\documentclass[preprint,12pt]{aastex62}

\shorttitle{ALMA Perseus Multiplicity}
\shortauthors{Tobin et al.}

\newcommand{\cateo}{\mbox{C$^{18}$O}}
\newcommand{\thco}{\mbox{$^{13}$CO}}
\newcommand{\twco}{\mbox{$^{12}$CO}}

\newcommand{\kms}{\mbox{km s$^{-1}$}}

\begin{document}

\title{The VLA/ALMA Nascent Disk and Multiplicity (VANDAM) Survey of Perseus Protostars. VI. Characterizing the Formation Mechanism for Close Multiple Systems}
\author{John J. Tobin}
\affiliation{Current address: National Radio Astronomy Observatory, 520 Edgemont Rd., Charlottesville, VA 22903, USA}
\affiliation{Homer L. Dodge Department of Physics and Astronomy, University of Oklahoma, 440 W. Brooks Street, Norman, OK 73019, USA}
\affiliation{Leiden Observatory, Leiden University, P.O. Box 9513, 2300-RA Leiden, The Netherlands}
\author{Leslie W. Looney}
\affiliation{Department of Astronomy, University of Illinois, Urbana, IL 61801}
\author{Zhi-Yun Li}
\affiliation{Department of Astronomy, University of Virginia, Charlottesville, VA 22903}
\author{Sarah I. Sadavoy}
\affiliation{Harvard-Smithsonian Center for Astrophysics, 60 Garden St, MS 78, Cambridge, MA 02138}
\author{Michael M. Dunham}
\affiliation{Department of Physics, State University of New York Fredonia, Fredonia, New York 14063, USA}
\affiliation{Harvard-Smithsonian Center for Astrophysics, 60 Garden St, MS 78, Cambridge, MA 02138}
\author{Dominique Segura-Cox}
\affiliation{Department of Astronomy, University of Illinois, Urbana, IL 61801}
\author{Kaitlin Kratter}
\affiliation{University of Arizona, Steward Observatory, Tucson, AZ 85721}
\author{Claire J. Chandler}
\affiliation{National Radio Astronomy Observatory, P.O. Box O, Socorro, NM 87801}
\author{Carl Melis}
\affiliation{Center for Astrophysics and Space Sciences, University of California, San Diego, CA 92093}
\author{Robert J. Harris}
\affiliation{Department of Astronomy, University of Illinois, Urbana, IL 61801}
\author{Laura Perez}
\affil{Departamento de Astronom\'ia, Universidad de Chile, Camino El Observatorio 1515, Las Condes, Santiago, Chile}

\begin{abstract}
We present Atacama Large Millimeter/submillimeter Array (ALMA) 
observations of multiple protostar systems in the Perseus molecular cloud
previously detected by the Karl G. Jansky Very Large Array (VLA).
We observed 17 close ($<$600~AU separation) multiple systems 
at 1.3~mm in continuum and five molecular lines (i.e., \twco, \cateo, \thco, H$_2$CO, SO) 
to characterize the circum-multiple
environments in which these systems 
are forming.
We detect at least one 
component in the continuum for the 17 multiple systems. In three systems,
one companion is not detected, and for two systems the companions are
unresolved at our observed resolution.
We also detect circum-multiple dust emission toward 8 out of 9 Class 0 multiples. Circum-multiple
dust emission is not detected toward any of the 8 Class I multiples.
Twelve systems are detected in the dense gas tracers toward their disks/inner envelopes. 
For these 12 systems, we use the dense gas observations to characterize their formation 
mechanism.
The velocity gradients in the circum-multiple gas are clearly orthogonal to the outflow
directions in 8 out of the 12 systems, consistent with disk fragmentation.
Moreover, only two systems with separations $<$200~AU are \textit{inconsistent} with disk fragmentation,
in addition to the two widest systems ($>$500~AU). Our results suggest 
that disk fragmentation via gravitational instability is an important formation 
mechanism for close multiple systems, but further statistics are needed to better
determine the relative fraction formed via this method.
\end{abstract}

\section{Introduction}

Star formation typically occurs within dense clouds of molecular gas in the interstellar medium.
These clouds of dense gas collapse to form stars when their self-gravity 
dominates over other sources of support \citep[e.g., thermal pressure, magnetic fields, turbulence,][]{mckeeostriker2007}, forming either a single or multiple
star system. Nearly half of Sun-like stars (in terms of stellar mass)
are found in binary or higher-order multiple systems \citep{dm1991,raghavan2010} with typical
separations of $\sim$50~AU. The frequency
of stellar multiplicity strongly depends on stellar mass. Stars more massive than the Sun have
a higher fraction of multiplicity, and stars less massive than the Sun have a lower degree of multiplicity;
see \citet{duchene2013} for a recent review. Thus, it is clear that multiple star formation is a common
outcome of star formation at all masses, and a comprehensive understanding
of the star formation process must also account for multiplicity.

There are two favored routes to explain the formation of multiple star systems: disk fragmentation
due to gravitational instability \citep{adams1989,bonnell1994a,stamatellos2009,kratter2010} and turbulent fragmentation
of the molecular cloud \citep{padoan2002,offner2010}. Disk fragmentation will preferentially result
in the formation of close ($<$600~AU) multiple star systems and requires the existence 
of a rotationally-supported disk around the primary star. Turbulent fragmentation can result in the
formation of both wide and close multiple systems. The initial protostars form with separations
$\sim$1000~AU (or larger), and depending on their relative motions and masses they can migrate closer
together, remain at wide separations, or drift further apart \citep{offner2010,sadavoy2017}. 
Previously, it was also thought that envelope rotation could play a role in binary/multiple formation
\citep[e.g.,][]{bb1993},
but this is no longer favored given the lack of ordered rotation for most 
star forming cores \citep{tobin2011,storm2014,fernandez2014} and the apparently random
orientations of angular momenta (traced by outflow directions) for the wide companion protostars \citep{lee2016}.
A key prediction of turbulent fragmentation is that the seeds for multiplicity are produced in the 
pre-stellar phase. Early ALMA survey results suggest tentative agreement 
with these predictions \citep{pineda2015,dunham2016,kirk2017}, but better statistics are still needed.

Despite the excellent statistics offered by studying field stars, the properties of field star multiples
cannot alone reveal the origin of multiplicity. This is because these systems are observed in the
present epoch and represent the culmination of Myr to Gyr of dynamical evolution, and to understand
the origin of multiplicity, it must be characterized during or shortly after the onset of star formation
during the protostellar phase. 

Therefore, to gain a more clear picture of multiple star formation, protostars
in the earliest phase of the star formation
process, known as the Class 0 phase \citep{andre1993}, must be observed. During this phase
the forming protostar is enshrouded in a dense,
infalling envelope of gas and dust. The high levels of obscuration, however, made searching for multiplicity
toward the Class 0 protostars difficult in the infrared, especially at scales less than 1000~AU. Thus,
centimeter and millimeter interferometry were necessary to examine the formation of such systems.
Early searches detected a number of wide multiple systems ($>$600~AU) and a few close ($<$600~AU) multiple systems 
\citep{looney2000,rodriguez1998,brown2000}. Later surveys aimed to improve upon the statistics, and \citet{chen2013}
characterized a number of wide systems in the Class 0 phase, finding that most (67\%) Class 0 protostars
may begin their lives as part of a wide multiple system. \citet{maury2010, tobin2013,tobin2015b} aimed 
to expand the characterization of close multiples, but samples were still too small to provide
statistically significant results.

Several efforts were also made to characterize multiplicity
in the late protostellar phase \citep{connelley2008,duchene2007}, the so-called Class I protostars \citep{dunham2014}.
However, even these systems may be too evolved to retain signatures of their
formation mechanism given that enough enshrouding material has been accreted/dispersed to enable
their detection in the near-infrared.

The VLA/ALMA Nascent Disk and Multiplicity (VANDAM) Survey \citep{tobin2015a,tobin2016a,segura-cox2016}
unlocked the distribution of protostellar multiples on scales less than 
600~AU (down to $\sim$20~AU),
thereby sampling the peak of the field solar-type separation distribution. This survey conducted
with the NSF's Karl G. Jansky Very Large Array (VLA) observed all known
Class 0 and Class I protostars in the Perseus star forming region, 37 of which are Class 0 
protostars (FHSCs and VeLLOs included), 8 are Class 0/I protostars, and 37 are Class
I protostars (flat spectrum included) \citep{tobin2016a}. In addition, 12 Class II sources 
(pre-main sequence stars with disks) that were bright in the far-infrared were also observed. 
We adopt
a distance of 300~pc as an average distance for the Perseus molecular cloud \citep{ortiz2018}. 
There may be a distance gradient across the cloud \citep{ortiz2018,zucker2018} with the 
eastern part of the cloud (IC348) at $\sim$320~pc, the central part (NGC 1333) at $\sim$293~pc, and
the western part (L1448 and L1451) at $\sim$280~pc, but 300~pc is adopted for simplicity because these
distances are consistent with each other when considering the uncertainties of the current Gaia
measurements. We note that these revised distances are in disagreement with the maser parallaxes
from \citet{hirota2008,hirota2011} which indicated $\sim$230~pc. This 230~pc distance was used 
in \citet{tobin2016a,tobin2016b} to determine physical separations and convert continuum flux density 
to mass.
From this sample,
18 systems were identified as multiple with companion separations less than 
600~AU, and 16 of these were new detections by the VANDAM survey.

The VANDAM survey greatly increased the detection statistics for protostars on scales
less than 600~AU,
but the formation mechanism(s) of the multiple systems was difficult to quantify
because the VLA survey was continuum-only.
\citet{tobin2016a} argued that the systems with separations beyond $\sim$500~AU most likely 
formed from turbulent fragmentation \citep{offner2016}; this conclusion was further supported by 
observations of outflow directions by \citet{lee2016}. For the close multiples detected,
disk fragmentation was argued to be the most likely formation mechanism for 17 out of the 18\footnote{Only one close companion in the VANDAM sample 
(NGC 1333 IRAS2A/Per-emb-27) had clearly misaligned outflows, indicating that formation within
a circumbinary disk is unlikely \citep{tobin2015a}.}
based on their proximity, circumbinary material around some sources, and
the double-peaked nature of the separation distribution. Nevertheless, it remains a possibility that 
the close multiple systems initially formed at wide separations and migrated 
inward \citep{offner2010, sadavoy2017}.
However, definitive evidence in favor of disk fragmentation would be
finding them embedded within a common, circumbinary/multiple structure that has Keplerian rotation.
To search for the presence or lack of such a signature, we observed the sample of 17 (out of 18) close multiple 
sources in Perseus with the Atacama Large Millimeter/submillimeter Array 
(ALMA); an additional close multiple system (IRAS 03282+3035/Per-emb-5) 
was discovered after the program was accepted. 
ALMA enables simultaneous observation of both dust continuum to look for circum-multiple
structure and molecular lines to search for rotation signatures
in the surrounding gas \citep[e.g.,][]{takakuwa2014,takakuwa2015,tobin2016b}.

The 
results presented here reveal the circum-multiple environment around each close system, such
as whether or not there is rotation at small-scales and the presence or absence of surrounding 
material. This paper is structured as follows: we discuss the observations and data reduction
in Section 2, the results from dust continuum observations and molecular line mapping
in Section 3, the implications of the observations in Section 4, and present
our conclusions in Section 5.

\section{Observations and Data Reduction}

The Perseus multiple systems were observed with the ALMA during
Cycle 2 on 27 September 2015 with 33 antennas operating and sampling baselines between 32 - 2000 meters. 
The observations were executed within a 1.8 hour block
and the total time spent on each source was $\sim$2.9 minutes. 
The precipitable water vapor was $\sim$0.7 mm throughout the observing session.
The phase calibrator was J0319+4130 (3C84), the bandpass calibrator
was J0237+2848, and the amplitude and absolute flux calibrator was the monitored 
quasar J0238+166. The absolute flux calibration accuracy is expected to be better than 10\%; however,
we do not include this uncertainty in our calculations and utilize statistical
uncertainties throughout the paper.
The correlator was configured to observe a 2 GHz continuum band 
centered at 232.5 GHz and observed in TDM mode with 128 channels. The three
other basebands were allocated to 60 MHz windows with 1960 channels (0.083~\kms\ velocity resolution)
each and centered on the following molecular transitions: $^{12}$CO ($J=2\rightarrow1$), 
$^{13}$CO ($J=2\rightarrow1$), C$^{18}$O ($J=2\rightarrow1$), SO ($J_N = 6_5\rightarrow5_4$), 
and H$_2$CO ($J=3_{03}\rightarrow2_{02}$).

The raw visibility data were manually reduced
by the North American ARC staff using CASA version 4.5.0, and we additionally performed
self-calibration on the continuum data to increase the signal to noise ratio. 
When possible, we performed 2 rounds of phase self-calibration, first with solution
intervals that encompassed the length of an entire on-source scan, then the second 
round utilized either 12.1 or 6.05 second solution intervals. The shortest possible
solution interval was 6.05 seconds, corresponding to the length of a single integration.
Following phase self-calibration, we performed amplitude self-calibration using the
solution normalization option, which normalizes the amplitude solutions to around the factor
1.0 such that the flux density scaling is not altered.
Following the completion of self-calibration on the continuum, the
solutions were also applied to the 
spectral line bands. Self-calibration was possible for 16 out of 17 fields observed, only Per-emb-48 could
not be self-calibrated due to insufficient S/N. The resultant noise in the 1.3~mm continuum 
was $\sim$0.14~mJy~beam$^{-1}$ and $\sim$15~mJy~beam$^{-1}$ in 0.25~\kms\ channels for the spectral line observations.
The non-self-calibrated continuum images had a noise level of $\sim$1~mJy~beam$^{-1}$.

The 
data were imaged using the \textit{clean} task within CASA 4.5.0; the ALMA images shown 
in Figures 1 and 2 for each target were generated using Briggs weighting with a robust parameter of 0.5.
Within the \textit{clean} task, we only include data at $uv$-distances $>$ 50 k$\lambda$ to 
mitigate striping in the images from large-scale emission detected on the shortest baseline
that could not be properly imaged. The $^{13}$CO ($J=2\rightarrow1$) and C$^{18}$O ($J=2\rightarrow1$) 
images were generated with Natural weighting, tapering at 500~k$\lambda$, and also only using the data having
$uv$-distances $>$50~k$\lambda$. Tapering reduces the weight of longer 
baseline data in the deconvolution to facilitate the detection of larger, lower surface brightness
 structures. The typical beam sizes of the continuum and molecular line images are
0\farcs27$\times$0\farcs16 (81~AU~$\times$~48~AU) and 0\farcs35$\times$0\farcs25 (108~AU~$\times$~76~AU), respectively.

 We note that the phase 
was very unstable during the course of our observations, and
the phase calibrator (J0319+4130) was at $\sim$10\degr\ lower elevation than our science 
targets.
As such, there are systematic position offsets in each field up to $\sim$0\farcs1. Thus, the most
precisely measured positions for the observed protostars in Perseus 
are those listed in \citet{tobin2016a} from the VLA, despite the higher S/N offered
by the ALMA observations.

\subsection{Data Analysis}
With the observed sample of protostellar multiple systems, one of the key observables is the flux density of the
resolved sources and that of the surrounding extended structure. We measured the flux densities toward 
the binary and wide companion systems that fell within the primary beam using Gaussian fitting.
First, we fit Gaussians to all compact, individual components that were resolved and/or marginally resolved in the data.
The Gaussians were fit using the \textit{imfit} task of CASA 4.7.2. The integrated flux densities and
uncertainties, source positions, and Gaussian parameters 
from \textit{imfit} are given in Table 1. Some sources have clear non-Gaussian, 
extended components. In these cases, we used \textit{casaviewer} to draw a polygon region encompassing 
all the extended emission from the sources (out to $\sim$2\arcsec, where we still recover all flux) 
and measure the flux density within the polygon; this value is given as Extended Flux in Table 1. 
For binary systems, there were many cases where the extended flux was greater than the combined 
flux densities of the Gaussian fits to the compact sources. This indicates that there is a significant amount
of emission in extended, non-Gaussian components.

For the observed molecular lines, we made moment maps over the selected channel ranges where
there is spectral line emission. These images are used to assess whether or not there is rotation around a particular
multiple system and if there is an outflow.

\section{Results}

\subsection{Dust Continuum Imaging}

We detect a wide variety of
structures toward the sample of 17 multiple protostar systems observed by ALMA. 
The dust continuum images of the Class 0 systems are shown in Figure 1 and the Class I systems in Figure 2.
The ALMA observations had high enough angular resolution to resolve the companions in 15 out of 
the 17 systems. The systems with unresolved companions are Per-emb-2 and
Per-emb-18, which have separations of 0\farcs08 ($\sim$24~AU). For 
the 15 systems with resolvable companions, we detected all known components in 12. 
The companions were not clearly detected for Per-emb-40, Per-emb-55, and Per-emb-48. 
However, toward 
Per-emb-55, there is emission at the expected companion position at the 2$\sigma$ level, and
for Per-emb-40 there is extended emission toward the companion position that might reflect
emission from the companion. Per-emb-48 was among 
the faintest sources detected by ALMA and self-calibration 
was not possible for this field. The increased noise from phase decorrelation may have prevented the 
companion from being detected, but we cannot rule-out that its flux density is below
our detection limit.

The Class 0 systems all have some extended emission surrounding the 
companions. This is demonstrated quantitatively in Table 1, where
the total extended flux density (the sum of compact and extended emission) 
is greater than the flux density from the compact sources. 
Per-emb-2, Per-emb-18, Per-emb-17, L1448~IRS3B, SVS13A, 
L1448~IRS3C, and L1448~IRS2 all have clear emission surrounding and/or bridging the companion 
sources. The degree to which the total flux density is greater varies, sometimes the 
difference is as small as a few percent (e.g., Per-emb-17), but it can be greater than 
a factor of two (e.g., Per-emb-33).
This emission could be described as a circumbinary disk and/or envelope.

The extended structure surrounding the companions is quite prominent in L1448~IRS3B, SVS13A, Per-emb-18, 
and Per-emb-2. The extended emission can be described as either spirals or streamers toward
L1448~IRS3B, SVS13A, and Per-emb-2 (Figure 1). The spirals in L1448~IRS3B 
were previously reported in \citet{tobin2016b}. Finally, the extended structures surrounding Per-emb-2 and 
Per-emb-18 show evidence of being optically thick at 1.3~mm as 
evidenced by their emission peaks not being located toward what are assumed protostar positions, but off source, 
and having smooth surface brightness profiles across their central regions. In all these
sources, the VLA imaging in \citet{tobin2016a} detects the peak emission 
toward what are assumed to be the positions of the individual protostars,
and the ALMA imaging with higher surface brightness sensitivity is able to detect much 
more extended dust emission than detected by the VLA. Furthermore, the brightest feature in Per-emb-18 
is offset to the east; this was a weakly detected feature at 9~mm with the VLA \citep{tobin2016a}.

Finally, there is less-prominent extended emission toward NGC1333 IRAS2A and L1448~IRS2.
NGC1333 IRAS2A has extended features at low surface brightness, and asymmetric extended
emission to the east. Then the extended emission for 
L1448~IRS2 is also more diffuse and does not have as well-defined circumbinary structure like some of the others.
However, the larger structure surrounding the protostars may be impacted by spatial filtering
given that \citet{tobin2015b} detected a surrounding structure on larger 
scales in lower resolution data.

The Class I sources on the whole have little or no extended emission surrounding
the components of these multiple systems. The images show
this visually in Figure 2, and Table 1 shows that the flux densities for the Gaussian 
fits and the extended emission from a larger
area encompassing the two protostars are comparable. However,
some show resolved structure toward one component. Both NGC 1333 IRAS2B and L1448 IRS1 have at least one 
component that is dominant, with resolved structure in their dust emission. The other Class I binaries
appear consistent with point sources, and the components have similar flux densities.

While the close companions were the primary targets, we detected several wide companions in the
observed fields. In the field of SVS13A, we detect several additional sources. One source is
RAC1999 VLA20 \citep{rodriguez1999}, located northeast of SVS13A. This source was
 previously detected by the VLA, but has no counterpart at shorter wavelengths
 (i.e., mid-infrared and far-infrared) and has been hypothesized 
to be extragalactic \citep{tobin2016a}, see Figure \ref{SVS13}.
The wider companion to SVS13A, often called VLA3 or SVS13A2, 
is detected (Figure \ref{SVS13}), and appears marginally resolved. SVS13B is also detected and  
resolved as shown in Figure \ref{SVS13}). VLA 8~mm imaging of
 SVS13B indicated that it has a small embedded disk \citep{segura-cox2016}, but a larger, resolved 
structure is detected in the ALMA 1.3~mm data.
We also detected L1448 IRS3A in the field of L1448 IRS3B (Figure \ref{L1448IRS3}). 
The small-scale structure of L1448 IRS3A appears to be a resolved disk (Figure \ref{L1448IRS3}), 
consistent with the resolved emission detected in \citet{tobin2015b}. In the field of
Per-emb-55 (Figure \ref{per-emb-55}), we also detected 
Per-emb-8 which appears to have a large extended disk surrounding it
(Figure \ref{per-emb-55}). Per-emb-21 was detected in the field of Per-emb-18, 
but appears unresolved (Figure {\ref{per-emb-18}).

The resolved structures around Per-emb-8 and L1448 IRS3A were not 
ideal for fitting a single Gaussian, so we fit two
Gaussians to these sources and list the inner and outer Gaussian 
fits separately, referring to them as Inner Disk and
Outer Disk, in addition to the fit of a single Gaussian in Tables 1 and 2.

\subsection{Mass Estimates from Dust Continuum}

The mass of material found toward individual components of the multiple systems, as well
as surrounding all components, can be estimated using the flux density of the dust continuum emission. 
Under the assumption that the dust emission is optically thin and isothermal, the dust mass can be calculated
with the equation
\begin{equation}
\label{eq:dustm}
M_{dust} = \frac{D^2 F_{\nu} }{ \kappa_{\nu}B_{\nu}(T_{dust}) },
\end{equation}
where $D$ is the distance ($\sim$300~pc), $F_{\nu}$ is the observed flux density, $B_{\nu}$ is
the Planck function, $T_{dust}$ is the dust temperature, and $\kappa_{\nu}$ is the
dust opacity at the observed wavelength. $T_{dust}$ is assumed to be 30~K, consistent
with temperature estimates on $\sim$100~AU scales \citep{whitney2003a} and $\kappa_{\nu}$
is 0.899, taken from \citet{ossenkopf1994}, Table 1 column 5. We then multiply the resulting value of M$_{dust}$ by
100, assuming the canonical dust to gas mass ratio of 1:100 \citep{bohlin1978}. We caution that
the masses calculated may have systematic uncertainty because opacity will hide some mass, the
dust emission is not isothermal and
the dust opacity along with the dust to gas mass ratio are uncertain and may vary between components
within the same system.

The results from the mass calculations are given in Table 2. A number of sources have total masses greater
than 0.1~M$_{\sun}$. The mass calculations presented here tend to be about 20\% higher than the masses
reported for some of the same sources in \citet{tobin2015b} from observations with CARMA
(taking into account the different distances adopted),
despite observing at the same wavelength, slightly lower resolution (for CARMA), 
and making the same assumptions for the mass calculation. 
The minor discrepancy likely results from a combination 
of greater absolute flux uncertainty in the CARMA data, better sensitivity to low-surface brightness 
emission from ALMA, better uv-coverage with ALMA, and less decoherence in the self-calibrated ALMA data.

Most of the close companions have large flux density ratios between them, which could
be interpreted as a large circumstellar mass ratio; this ratio offers no 
information as to the mass ratio between the stellar components and only reflects the compact circumstellar
mass. Per-emb-17, EDJ2009-269, and
Per-emb-35 have ratios closest to unity, and L1448 IRS1, L1448 NW, Per-emb-12, 
Per-emb-22, Per-emb-27, Per-emb-33, Per-emb-36, Per-emb-40, Per-emb-44, Per-emb-49, 
and Per-emb-55 all have circumstellar mass ratios that are less than 0.63. 
Both Class 0 and Class I systems span a range of circumstellar mass ratios.
We note that several cases with large circumstellar mass ratios are in hierarchical and/or higher order systems.

\subsection{Spectral Indices}
Because most of the companions are well-enough resolved to enable their 1.3~mm
flux densities to be measured individually, We were able to calculate the spectral 
index of the emission from 1.3~mm to 9.1~mm. We calculated the
spectral index ($\alpha$) using the equation
\begin{equation}
\alpha = \frac{ln(F_{\nu,1.3mm})-ln(F_{\nu,9.1mm})}{ln(\nu_{1.3mm}) - ln(\nu_{9.1mm})}
\end{equation}
and the associated uncertainty using the equation
\begin{equation}
\sigma_{\alpha}^2 = \left(\frac{1}{ln(\nu_{1.3mm}) - ln(\nu_{9.1mm})}\right)^2\left(\frac{\sigma_{\nu,1.3mm}^2}{F_{\nu,1.3mm}^2}+\frac{\sigma_{\nu,9.1mm}^2}{F_{\nu,9.1mm}^2}\right)
\end{equation}
following \citet{chiang2012}. We use the 9.1~mm flux densities from \citet{tychoniec2018} to calculate $\alpha$ 
with respect to the flux densities at 1.3~mm. Because \citet{tychoniec2018} attempted to remove the free-free
emission that can contribute to the 9.1~mm flux densities, we calculated $\alpha$ with 
respect to both the corrected and uncorrected 9.1~mm flux densities. 
The spectral indices for the sources are given in Table 1. The statistical
uncertainty on $\alpha$ is generally quite small $\sim$0.1 due to the large different in wavelength between
the two bands. We do assume a 10\% uncertainty in the flux calibration for each flux density and add this
in quadrature to the statistical uncertainty on the flux density. The uncertainty on $\alpha$ is
generally smaller than the range of $\alpha$ from the corrected and uncorrected flux densities. 

Most of the values for $\alpha$ are between 2 and 3; optically thin thermal dust emission, where the dust
opacity spectral index $\kappa_{\nu}$~$\propto$~$\nu^{\beta}$, is expected to have $\alpha$~=~2+$\beta$.
Thus, for thermal dust emission alone (optically thin or partially optically thin), 
the spectral indices should be $\ge$2; the spectral index would be 2 for
optically thick emission. Most of the values of $\alpha$ are $>$2 even when the 9.1~mm emission was not corrected
for free-free emission, thus it is likely that the 9.1~mm mostly reflects dust emission
as also found by \citet{tychoniec2018}. There are some sources for which $\alpha$~$\sim$~2, 
this could indicate that either the   properties of the dust grains are such 
that $\beta$ is quite small, the dust emission is becoming
optically thick, or there is a some amount of free-free emission to the 9.1~mm 
flux density that could not be removed. These three possibilities are difficult to disentangle with the data
in-hand, limiting the possible interpretations of these spectral index data. Therefore, the 1.3~mm emission
presented here is most likely to \textit{only} reflect dust emission while the 9.1~mm emission might
reflect dust and free-free emission.

\subsection{Molecular Line Kinematics}

The dust continuum emission provides crucial evidence of the presence or absence of material surrounding the 
close binary/multiple stars, which will help with the interpretation of their formation mechanism.
However, the continuum data alone are not enough to fully understand how
these multiple systems formed. Therefore, we also observed five molecular tracers toward each system
along with the dust continuum. We expect that \cateo\ will generally trace the kinematics of the
dense circum-multiple material because this line has tended to be the most reliable
tracer of protostellar disks \citep{ohashi2014,yen2014,aso2015}. Also, \thco\ should 
generally trace the circum-multiple material depending \citep{takakuwa2012,tobin2012}, but we 
use \thco\ with caution given that it can also trace outflowing
gas. H$_2$CO is also expected to trace the circum-multiple environment 
because this is a typical high-density tracer \citep[e.g.,][]{mangum1993}, and it has been found to trace
the Class 0 disk and inner envelope of L1527 \citep{sakai2014}. Similarly, SO has also been found to
trace the disk toward some protostars \citep{sakai2014,yen2014}. However, there are instances
where both H$_2$CO and SO have been shown to trace outflowing material, in addition
to circumstellar/multiple material \citep[e.g.,][]{wakelam2005}. Finally, \twco\ is expected to trace
the outflow toward most protostars in the sample given their youth, but \twco\ could trace the disk kinematics 
of some of the more evolved multiple systems in our sample \citep[e.g.,][]{simon2000}.
The figures showing the molecular line detections toward circum-multiple and circumstellar structures 
are shown in Figures 7 to 22. The intervals over which the red and blue shifted emission are integrated
are given in Table 3.

Due to the short observations toward each source, the 
S/N is not very high in each channel for these high-angular 
resolution spectral line maps. Thus, we
examine the integrated intensity maps constructed for 
emission red- and blue-shifted with respect to the line center.
The detection rates for the lines vary from source to source, some only have one 
line detected, whereas others have detections in all five lines. We examine 
the integrated intensity maps of each molecule, relative to the outflow 
directions and the extended continuum structure. The larger-scale outflows 
have previous characterization of their orientations, when possible, from \citet{stephens2017}, 
therefore we only show the outflow maps in Appendix A.

\subsubsection{Characterization of Kinematics in Multiples}

We use the red- and blue-shifted integrated intensity maps of each molecule to
 determine whether or not the velocity patterns observed toward the 
multiple systems reflect rotation, outflow, or an indistinct process. 
We do this by comparing the
orientation of the velocity gradient relative to the outflow direction (and extended
continuum if possible). If there is a velocity gradient in the circum-multiple structure, that
is within 30\degr\ of orthogonal to the outflow direction, we classify it as being consistent with 
disk rotation. To quantify the relative orientations of the outflow and velocity gradient directions,
we measure the position angle of the outflow and velocity gradient and then take the difference. These
values are tabulated in Table 4. To measure the outflow position angle, we draw a line bisecting the
outflow and measure its position angle. We follow the same procedure for the velocity
gradients, but measure this from the emission peaks in the blue- and red-shifted integrated intensity maps
tracing rotation, generally \cateo. The position angles do have uncertainty and we estimate that this
is $\pm$10\degr.

Molecular line emission is detected
with sufficiently high S/N in 12 systems to characterize the kinematics as rotation or not.
This classification is most firm for the 9 Class 0 systems with the strongest detections and 
seven of these Class 0 systems have velocity gradients that indicate rotation in the
circum-multiple gas. If we include those with tentative
classifications, we find that 8 out of 12 systems are consistent with the circum-multiple 
emission tracing rotation orthogonal to the outflow direction. The systems classified
as indistinct generally have disorganized kinematic 
structure on the scales and at the sensitivity we observed. Also, certain molecules
may better trace different kinematic components of protostellar systems \citep[e.g.,]{sakai2014}.
The detection and characterization of the molecular line data are summarized in Table 4.
Each source is discussed in more detail in Section 4.1 with respect to the data
presented in this paper and to interpret previous results in the context of the new
ALMA data.

\subsubsection{Detections of Individual Rotating Disks}

In addition to the kinematic structure of circum-multiple material, we were also able to 
examine the kinematic structure of both some individual circumstellar disks either as part of a close
multiple system or as a wide companion. Per-emb-8 shows the clearest example of rotation in 
all observed molecular lines; \twco\ interestingly shows a combination of rotation and outflow. 
L1448 IRS3A also shows rotation in all molecular lines and is oriented orthogonal to the position angle
of L1448 IRS3B. The individual disks toward Per-emb-35-A and -B appear to have molecular line emission as well,
but the line emission is not well-resolved spatially. Only the H$_2$CO appears to have a clear indication
of rotation. The characterization of the observed kinematics from all lines are given in Table 4. 

\section{Discussion}

The ALMA 1.3~mm continuum and molecular line emission observed toward the close multiple
systems in Perseus dramatically
increases our knowledge of the spatial and kinematic structure of dense gas on a few hundred
AU scales around these multiple protostar systems. The ALMA dust continuum traces 
more tenuous emission surrounding the companions as compared to the previous VLA observations.
While the extended emission is more diffuse than the compact sources detected by both the VLA and ALMA, this 
emission can account for a significant fraction of the mass in the systems. The molecular maps show the gas motion, revealing how well (or not) the 
motion of the extended dense gas surrounding the protostar(s) is organized. The molecular line maps provide evidence
for rotation in several systems, but not universally, as some systems have unclear kinematic structure
with the current S/N of the ALMA data. Nevertheless, these ALMA data now enable us to better interpret
how multiple star systems form on a case-by-case basis and as an ensemble. 
We consider each system individually and summarize our findings for the sample as a whole.

\subsection{Individual Systems}
\subsubsection{Class 0 and 0/I Systems}

\textbf{Per-emb-2}-- The continuum structure is extended with a very
smooth surface brightness distribution, in contrast to the VLA data with a central 
intensity peak toward the protostars, forming a binary system with a separation of 
$\sim$24~AU. Only
\cateo\ and \thco\ were detected toward this system, and we detect a 
clear velocity gradient orthogonal to the outflow direction,
as shown in Figure \ref{per-emb-2-lines}. The
\cateo\ emission is not fully coincident with the continuum image, 
possibly due to dust opacity and/or spatial filtering of low-velocity
emission near line-center. Nevertheless, the \cateo\ and \thco\ kinematics still exhibit
evidence for rotation on the scale of the circumbinary continuum.

\textbf{Per-emb-12}-- Also known as NGC 1333 IRAS4A, this system is the most massive overall in terms 
of dust emission out to very large scales. It was also known to be a binary prior to the VANDAM
survey with a separation of $\sim$549~AU \citep{looney2000}. The dust emission 
toward IRAS4A has significant opacity, which prevents 
tracing kinematics toward the protostars; only some emission around the 
protostars in \cateo\ and \thco\ is revealed in Figure \ref{per-emb-12-lines}.

\textbf{Per-emb-17}-- There were detections in multiple lines 
toward Per-emb-17, and the material surrounding the $\sim$83~AU binary system
forms a flattened structure in dust continuum that is similarly flat in \thco\ and \cateo. The
molecular line emission exhibits velocity
gradients orthogonal to the outflow direction, see Figure \ref{per-emb-17-lines}. The rotation
center of the system is not clearly centered on a particular component, so it is unclear
which protostar is more massive.
H$_2$CO and SO, however, seem to be more associated with the outflow in this 
source. SO in particular is strong toward the B component and extended along the outflow.

\textbf{Per-emb-18}-- This source has a flattened structure in the dust continuum surrounding
 the companions separated by $\sim$26~AU and detections in multiple lines. 
The continuum might be tracing dust emission with high optical depth, given that the emission is
not very peaked toward the main pair of protostars. The \thco, \cateo, SO, and H$_2$CO trace 
velocity gradients orthogonal to the outflow direction, see Figure \ref{per-emb-18-lines}. The 
fact that all molecules have gradients in the same direction lends confidence to interpreting
the kinematics as rotation.

\textbf{Per-emb-22}-- Also known as L1448 IRS2, this binary system has a separation of
$\sim$225~AU.
Continuum emission was mainly detected toward individual components, with some
emission in between, exhibiting an apparent `bridge' between them. The molecular lines have evidence
 of velocity gradients orthogonal to the
outflow (Figure \ref{L1448IRS2-lines}), but much of the molecular line emission appears to 
be filtered out and/or have very low surface brightness. 
\citet{tobin2015b} previously found evidence for velocity gradients/rotation using combined CARMA and SMA data
that had lower resolution, in addition to a larger circumbinary structure. 

\textbf{Per-emb-27}-- Also known as NGC 1333 IRAS2A, this system does not have much 
material surrounding the 
$\sim$186~AU binary, but there is some extended material associated 
with the brighter source. There is weak evidence for a velocity gradient in \cateo\ on scales 
larger than the binary system. SO is associated with the outflow, but H$_2$CO is compact
toward the brighter component with a small extension toward the companion, see Figure \ref{per-emb-27-lines}.

\textbf{Per-emb-33/L1448 IRS3B}-- This source was shown in \citet{tobin2016b} to 
have clear velocity gradients associated with its extended circumbinary material 
in \thco, \cateo\, and H$_2$CO. The separations of the companions from Per-emb-33-A 
are $\sim$79 and $\sim$238~AU.
Despite being published previously, we show this source for the sake of
comparison to the full sample (Figure \ref{L1448IRS3B-lines}).

\textbf{L1448 IRS3C}-- The two companions in this system are separated by just 
$\sim$75~AU, and they 
appear blended in the image. However, the emission toward each component is measured using a 
two component Gaussian fit. There is low-level dust continuum emission surrounding the two blended
protostars, and the circumbinary material also shows a velocity gradient in \cateo\ orthogonal to 
the outflow, see Figure \ref{L1448IRS3C-lines}.

\textbf{Per-emb-44}-- Also known as SVS 13A,  this binary system has a companion separation of
$\sim$90~AU and is one of the most prominent protostars within NGC1333.
The 1.3~mm continuum also shows a 1-armed spiral pattern,
almost as prominent as that of L1448 IRS3B. Both \thco\ and \cateo\ are detected toward this 
protostar and exhibit velocity gradients that are at least partially orthogonal to the outflow, 
see Figure \ref{per-emb-44-lines}. This source might be viewed more face-on making 
rotation harder to distinguish in this complex environment. The \cateo\ emission 
also follows the spiral through multiple velocity channels.
Furthermore, both SO and H$_2$CO are detected at smaller radii 
near the close pair and may also show a velocity gradient, 
but it is uncertain due to blending.  This protostar was 
previously known to have a close companion \citep{anglada2004}, and the original study
indicated that the disk around the western companion (Per-emb-44-B; VLA 4A) was not very prominent given 
its low flux density relative to the eastern component (Per-emb-44-A; VLA 4B) at a wavelength of 7~mm.
However, their masses (flux densities) differ by only a 
factor of 1.6 at 1.3~mm, which may have more to do with either the amount of 
large grains and/or variable free-free emission because
\citet{anglada2004} used data at a wavelength of 7~mm from the VLA. 
However, \citet{tychoniec2018b} also finds that the
9~mm flux densities (with estimated free-free contribution 
removed) have a flux ratio of 2.8, 
and the spectral indices 
are quite different $\sim$2.3 for the eastern component and $\sim$3 for the
western component. 
This could indicate
more large grains in the eastern source to explain the 
increased flux density, or the 1.3~mm is optically thick
and the 8~mm emission is less opaque, enabling more dust 
emission to be detected.

\subsubsection{Class I \& II Systems}

\textbf{Per-emb-36 (NGC 1333 IRAS2B)} -- The two sources detected toward Per-emb-36 are marginally
resolved in the ALMA 1.3~mm data with a separation of $\sim$93~AU,
and it has multiple detected emission lines
(Figure \ref{per-emb-36-lines}). However, the line emission does not 
obviously trace velocity gradients across the source, and the emission is 
mainly found west of the protostars rather than around them.

\textbf{Per-emb-35} -- These protostars are separated by 
$\sim$572~AU, and they are the source
of an S-shaped outflow detected by \textit{Spitzer} \citep{gutermuth2008}. These sources do not
have obvious dust continuum surrounding the compact continuum sources, 
but we cannot rule-out diffuse circumbinary emission
below our detection limits. H$_2$CO and SO are strongly detected toward both of the protostars and
may show evidence of velocity gradients orthogonal to their outflow directions (Figure \ref{per-emb-35-lines}). Furthermore,
there appears to be emission in these molecules between the two protostars. Both protostars
are driving outflows, and they appear parallel as observed in the plane of the sky. \thco\ is detected
toward both sources, but it is not obviously tracing rotation, and \cateo\ is only weakly detected, 
also see Figure \ref{per-emb-35-lines}.

\textbf{Per-emb-40} -- Only the primary component of this system is well-detected, but the companion
is tentatively detected. The molecules \thco\ and \cateo\ are only weakly detected (Figure \ref{per-emb-40-lines}). \thco\ does have
blue- and red-shifted components, but not obviously tracing rotation.

\textbf{L1448 IRS1} -- This protostar appears to be the most evolved source in L1448, and it
is detectable at visible wavelengths. The separation of the two protostars is 
$\sim$427~AU. Unlike
the other protostellar systems, the rotation in L1448 IRS1 is traced by \twco\ that
is extended toward the companion, see Figure \ref{L1448IRS1-lines}. 
In this particular case, \twco, is not tracing outflowing emission but rather
the disk because of its advanced evolutionary state. We know this 
because the scattered light nebula from this source \citep{foster2006}, which is
an alternative indicator of the outflow direction, is oriented orthogonal to the 
velocity gradient in \twco.

\textbf{EDJ2009-269} -- This system is a likely Class II binary in Perseus,
separated by $\sim$157~AU. Its only detected
emission line is \twco\ toward one of the protostars 
(Figure \ref{EDJ2009-269-lines}). In this more-evolved system,
the \twco\ also appears to trace kinematics of gas toward the 
individual circumstellar disks rather than
outflowing material.

\textbf{Per-emb-48, Per-emb-49, and Per-emb-55}-- These protostars do not show strong evidence for molecular line
emission surrounding the companion stars, thus we cannot examine whether or
not there are velocity gradients in the circum-multiple emission. 
These are all Class I systems and may simply 
have a much smaller reservoir of circumbinary material, and/or the molecular line emission could
not be detected with our uv-coverage and/or sensitivity. Per-emb-55, however, does appear to have an
outflow associated with it in \twco\ emission, see Appendix Figure \ref{outflows-2}.

\subsubsection{Wide Companions with Disks}

\textbf{L1448 IRS3A}-- This is the wide companion to L1448 IRS3B, separated by 7\farcs3 
($\sim$2195~AU), and has extended dust emission with disk-like morphology. The protostar
exhibits a weak outflow in \twco, but rotation in all the other observed lines, see 
Figure \ref{L1448IRS3A-lines}.

\textbf{Per-emb-8}-- This protostar is clearly detected with extended dust
emission that shows signatures of rotation from the molecular line emission 
(Figure \ref{per-emb-8-lines}). 
It is separated from Per-emb-55 by $\sim$2867~AU. Per-emb-8 was
previously identified as a disk candidate by \citet{segura-cox2016}, but with
a very different position angle than is observed with ALMA. Since this source
appears to have strong, extended free-free emission \citep{tychoniec2018a},
the geometric parameters from the VLA data may not solely trace the dusty disk.

\subsection{The Origin of Close Protostellar Multiples}

The ALMA continuum and molecular line maps are key to determining the most likely
origin of the multiple star systems. Systems that have formed via disk fragmentation are 
expected to have circum-multiple dust emission and this circum-multiple material should be
rotating in an organized manner. Furthermore, if fragmentation happened in a disk, the
disk should be larger than the separation of the companions.
Systems that result from turbulent fragmentation have no
such expectations for their circum-multiple emission and may not have well-organized
kinematics surrounding the system. We discuss our interpretation on the relative
frequency of these mechanisms based on our ALMA molecular line and continuum data.

\subsubsection{The Likelihood of Disk Fragmentation}

The ALMA observations of close multiple protostar systems reveal several features that were 
previously unclear from the VLA-only dataset. First, the continuum emission from 
extended circumbinary (or triple) material is detected in 9 out of the 18 observed sources. 
The Class 0s and most deeply embedded Class Is have the highest tendency
for circum-multiple dust continuum. Furthermore, this circum-multiple continuum 
emission is almost always extended
orthogonal to the outflows (L1448 IRS3B, L1448 IRS3C, L1448 IRS2, 
Per-emb-12, Per-emb-17, Per-emb-18), barring a few cases that appear to 
have inclinations that are nearly
face-on (SVS13A, Per-emb-2). Second, the molecular line emission 
detected toward these protostars and their 
circum-multiple disks strongly tends to have velocity gradients orthogonal to the outflow 
direction (Per-emb-2, L1448 IRS3B, L1448 IRS3C, L1448 IRS2, Per-emb-17, 
Per-emb-18, SVS13A, L1448 IRS1). 
The line emission is spatially coincident
with the circum-multiple dust emission (taking into account the
possible dust opacity and spatial 
filtering in some sources), and the velocity gradients orthogonal to the outflows 
on these scales are interpreted as rotation because the mass of the protostar and its 
disk are dominant at these scales ($<$~500~AU) and do not appear to be significantly confused 
with large-scale effects in the clouds. 

Therefore, it is clear that for many of the youngest multiple systems 
we have circum-multiple dust emission oriented in a manner that 
is consistent with a disk, and we detect molecular line kinematics that are consistent 
with rotation on the same scale as the circum-multiple material. We detail the
molecular line detections and their likely kinematic origin for each source in Table 4.
The observational results of rotating, circum-multiple gas align with the expectations for
companion formation via disk fragmentation. 
Direct evidence for the circum-multiple material in only the younger sources makes 
sense, because as they evolve, the circum-multiple
material is accreted rapidly \citep{bate2018}. Thus, about 7 protostars in the sample appear 
to have recently undergone disk fragmentation (those listed above with velocity
gradients orthogonal to the outflow and extended circum-multiple mass). Many of the Class I close systems 
without circum-multiple dust or gas may have formed earlier and/or consumed/dispersed their surrounding mass.
The widest systems that appear convincingly consistent with disk fragmentation are 
Per-emb-22 (225~AU) and Per-emb-33 (79, 238 AU)
and there is circum-multiple continuum detected in both cases.
L1448 IRS1 is possibly consistent with disk fragmentation, 
but the companion is at a wider separation (427~AU)
and the disk would have had to be quite large initially to form 
the companion, unless there was a three-body interaction
that moved the companion to a larger orbit \citep{reipurth2012}. 
This would then require that L1448 IRS1-A 
has an unresolved companion.

These ALMA observations provide important evidence for the possibility
of fragmentation via gravitational instability in these systems. Additional follow-up 
with higher S/N will be required to confirm that these apparent disks are rotationally supported
and to determine the outer radius of the Keplerian disk, ensuring that it encompasses the companion.
The short observations presented here do not have the S/N to robustly fit a rotation curve
that can distinguish Keplerian rotation from other velocity profiles. Therefore, 
at this point, we can conclude that there is evidence in favor of these systems 
forming via fragmentation in a gravitationally unstable disk, like L1448 IRS3B. 
There is rotation, circum-multiple dust, and the companions are 
most often orthogonal to the outflow and within the circum-multiple material.

\subsubsection{Evidence for Turbulent Fragmentation in Some Systems}

There are a few systems where disk fragmentation is not obviously consistent with the 
observations and turbulent fragmentation 
is more likely \citep[i.e., Per-emb-27; ][]{tobin2015a}.
 Per-emb-27 in particular has some evidence for a 
velocity gradient in \cateo\, but not in other tracers. H$_2$CO in fact
appears very centrally peaked toward Per-emb-27-A and may be tracing the
hot corino \citep[e.g.,][]{taquet2015}. Furthermore, the disk toward Per-emb-27-A
appears to be very small, $\sim$10~AU in radius \citep{seguracox2018}. Thus,
Per-emb-27 seems unlikely to have any characteristics that make disk fragmentation
a likely possibility for this system.

The source Per-emb-12 (NGC 1333 IRAS4A) is borderline for consideration as a close companion
with its $\sim$549~AU
separation. We see a significant amount of material surrounding the two
sources, but this is not clearly organized as a disk and is perhaps more likely a circumbinary
envelope. The outflows from these two sources are roughly aligned, indicating that envelope
rotation could have played a role in their formation, but the kinematics are unclear because the line 
emission is so complex due to the high opacity and spatial filtering that we cannot clearly 
say if there is organized rotation or
if the inner envelope kinematics appear random.

The source Per-emb-36 (NGC 1333 IRAS2B) has extended gas emission surrounding it, but
not obvious circum-multiple dust emission. Some of the gas appears off to one side and the
companion is located toward the western side where much of the gas emission tends to originate.
The kinematic structure of this system is unclear and the outflow is quite wide \citep{plunkett2013}.
Thus, we cannot firmly put this binary system in either category. Given that Per-emb-36 is 
a Class I protostar, it is possible that much of the original gas has been accreted and there
is no longer a prominent circum-multiple disk. This could be a case of turbulent fragmentation, but it is hard to be
more definitive with the more-evolved state of Per-emb-36.

Per-emb-35 is another case that is not obviously disk fragmentation given its wide separation of 
572~AU. However,
unlike Per-emb-12, it is much less embedded and the continuum emission is dominated by the compact circumstellar
components that may exhibit rotation in some molecular species (Figure \ref{per-emb-35-lines}). While there is not
much evidence for circumbinary material, the outflows (and possibly rotation axes traced by molecular lines) 
appear roughly parallel. Furthermore, the outflows are inclined in the same direction with respect to the 
plane of the sky. This case is also not obviously turbulent fragmentation or disk fragmentation. On the other hand the fragmentation mechanism in this system could
be related to the overall rotation of the envelope \citep[e.g.,][]{bb1993}, despite this mechanism not
being favored. However, turbulent fragmentation does not rule-out some outflows being parallel, just that
the distribution over a larger sample should be random.

\subsubsection{Which is the Most Common Mechanism?}

In total, we can make reasonably strong statements about the formation mechanisms for 9 out
of the 17 protostars in our sample based on the extended dust emission and gas kinematics. 
Seven out of the 9  (Per-emb-2, L1448 IRS3B, L1448 IRS3C, L1448 IRS2, Per-emb-17, Per-emb-18, SVS13A)
appear to be consistent with fragmentation of a gravitationally unstable disk given
their observed morphologies and velocity gradients. There are only two systems
(Per-emb-12, Per-emb-27) for which we can state with reasonable confidence that they are inconsistent with
disk fragmentation and that turbulent fragmentation is more likely. With the caveats of having a large
separation and only a hint of rotation in \twco\, we could 
add L1448 IRS1 to the disk fragmentation group. Then Per-emb-36 and Per-emb-35 could also
be added to the turbulent fragmentation group, 
bringing the total to 8 systems likely forming via disk fragmentation to 4 systems likely 
forming via turbulent fragmentation.
The raw numbers indicate that for multiples on scales
$<$600~AU, $\sim$67\% form via disk fragmentation and $\sim$33\% form via turbulent fragmentation.
If we only consider systems with $<$300~AU separations, more in line the sizes of disks, then
there are a total of 10 systems with only 2 systems inconsistent with disk fragmentation.
The smaller separations systems, further indicates that the closer systems may be more
frequently formed via disk fragmentation. However, these percentages obviously represent 
small number statistics making, prohibiting us from making a strong, general conclusion on 
the typical origin of close multiple systems.

A dominant contribution 
from turbulent fragmentation would not necessarily result in the 
presence of circumbinary disks within which the companions are embedded, and outflows 
would not tend to appear aligned orthogonal to the companion position angles
and/or velocity gradients.

There is not currently a clear prediction for the relative fractions from theory/simulations. 
\citet{offner2010} indicate that most close companions should result from turbulent fragmentation
because radiative heating of the protostellar disk should suppress fragmentation. On the other hand, 
the simulations from \citet{bate2018} include radiative feedback and do find disk fragmentation, but
a study of the relative population of companions in different ranges of separations has not been conducted.
However, it should be noted that \citet{offner2010} and \citet{bate2012} employ different
initial cloud conditions (turbulence and volume density) and different assumptions
for the radiative heating. Both of these factors are likely to influence 
the fragmentation on cloud and disk scales in both simulations. Even in the
presence of radiative heating, fragmentation can still occur for systems with total masses (star+disk) of
$>$1~M$_{\sun}$ \citep{kratter2008,kratter2010}. The \citet{offner2010} simulations may be underpredicted
disk fragmentation because most of the systems in that simulation were $<$1~M$_{\sun}$, and the different
treatment of radiative heating in \citet{bate2012,bate2018} could overpredict fragmentation in this mass range.

While we suggest that our results favor disk fragmentation 
as a formation mechanism for these systems,
there is an inescapable caveat. Even if the disks are Keplerian, there remains the possibility
that initially widely separated companions could have migrated closer together \citep{offner2010,sadavoy2017}. 
In the process, the original disks may have been destroyed and a new disk 
reformed around the companions in the same orbital plane. While the outflows
could be misaligned initially \citep{offner2016}, they could be brought 
into alignment and appear as a single outflow as the protostars
accrete from the disk surrounding them. However, there is not evidence for 
past multiple outflows in different directions
for the systems where we would have the possibility of determining this. 
But, the timescales over which we could detect
remnant outflows might be too short. Further study of circum-multiple material 
surrounding systems formed via initially turbulent fragmentation may provide clues to more
uniquely distinguish between primordial disk fragmentation and turbulent fragmentation with a 
re-formed disk.
Therefore, our main result finding that 8 out of 12 sources are consistent with disk
fragmentation could be rephrased to say that 66\% is an upper limit on the frequency disk 
fragmentation for the production of close multiples.

\subsection{Non-detections of Companions}

In our ALMA follow-up of the VLA-detected close multiples, we detect the companions in 
12 out of 15 multiple systems that we were able to resolve. We left Per-emb-2 and Per-emb-18 
out of these numbers because it is not possible to resolve their companions in these
observations. All Class 0 companions are detected, and only the Class I companions 
have some non-detections. However, in the
case of the non-detections, we do marginally detect one of the components in Per-emb-40,
Per-emb-55 has a tentative detection toward the companion position.
Finally, Per-emb-48 could also not be self-calibrated, reducing 
its likelihood to detect the companion.

The fact that most of the non-detections were more-evolved 
Class I systems makes sense given that the VLA is detecting a 
combination of dust and free-free emission in most cases. 
The non-detected component of Per-emb-48 was Per-emb-48-B;  Tychoniec et al. (2018) detected Per-emb-48-B
with a declining spectral index from 6.4~cm to 8~mm, consistent with gyro-synchrotron 
from a stellar corona. Per-emb-55-B, the very tentative detection, had a spectral index 
that was rising (0.9) but more flat than expected from
dust emission alone ($\ge$2). Assuming a spectral index of 3 (implicitly assumes a dust opacity
spectral index of 1), we extrapolate the 9~mm flux density of 0.07~mJy to 1.3~mm. In order to have
a flux density consistent with the observations, a flux density of less than 1~mJy, 
$\sim$95\% of the 9~mm emission from Per-emb-55-B must be from free-free emission. \citet{tychoniec2018}
found that 79\% of the 9~mm emission was from free-free emission, this would also point to
the spectral index of the source emission being shallower than 3.
Per-emb-40-B was quite faint and also appeared to have a declining spectral index in 
8~mm to 1~cm; the 4.1~cm and 6.4~cm were much fainter than expected, possibly due to 
source variability \citep[e.g.,][]{dzib2013}. However, Per-emb-40-B has extended dust toward it, so it is probably 
real, but not detected at our level of sensitivity. 

The faintness of companions toward
some of these close Class I protostars is consistent with the results of \citet{harris2012}. While
that study examined Class II stellar systems, they found that the systems with close separations ($<$300~AU)
had a factor of 5 lower millimeter emission compared to single systems and systems with $>$300~AU separations.
Moreover, the primary was always brighter than the secondary in the cases where both components
of the system were detected at millimeter wavelengths. Thus, evolution may drive the companion(s) to be significantly fainter
than the primary, and the flux density of close ($<$300~AU) systems to be lower than more widely separated and/or single systems.

It is also worth discussing the cases of some other companion sources reported
toward protostars in Perseus. A possible new companion source was reported 
near Per-emb-12 (NGC 1333 IRAS4A), referred to as A3 in \citet{santangelo2015}, with 
a 1.3~mm flux density of 187 mJy. This source A3 is not detected with the 
VLA in VANDAM nor with our ALMA 1.3~mm data; we only detect the two 
known protostars and the extended emission around them. \citet{santangelo2015} 
also did not detect this source at a wavelength of 3~mm. This case of A3 
toward Per-emb-12 is very similar to the case of the companions (denoted MM2 and MM3)
around Per-emb-27 (NGC 1333 IRAS2A) reported by \citet{maury2014} from PdBI
observations. These companions were not detected 
by the VLA, CARMA, or the SMA, as detailed in \citet{tobin2015a} or by
the more sensitive ALMA data presented here. It is most likely that A3 toward Per-emb-12 and MM2 and MM3 toward Per-emb-27
are either artifacts from the deconvolution process and/or additional noise resulting from dynamic range
limitations of the PdBI as was previously discussed in \citet{tobin2015a}.

In a similar vein, there have been hypotheses put forward that some clearly detected
continuum sources were features produced by outflow interactions \citep{maury2010, maury2012}.
However, further study over the past several years has found that, the continuum features that
were suggested to be outflow knots in VLA 1623, are definitively
real protostars \citep{murillo2013,murillo2013b, sadavoy2018, harris2018}.
The other cases where continuum sources were 
suggested to be outflow features \citep[i.e., L1448-mm, 
NGC 1333 IRAS2A, NGC 1333 IRAS4A,][]{maury2010,maury2014,santangelo2015}
seem to be artifacts in those data given their lack of detection here with ALMA and in
other works at the same and other wavelengths \citep{tobin2015a,tobin2015b}.

While it seems unlikely that outflow features are significantly contributing 
to the detection of false companions, 
we note that none of the companions without detections are found along the outflow
direction. Moreover, the host protostars of
the three non-detections are not driving prominent outflows as viewed from the ALMA \twco\ data. Therefore
it seems unlikely that companions are being mimicked by outflow features in our case.

We had an 80\% detection rate with ALMA toward the VLA-detected
close multiples in Perseus; 100\% for the Class 0s and 62.5\% for the Class Is. The Class I
non-detections based on the VANDAM survey are likely to be real companions 
but were below our detection limits. For these
sources in particular, it might be useful to search for their companions in the near-infrared since 
they are observed toward sources that are not highly embedded. Thus, we can conclude 
that VLA-detected companions are quite robust.

\section{Conclusions}

We have conducted ALMA observations toward 17 VLA-detected proto-multiple systems from
the VANDAM survey that are separated by less than 600~AU. We observed these systems in 
dust continuum at 1.3~mm and the
\twco, \thco, \cateo, SO, and H$_2$CO molecular lines, enabling
us to examine the circumbinary environments. Our main results are as follows:

\begin{itemize}
\item Of the 15 companions that could be resolved in our observations, we detect 12 
confidently, with tentative detections for 2 additional sources. We conclude that the 
VANDAM multiplicity detections from \citet{tobin2016a} had very high reliability. 
Though they were not the focus of this study, we also detected all wide companions 
in the higher-order systems around L1448 IRS3B, Per-emb-55, and Per-emb-44.

\item We detect circum-multiple dust emission in 8 out of the 9 deeply embedded Class 0
sources in the sample, but none in the remaining 8 more-revealed Class I systems. For
the sources that are closer to edge-on, this material is oriented orthogonal to the outflow direction and the 
companion stars are embedded within the circum-multiple material.  This is consistent with the expectations for
their formation within a circumbinary/multiple disk.

\item We detect molecular line emission associated with the circum-multiple material 
in 12 systems (all Class 0 systems and 3 out of 8 Class I systems). The molecular lines with
discernible velocity gradients were orthogonal to the outflow 
directions in 8 out of 12 cases (EDJ2009-269 had too
tentative of a detection in \twco). Velocity gradients orthogonal to the outflow
on these scales ($<$600~AU) are most likely due to
rotation. Therefore, 8/12 systems are consistent with disk fragmentation and
4/12 appear inconsistent with disk fragmentation. However, the overall numbers are
small and may not reflect a general ratio for the origin of close companions.

\item Most of the less-embedded Class I systems in our sample do not show circumbinary dust or gas emission.
These systems are likely near the end point of their evolution where the circum-multiple 
disk is mostly accreted or dispersed.

\item We identified two additional strong disk candidates, L1448 IRS3A and 
Per-emb-8. Both show extended dust emission
and have molecular line emission that is consistent with rotation. 
Per-emb-8 now has a better defined position angle
compared to what was known previously from the VLA.

\end{itemize}

We acknowledge fruitful discussions with D. Harsono, P. Sheehan, 
P. Hennebelle, and N. Reynolds. We also thank the anonymous referee for comments that 
improved the quality of the manuscript.
J.J.T. acknowledges support from NSF grant AST-1814762, the Homer L. Dodge Endowed Chair,
and grant 639.041.439 from the Netherlands Organisation for Scientific Research (NWO).
This paper makes use of the following ALMA data: ADS/JAO.ALMA\#2013.1.00031.S.
Z.-Y.L.  is supported in part by NSF AST-1616636 and AST-1716259 and NASA NNX14AB38G.
ALMA is a partnership of ESO (representing its member states), NSF (USA) and 
NINS (Japan), together with NRC (Canada), NSC and ASIAA (Taiwan), and 
KASI (Republic of Korea), in cooperation with the Republic of Chile. 
The Joint ALMA Observatory is operated by ESO, AUI/NRAO and NAOJ.
The National Radio Astronomy 
Observatory is a facility of the National Science Foundation 
operated under cooperative agreement by Associated Universities, Inc.
This research made use of APLpy, an open-source plotting package for Python 
hosted at http://aplpy.github.com. This research made use of Astropy, 
a community-developed core Python package for 
Astronomy (Astropy Collaboration, 2013) http://www.astropy.org.

 \facility{ALMA}, \facility{VLA}

\software{Astropy \citep[http://www.astropy.org; ][]{astropy2013,astropy2018}, APLpy \citep[http://aplpy.github.com; ][]{aplpy}, CASA \citep[http://casa.nrao.edu; ][]{mcmullin2007}, The IDL Astronomy User's Library (https://idlastro.gsfc.nasa.gov/)}

\appendix

\section{Outflow Observations}

Our observations toward the Perseus multiple systems also encompassed \twco, which 
typically traces outflowing gas toward protostars. We detected outflow signatures toward 
13 out of the 17 protostars observed. These detections are shown in 
Figures \ref{outflows-1}, \ref{outflows-2}, \& \ref{outflows-3}. 
Of the remaining 4 systems that do not appear to have outflows
traced by \twco, rotation
in the disk was found toward L1448 IRS1 (Figure \ref{L1448IRS1-lines}), two traced the continuum disk without a 
clear velocity gradient (Per-emb-48, EDJ2009-269; Figures \ref{EDJ2009-269-lines}), and one was not detected (Per-emb-49). In most cases, the close multiples only have one outflow that is clearly distinguished, there 
are a few exceptions, however. Per-emb-12 shows two approximately parallel outflows,
Per-emb-27 shows two orthogonal outflows, and Per-emb-35 shows approximately parallel outflows.

\section{Companion Protostar Separation Distributions}

The recently revised distance of $\sim$300~pc to the Perseus molecular cloud \citep{zucker2018,ortiz2018}
has resulted in the angular separations of the detected multiple systems corresponding to larger physical 
separations. We have generated a revised separation distribution using the updated distance for all the
the combined sample, and the classes individually and plot the distributions in Figure \ref{separations}.
We have provided the updated separations in Table 5, 6, and 7. The rest of the information contained
in the tables is the same as in \citet{tobin2016a}.

\begin{small}
\bibliographystyle{apj}
\bibliography{ms}
\end{small}

\clearpage

\begin{figure}
\begin{center}
\includegraphics[scale=0.3]{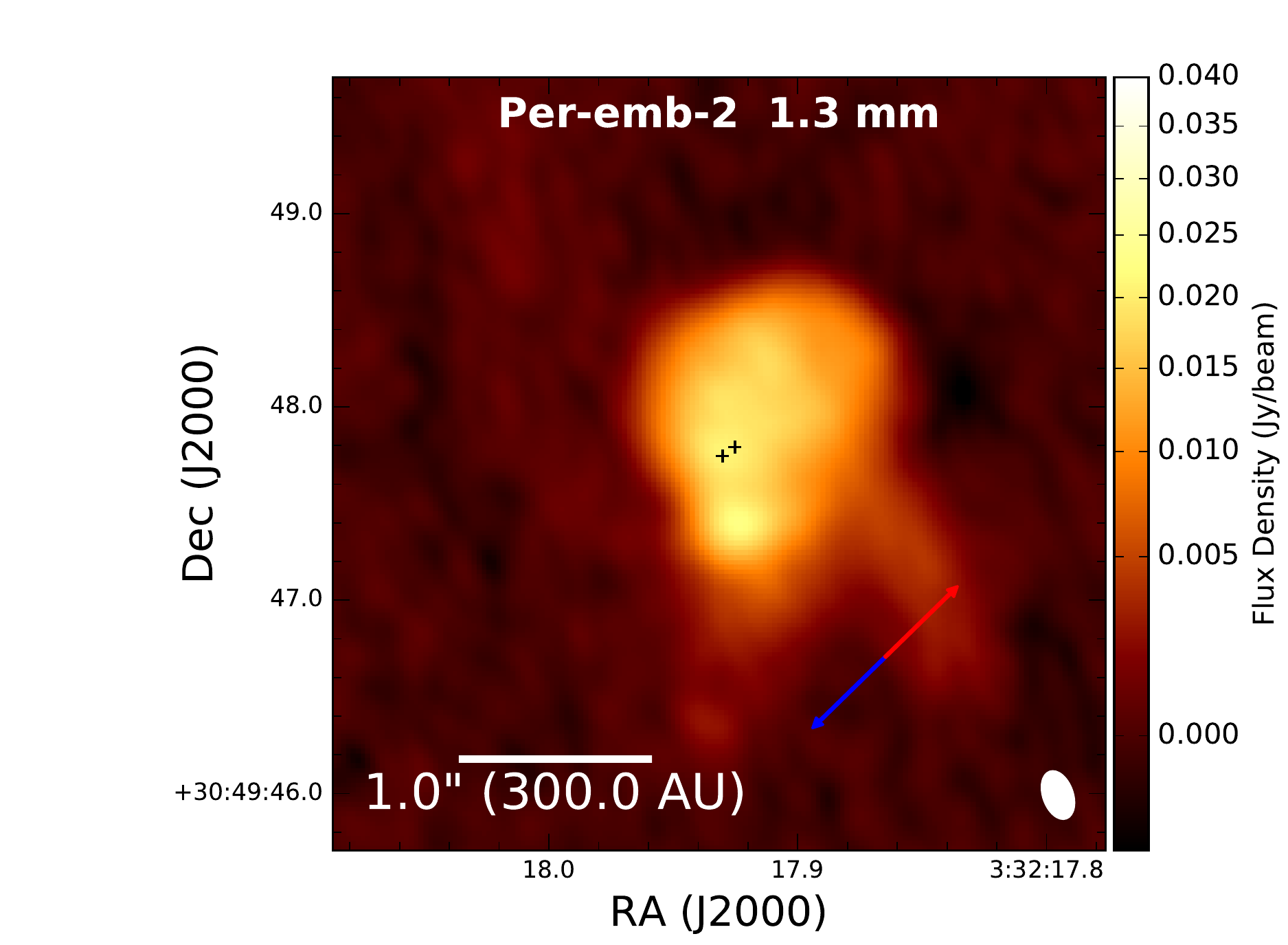}
\includegraphics[scale=0.3]{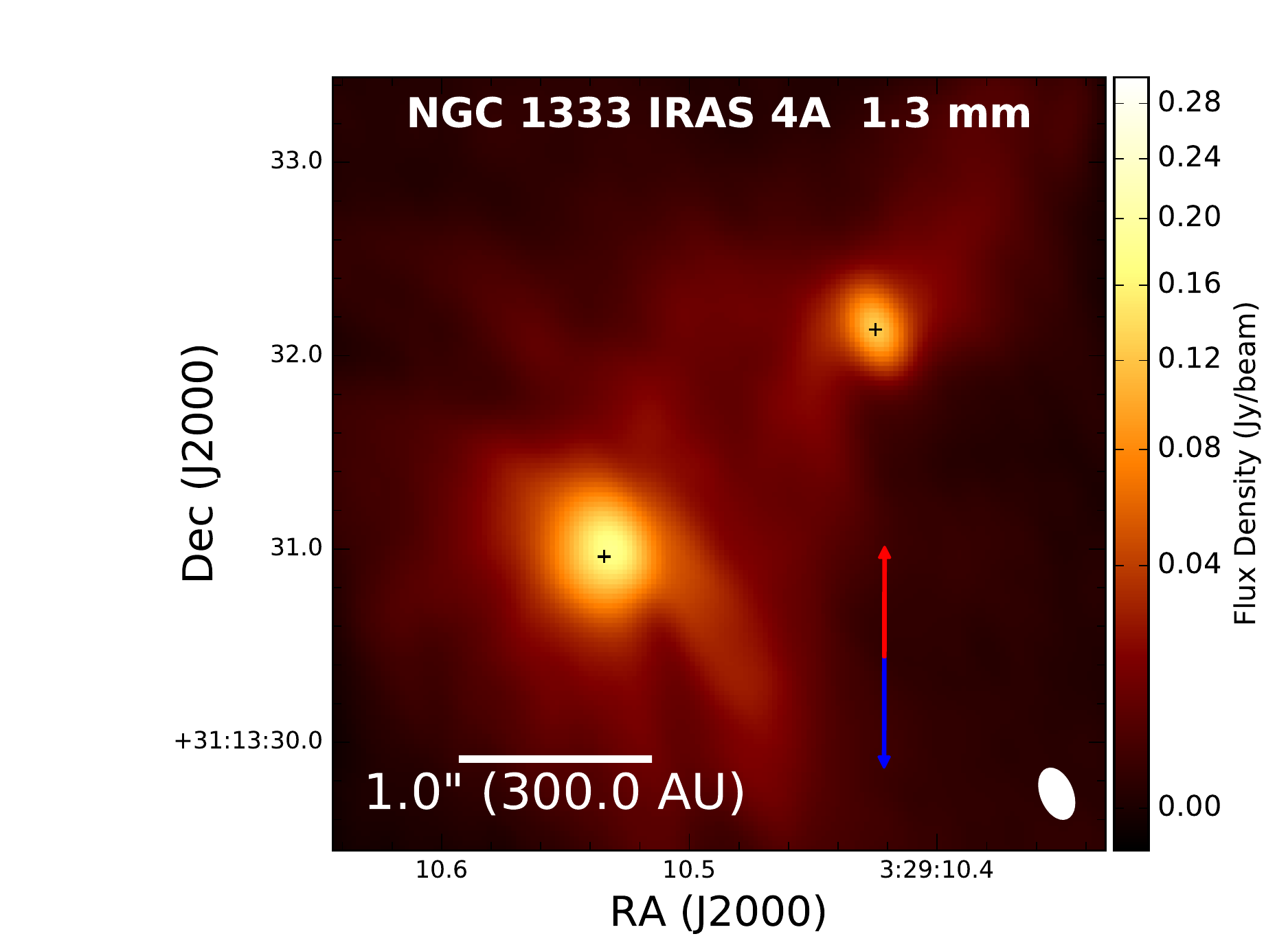}
\includegraphics[scale=0.3]{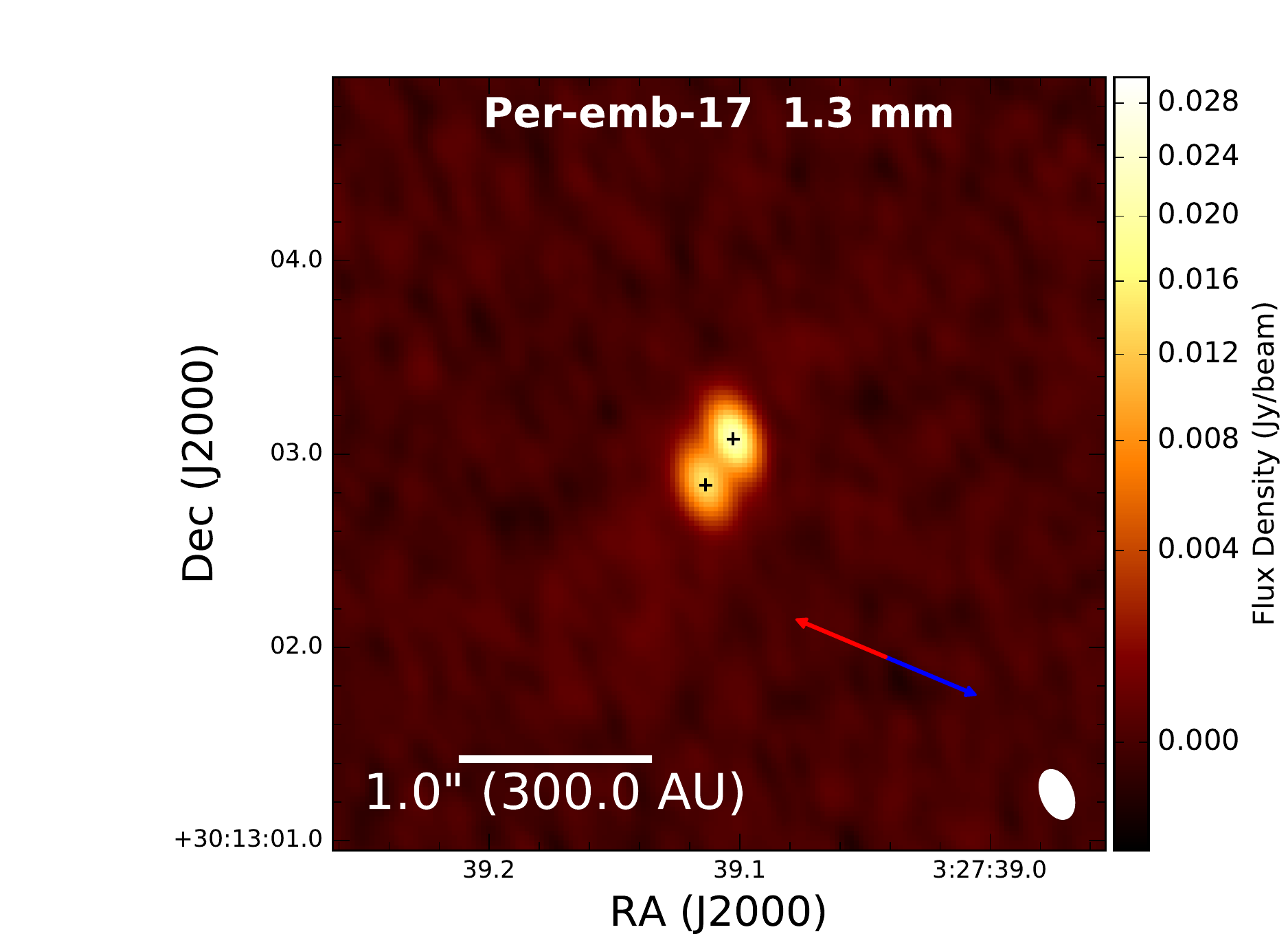}
\includegraphics[scale=0.3]{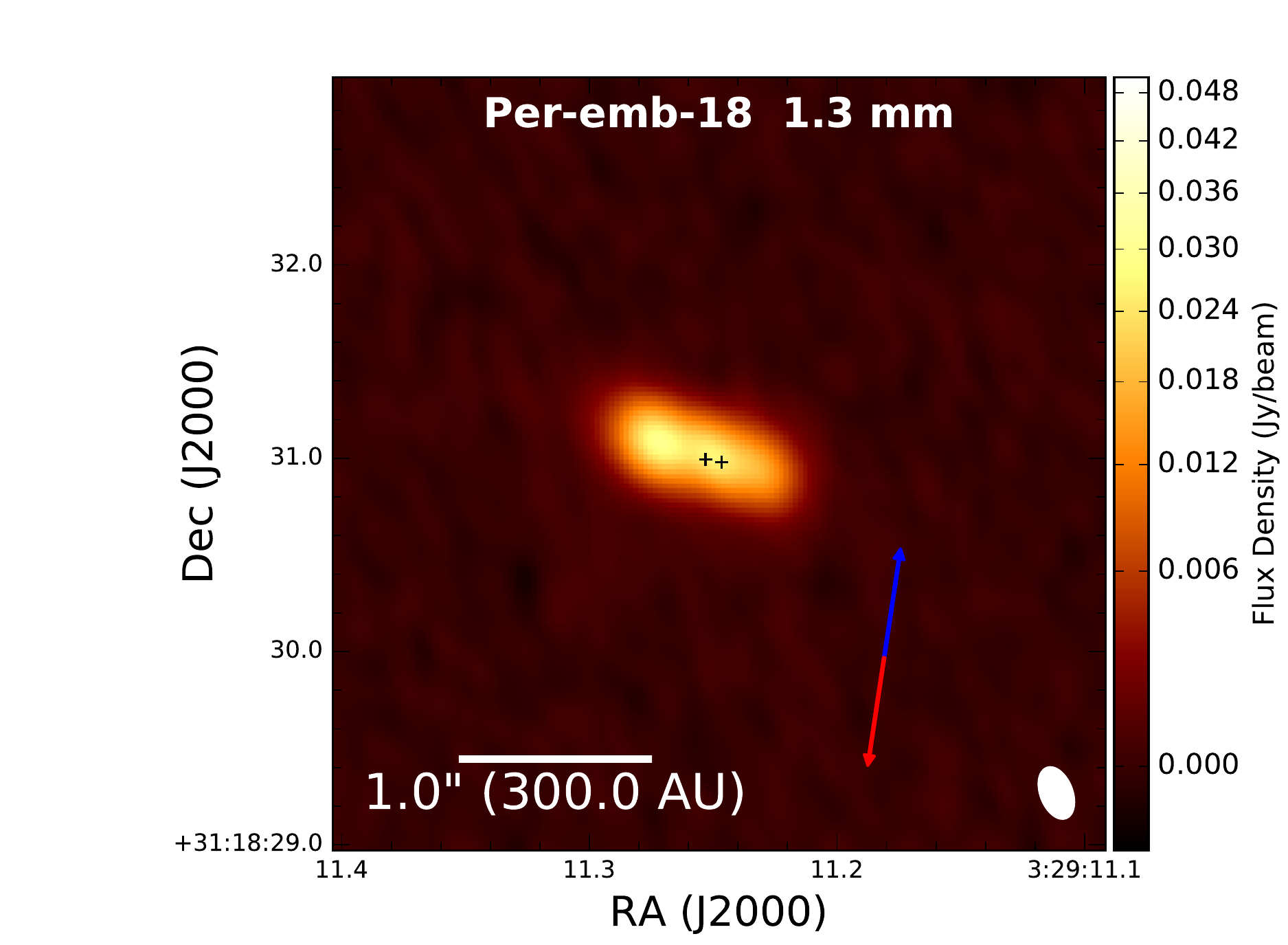}
\includegraphics[scale=0.3]{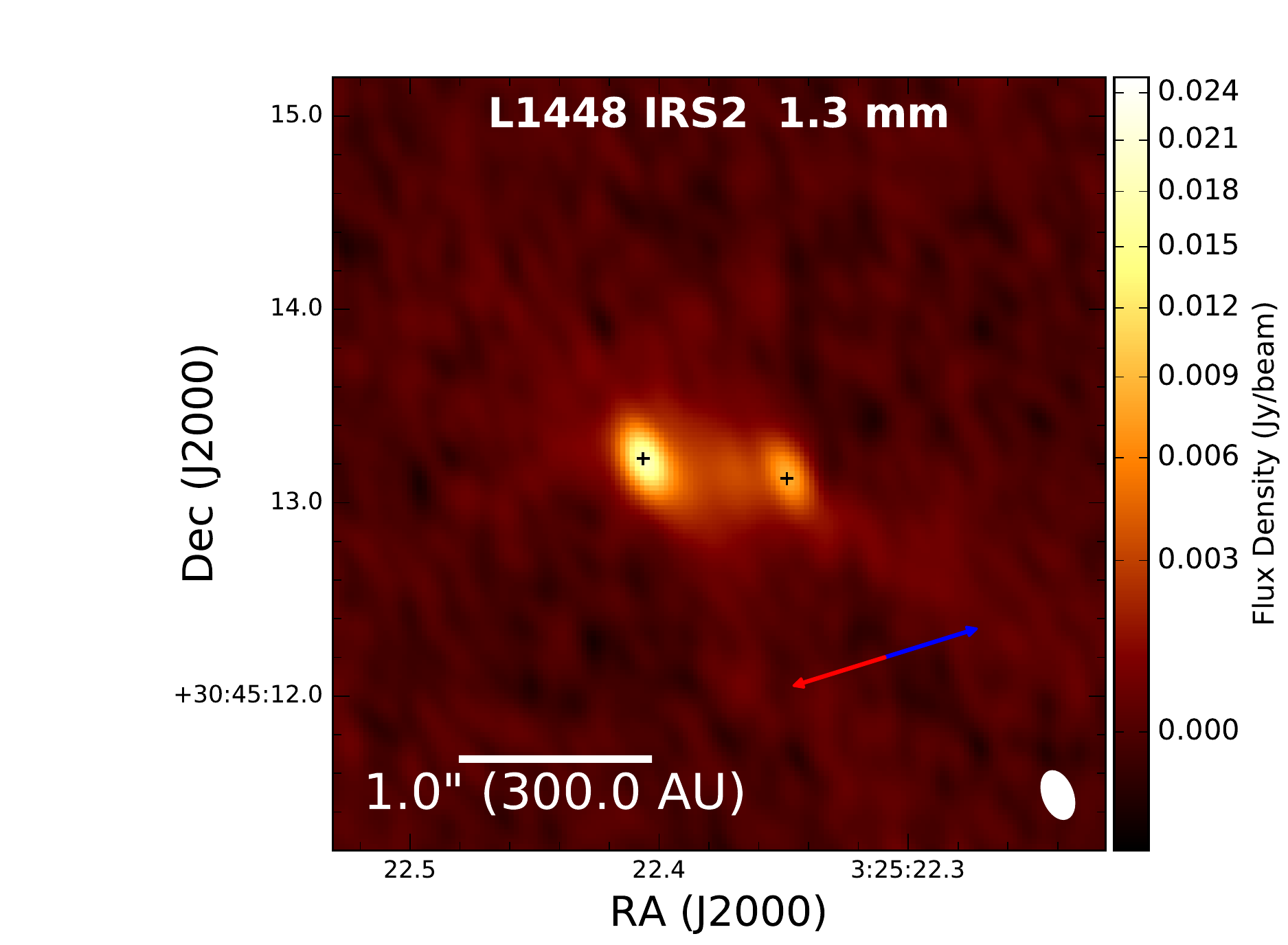}
\includegraphics[scale=0.3]{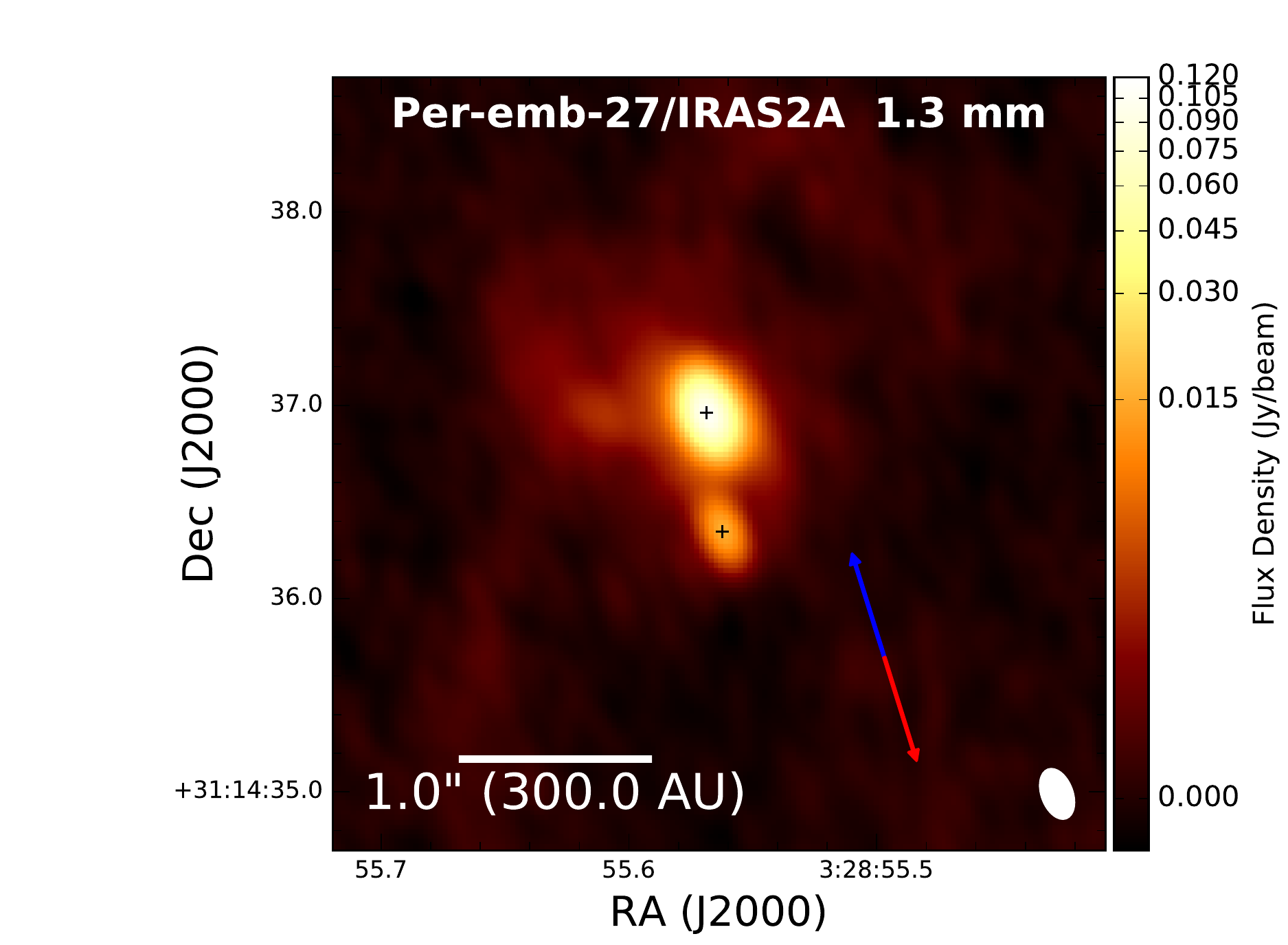}
\includegraphics[scale=0.3]{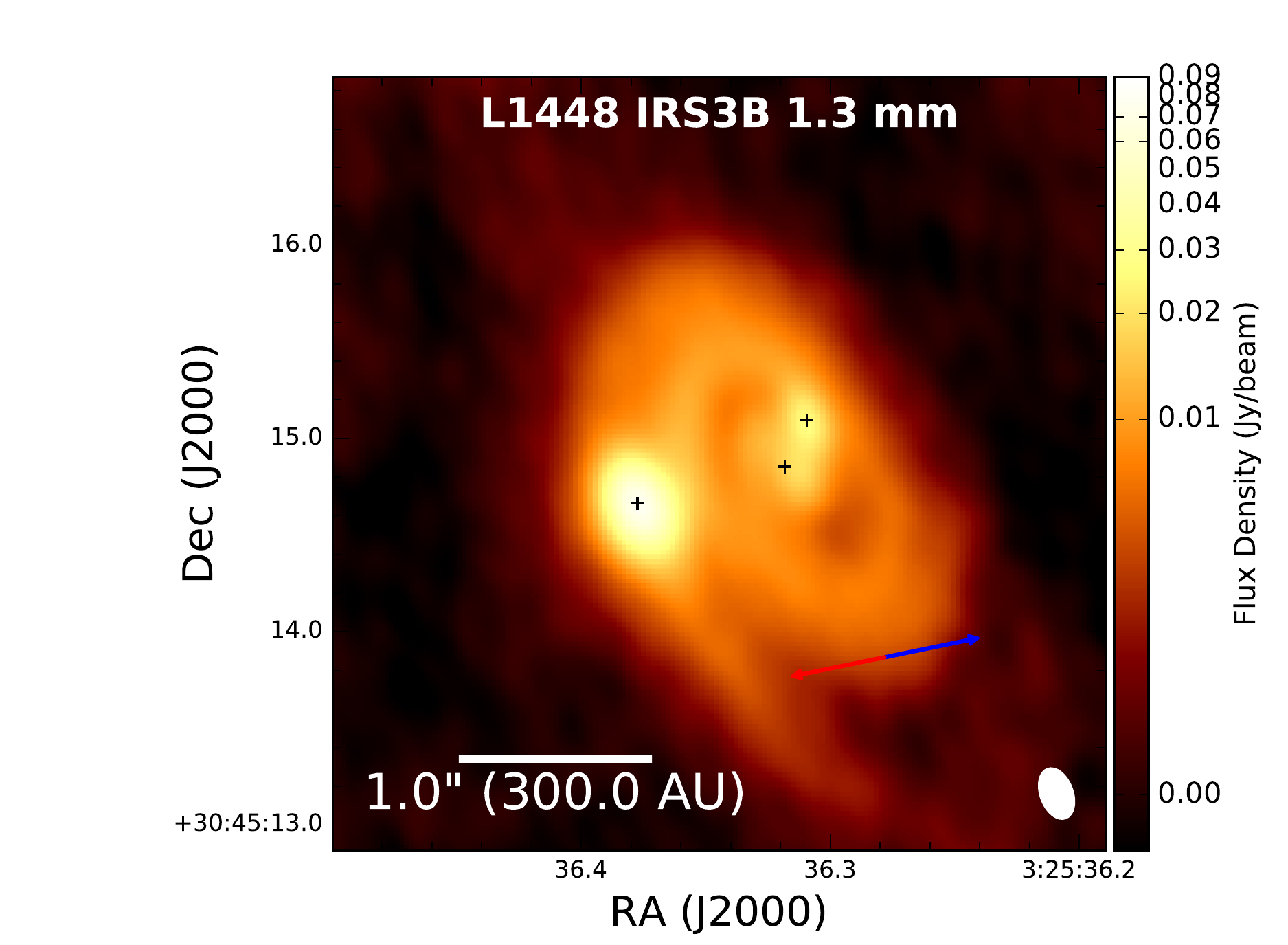}
\includegraphics[scale=0.3]{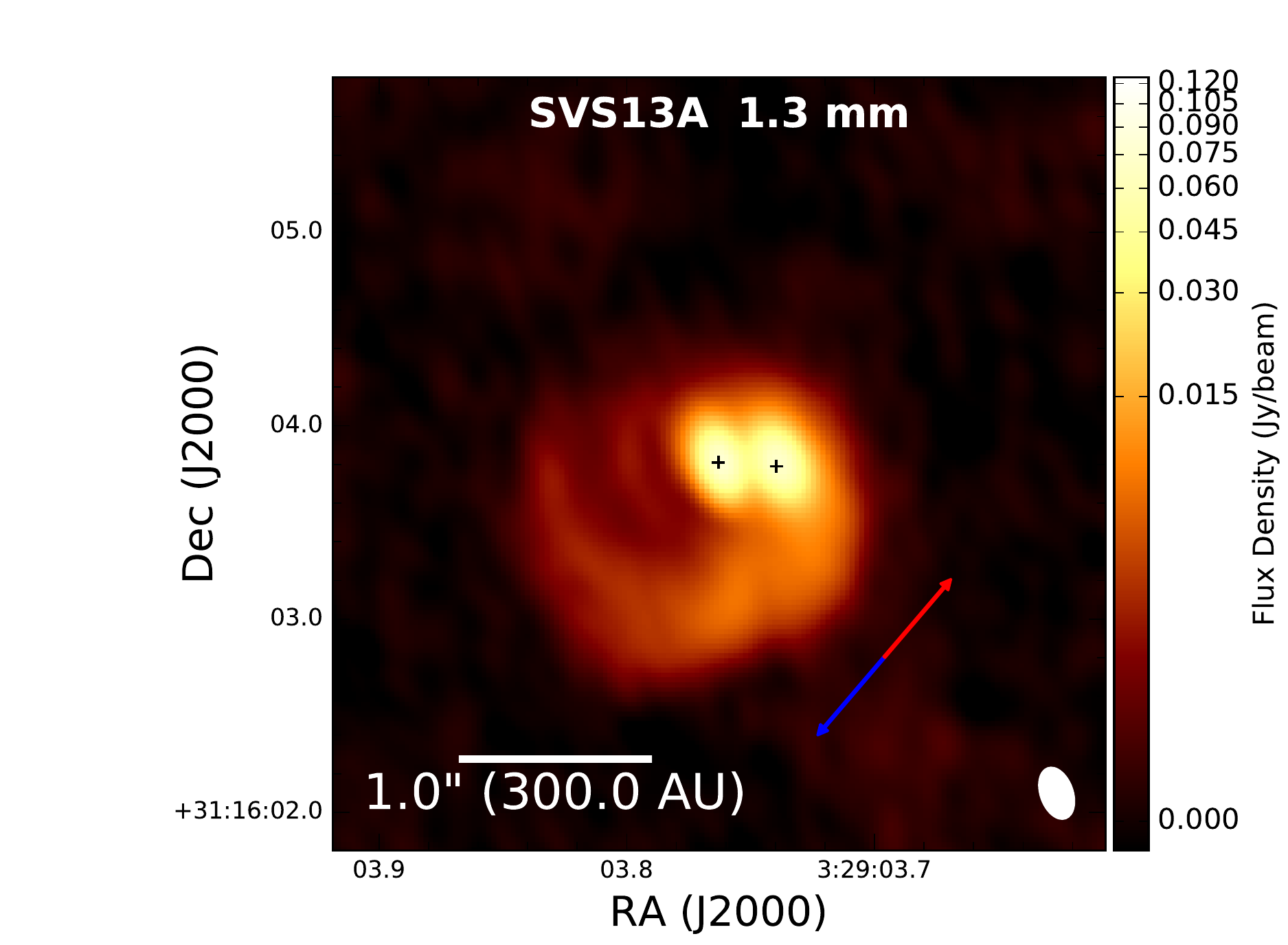}
\includegraphics[scale=0.3]{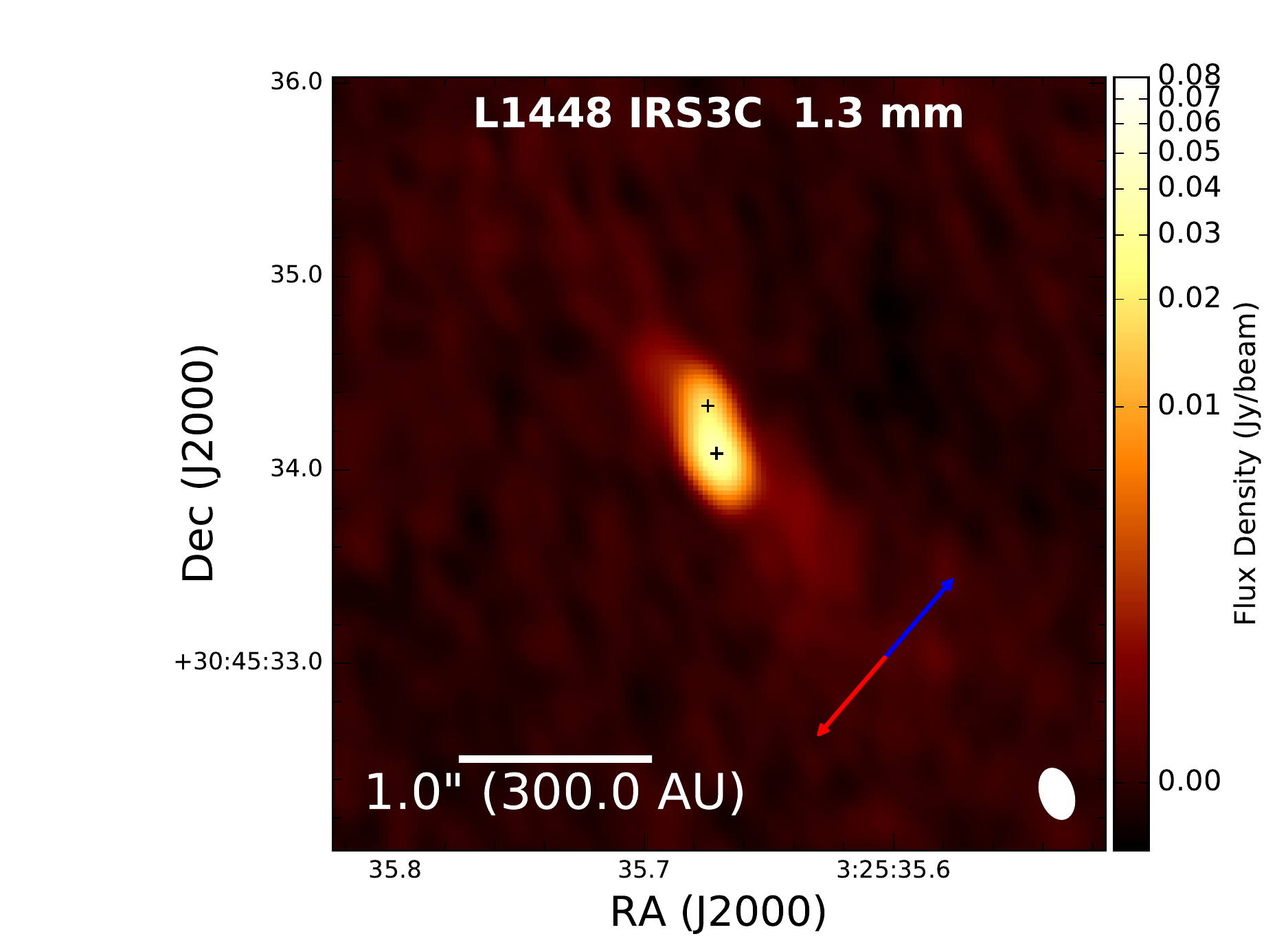}

\end{center}
\caption{ALMA images of Class 0 multiple protostar systems in Perseus at 1.3~mm. The
white or black crosses mark the VLA source positions in each image. A 1\arcsec\
scale bar is also drawn in each panel denoting 
300~AU. The beam of each image
is drawn in the lower right corner, corresponding to 
approximately 0\farcs27$\times$0\farcs17 (81~AU $\times$ 51~AU).
The noise level in each image is approximately 0.14 mJy~beam$^{-1}$, but this
varies somewhat between sources depending on dynamic range limits. The approximate outflow 
directions (when known) are drawn in the lower right corner with the red and blue arrows 
directions corresponding to the orientation of the outflow. Note that
the outflows apparently originate from the bright continuum peaks, but the arrows are 
drawn offset for clarity.
}
\label{class0}
\end{figure}

\begin{figure}
\begin{center}
\includegraphics[scale=0.3]{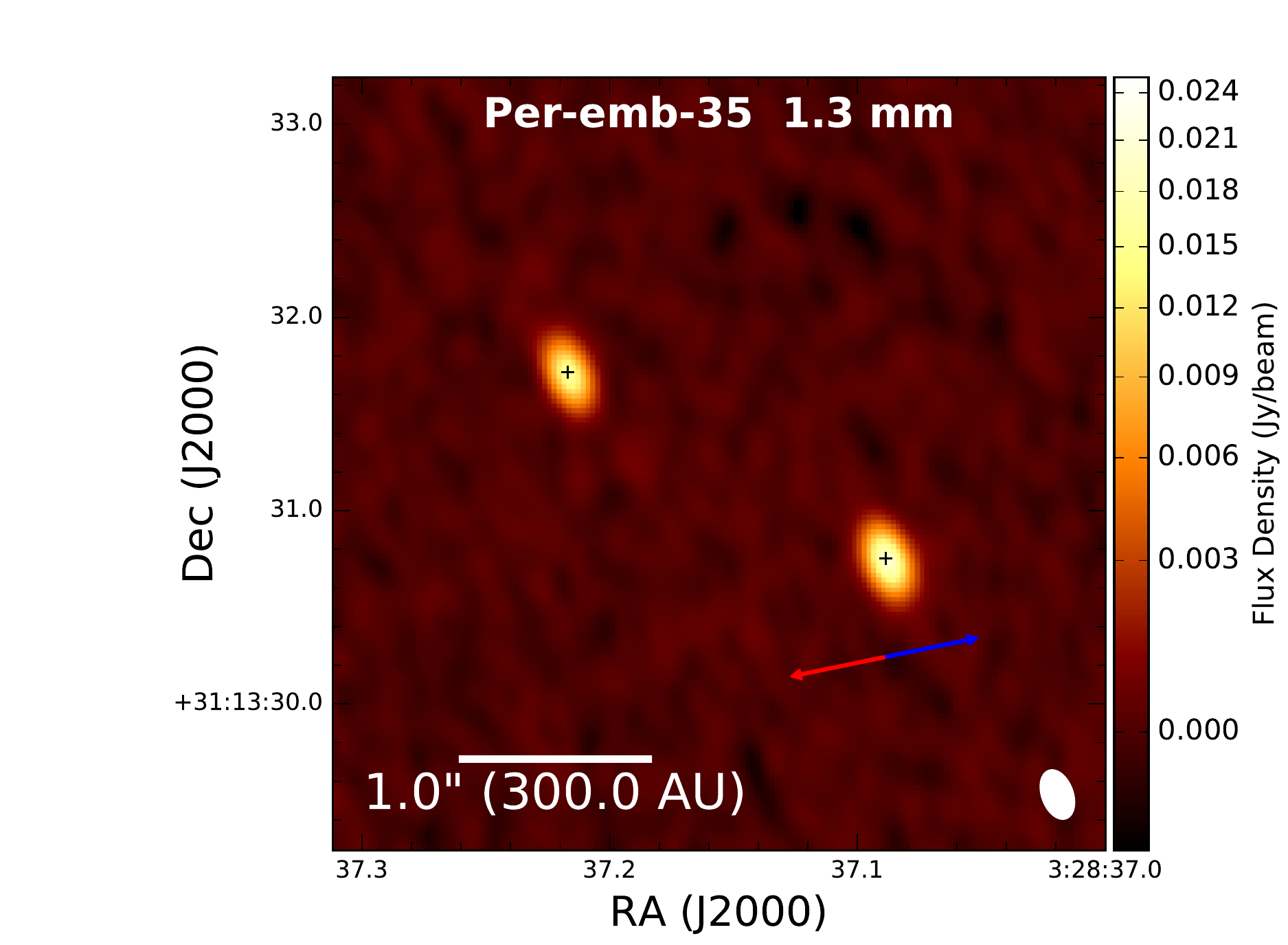}
\includegraphics[scale=0.3]{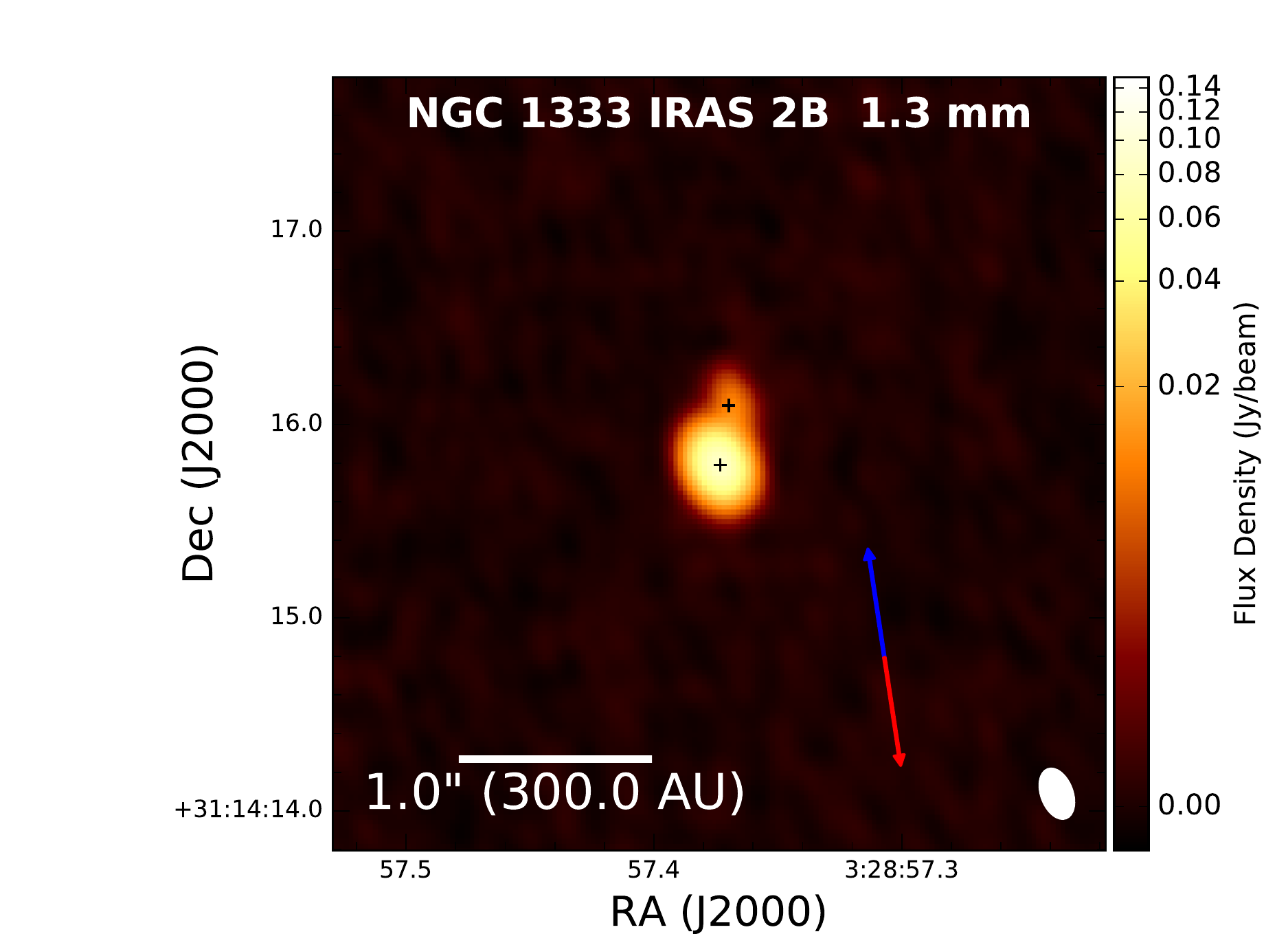}
\includegraphics[scale=0.3]{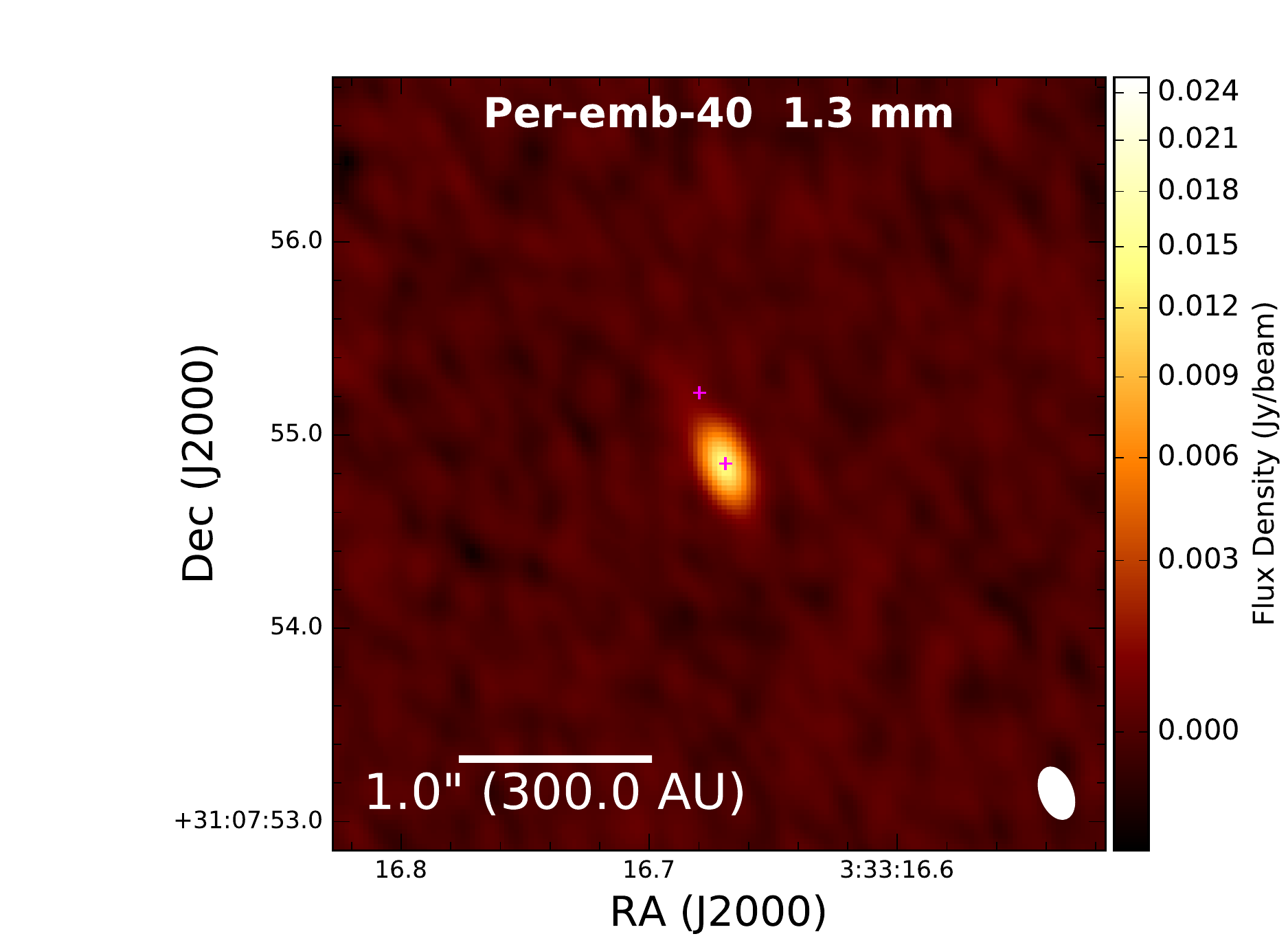}
\includegraphics[scale=0.3]{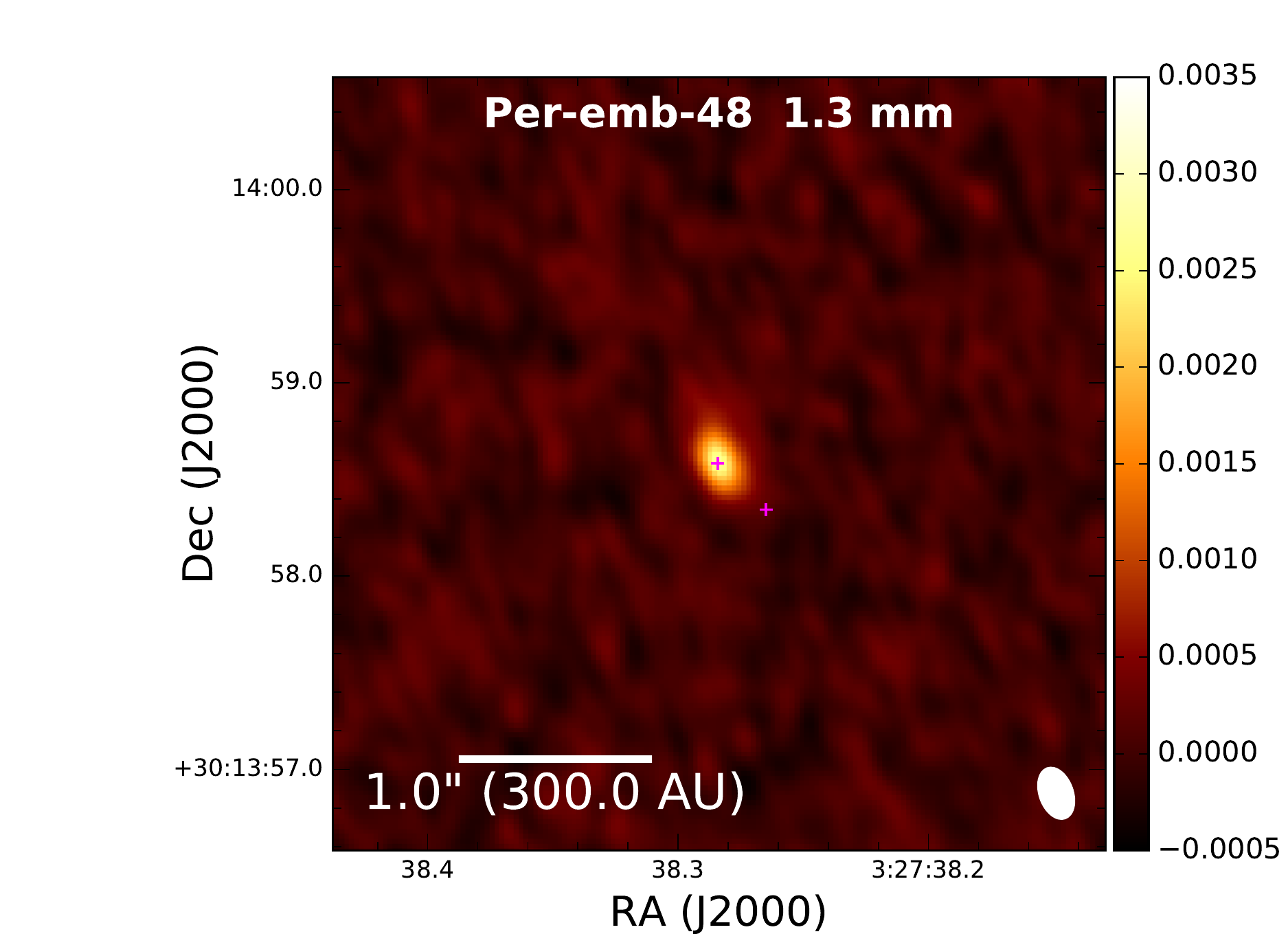}
\includegraphics[scale=0.3]{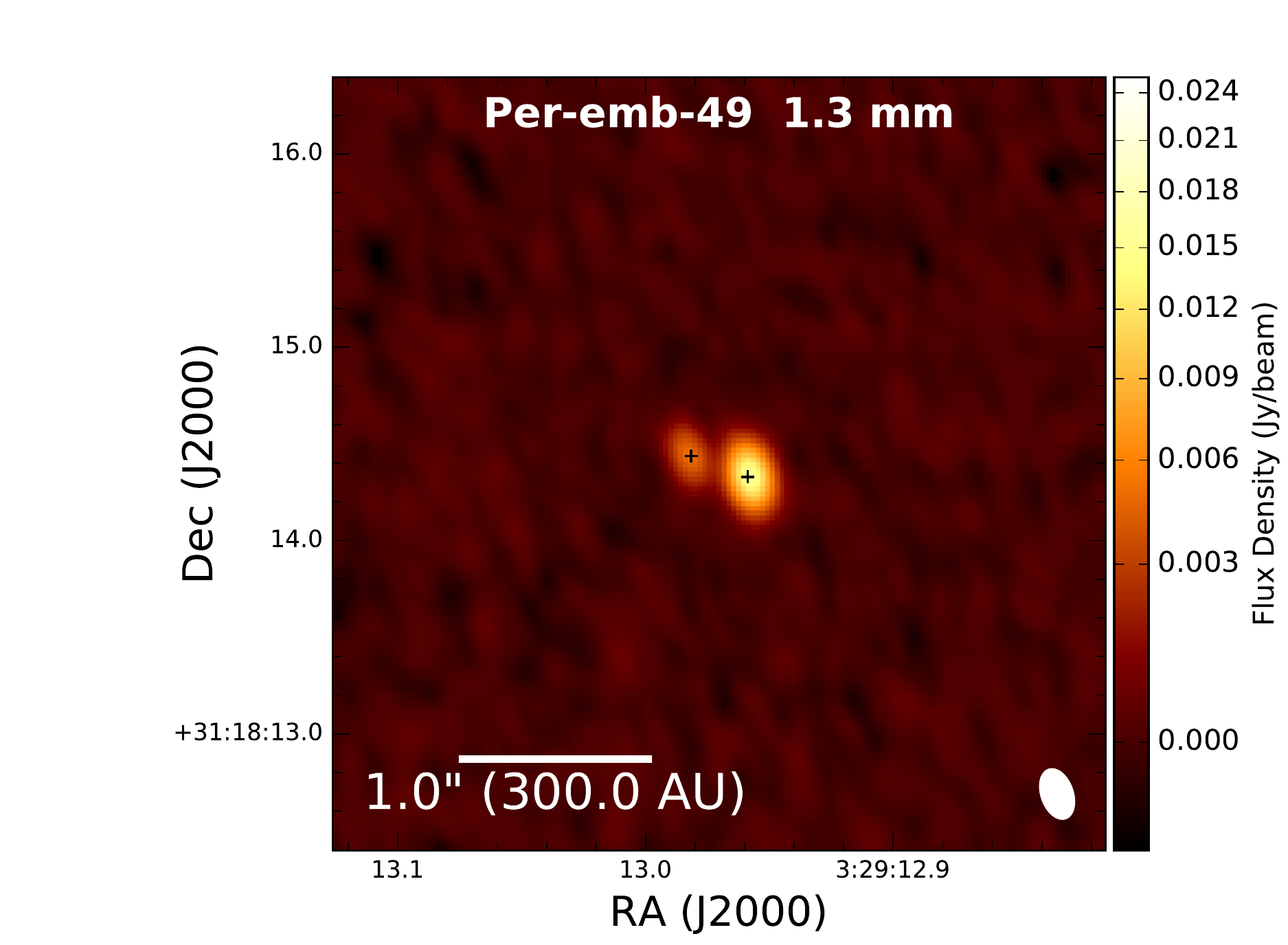}
\includegraphics[scale=0.3]{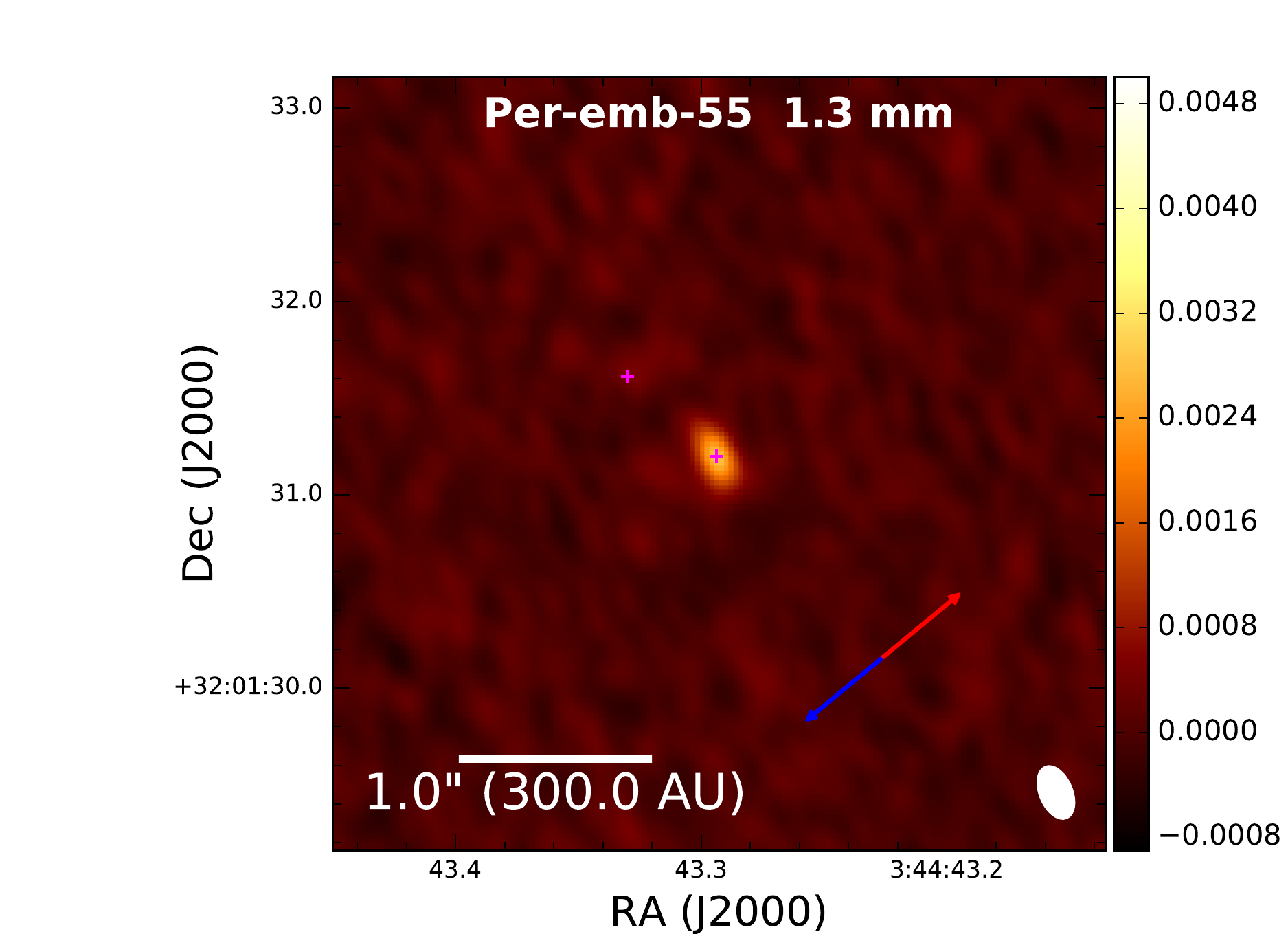}
\includegraphics[scale=0.3]{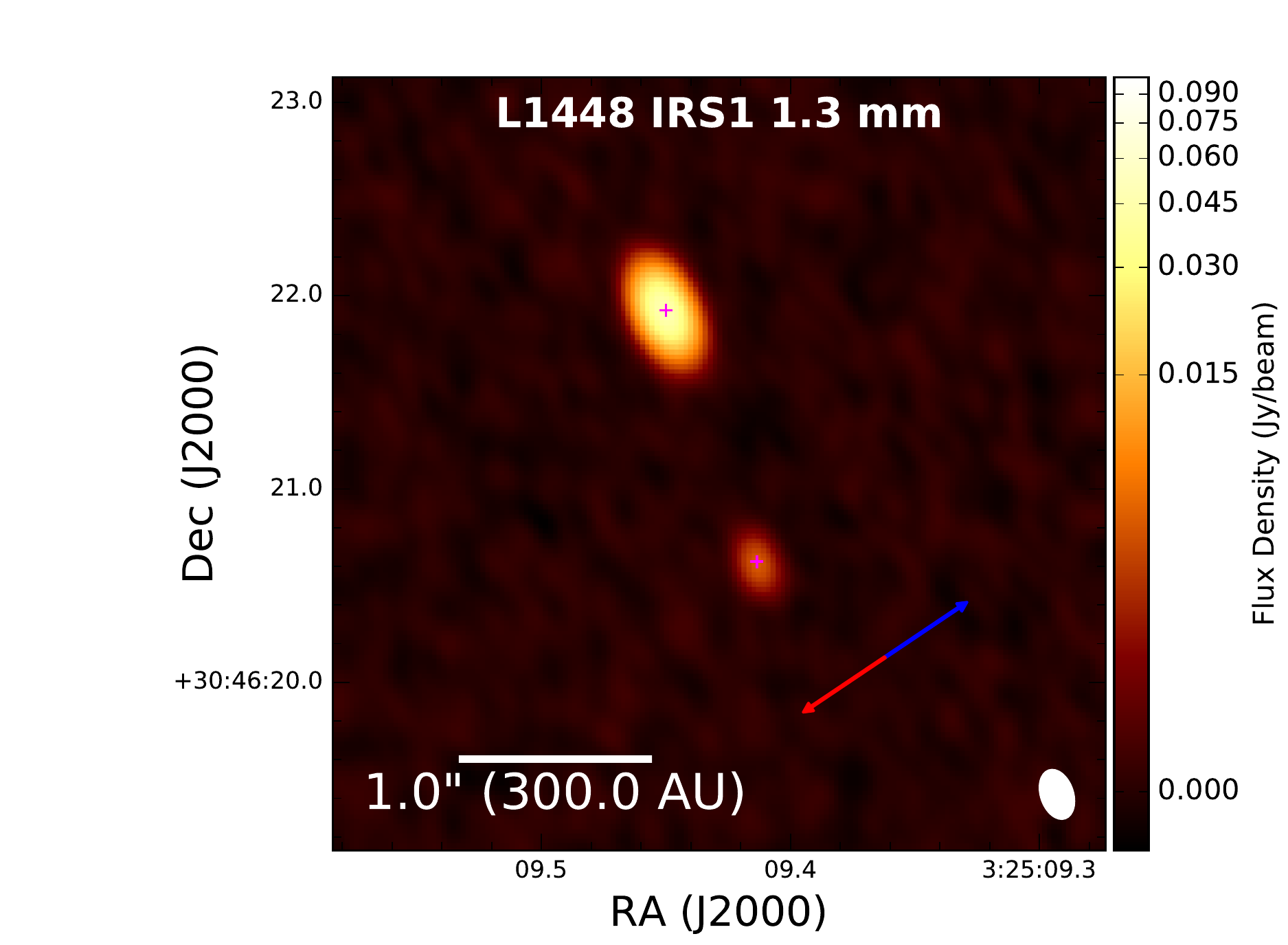}
\includegraphics[scale=0.3]{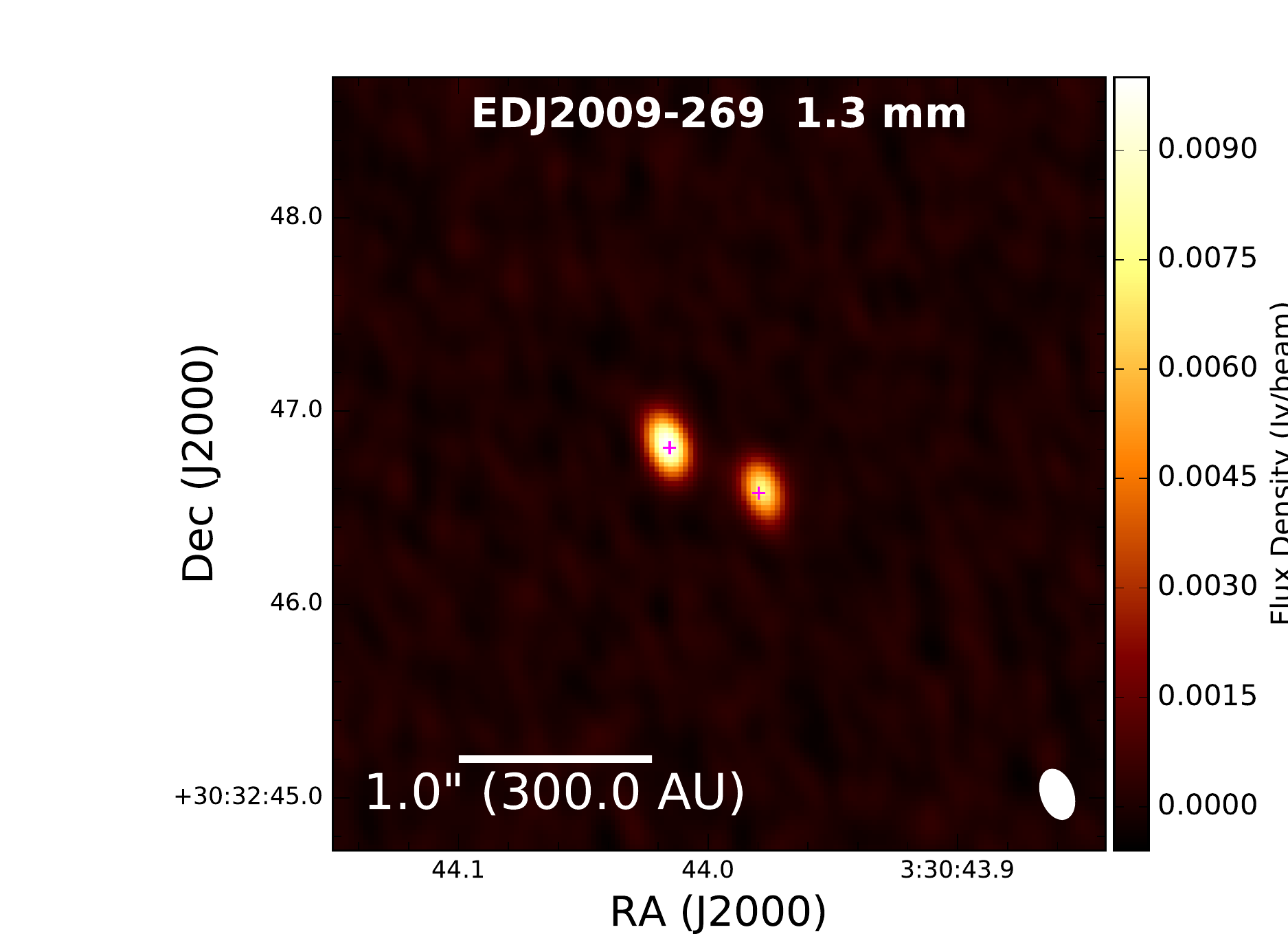}
\end{center}
\caption{Same as Figure 1, except for Class I multiple systems in Perseus.}
\label{class1}
\end{figure}

\begin{figure}
\begin{center}
\includegraphics[scale=0.45]{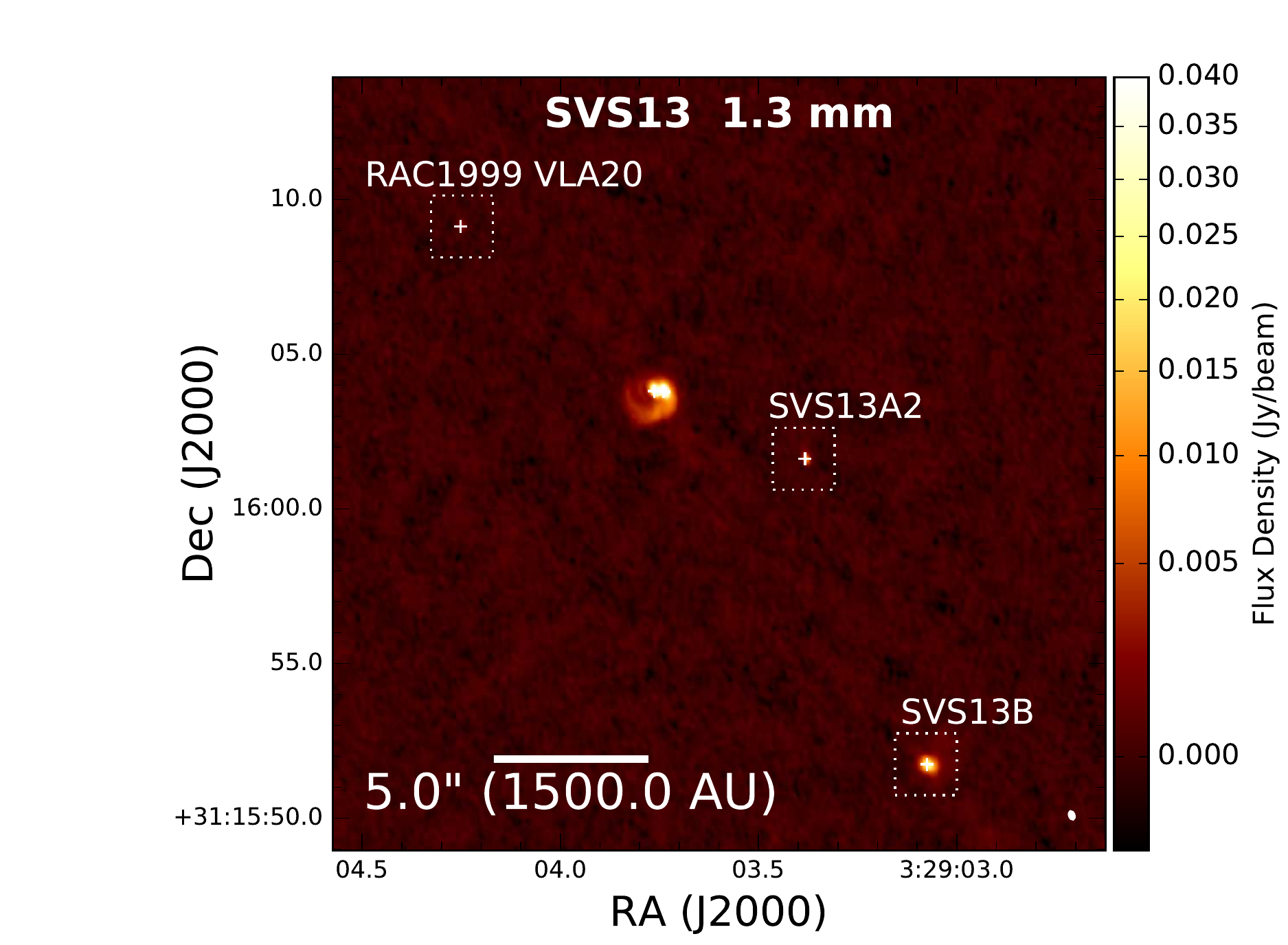}
\includegraphics[scale=0.45]{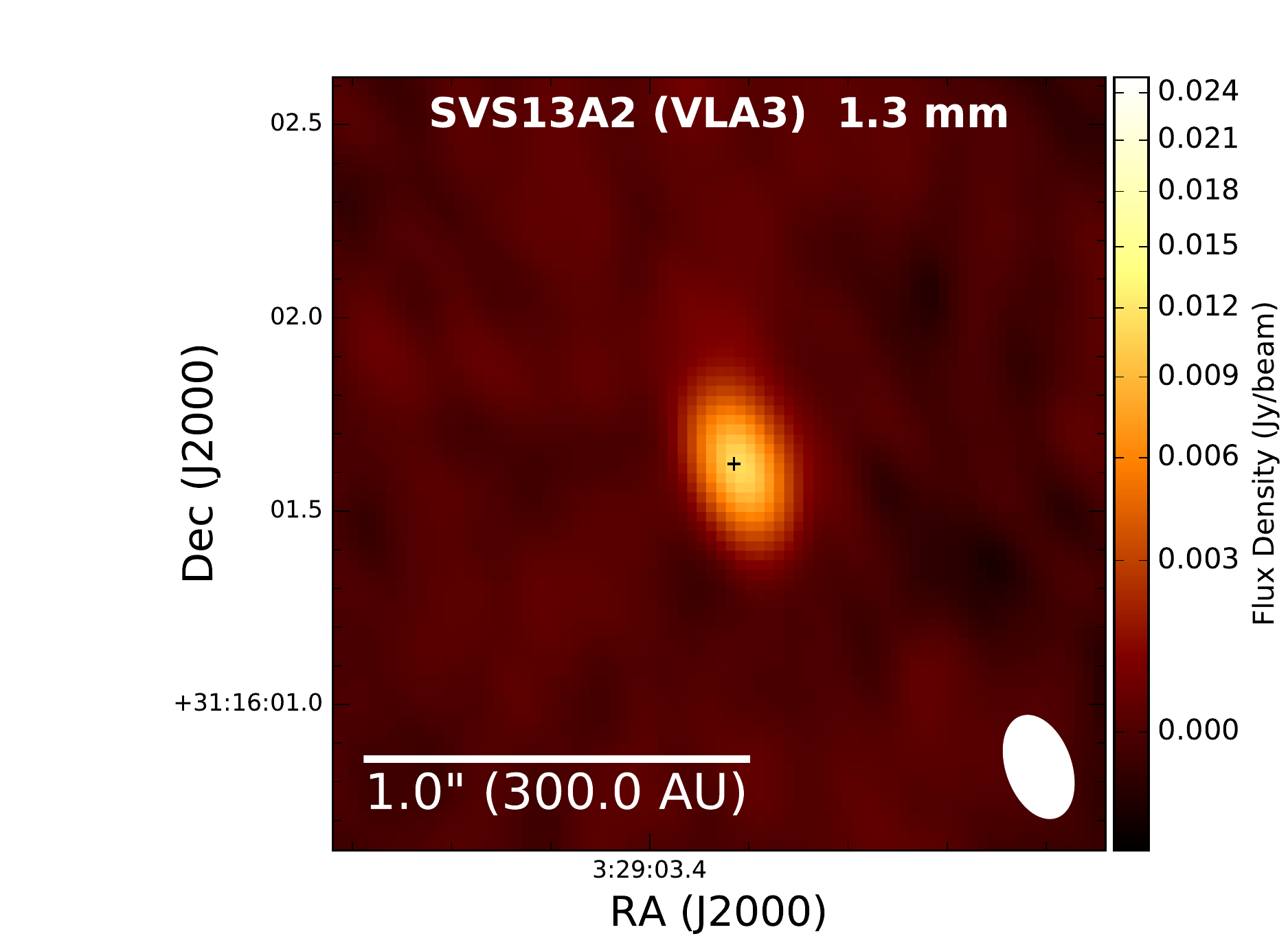}
\includegraphics[scale=0.45]{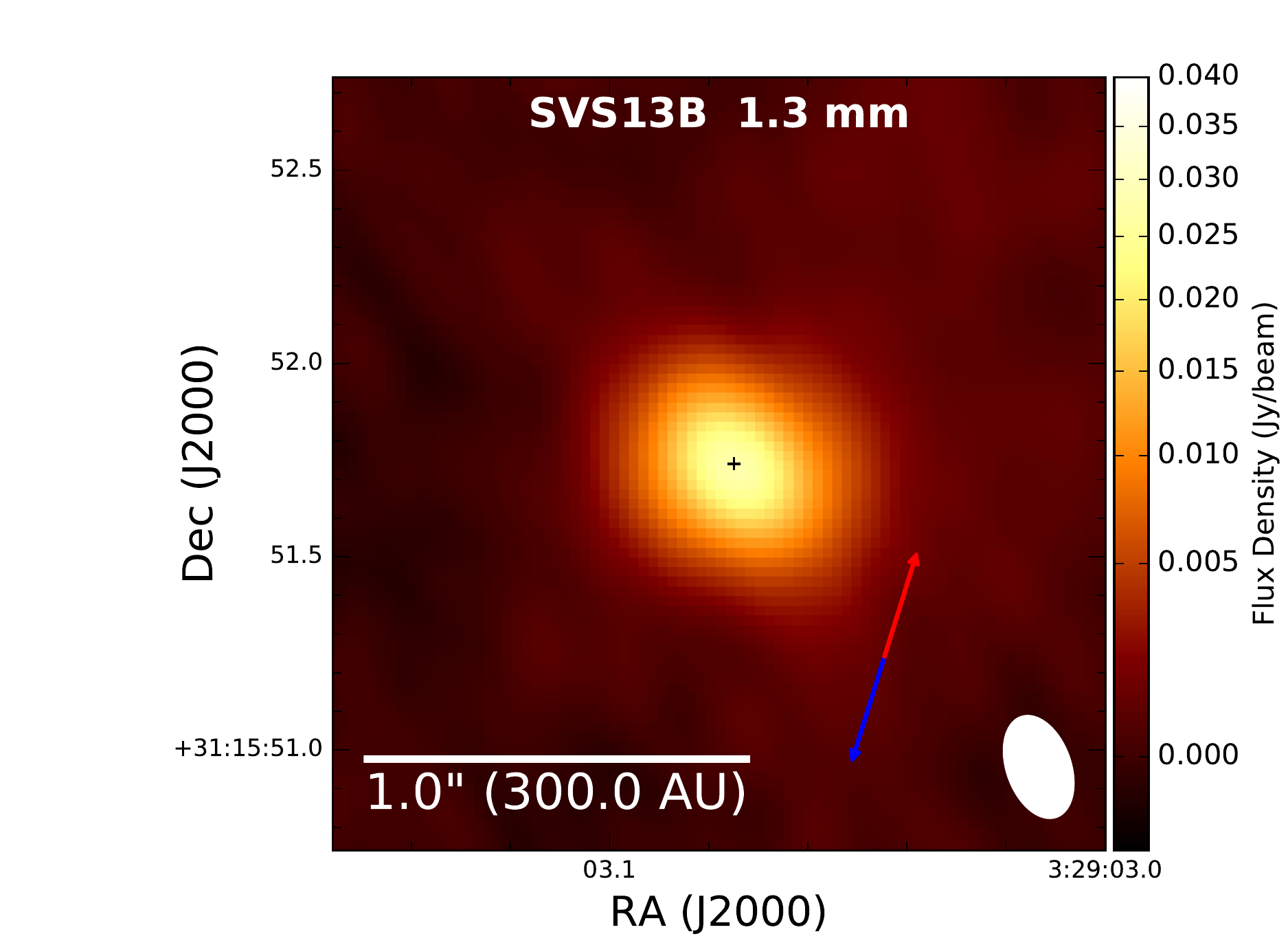}
\includegraphics[scale=0.45]{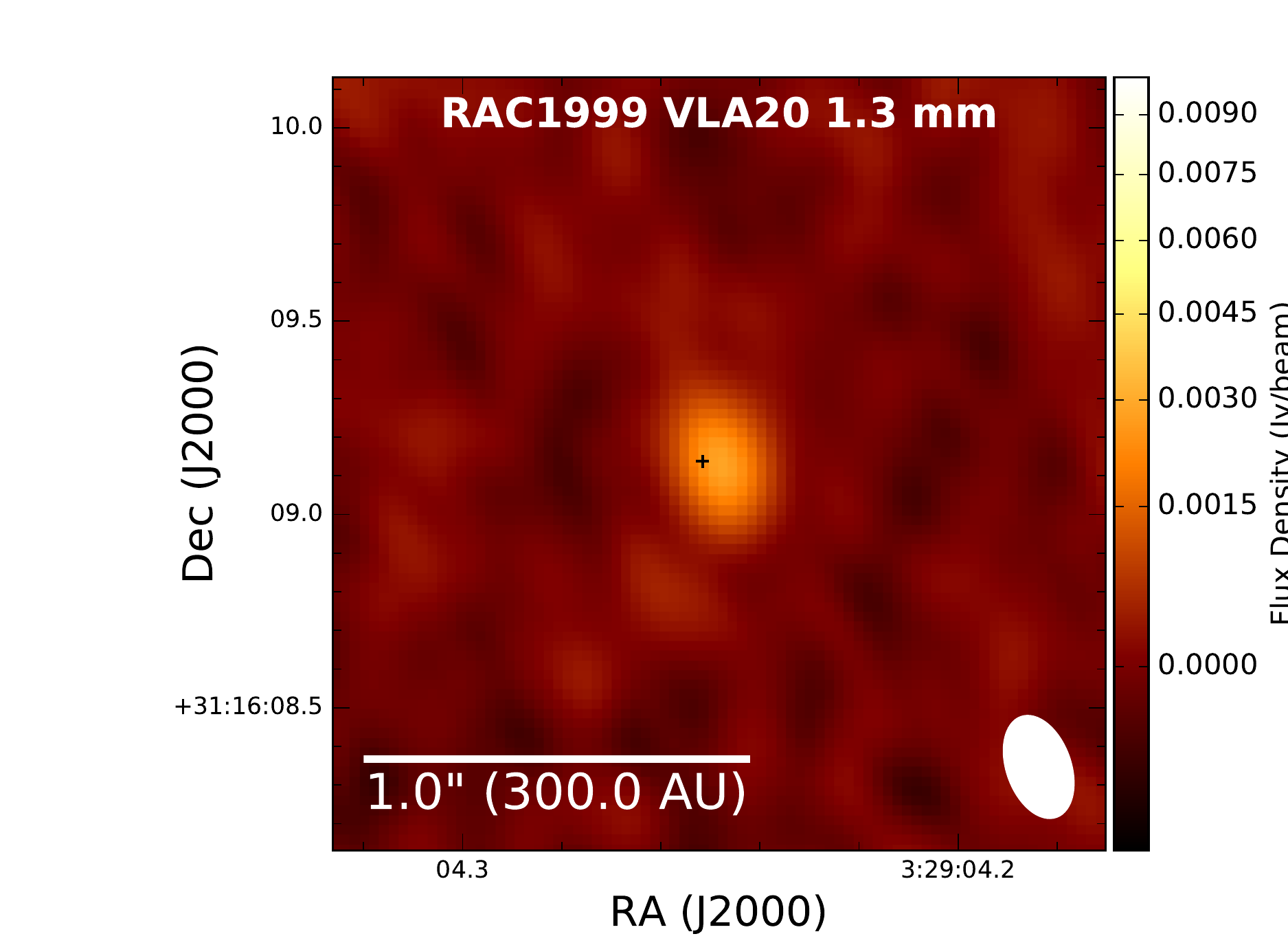}

\end{center}
\caption{ALMA 1.3~mm images of the SVS13 (Per-emb-44) region. A wide view
of the region encompassing all
the sources within the primary beam is shown in the top left and dashed boxes mark
the regions zoomed-in on in the subsequent panels. The northeastern-most source (RAC1999 VLA20) is most likely 
a background extragalactic source from the VLA imaging results. The two sources south of 
SVS13A are SVS13A2 (VLA3) and SVS13B. We show the zoom-in on SVS13A2 in the top right panel,
the zoom-in on SVS13B in the bottom left panel, and RAC1999 VLA20 in the bottom right panel. 
SVS13A2 and RAC1999 VLA20 do not appear highly resolved, while SVS13B is clearly resolved with respect
to the beam.}
\label{SVS13}
\end{figure}

\begin{figure}
\begin{center}
\includegraphics[scale=0.45]{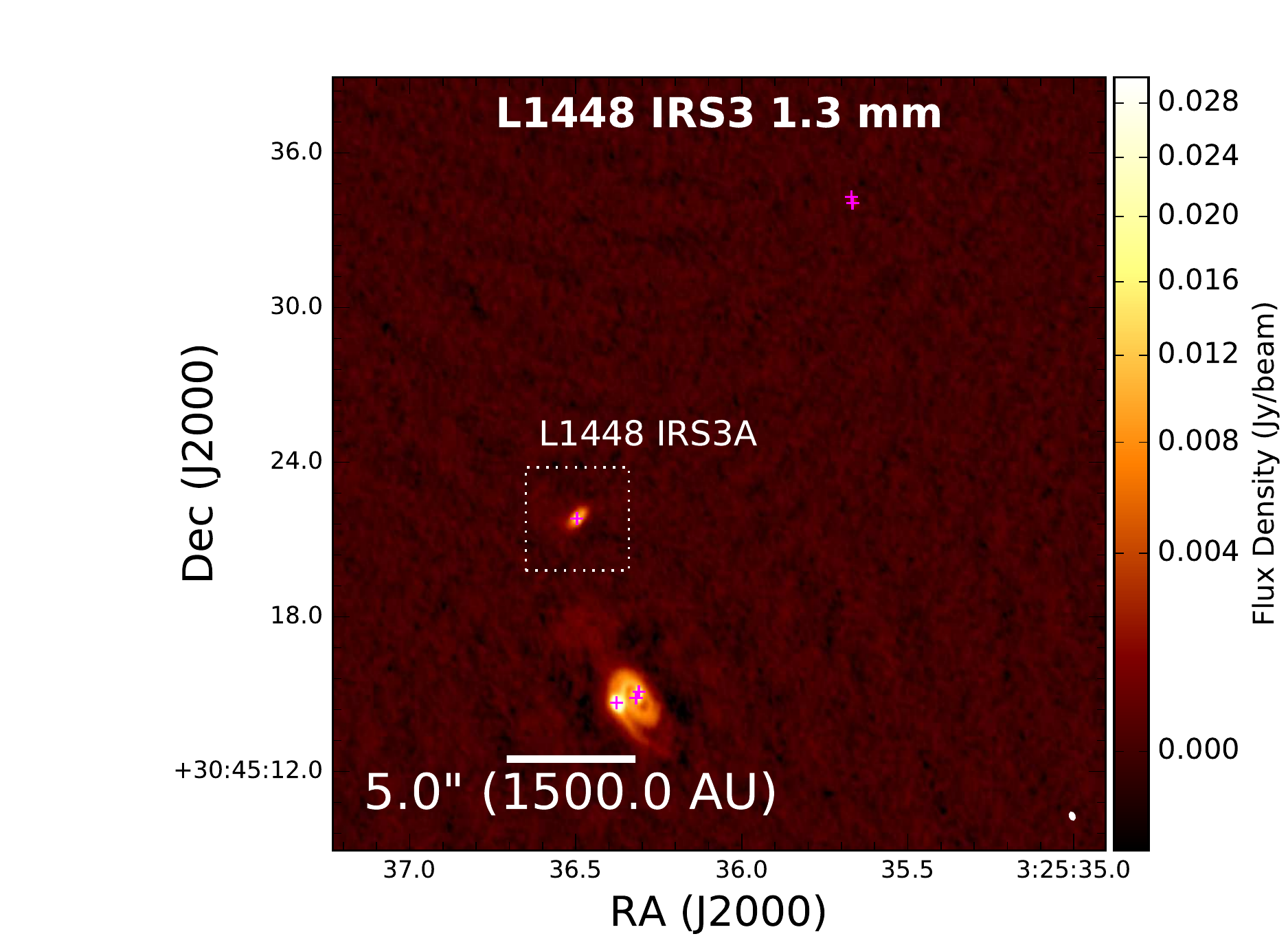}
\includegraphics[scale=0.45]{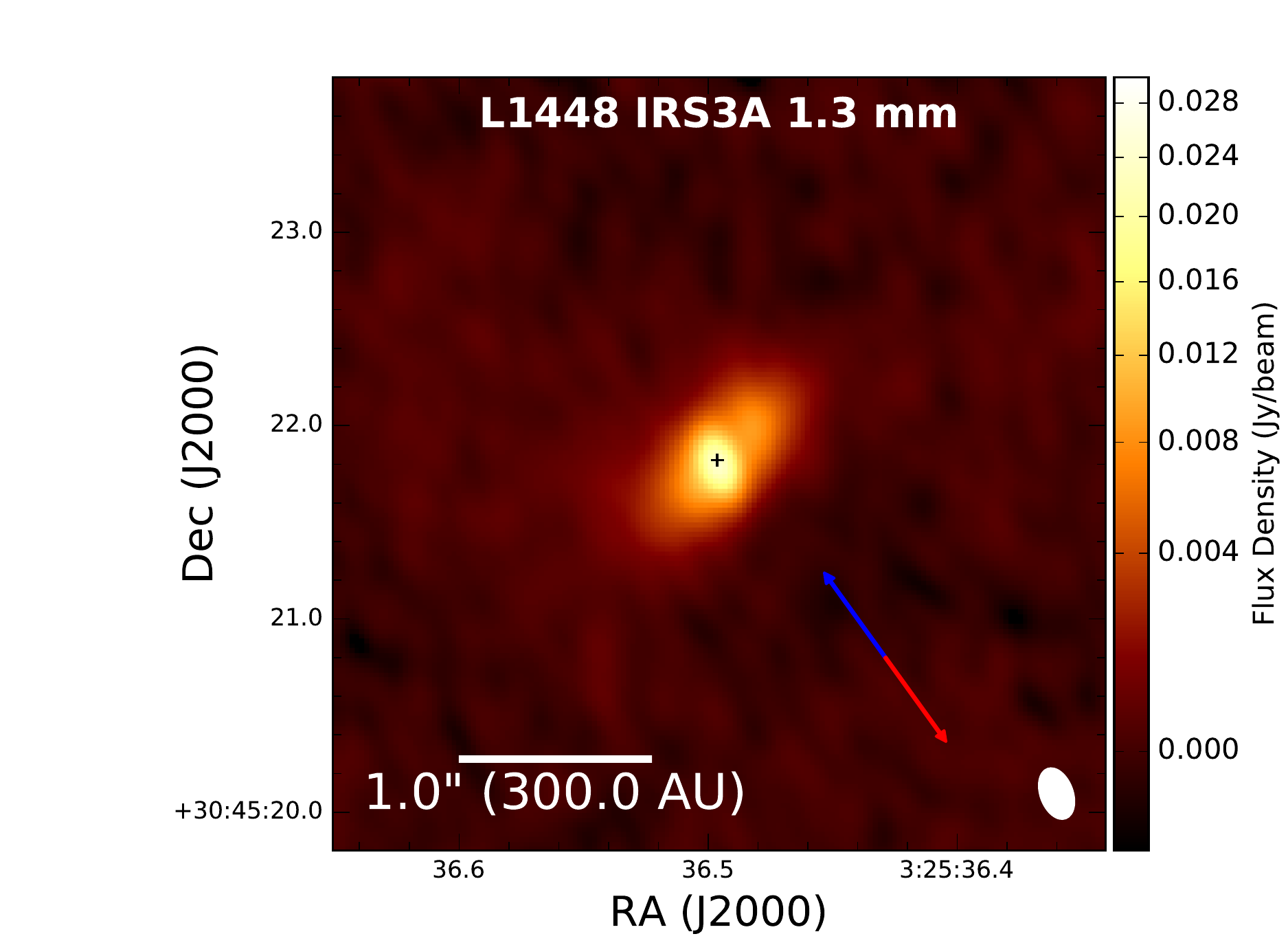}
\end{center}
\caption{ALMA 1.3~mm images of the L1448 IRS3 (Per-emb-33) region encompassing all
the sources within the primary beam; the dashed box marks
the regions zoomed-in on in the right panel. The observation was centered on L1448 IRS3B
largest and brightest source near the bottom of the image. Nonetheless, L1448 IRS3A is detected,
source in the middle-left, and L1448 IRS3C (L1448 NW) is marginally detected in this image toward
the upper right. L1448 IRS3C was also observed with a separate pointing given its large
angular distance from field center. Right: ALMA 1.3~mm zoom-in on L1448 
IRS3A showing the resolved disk structure toward
this protostar.}
\label{L1448IRS3}
\end{figure}

\begin{figure}
\begin{center}
\includegraphics[scale=0.45]{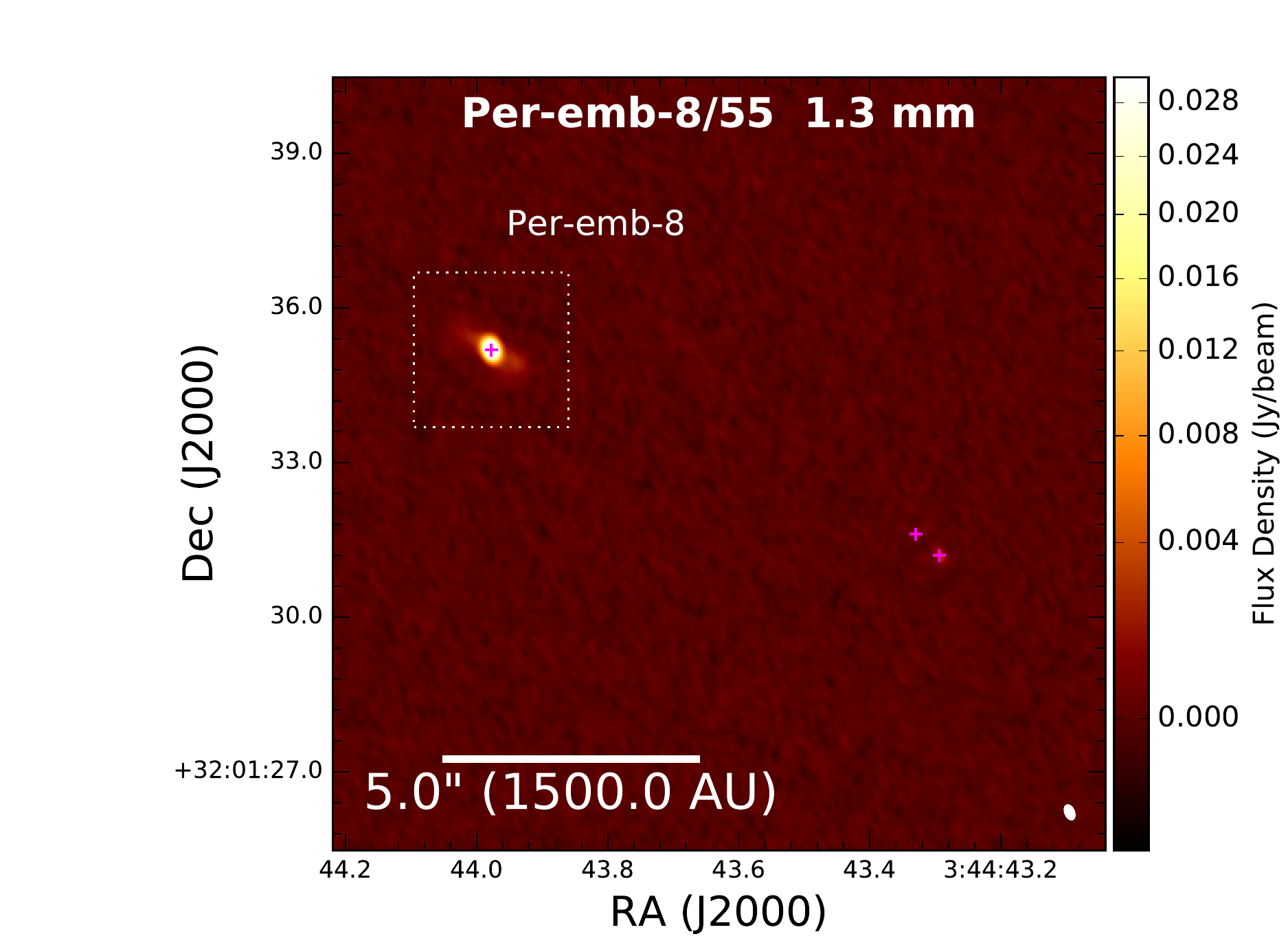}
\includegraphics[scale=0.45]{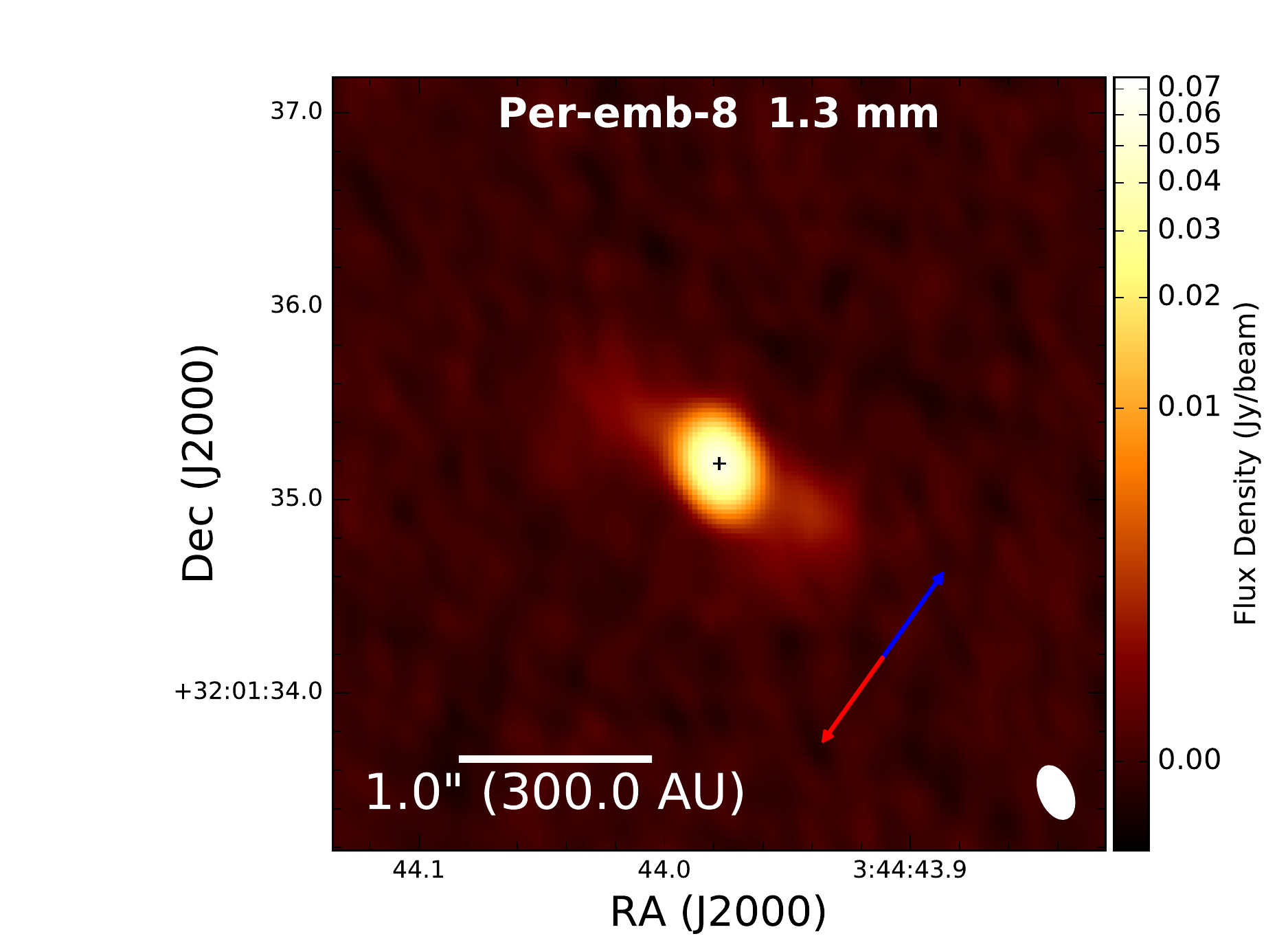}
\end{center}
\caption{ALMA 1.3~mm image of the Per-emb-8/Per-emb-55 region the two sources
within the primary beam; the dashed box marks
the regions zoomed-in on in the right panel. Per-emb-8 is located in the eastern portion of the image 
and Per-emb-55 is located in the western portion. Per-emb-8 is clearly
resolved and Per-emb-55 is a binary, but only one component is
strongly detected. Right: ALMA 1.3~mm image of Per-emb-8, an apparently single source with a surrounding disk.
The central region of Per-emb-8 is very prominent with respect to the extended disk.}
\label{per-emb-55}
\end{figure}

\begin{figure}
\begin{center}
\includegraphics[scale=0.45]{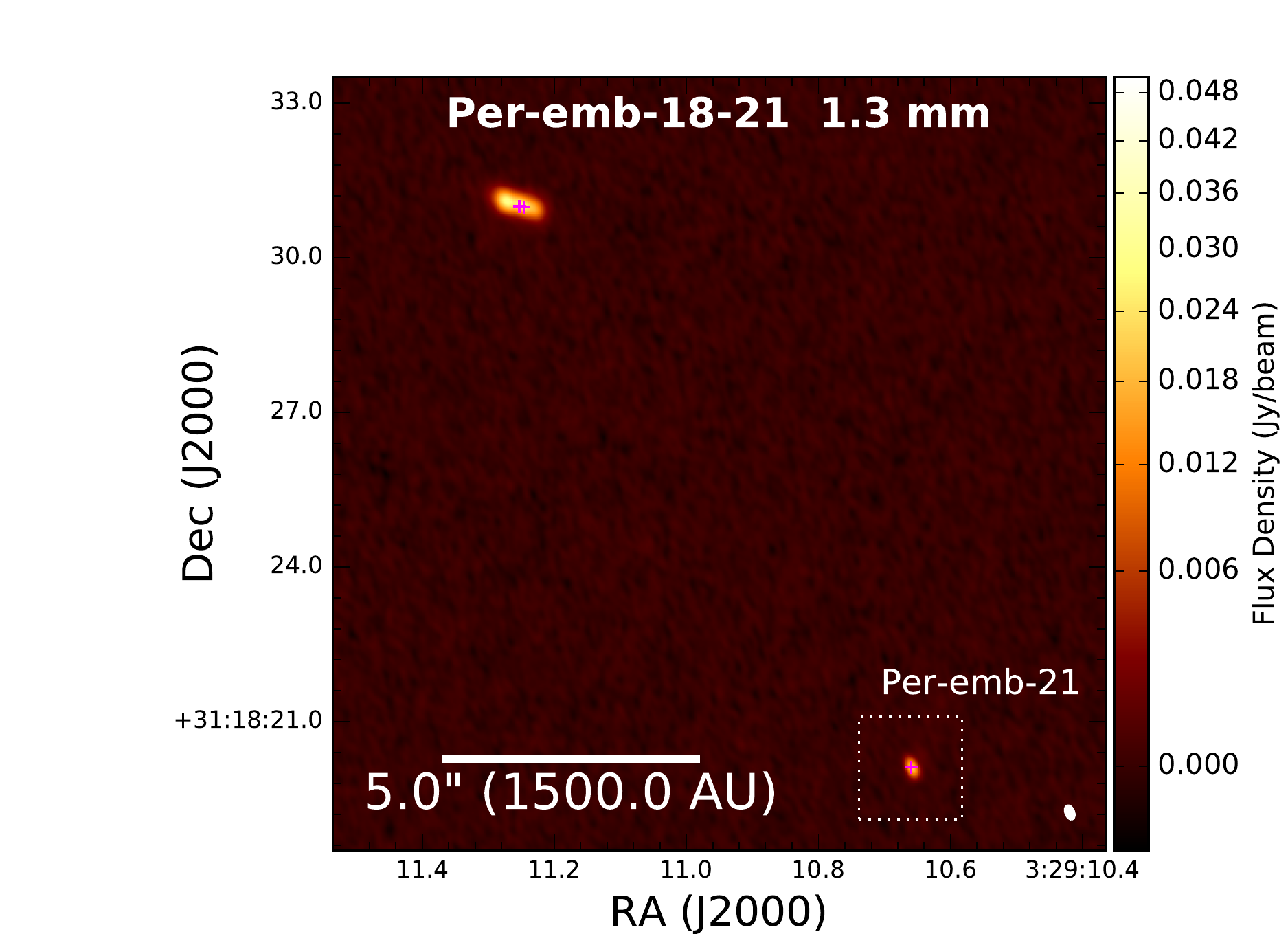}
\includegraphics[scale=0.45]{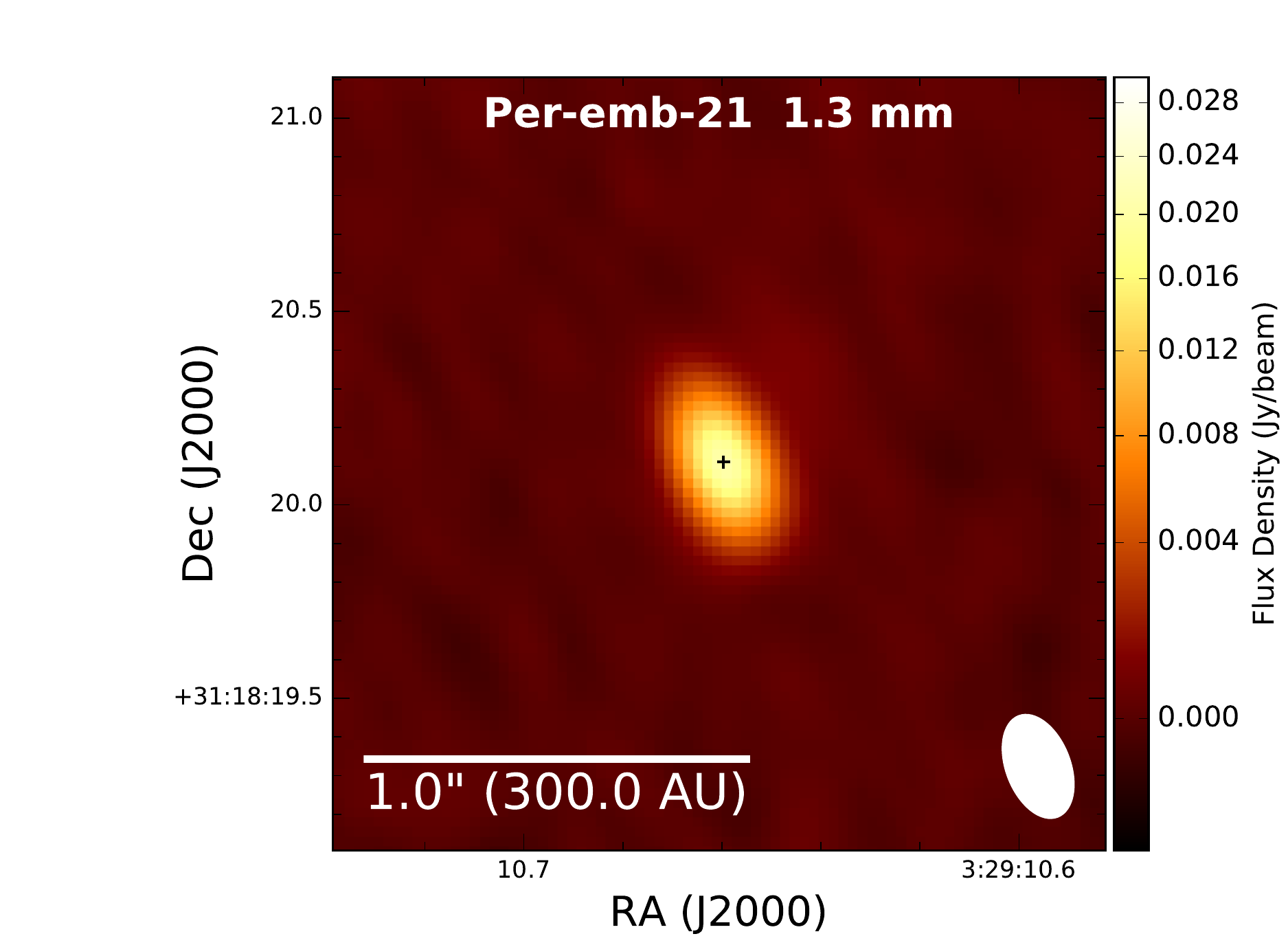}
\end{center}
\caption{ALMA 1.3~mm image of the Per-emb-18/Per-emb-21 region encompassing 
the two protostars within the primary beam; the dashed box marks
the regions zoomed-in on in the right panel. Per-emb-18 is located in the northern 
portion of the image and Per-emb-21 is located in the southern portion. 
Right: ALMA 1.3~mm image zooming-in on Per-emb-21, which appears unresolved.}
\label{per-emb-18}
\end{figure}

\begin{figure}
\begin{center}
\includegraphics[scale=0.425]{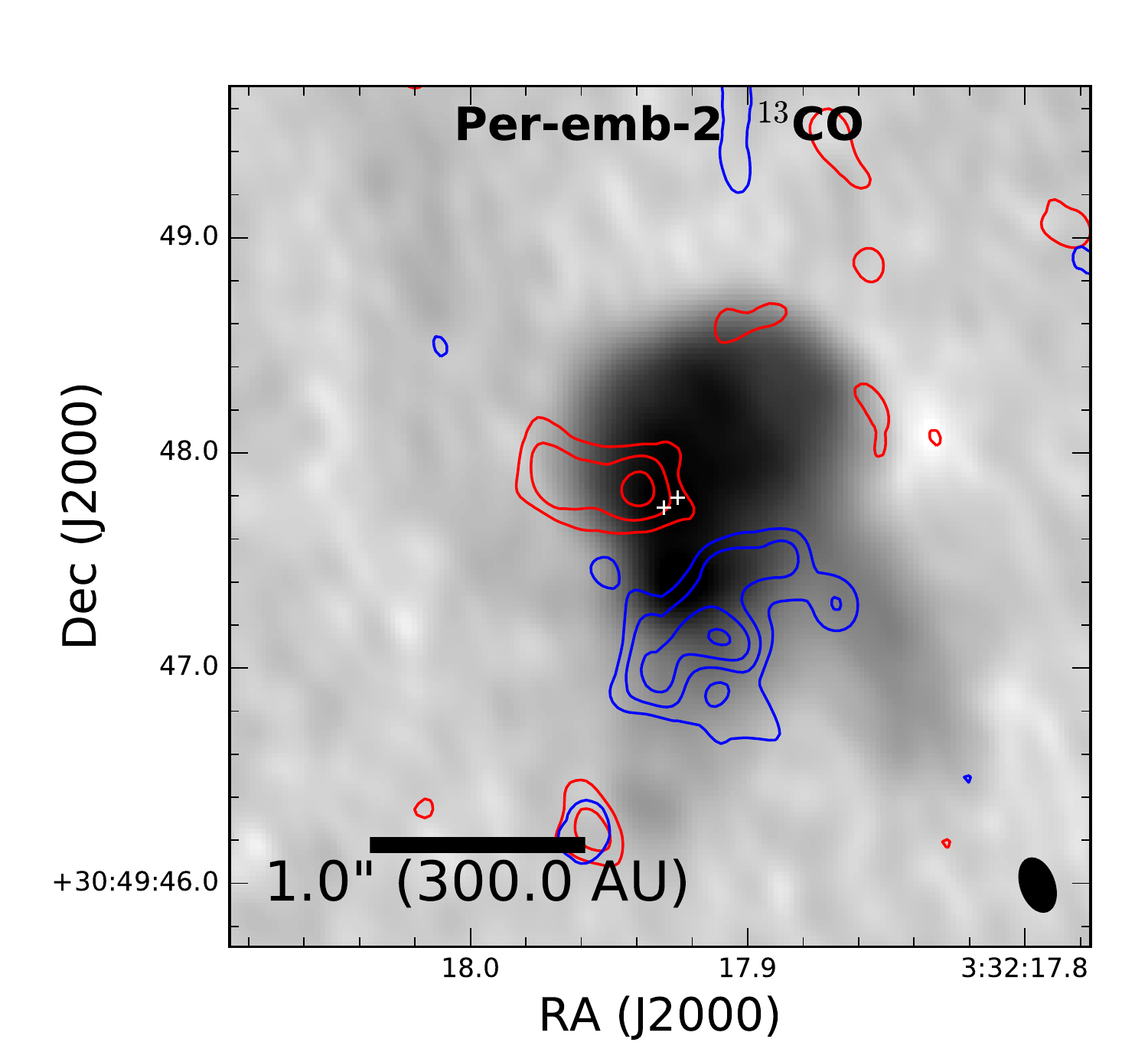}
\includegraphics[scale=0.425]{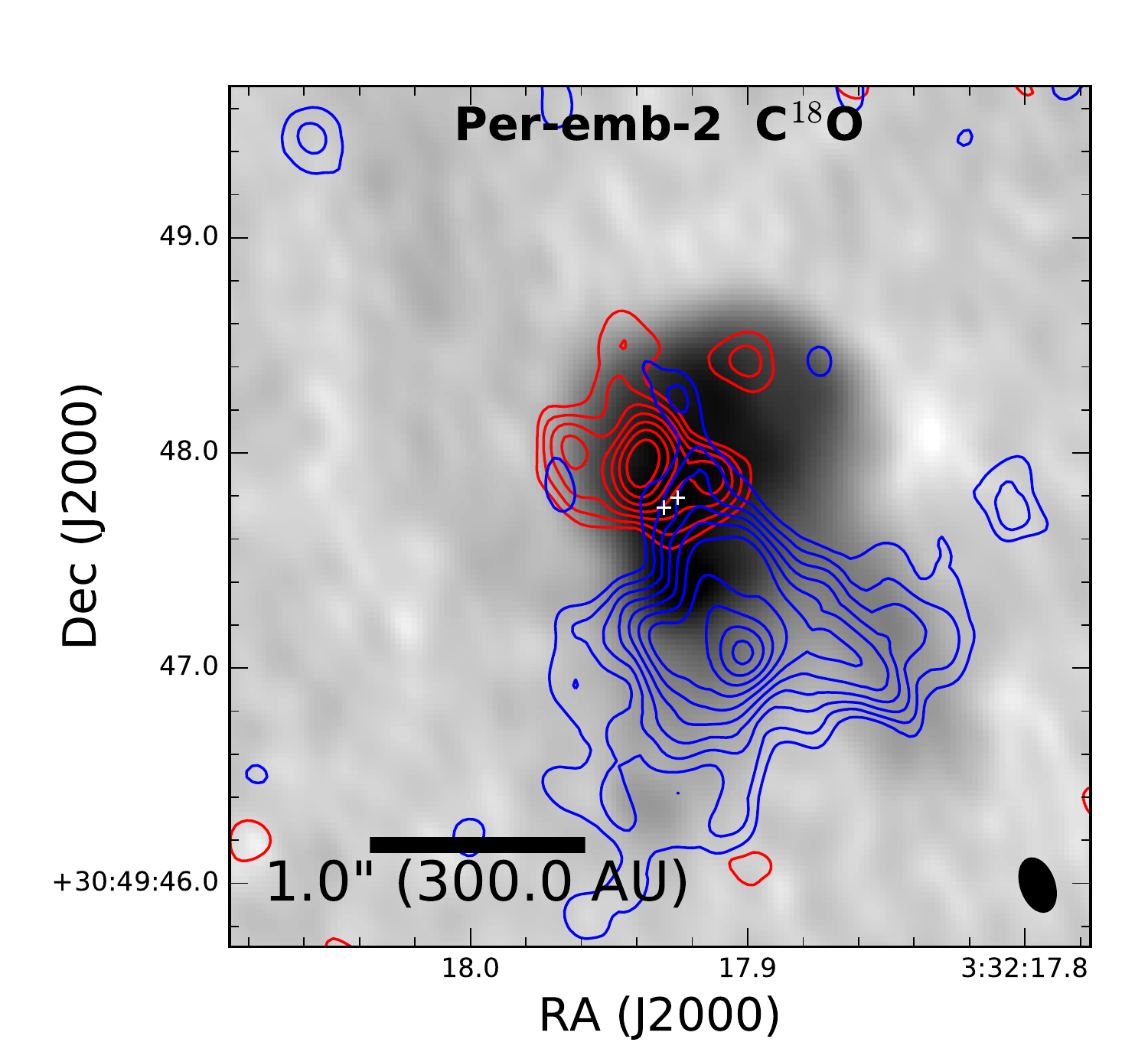}
\end{center}

\caption{Integrated intensity maps of \thco\ (left panel) and \cateo\ (right panel)
toward Per-emb-2. The integrated intensity maps are displayed as red and blue contours
corresponding to the integrated intensity of line emission red and blue-shifted with respect
to the system velocity. The contours are overlaid on the 1.3~mm continuum image. The
line emission shows evidence for a velocity gradient consistent with rotation.
The red-shifted contours start at (4,3)$\sigma$ and increase in (1,1)$\sigma$ increments, and
the blue-shifted contours start at (3,3)$\sigma$ and increase in (1,1)$\sigma$ increments. 
The values inside the parentheses in the previous sentence correspond to the \thco, \cateo, SO, and H$_2$CO integrated intensity maps, respectively. The
values for $\sigma_{red}$ and $\sigma_{blue}$ and velocity ranges over which the line
emission was summed can be found in Table 3.
The beam in the images is approximately 0\farcs36$\times$0\farcs26.
} 
\label{per-emb-2-lines}
\end{figure}

\clearpage

\begin{figure}
\begin{center}
\includegraphics[scale=0.425]{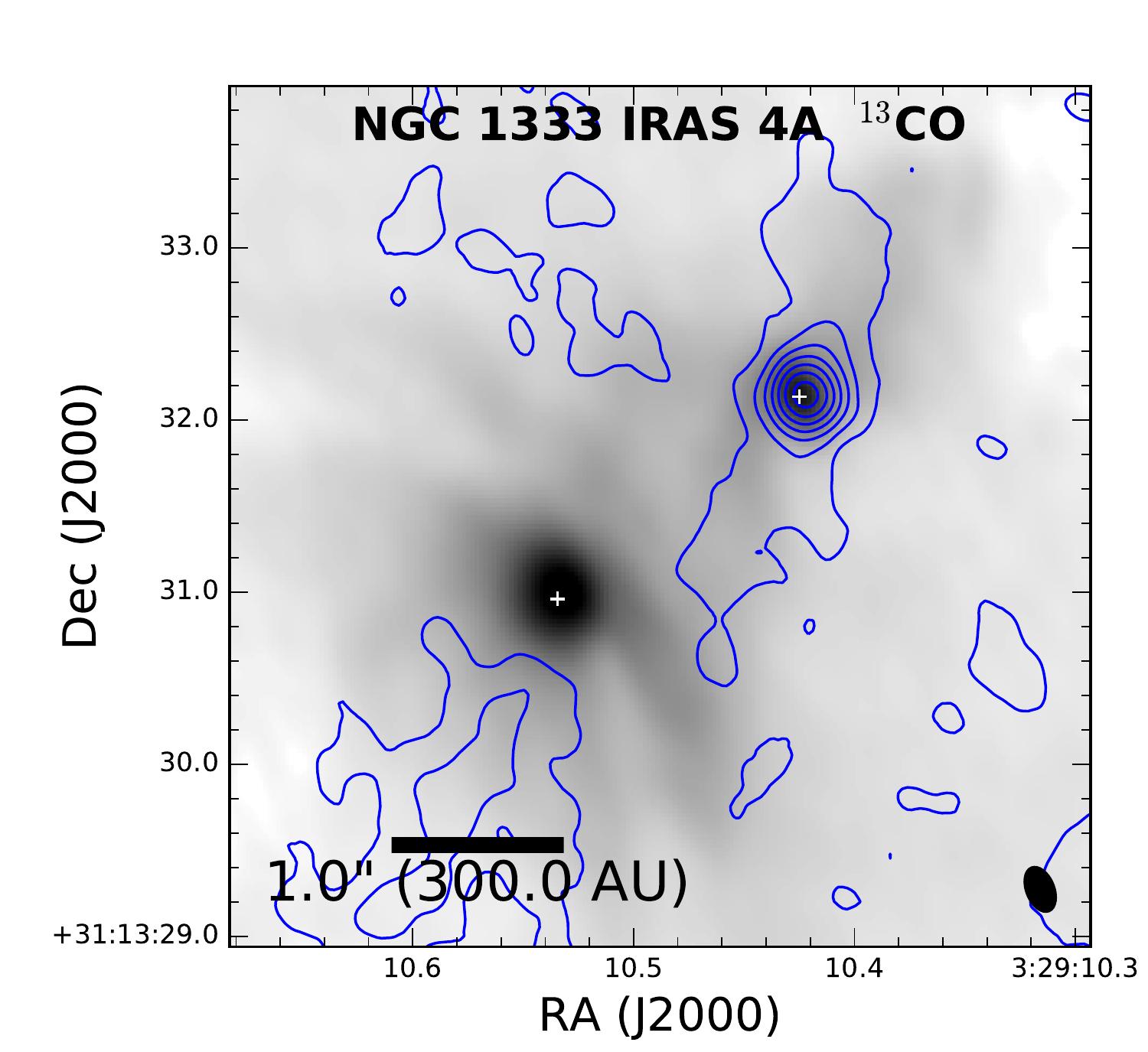}
\includegraphics[scale=0.425]{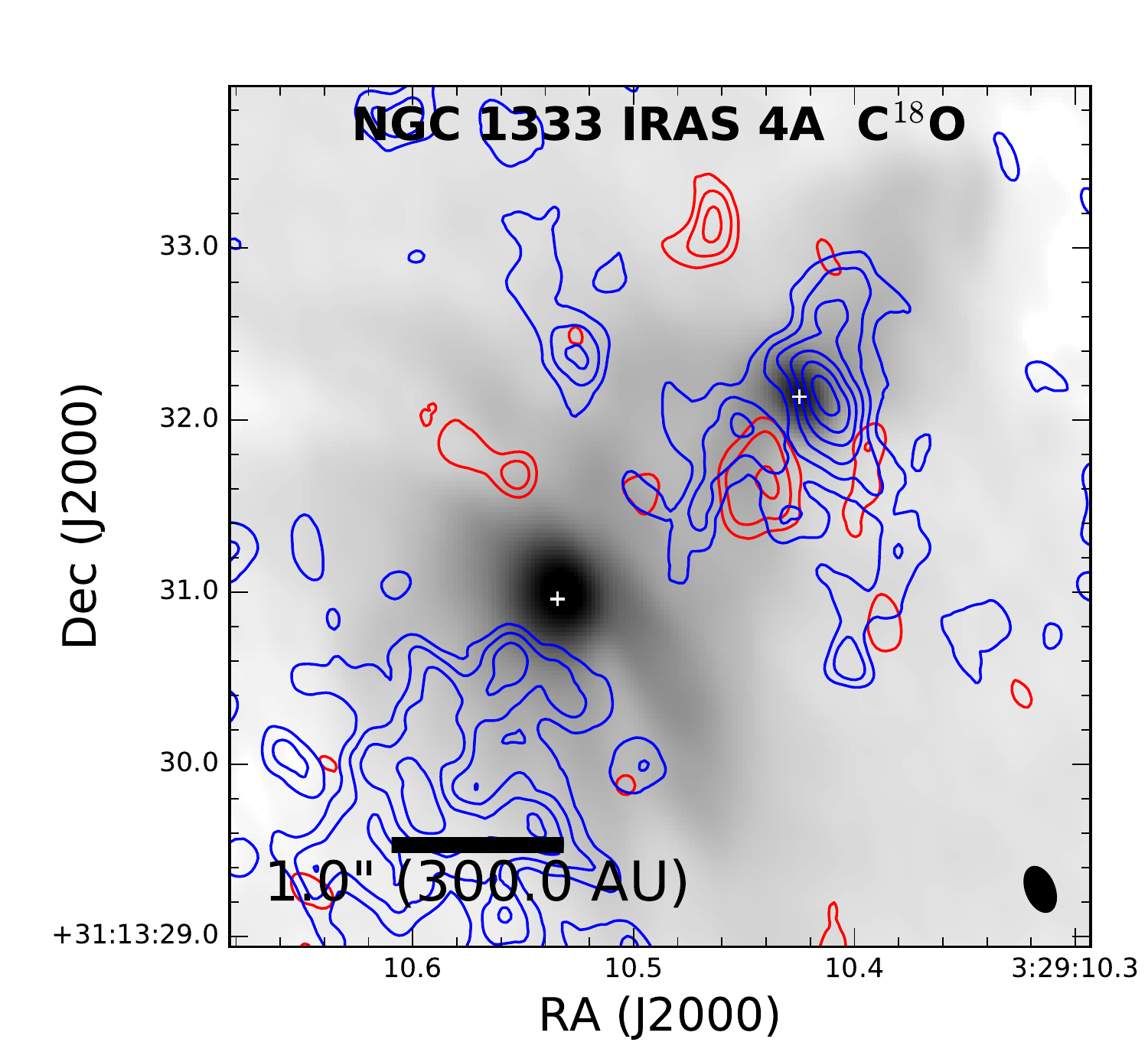}
\includegraphics[scale=0.425]{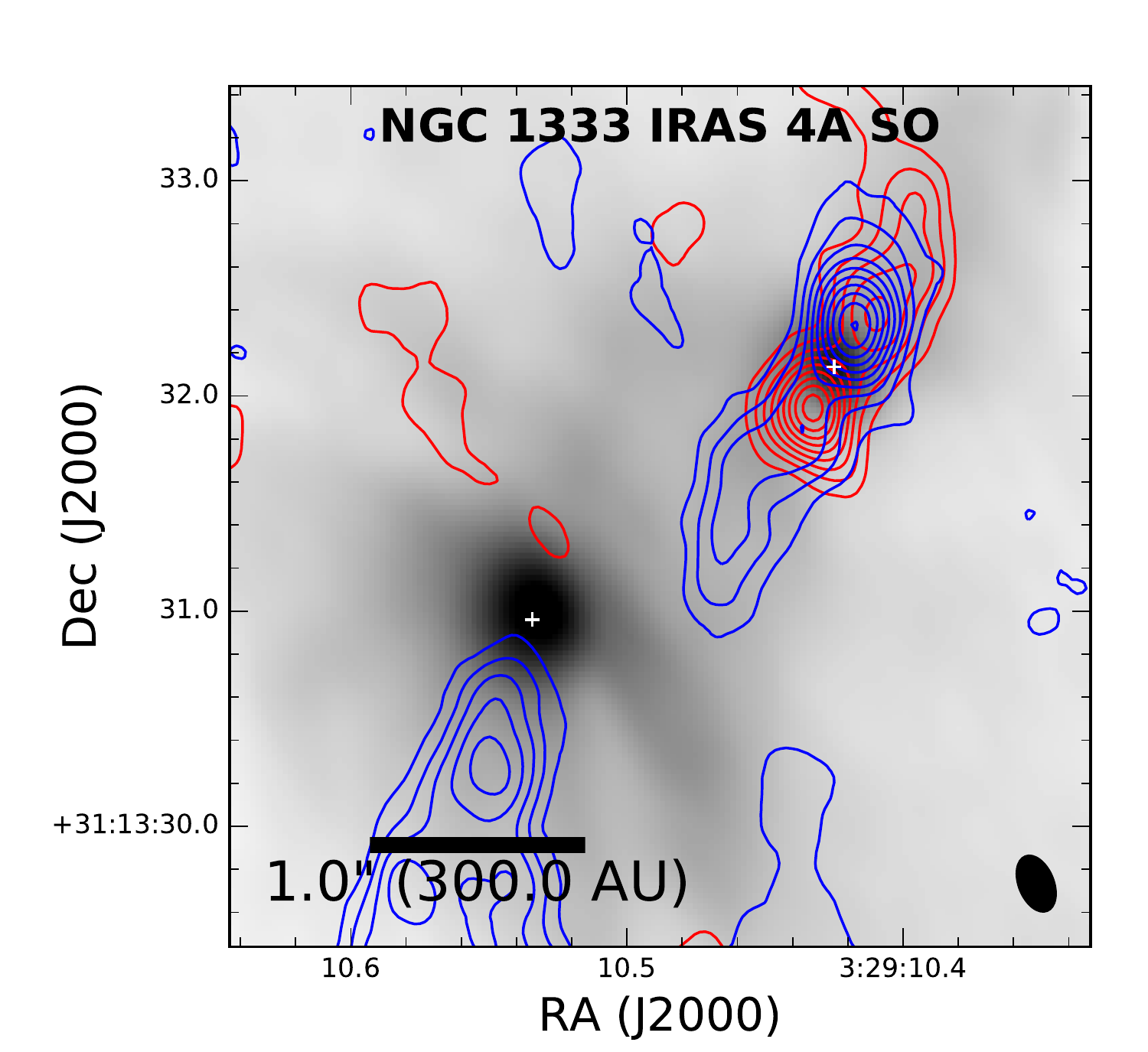}
\includegraphics[scale=0.425]{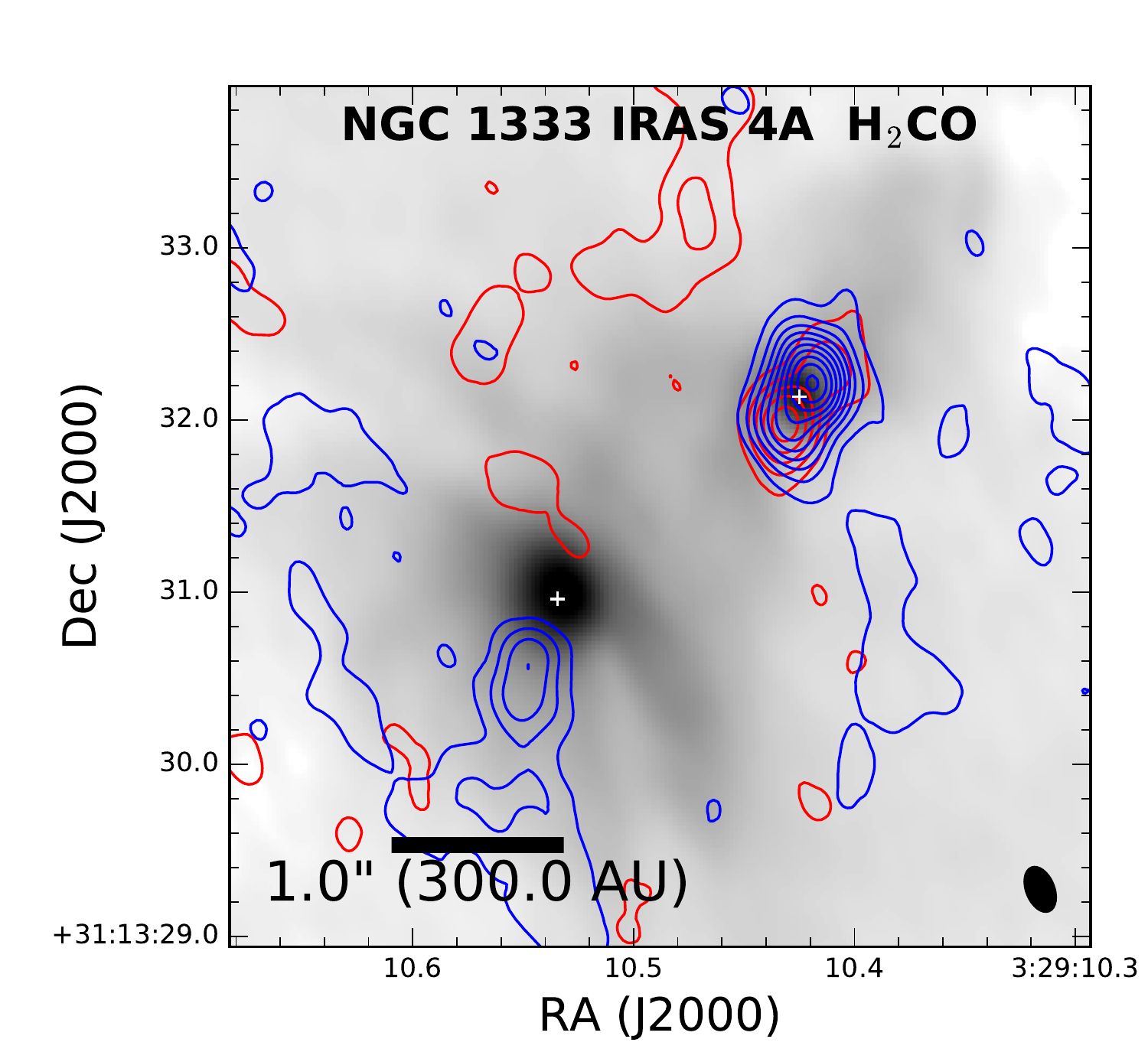}
\end{center}

\caption{Integrated intensity maps of \thco\ (top left panel), \cateo\ (top right panel), SO (bottom left panel), and H$_2$CO (bottom right panel)
toward Per-emb-12 (NGC 1333 IRAS4A). The integrated intensity maps are displayed as red and blue contours
corresponding to the integrated intensity of line emission red and blue-shifted with respect
to the system velocity. The contours are overlaid on the 1.3~mm continuum image. The
line emission shows evidence for a velocity gradient consistent with rotation.
The red-shifted contours start at (3,3,4,3)$\sigma$ and increase in (3,1,2,3)$\sigma$ increments, and
the blue-shifted contours start at (3,3,4,3)$\sigma$ and increase in (3,1,2,3)$\sigma$ increments.
The values inside the parentheses in the previous sentence
correspond to the \thco, \cateo, SO, and H$_2$CO integrated intensity maps, respectively. The
values for $\sigma_{red}$ and $\sigma_{blue}$ and velocity ranges over which the line
emission was summed can be found in Table 3.
The beam in the images is approximately 0\farcs36$\times$0\farcs26.}
\label{per-emb-12-lines}
\end{figure}

\clearpage
\begin{figure}
\begin{center}
\includegraphics[scale=0.425]{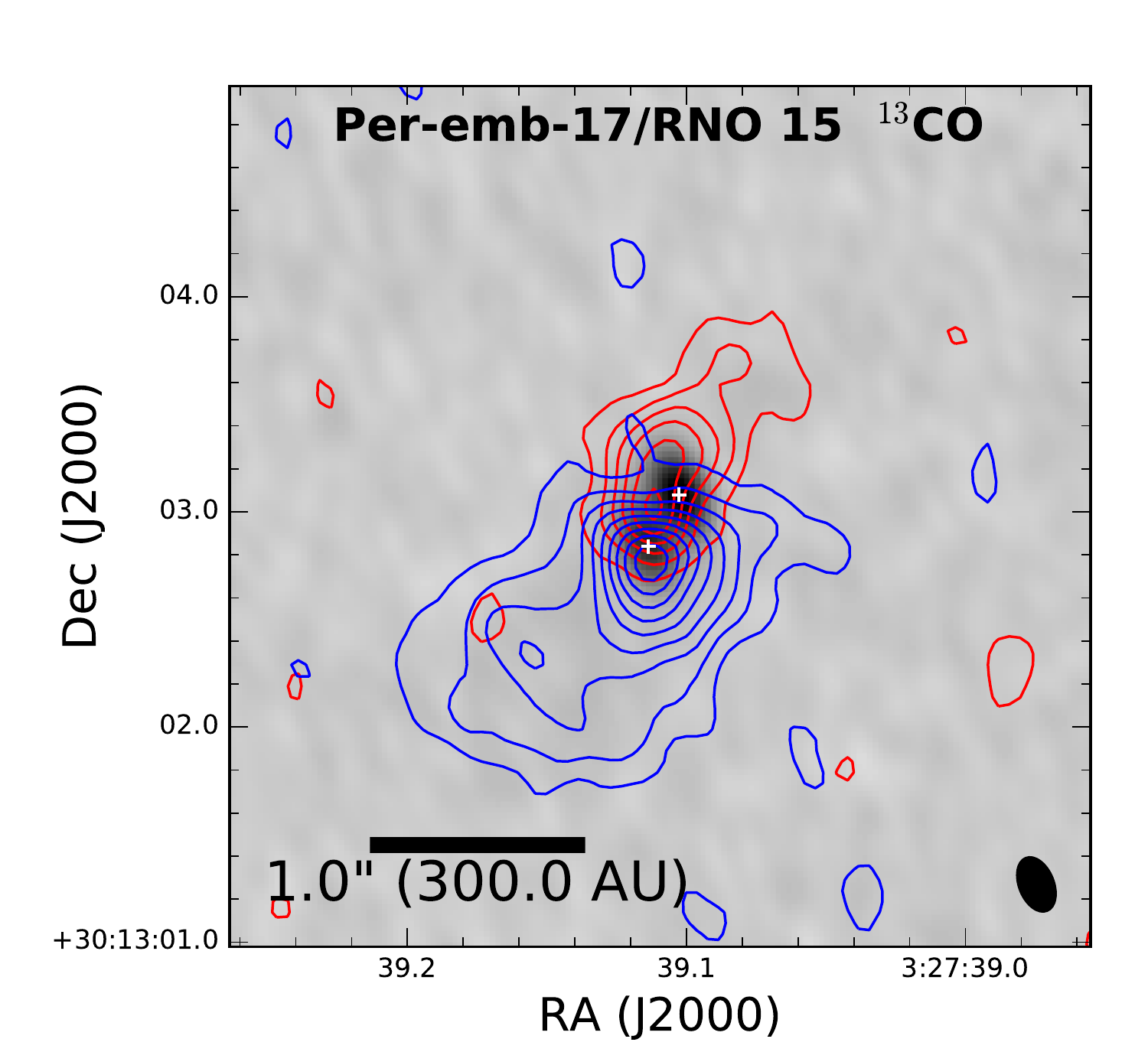}
\includegraphics[scale=0.425]{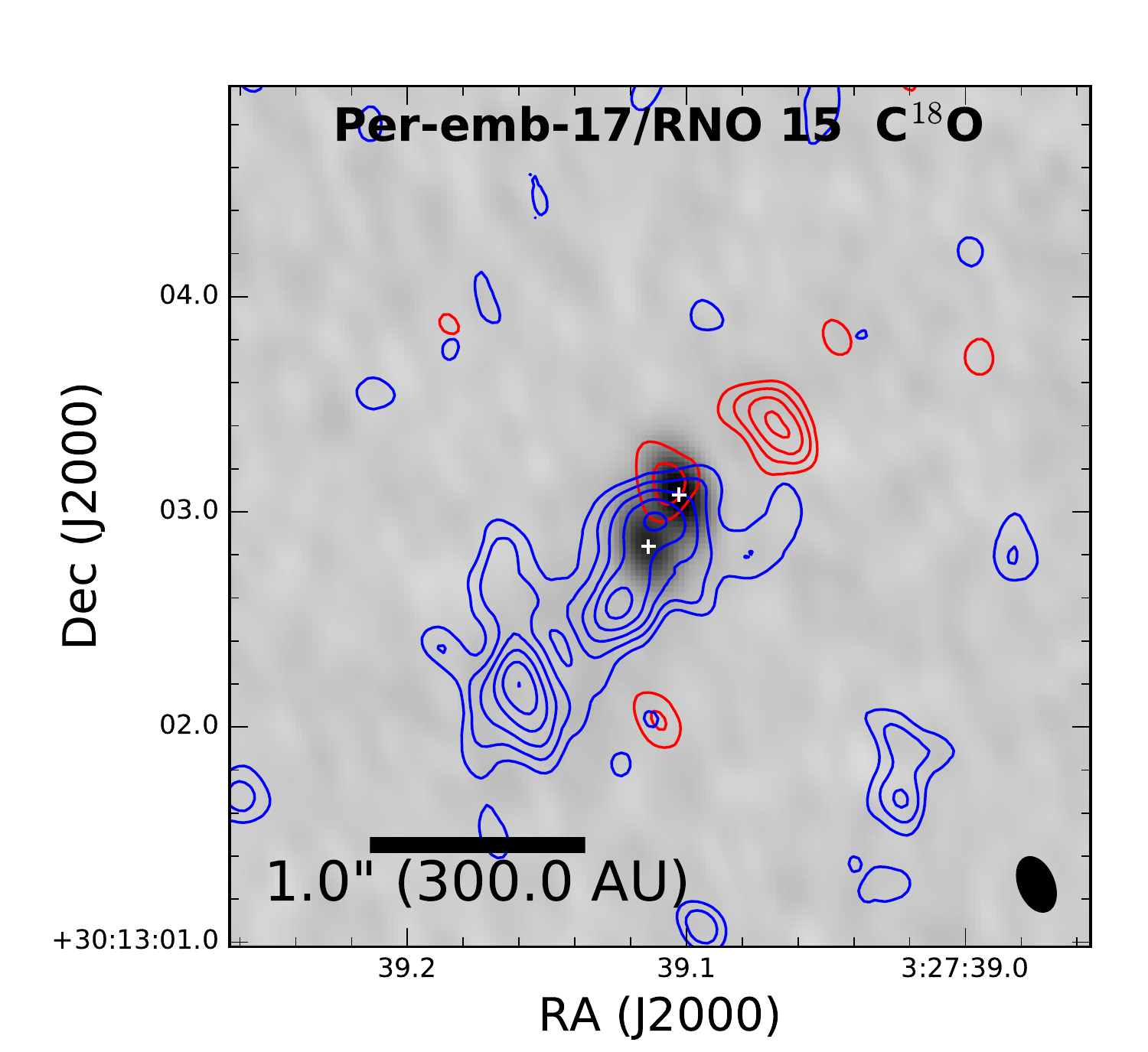}
\includegraphics[scale=0.425]{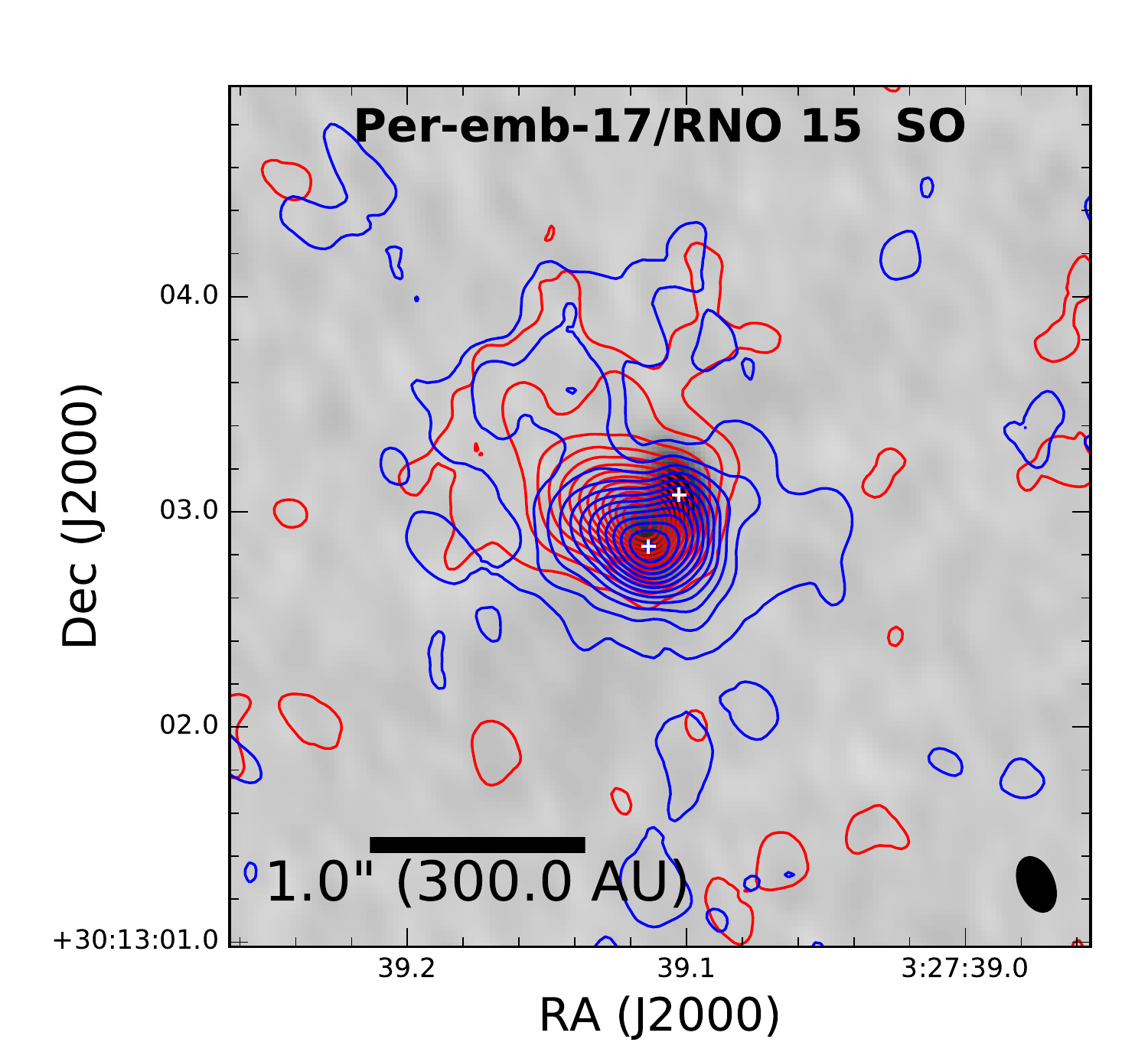}
\includegraphics[scale=0.425]{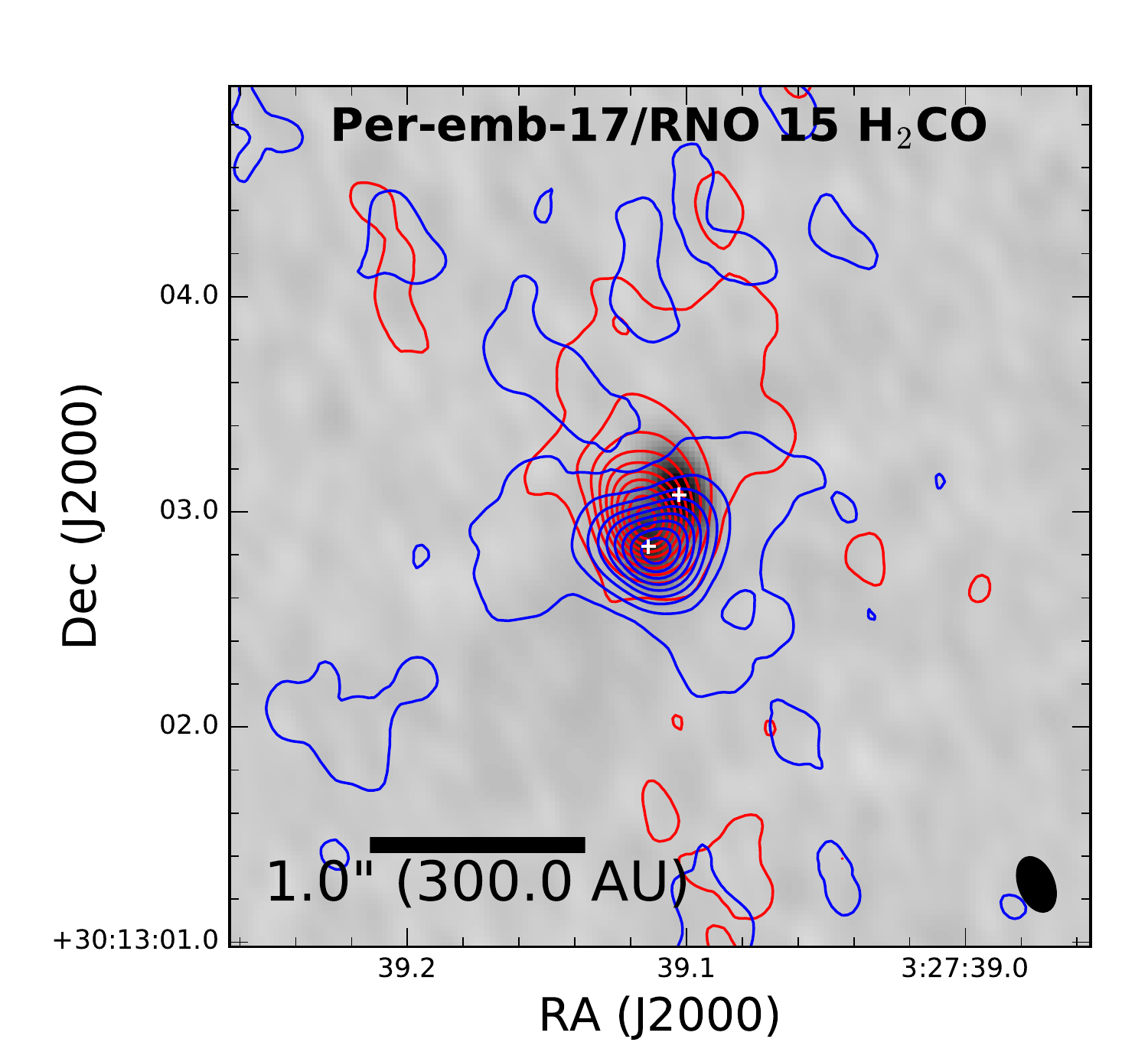}
\end{center}

\caption{Integrated intensity maps of \thco\ (top left panel), \cateo\ (top right panel), SO (bottom left panel), and H$_2$CO (bottom right panel)
toward Per-emb-17. The integrated intensity maps are displayed as red and blue contours
corresponding to the integrated intensity of line emission red and blue-shifted with respect
to the system velocity. The contours are overlaid on the 1.3~mm continuum image. The
line emission shows evidence for a velocity gradient consistent with rotation.
The red-shifted contours start at (3,4,3,3)$\sigma$ and increase in (2,1,2,3)$\sigma$ increments, and
the blue-shifted contours start at (3,3,3,3)$\sigma$ and increase in (2,1,2,3)$\sigma$ increments.
The values inside the parentheses in the previous sentence
correspond to the \thco, \cateo, SO, and H$_2$CO integrated intensity maps, respectively. The
values for $\sigma_{red}$ and $\sigma_{blue}$ and velocity ranges over which the line
emission was summed can be found in Table 3.
The beam in the images is approximately 0\farcs36$\times$0\farcs26.}
\label{per-emb-17-lines}
\end{figure}

\begin{figure}
\begin{center}
\includegraphics[scale=0.425]{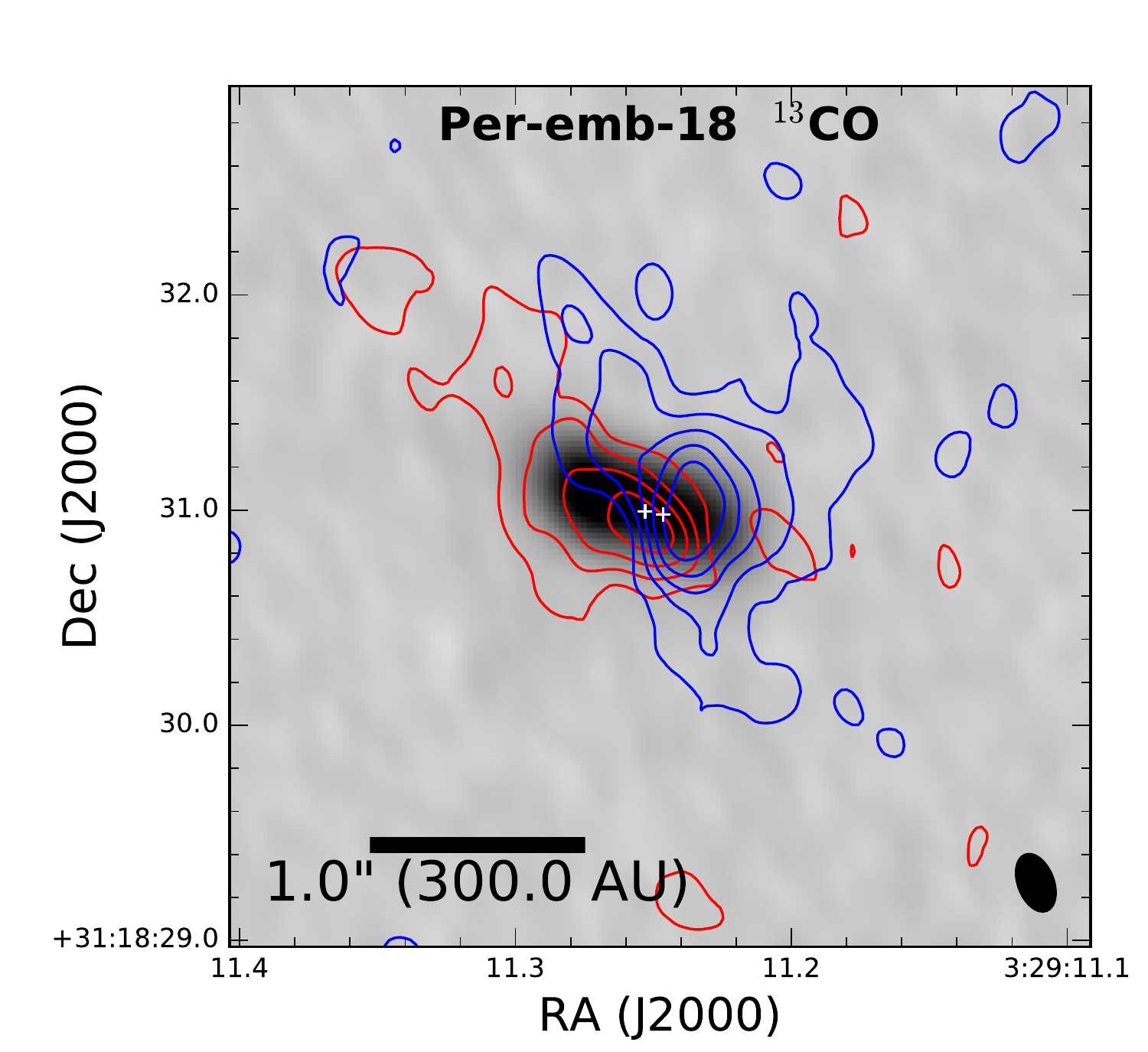}
\includegraphics[scale=0.425]{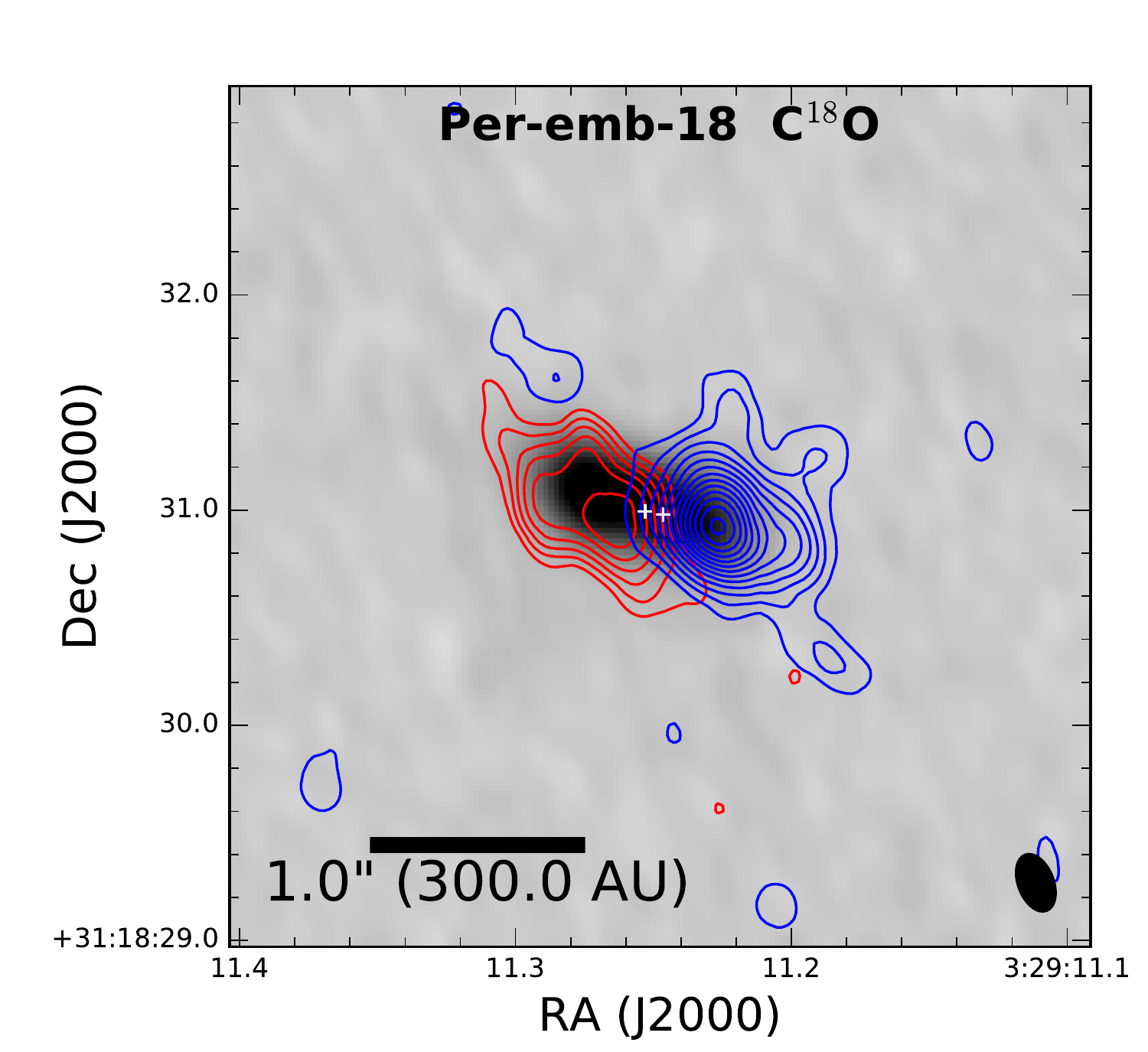}
\includegraphics[scale=0.425]{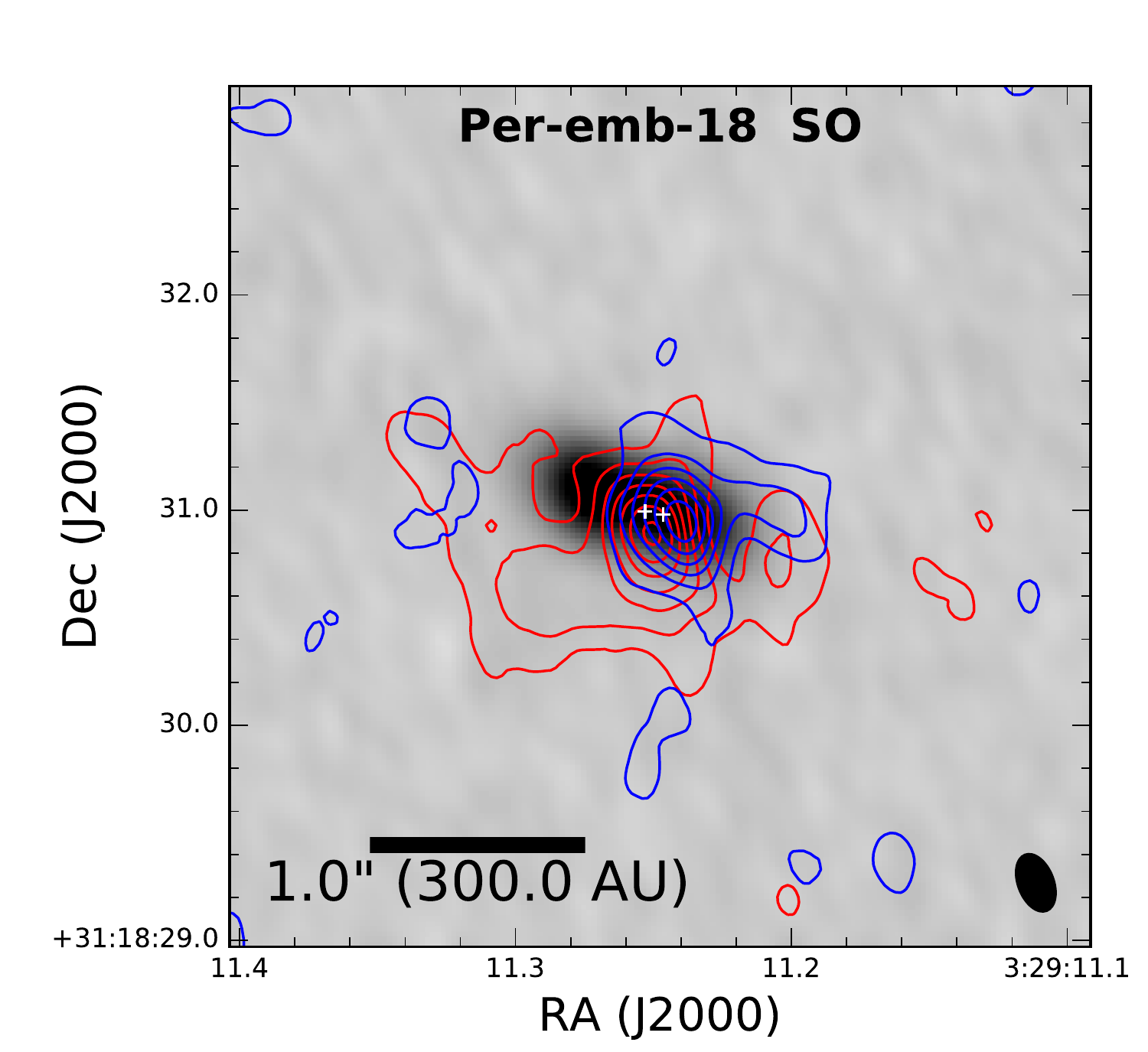}
\includegraphics[scale=0.425]{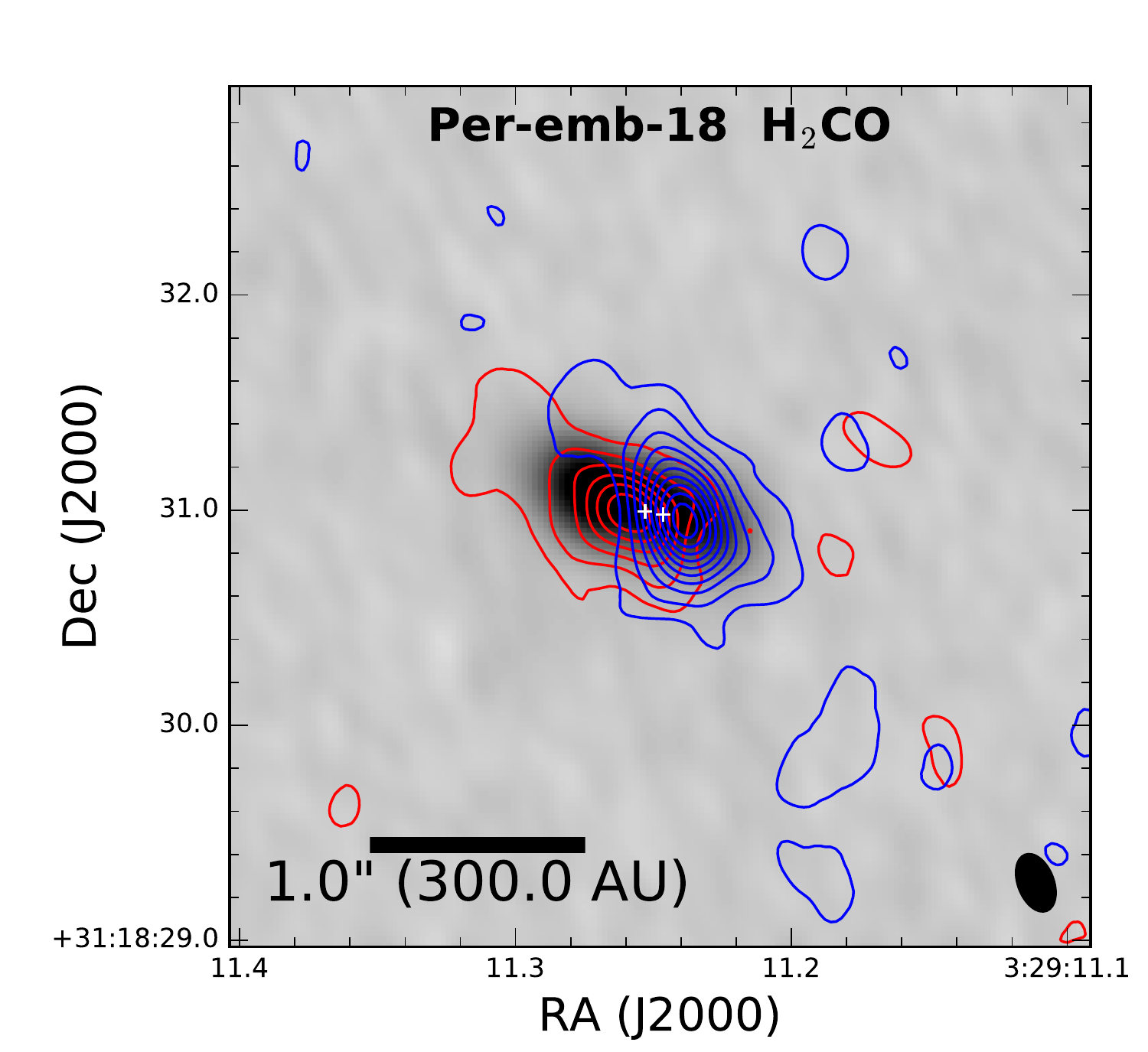}
\end{center}

\caption{Integrated intensity maps of \thco\ (top left panel), \cateo\ (top right panel), SO (bottom left panel), and H$_2$CO (bottom right panel)
toward Per-emb-18. The integrated intensity maps are displayed as red and blue contours
corresponding to the integrated intensity of line emission red and blue-shifted with respect
to the system velocity. The contours are overlaid on the 1.3~mm continuum image. The
line emission shows evidence for a velocity gradient consistent with rotation.
The red-shifted contours start at (3,3,3,3)$\sigma$ and increase in (2,1,2,2)$\sigma$ increments, and
the blue-shifted contours start at (3,4,3,3)$\sigma$ and increase in (2,1,2,2)$\sigma$ increments.
The values inside the parentheses in the previous sentence
 correspond to the \thco, \cateo, SO, and H$_2$CO integrated intensity maps, respectively. The
values for $\sigma_{red}$ and $\sigma_{blue}$ and velocity ranges over which the line
emission was summed can be found in Table 3.
The beam in the images is approximately 0\farcs36$\times$0\farcs26.}
\label{per-emb-18-lines}
\end{figure}

\clearpage

\begin{figure}
\begin{center}
\includegraphics[scale=0.425]{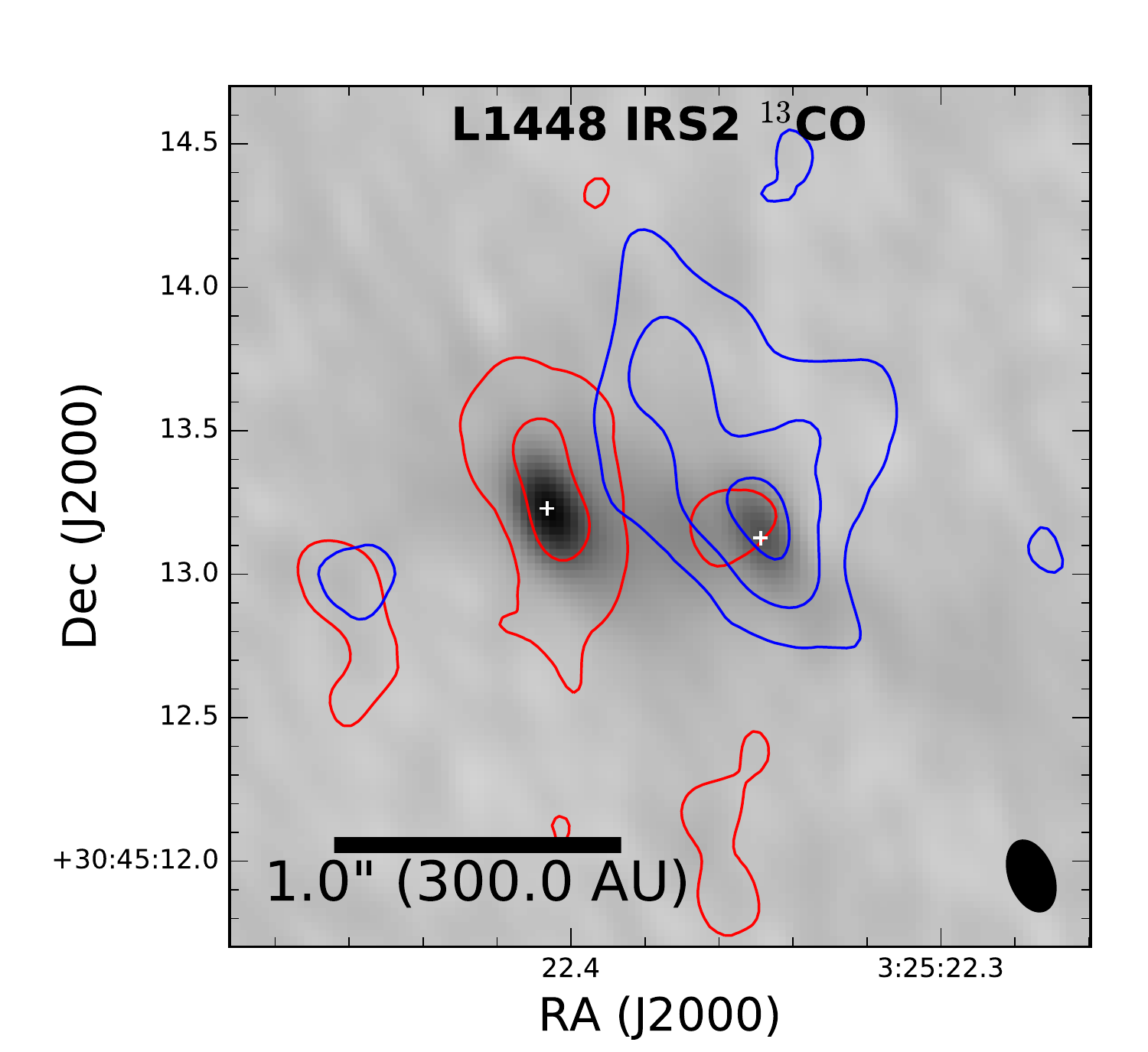}
\includegraphics[scale=0.425]{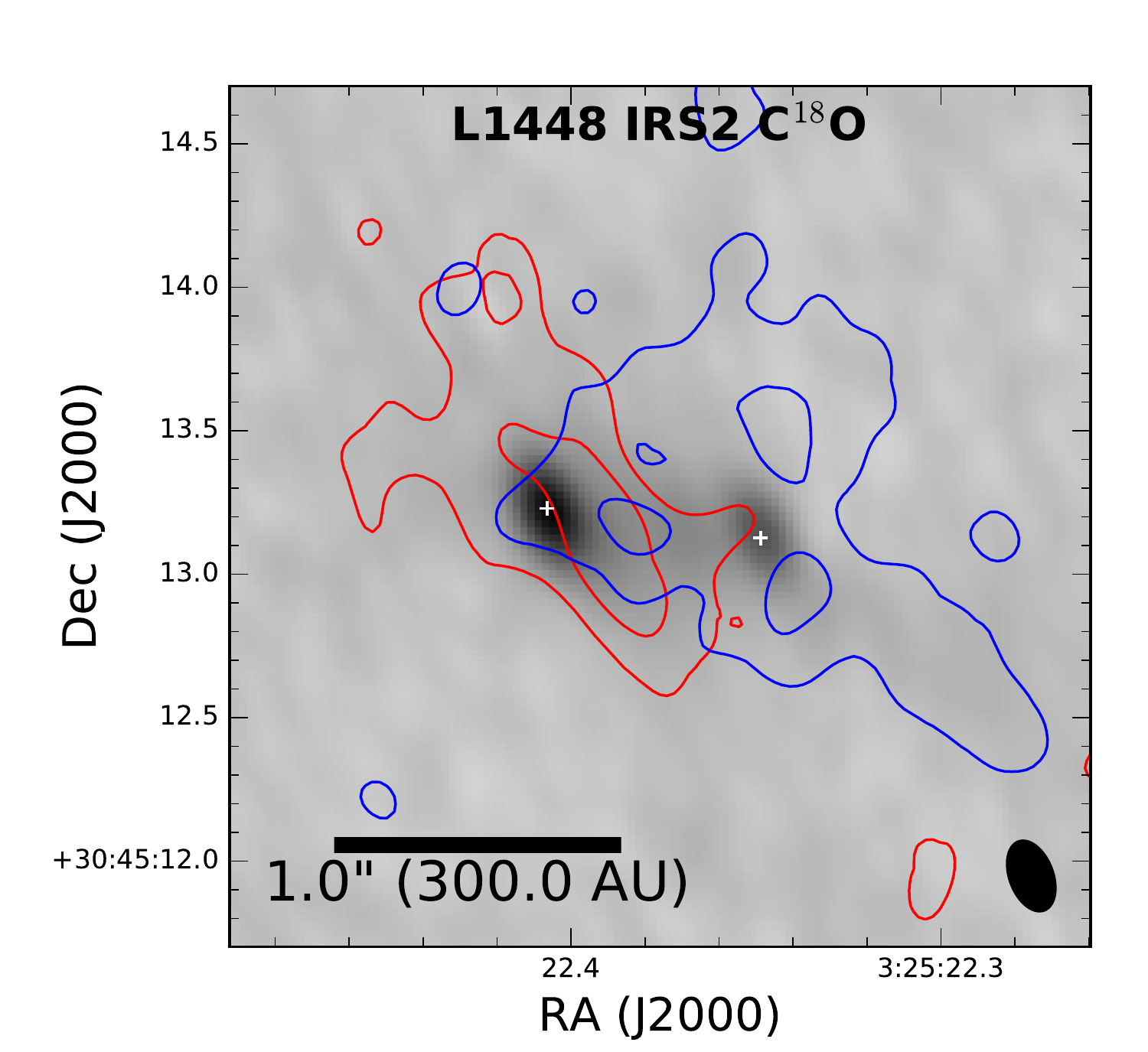}
\includegraphics[scale=0.425]{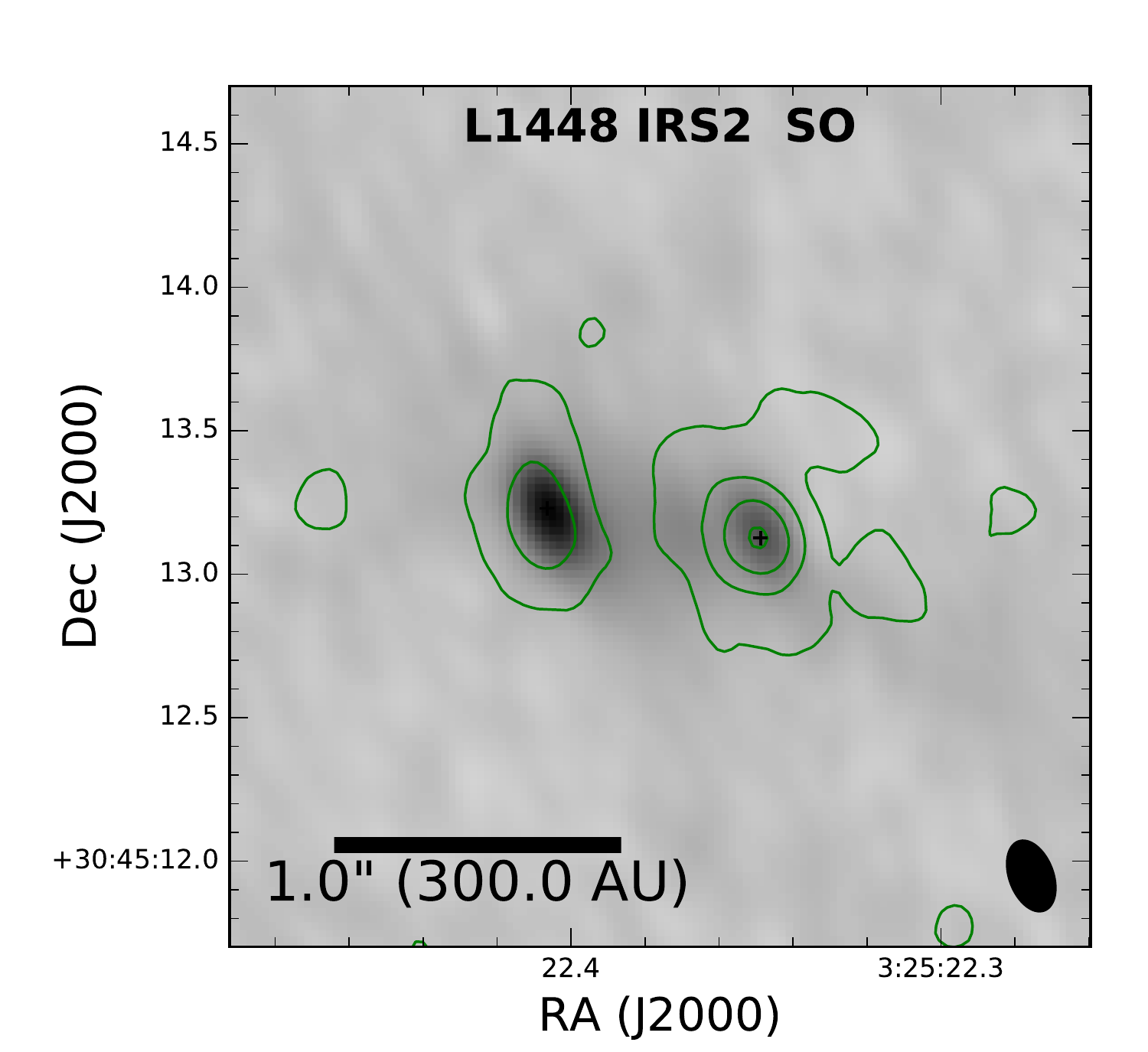}
\includegraphics[scale=0.425]{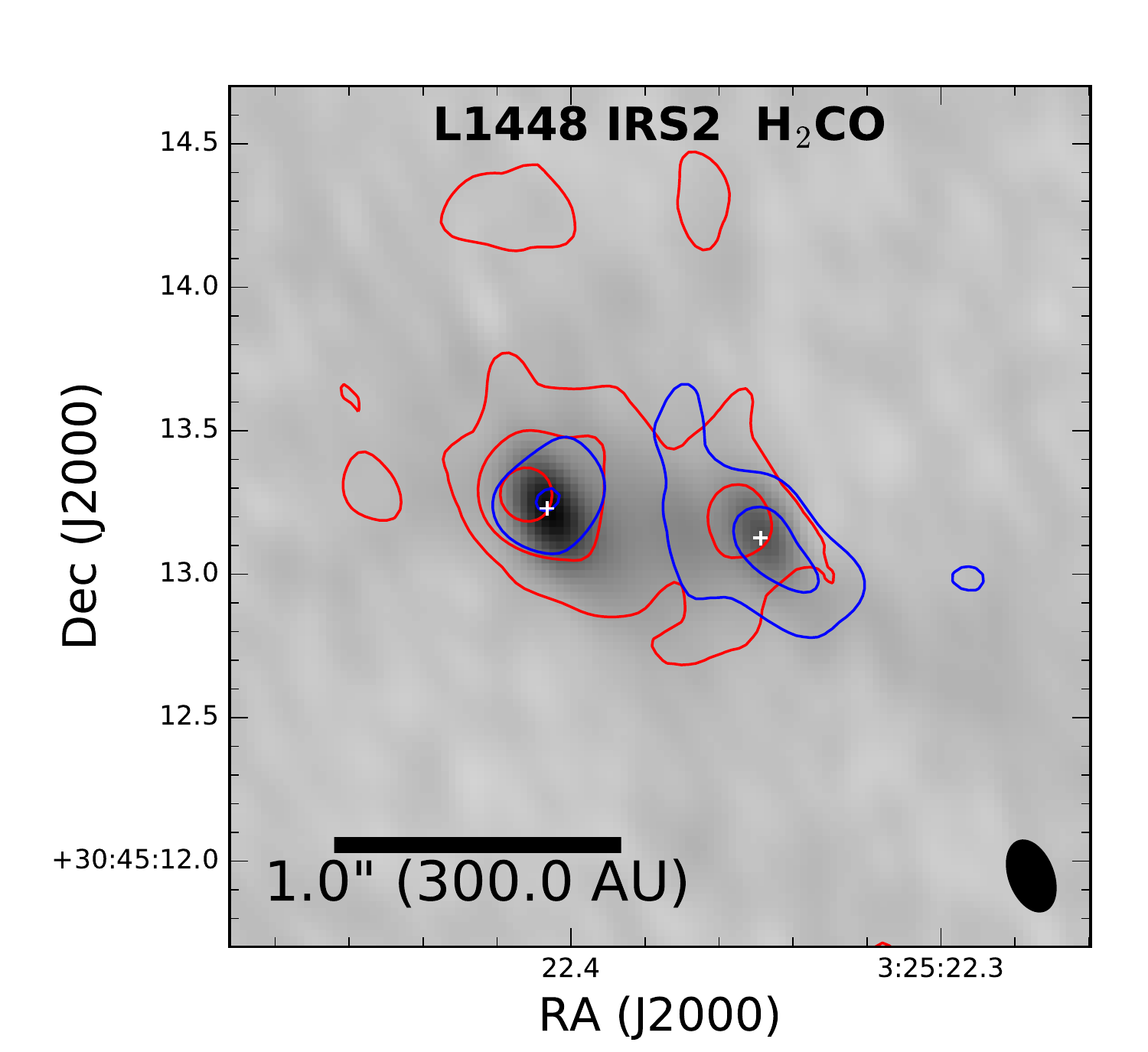}
\end{center}
\caption{Integrated intensity maps of \thco\ (top left panel), \cateo\ (top right panel), SO (bottom left panel), and H$_2$CO (bottom right panel)
toward Per-emb-22 (L1448 IRS2). The integrated intensity maps are displayed as red and blue contours
corresponding to the integrated intensity of line emission red and blue-shifted with respect
to the system velocity. The green contours for the SO emission correspond to emission integrated across the 
entire line because there was not a clear velocity gradient in this molecule due to low S/N.
The contours are overlaid on the 1.3~mm continuum image. The
line emission shows evidence for a velocity gradient consistent with rotation.
The red-shifted contours start at (3,3,3,4)$\sigma$ and increase in (3,3,3,2)$\sigma$ increments, and
the blue-shifted contours start at (3,3,3,4)$\sigma$ and increase in (3,3,3,2)$\sigma$ increments.
The values inside the parentheses in the previous sentence correspond to the \thco, \cateo, SO, and H$_2$CO integrated intensity maps, respectively. The
values for $\sigma_{red}$ and $\sigma_{blue}$ and velocity ranges over which the line
emission was summed can be found in Table 3.
The beam in the images is approximately 0\farcs36$\times$0\farcs26.}
\label{L1448IRS2-lines}
\end{figure}

\clearpage
\
\begin{figure}
\begin{center}
\includegraphics[scale=0.425]{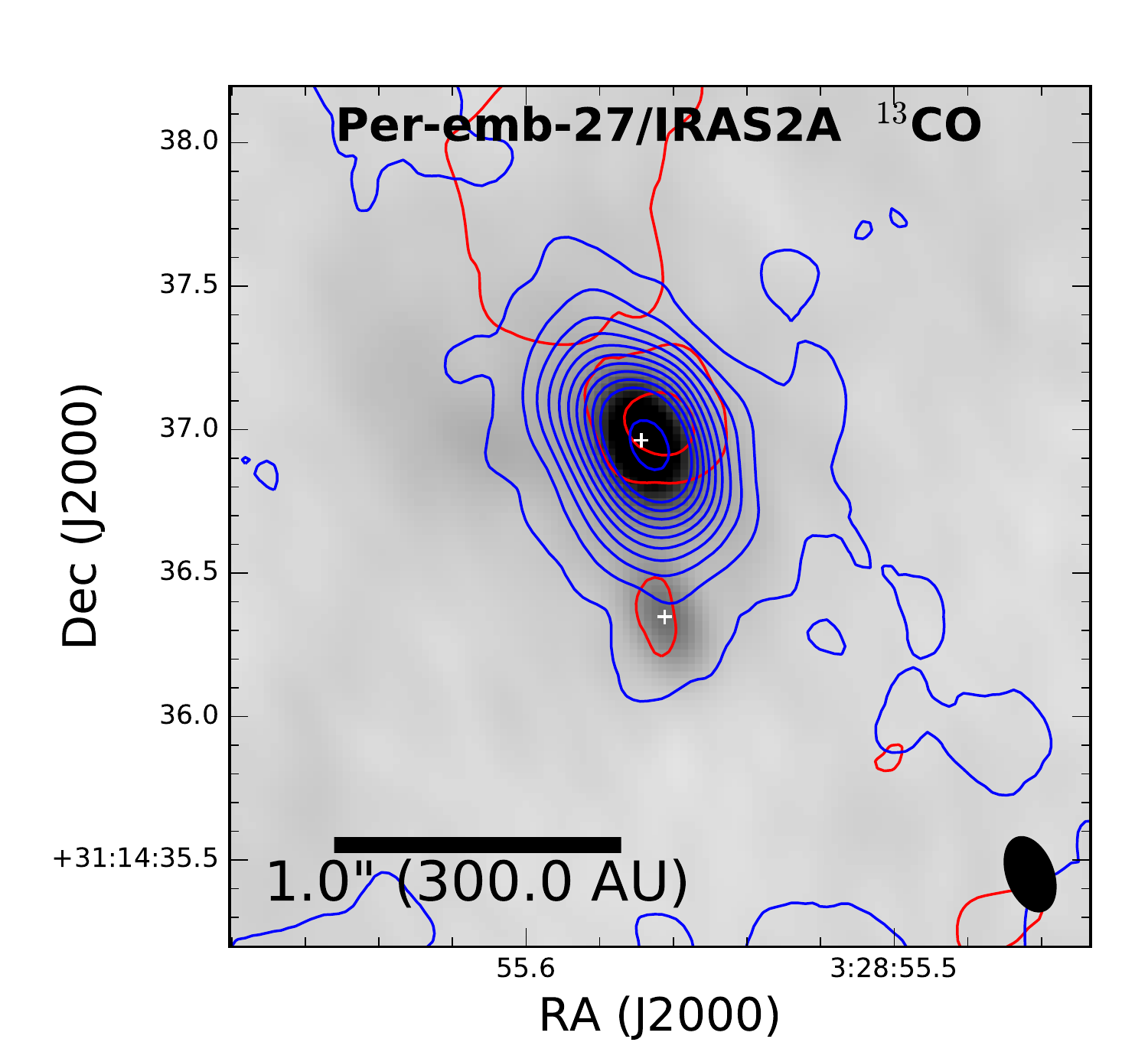}
\includegraphics[scale=0.425]{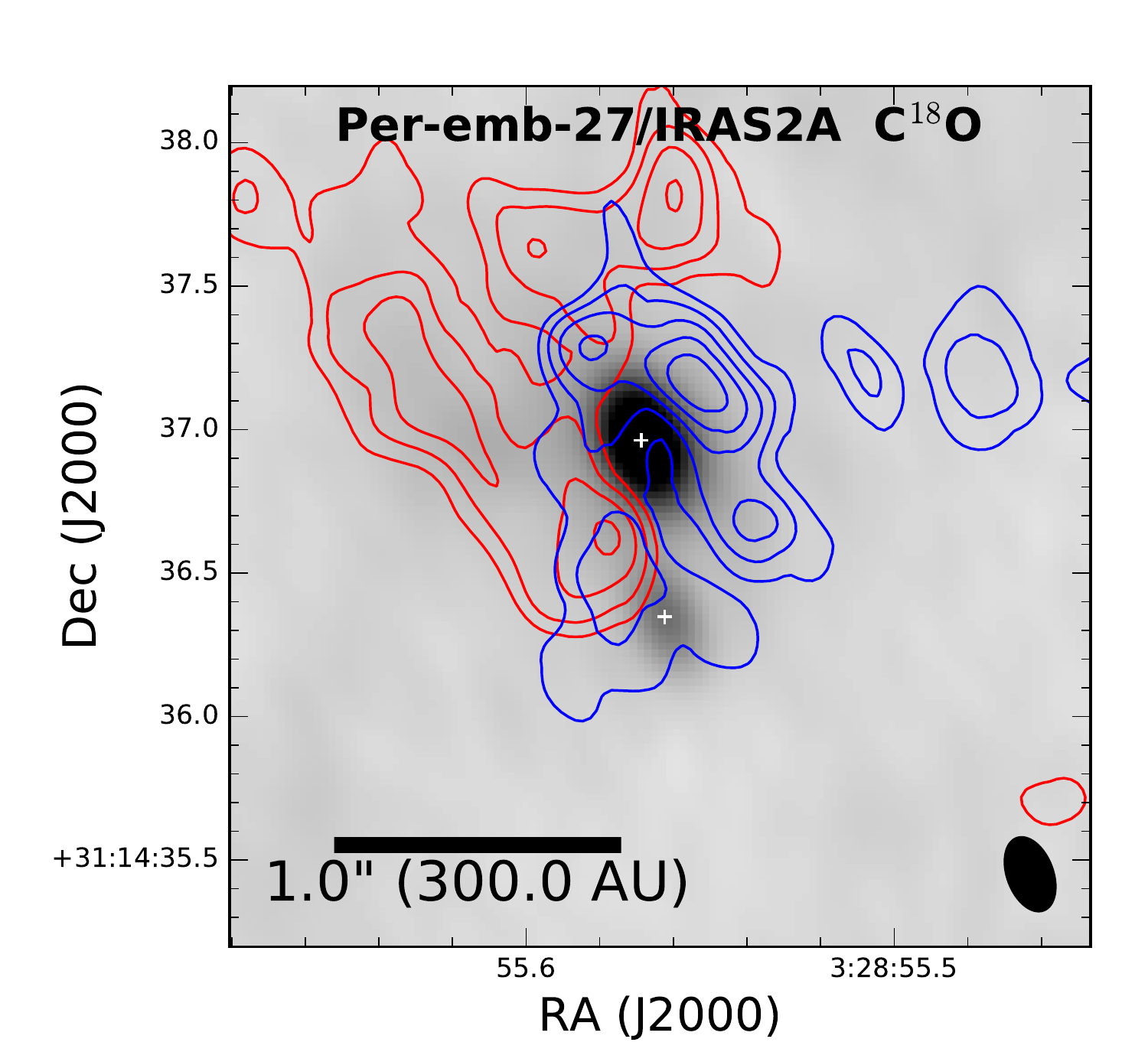}
\includegraphics[scale=0.425]{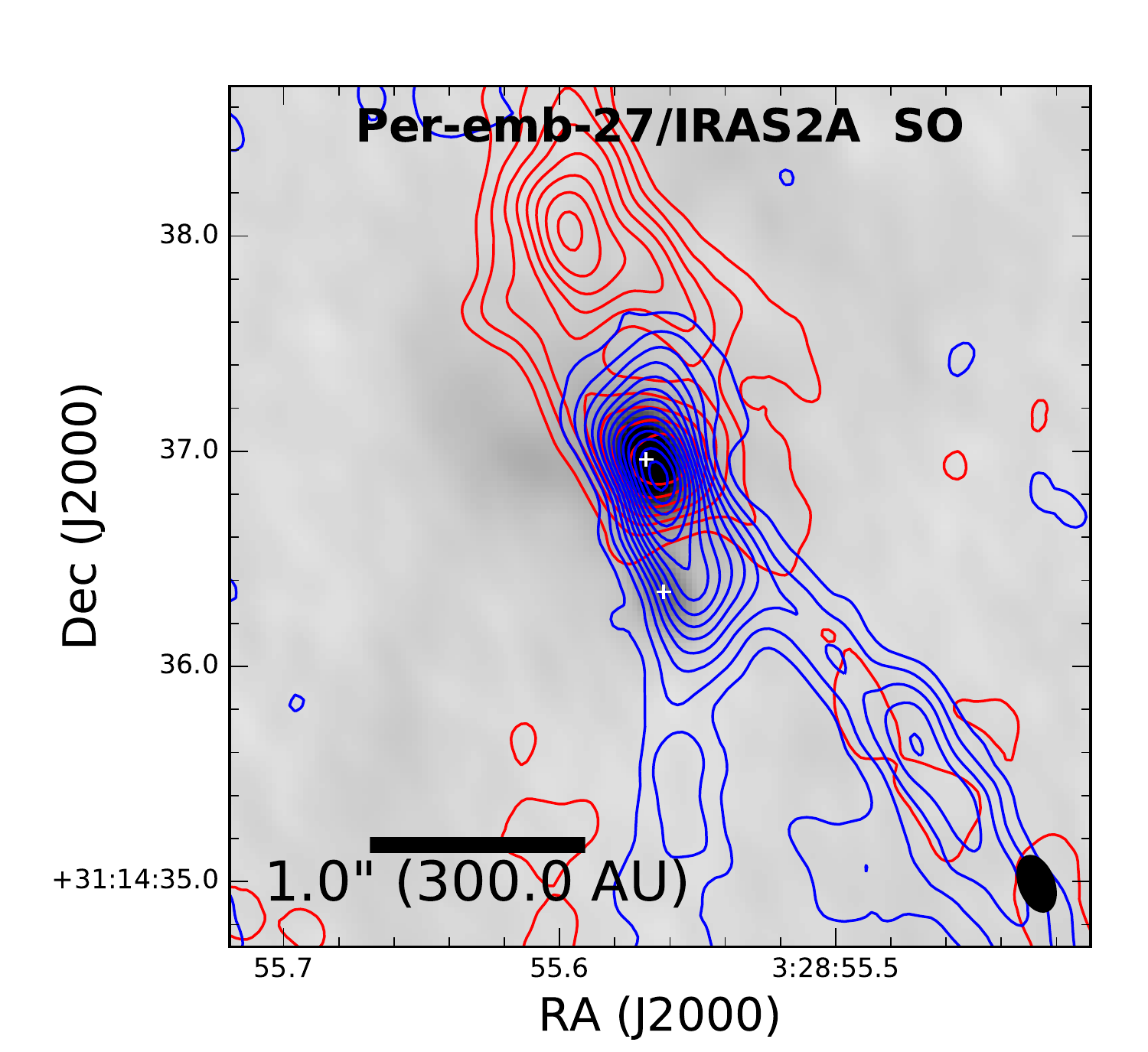}
\includegraphics[scale=0.425]{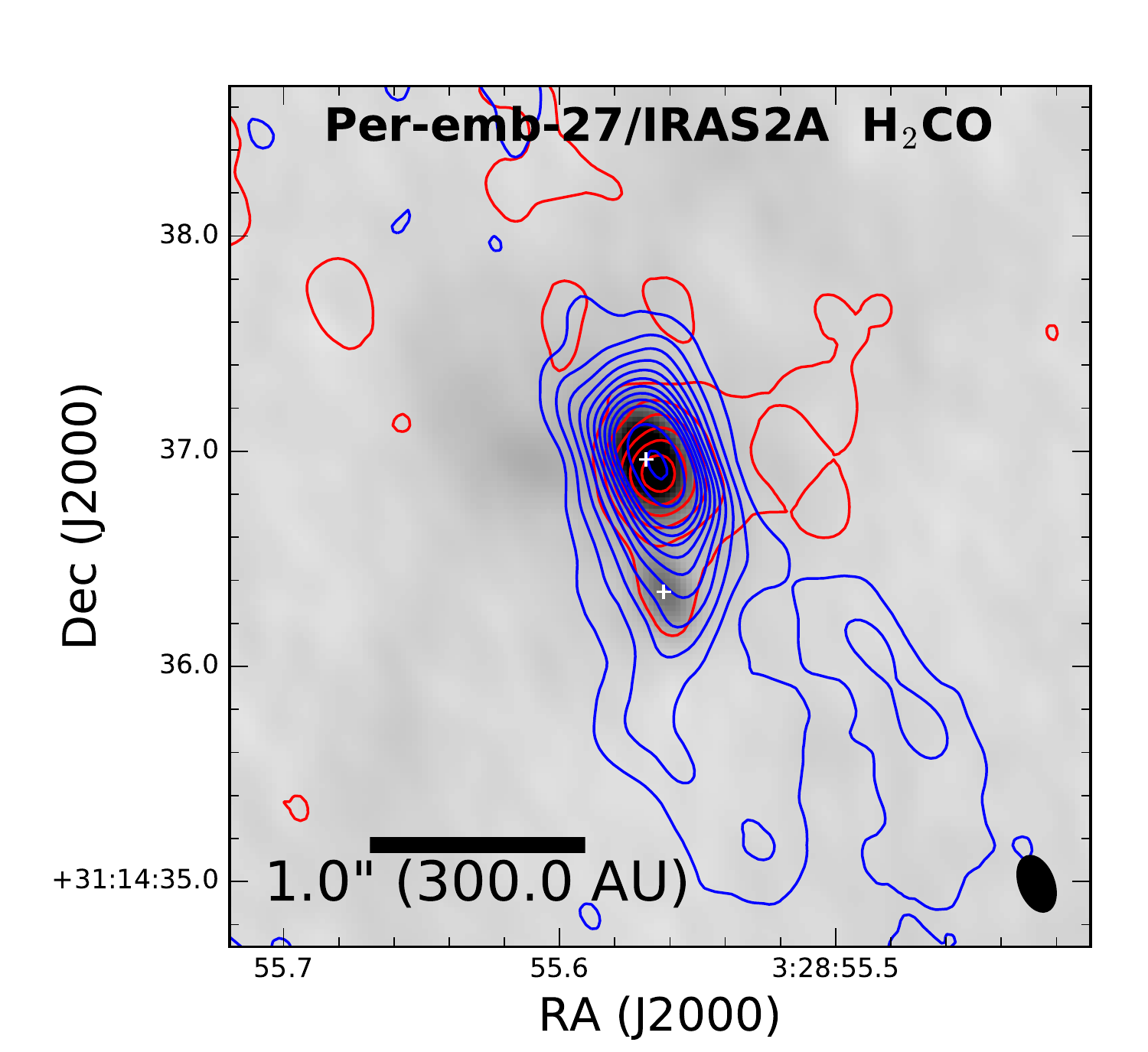}
\end{center}

\caption{Integrated intensity maps of \thco\ (top left panel), \cateo\ (top right panel), SO (bottom left panel), and H$_2$CO (bottom right panel)
toward Per-emb-27 (NGC 1333 IRAS2A). The integrated intensity maps are displayed as red and blue contours
corresponding to the integrated intensity of line emission red and blue-shifted with respect
to the system velocity. The contours are overlaid on the 1.3~mm continuum image. The
line emission shows evidence for a velocity gradient consistent with rotation.
The red-shifted contours start at (3,3,3,3)$\sigma$ and increase in (3,1,2,3)$\sigma$ increments, and
the blue-shifted contours start at (3,3,3,3)$\sigma$ and increase in (3,1,2,3)$\sigma$ increments.
The values inside the parentheses in the previous sentence correspond to the \thco, \cateo, SO, and H$_2$CO integrated intensity maps, respectively. The
values for $\sigma_{red}$ and $\sigma_{blue}$ and velocity ranges over which the line
emission was summed can be found in Table 3.
The beam in the images is approximately 0\farcs36$\times$0\farcs26.}
\label{per-emb-27-lines}
\end{figure}

\begin{figure}
\begin{center}
\includegraphics[scale=0.425]{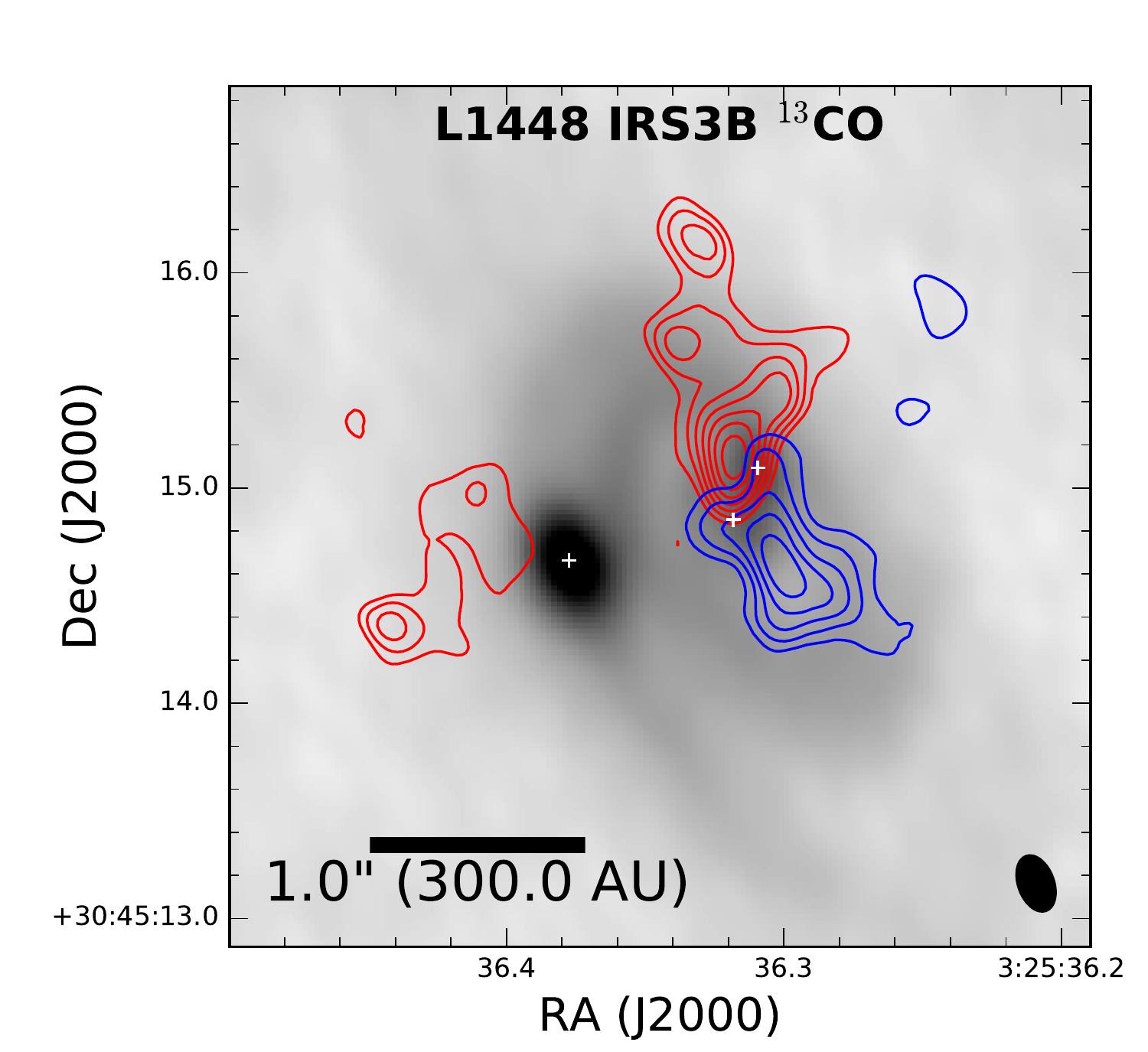}
\includegraphics[scale=0.425]{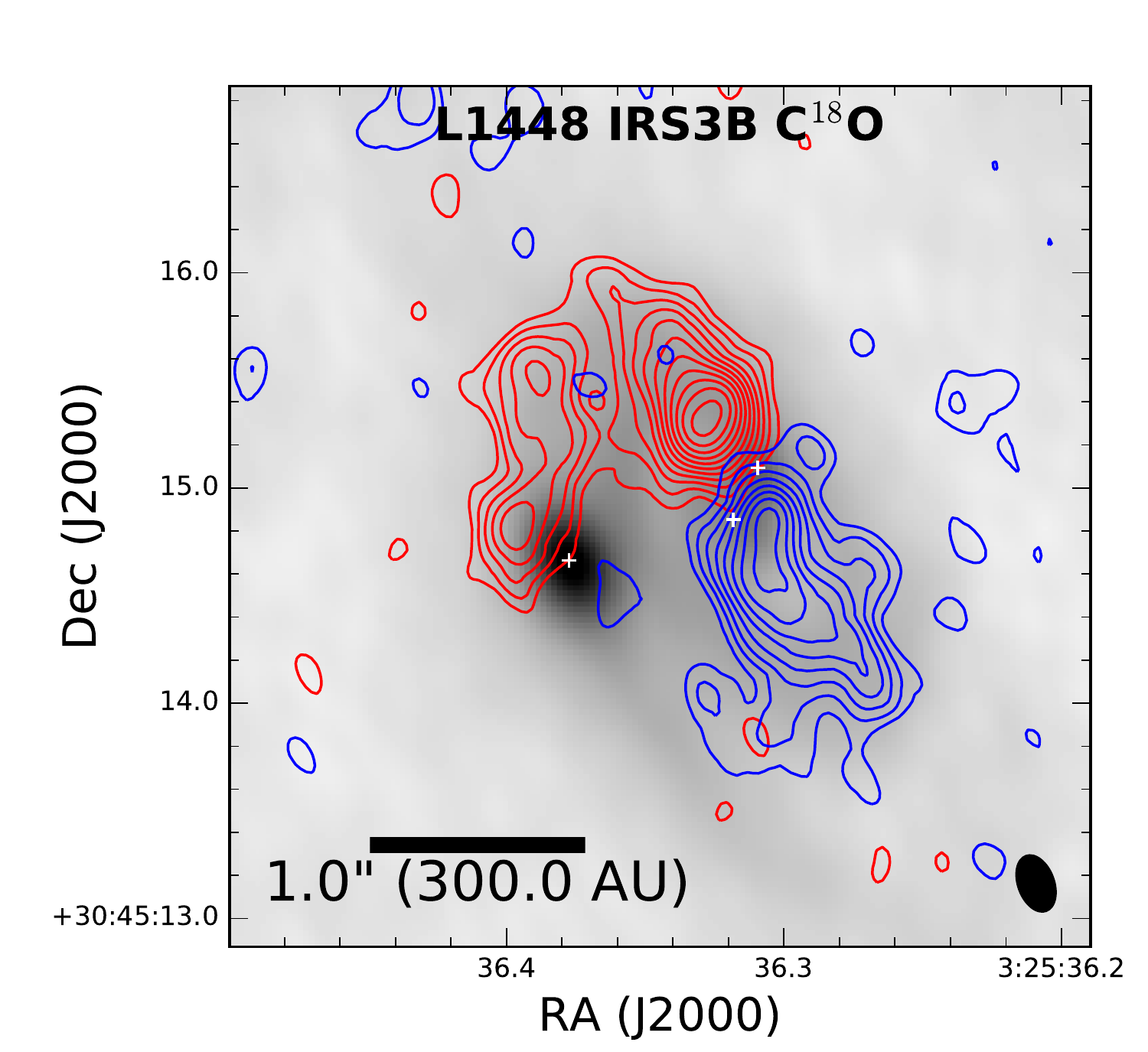}
\includegraphics[scale=0.425]{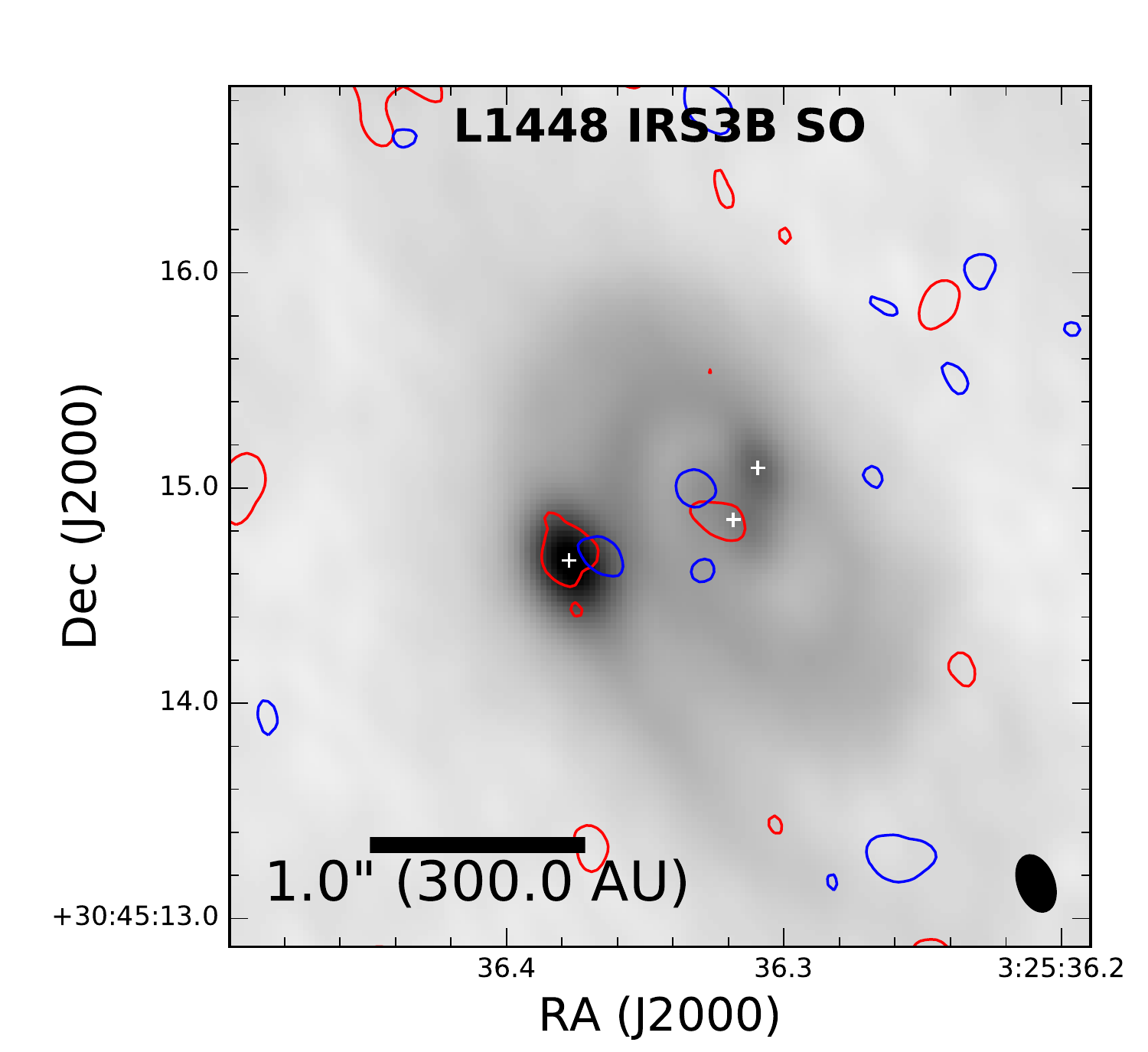}
\includegraphics[scale=0.425]{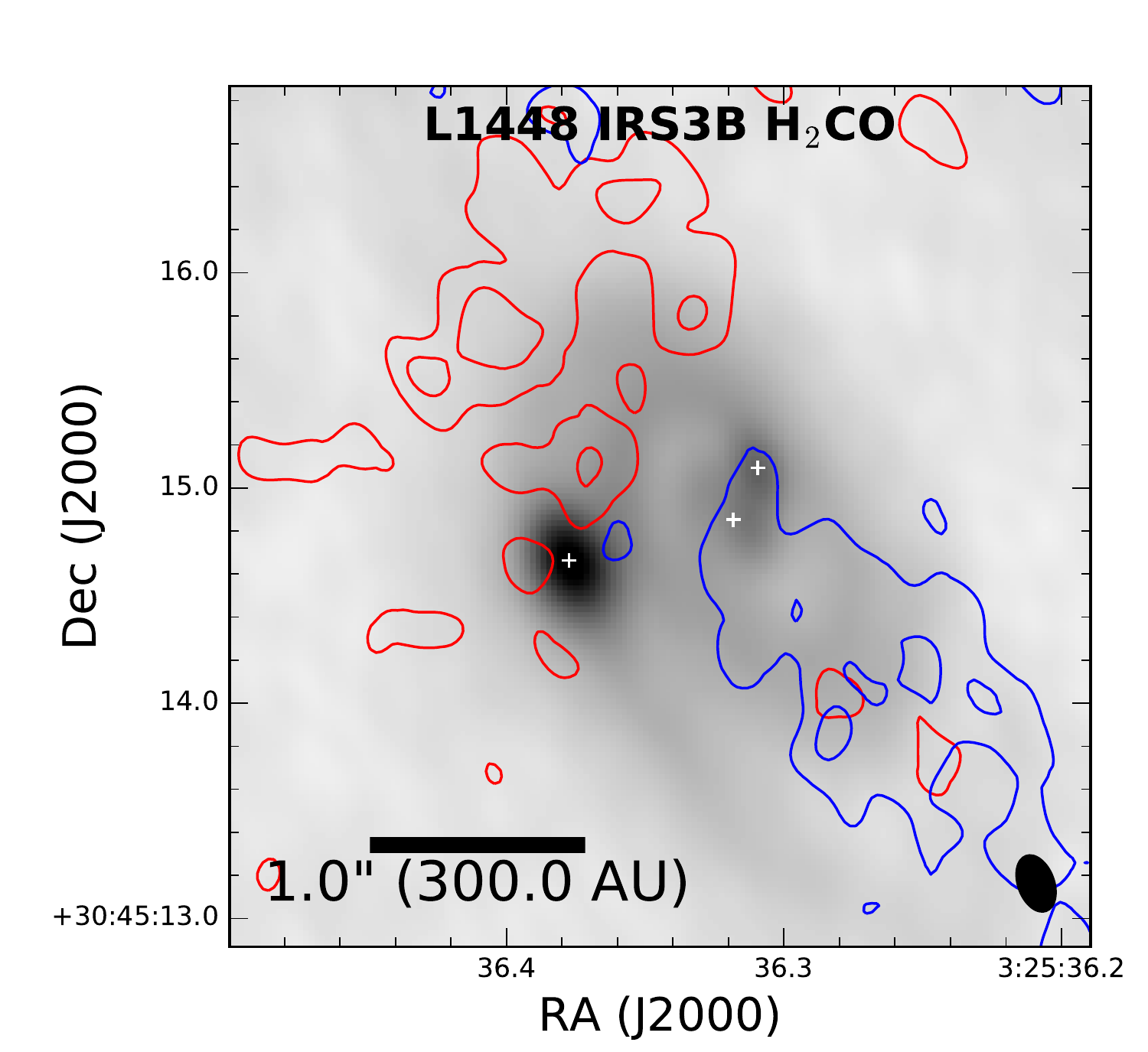}
\end{center}
\caption{Integrated intensity maps of \thco\ (top left panel), \cateo\ (top right panel), SO (bottom left panel), and H$_2$CO (bottom right panel)
toward Per-emb-33 (L1448 IRS3B). The integrated intensity maps are displayed as red and blue contours
corresponding to the integrated intensity of line emission red and blue-shifted with respect
to the system velocity. The contours are overlaid on the 1.3~mm continuum image. The
line emission shows evidence for a velocity gradient consistent with rotation.
The red-shifted contours start at (4,4,3,3)$\sigma$ and increase in (1,1,2,2)$\sigma$ increments, and
the blue-shifted contours start at (4,4,3,3)$\sigma$ and increase in (1,1,2,2)$\sigma$ increments.
The values inside the parentheses in the previous sentence correspond to the \thco, \cateo, SO, and H$_2$CO integrated intensity maps, respectively. The
values for $\sigma_{red}$ and $\sigma_{blue}$ and velocity ranges over which the line
emission was summed can be found in Table 3.
The beam in the images is approximately 0\farcs36$\times$0\farcs26.}
\label{L1448IRS3B-lines}
\end{figure}

\begin{figure}
\begin{center}
\includegraphics[scale=0.425]{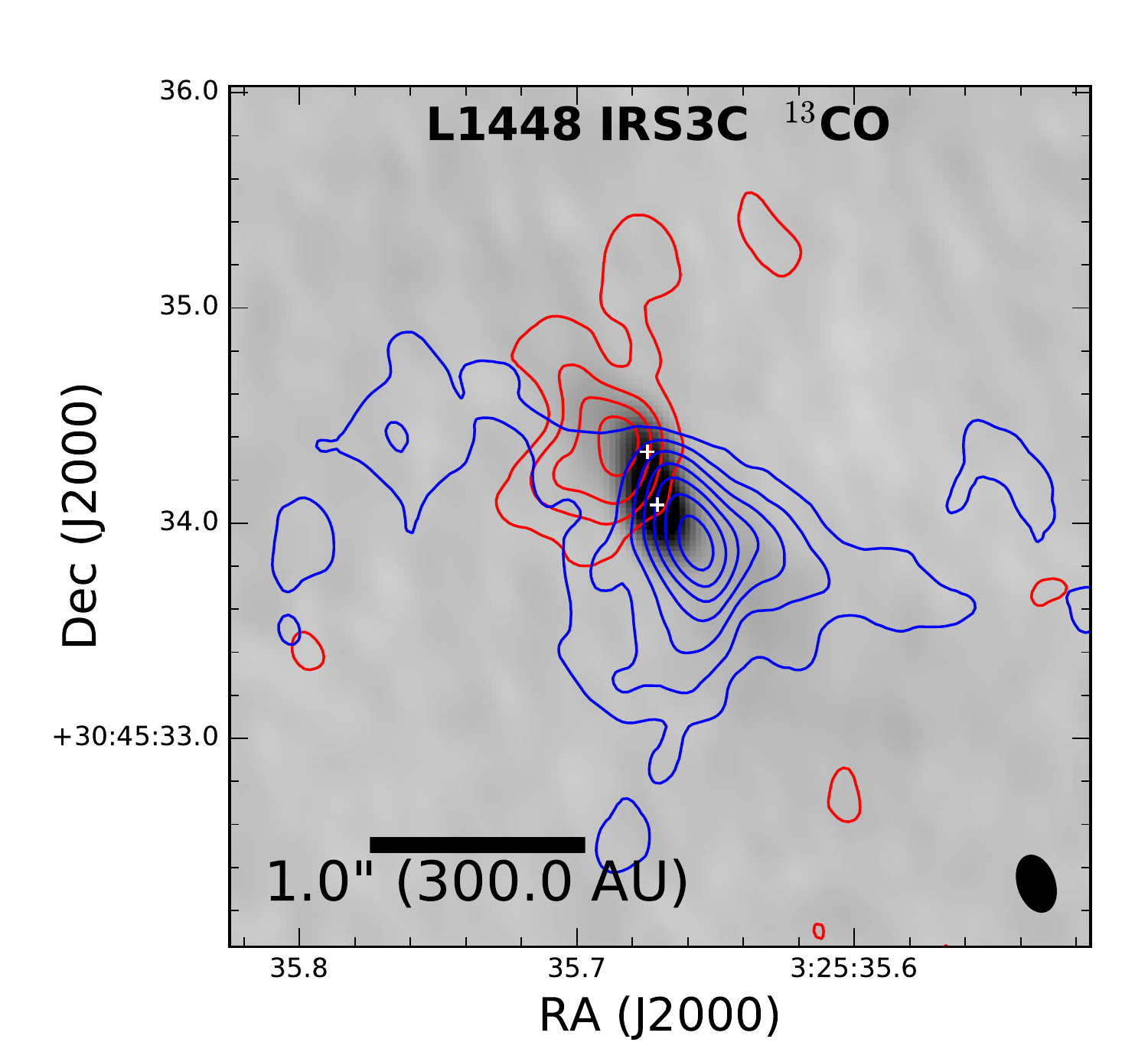}
\includegraphics[scale=0.425]{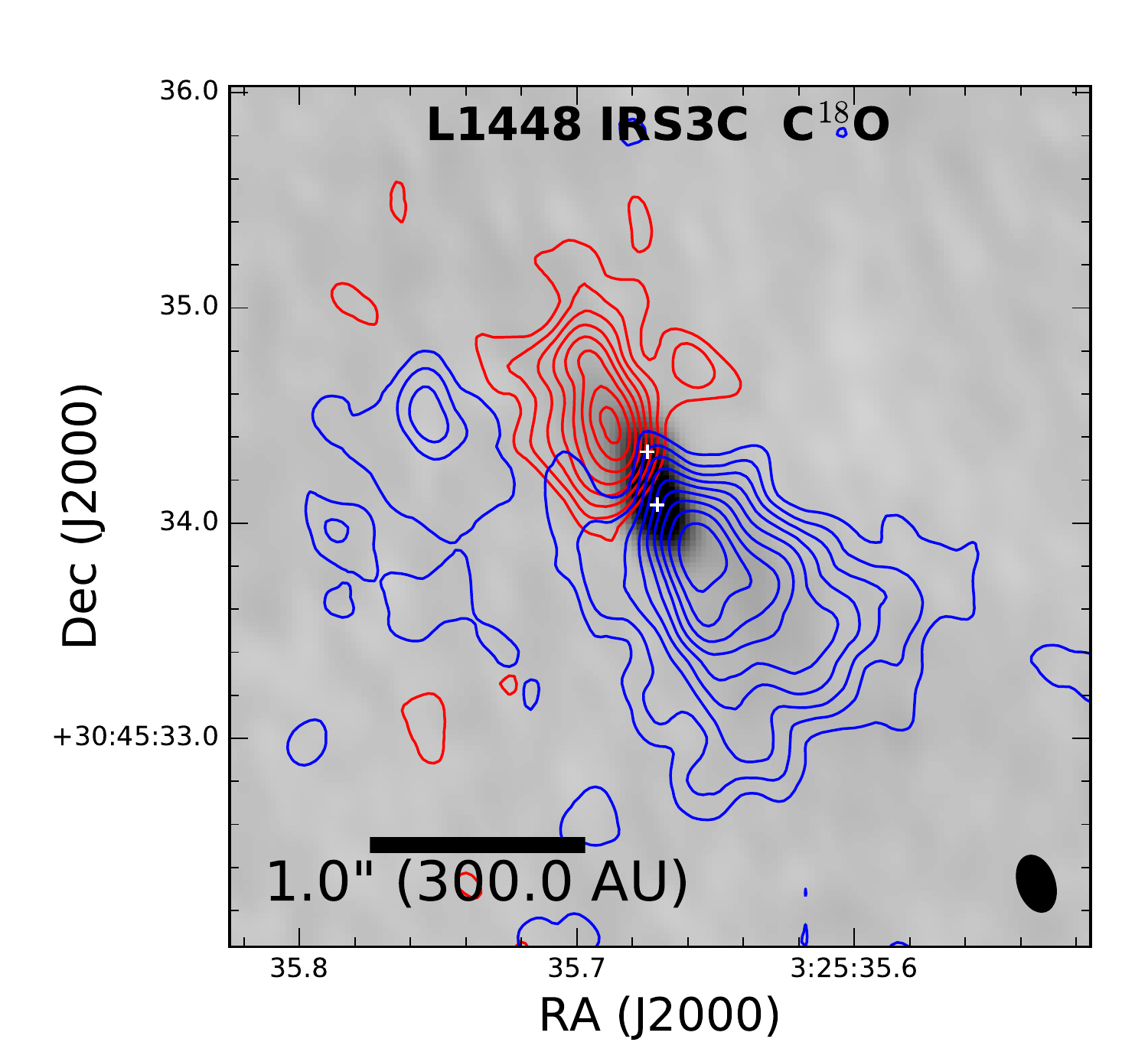}
\includegraphics[scale=0.425]{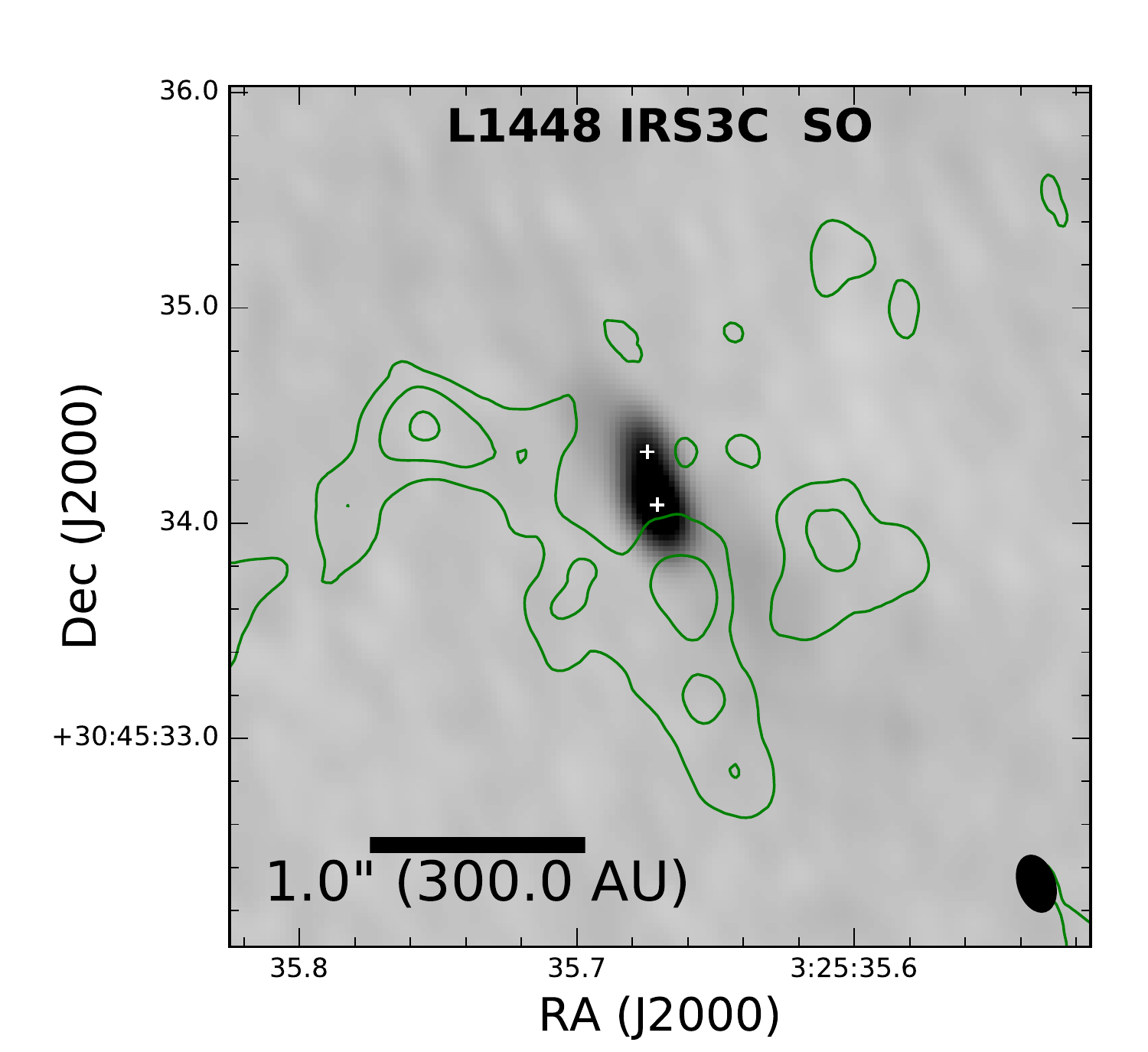}
\includegraphics[scale=0.425]{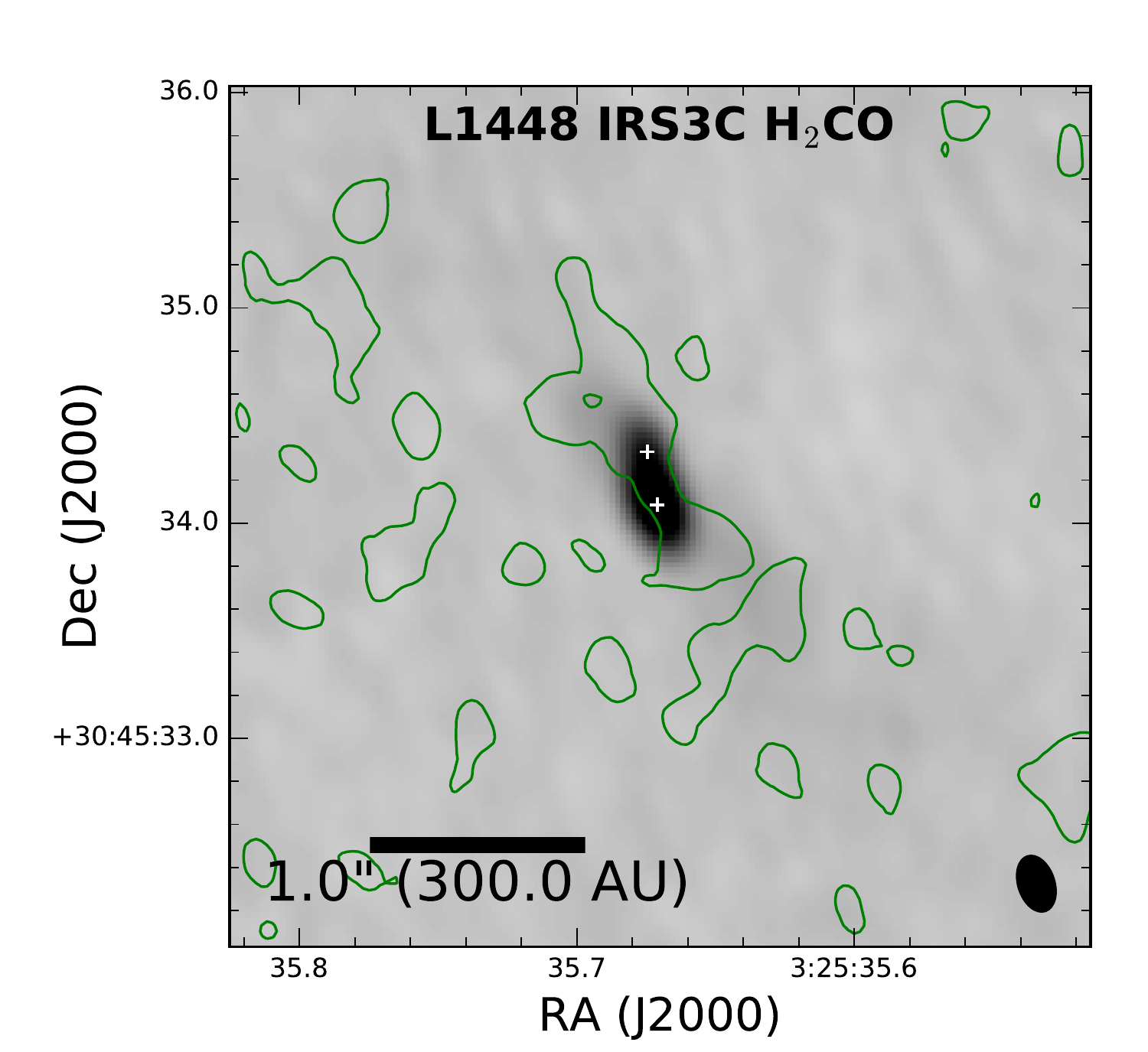}
\end{center}

\caption{Integrated intensity maps of \thco\ (top left panel), \cateo\ (top right panel), SO (bottom left panel), and H$_2$CO (bottom right panel)
toward L1448 IRS3C (L1448 NW). The integrated intensity maps are displayed as red and blue contours
corresponding to the integrated intensity of line emission red and blue-shifted with respect
to the system velocity. The contours are overlaid on the 1.3~mm continuum image. The
line emission shows evidence for a velocity gradient consistent with rotation.
The red-shifted contours start at (3,4)$\sigma$ and increase in (2,1)$\sigma$ increments, and
the blue-shifted contours start at (3,3)$\sigma$ and increase in (2,1)$\sigma$ increments.
The contours for SO and H$_2$CO start (3,3)$\sigma$ and increase in (2,3)$\sigma$, respectively
for a single interval.
The values for $\sigma$, $\sigma_{red}$ and $\sigma_{blue}$ and velocity ranges over which the line
emission was summed can be found in Table 3.
The beam in the images is approximately 0\farcs36$\times$0\farcs26.}
\label{L1448IRS3C-lines}
\end{figure}

\begin{figure}
\begin{center}
\includegraphics[scale=0.425]{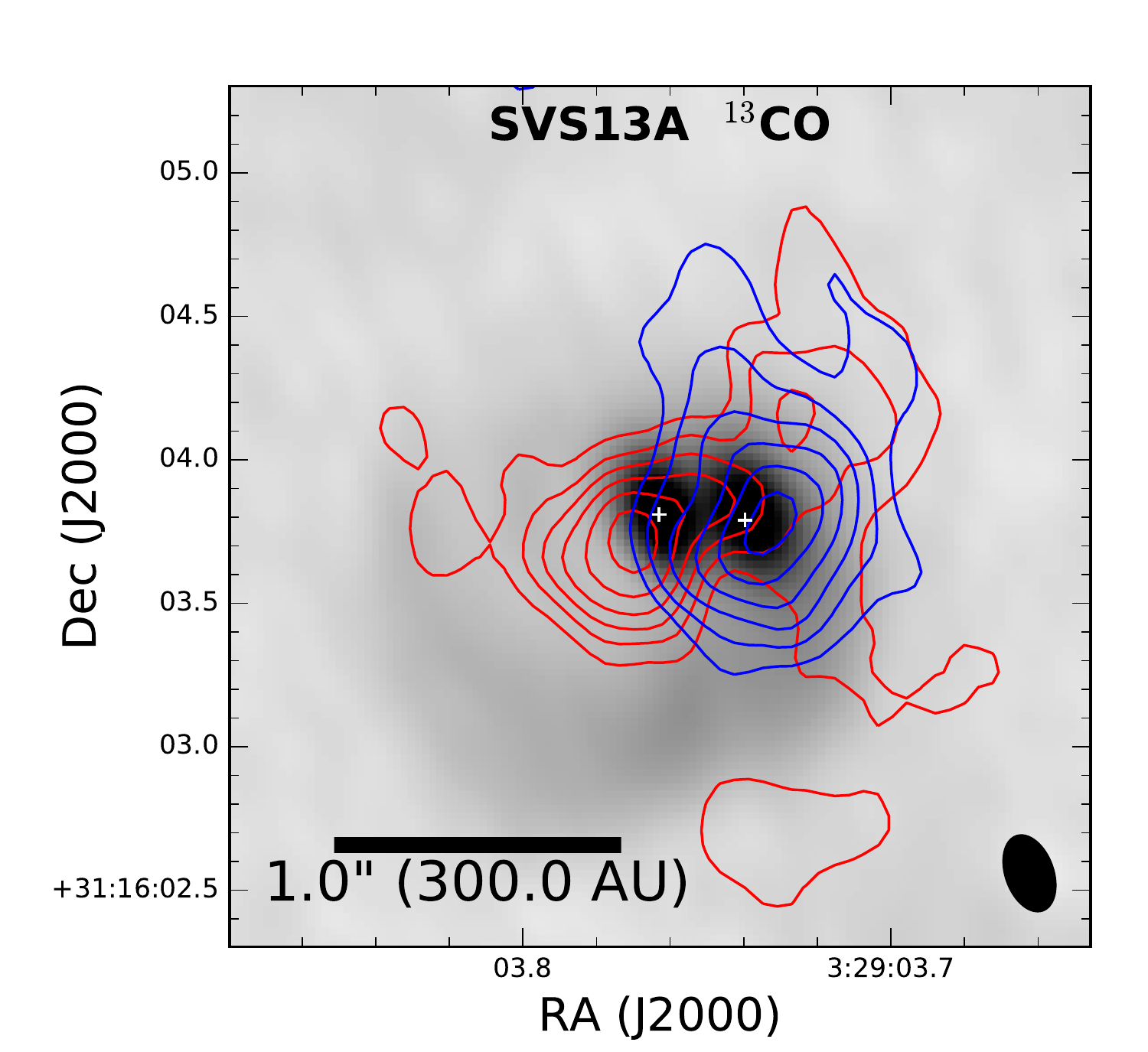}
\includegraphics[scale=0.425]{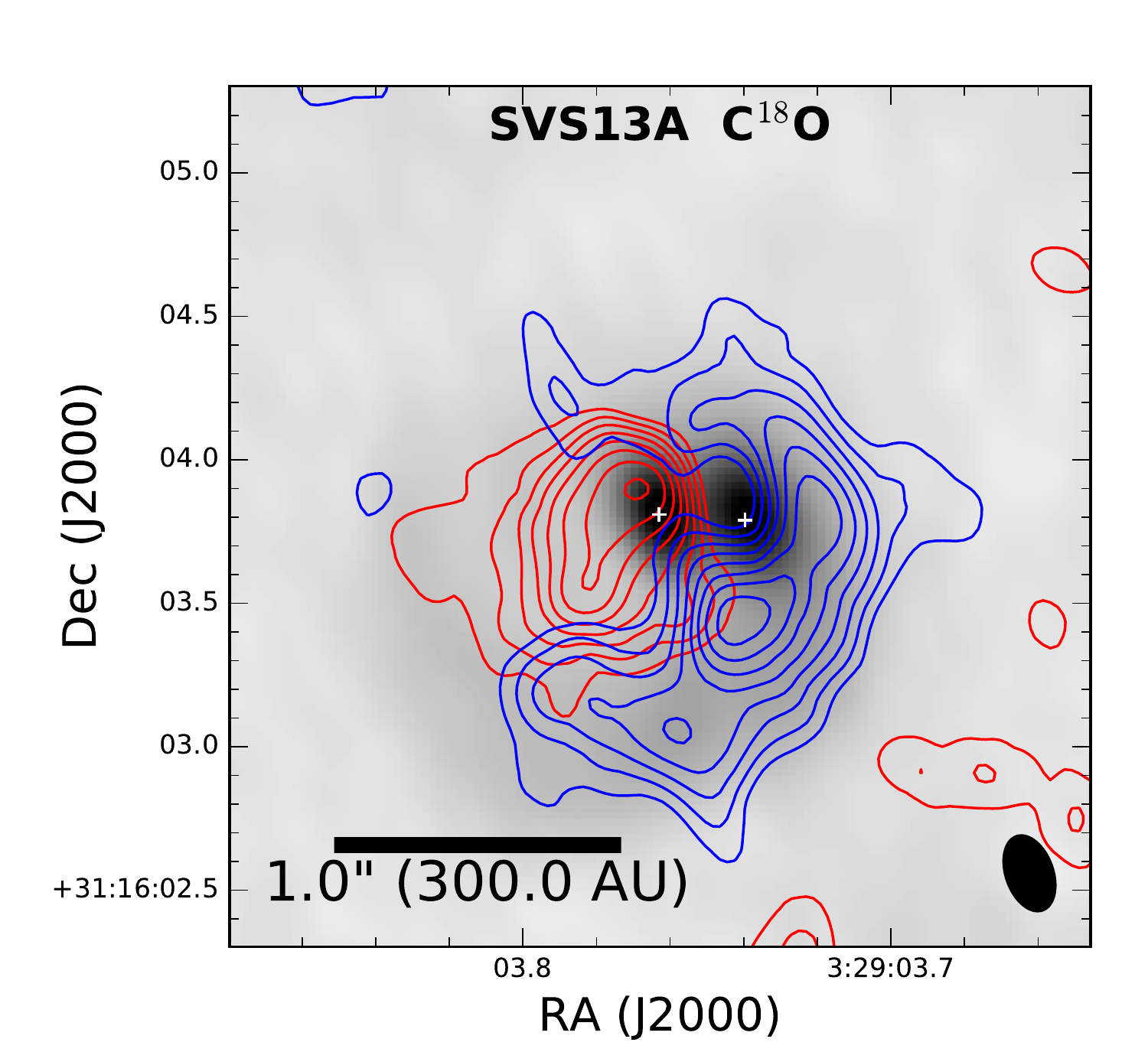}
\includegraphics[scale=0.425]{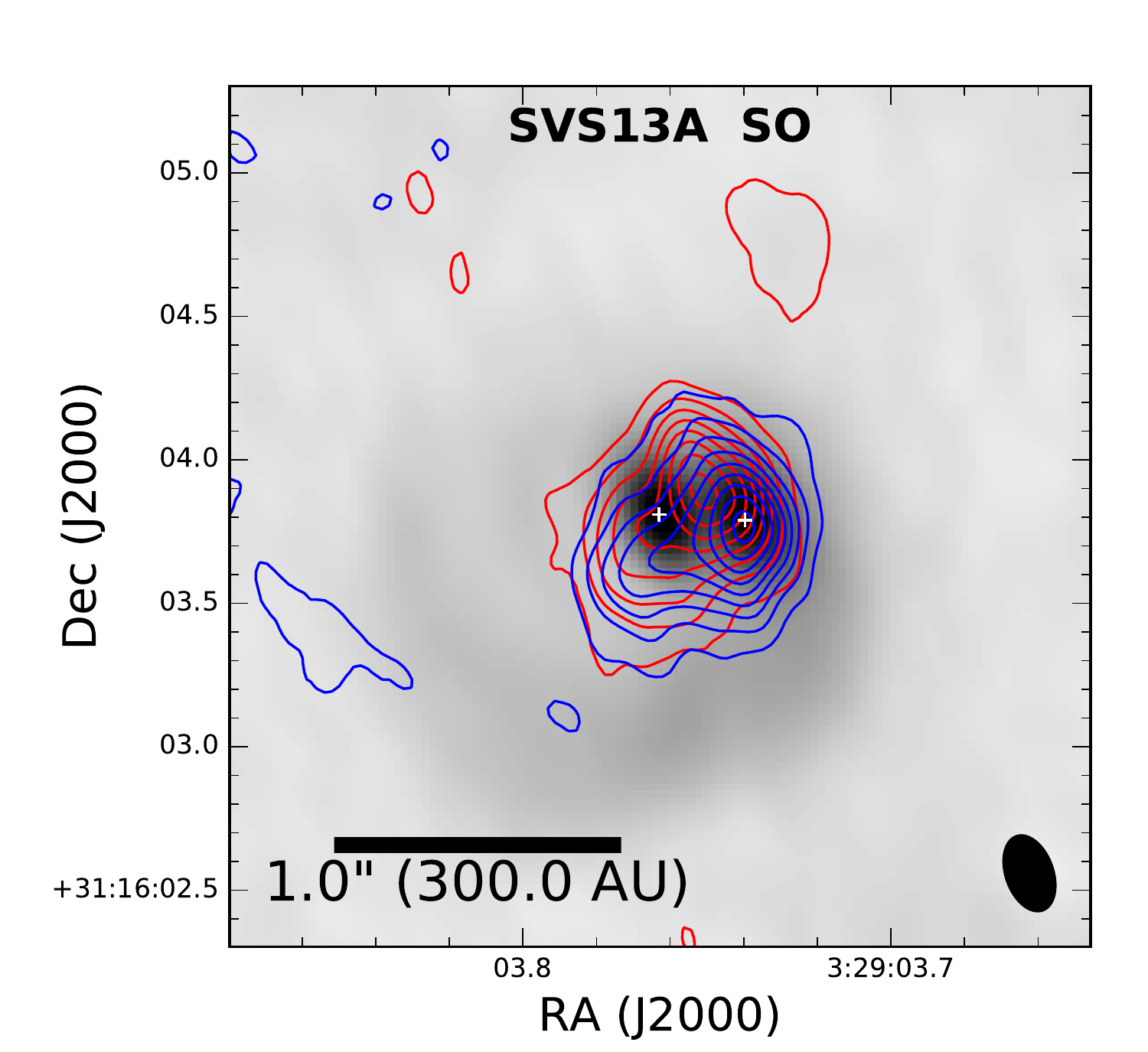}
\includegraphics[scale=0.425]{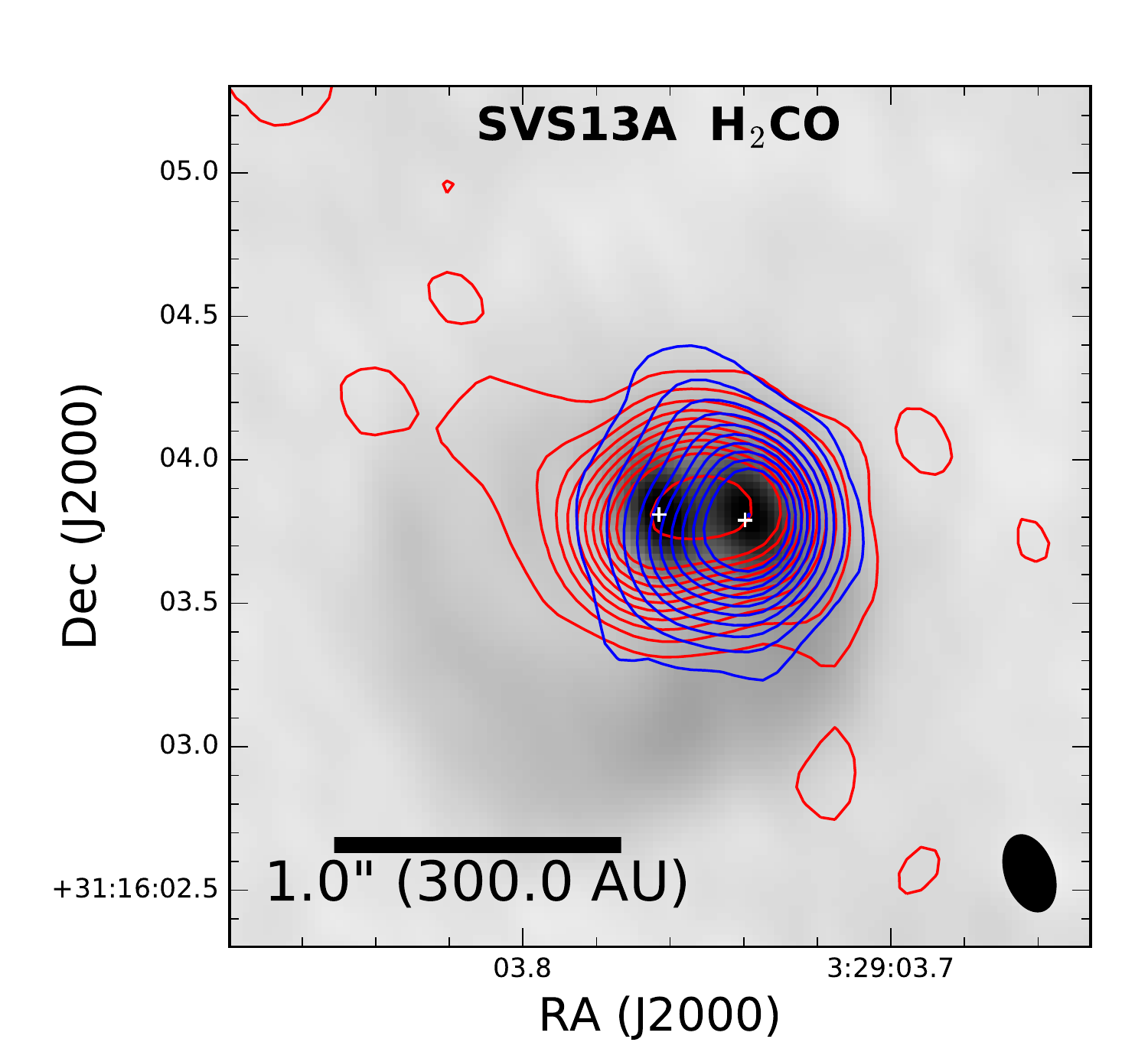}
\end{center}

\caption{Integrated intensity maps of \thco\ (top left panel), \cateo\ (top right panel), SO (bottom left panel), and H$_2$CO (bottom right panel)
toward Per-emb-44 (SVS13A). The integrated intensity maps are displayed as red and blue contours
corresponding to the integrated intensity of line emission red and blue-shifted with respect
to the system velocity. The contours are overlaid on the 1.3~mm continuum image. The
line emission shows evidence for a velocity gradient consistent with rotation.
The red-shifted contours start at (4,3,5,5)$\sigma$ and increase in (2,1,5,5)$\sigma$ increments, and
the blue-shifted contours start at (4,3,5,5)$\sigma$ and increase in (2,1,5,5)$\sigma$ increments.
The values inside the parentheses in the previous sentence correspond to the \thco, \cateo, SO, and H$_2$CO integrated intensity maps, respectively. The
values for $\sigma_{red}$ and $\sigma_{blue}$ and velocity ranges over which the line
emission was summed can be found in Table 3.
The beam in the images is approximately 0\farcs36$\times$0\farcs26.}
\label{per-emb-44-lines}
\end{figure}

\clearpage

\begin{figure}
\begin{center}
\includegraphics[scale=0.425]{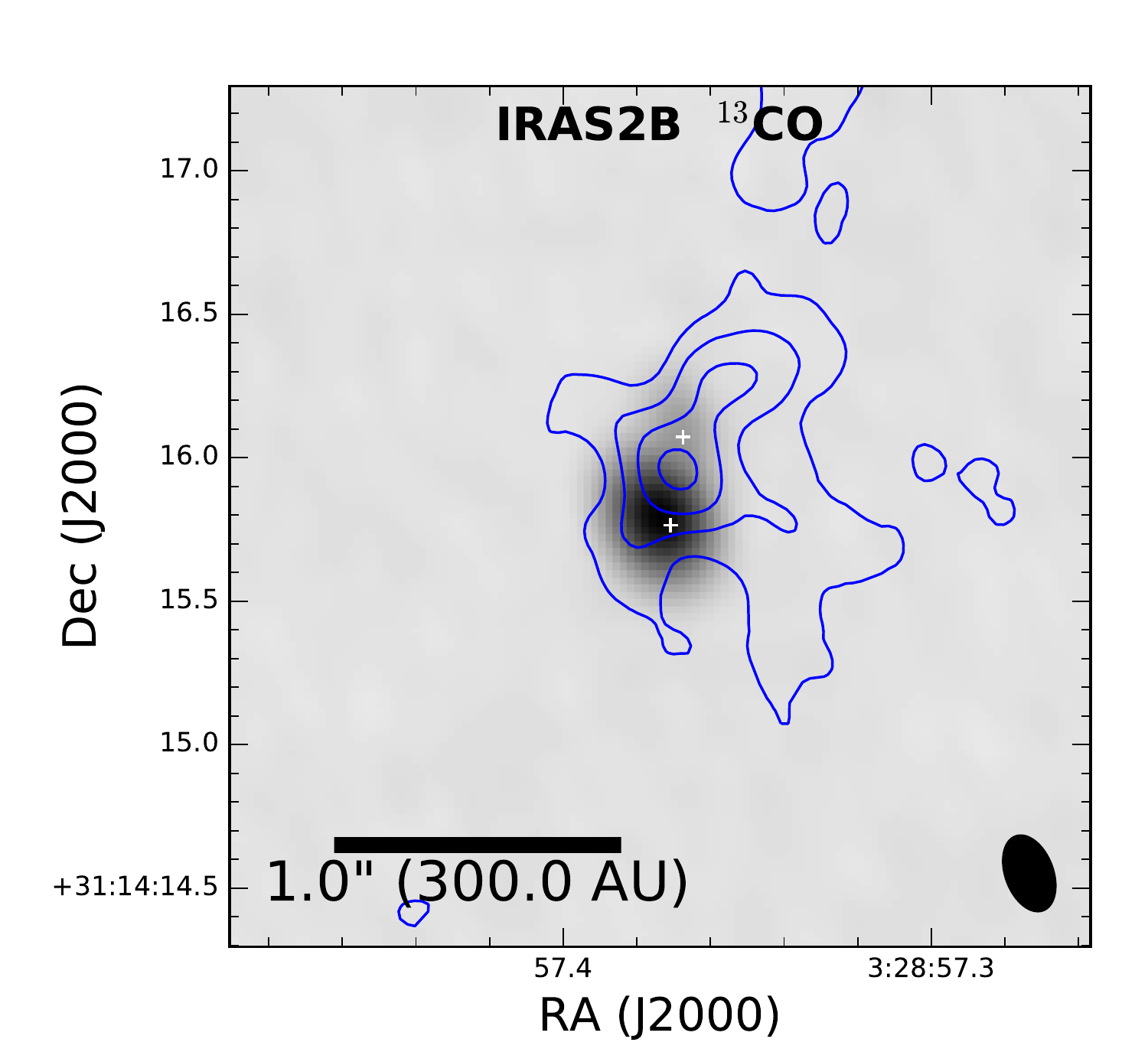}
\includegraphics[scale=0.425]{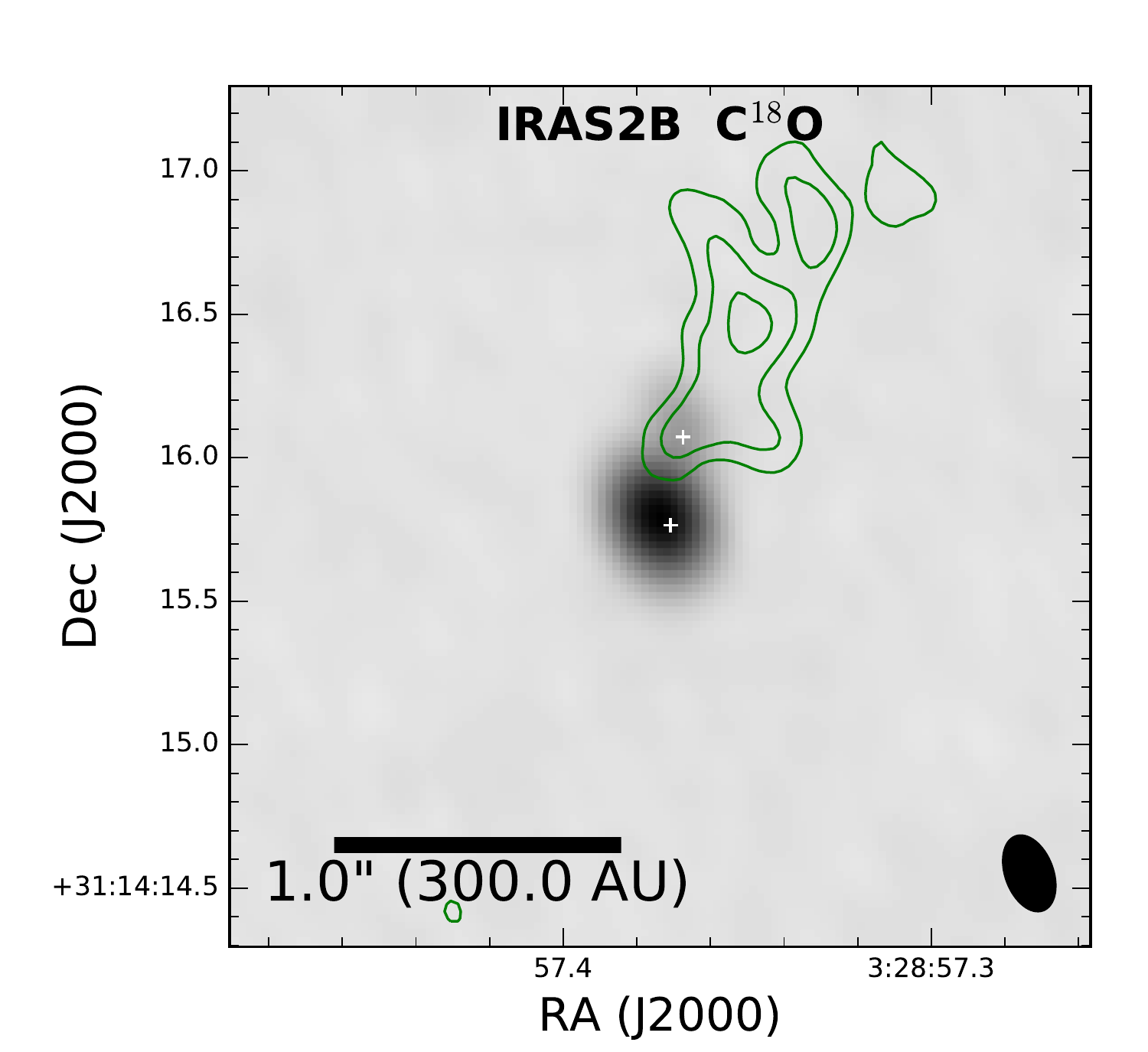}
\includegraphics[scale=0.425]{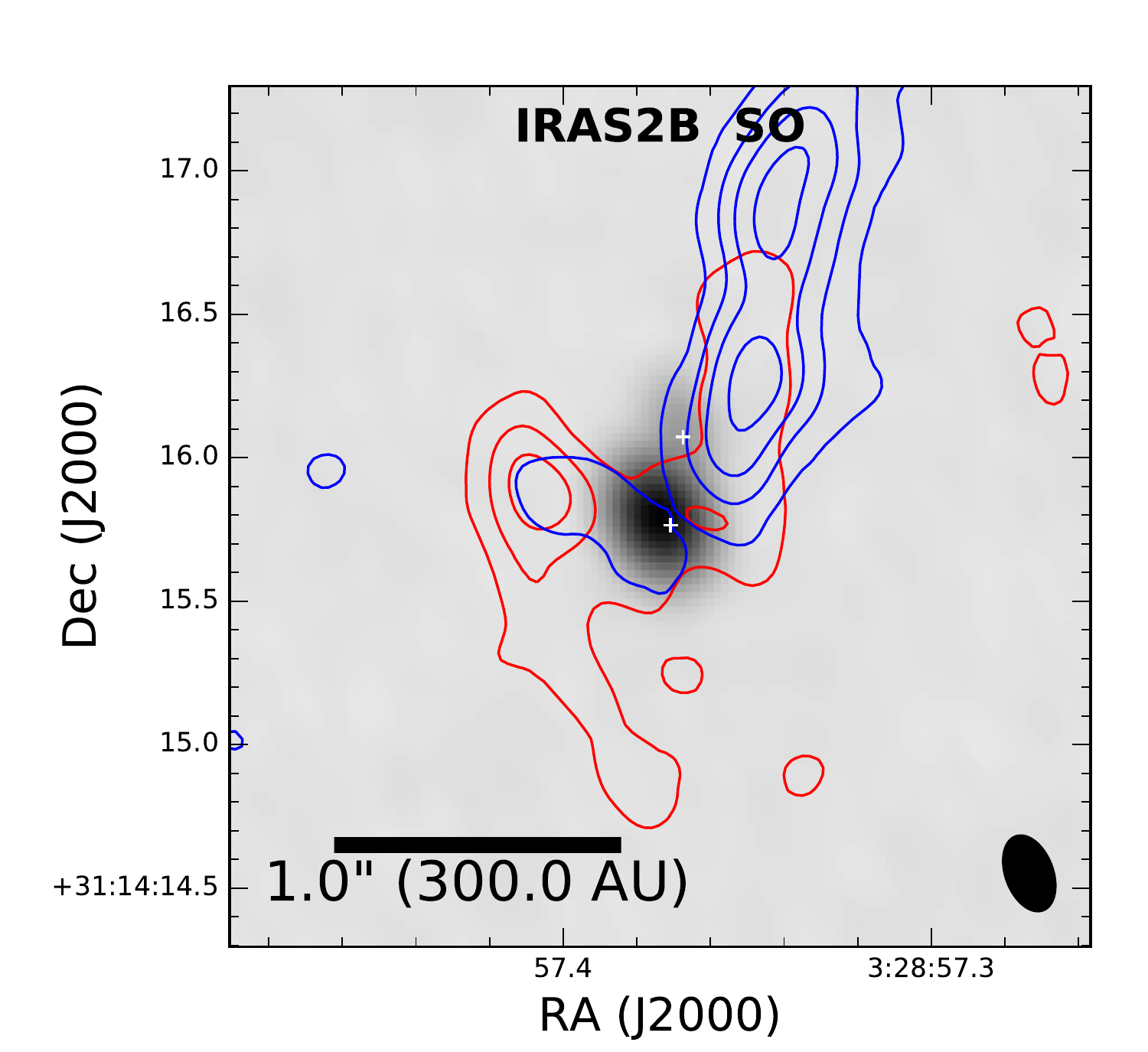}
\includegraphics[scale=0.425]{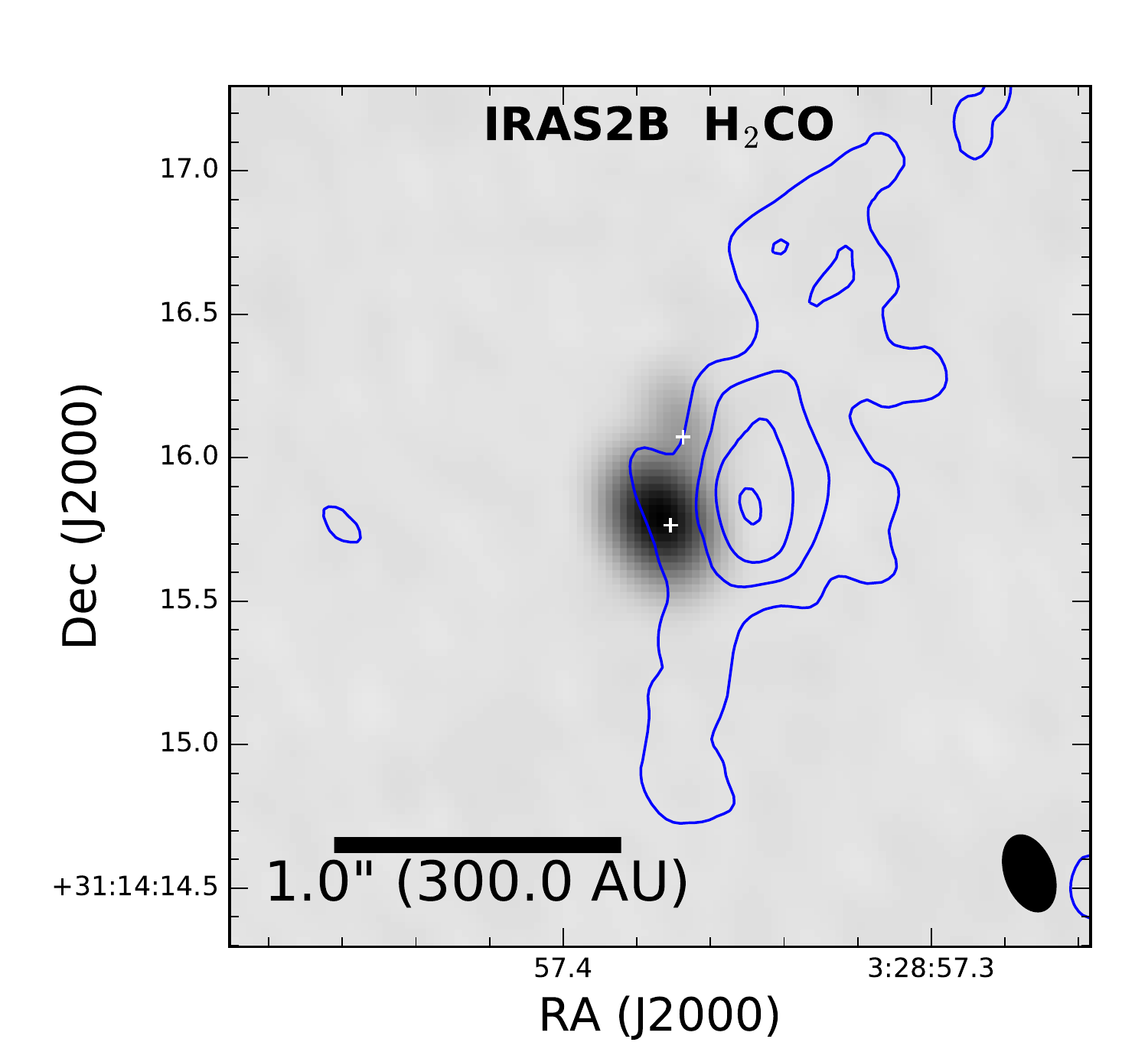}
\end{center}
\caption{Integrated intensity maps of \thco\ (top left panel), \cateo\ (top right panel), SO (bottom left panel), and H$_2$CO (bottom right panel)
toward Per-emb-36 (NGC 1333 IRAS2B). The integrated intensity maps are displayed as red and blue contours
corresponding to the integrated intensity of line emission red and blue-shifted with respect
to the system velocity. The contours are overlaid on the 1.3~mm continuum image. The
line emission shows evidence for a velocity gradient consistent with rotation.
The red-shifted contours start at (4,3)$\sigma$ and increase in (2,3)$\sigma$ increments for \thco\ and SO,respectively, and
the blue-shifted contours start at (3,3,3)$\sigma$ and increase in (2,3,2)$\sigma$ increments for \thco, SO, and H$_2$CO, respectively.
The line-center contours for \cateo\ start at 3$\sigma$ and increase by 1$\sigma$ increments.
The values for $\sigma_{red}$ and $\sigma_{blue}$ and velocity ranges over which the line
emission was summed can be found in Table 3.
The beam in the images is approximately 0\farcs36$\times$0\farcs26.}
\label{per-emb-36-lines}
\end{figure}
\clearpage

\clearpage

\begin{figure}
\begin{center}
\includegraphics[scale=0.425]{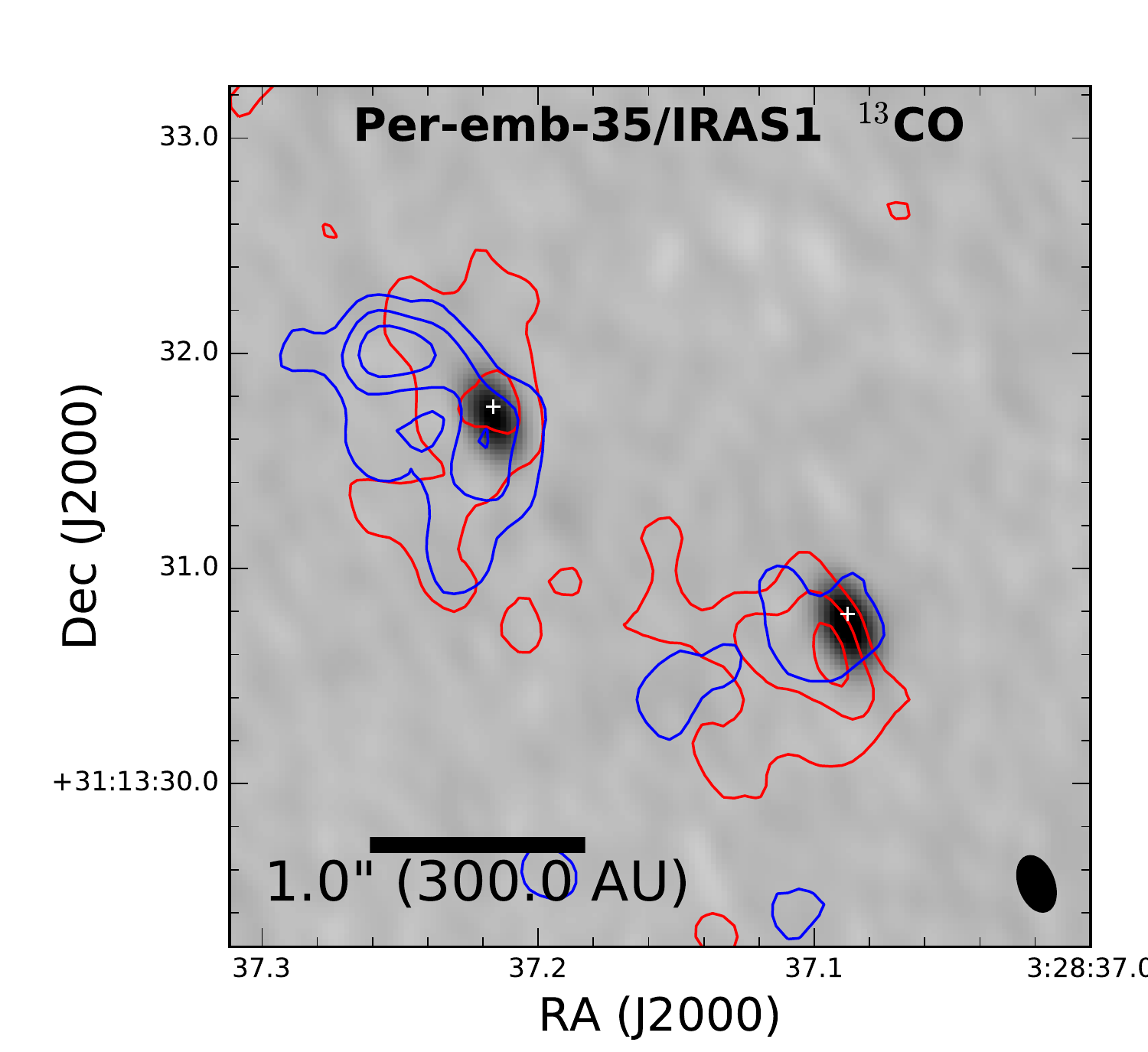}
\includegraphics[scale=0.425]{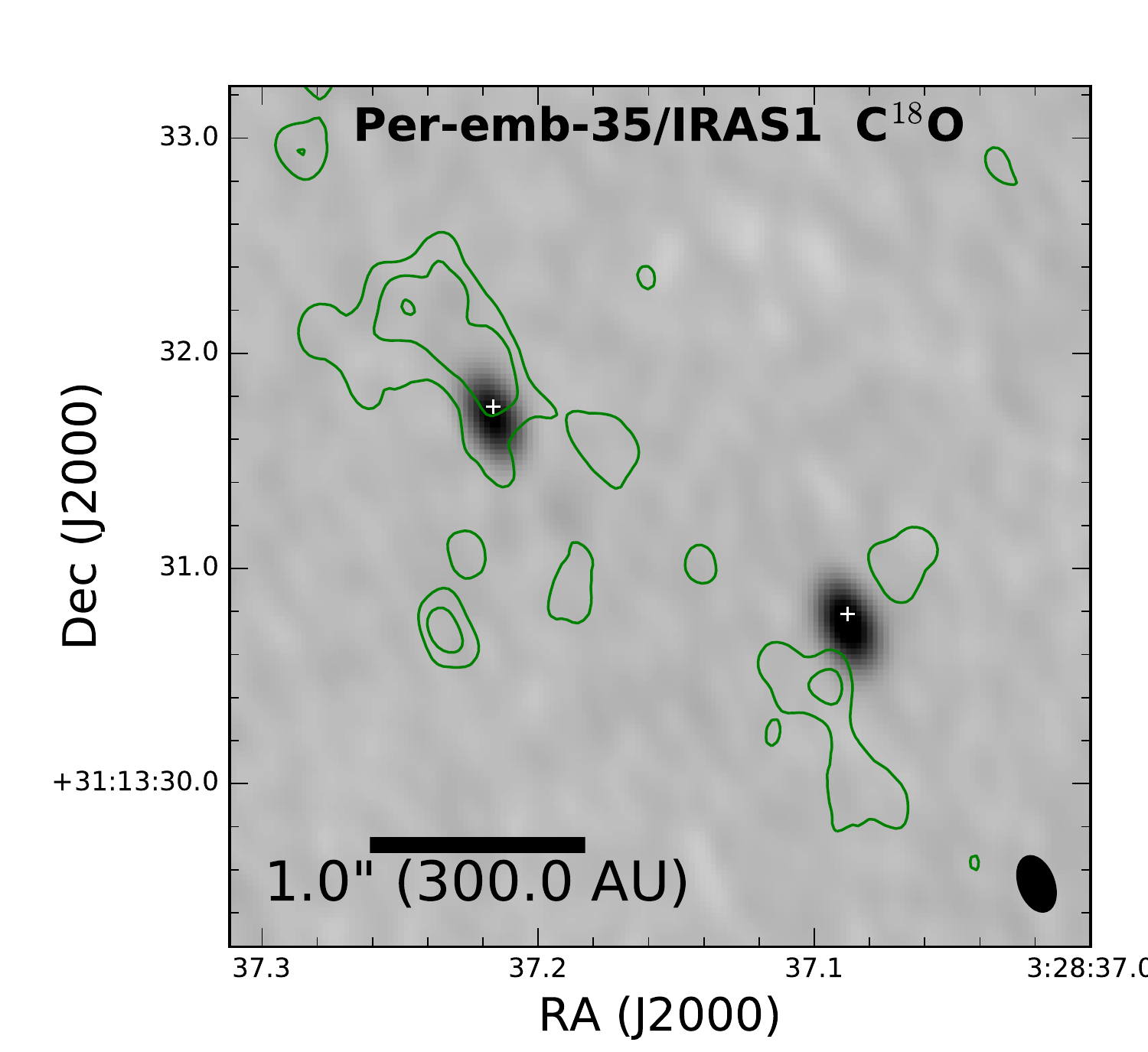}
\includegraphics[scale=0.425]{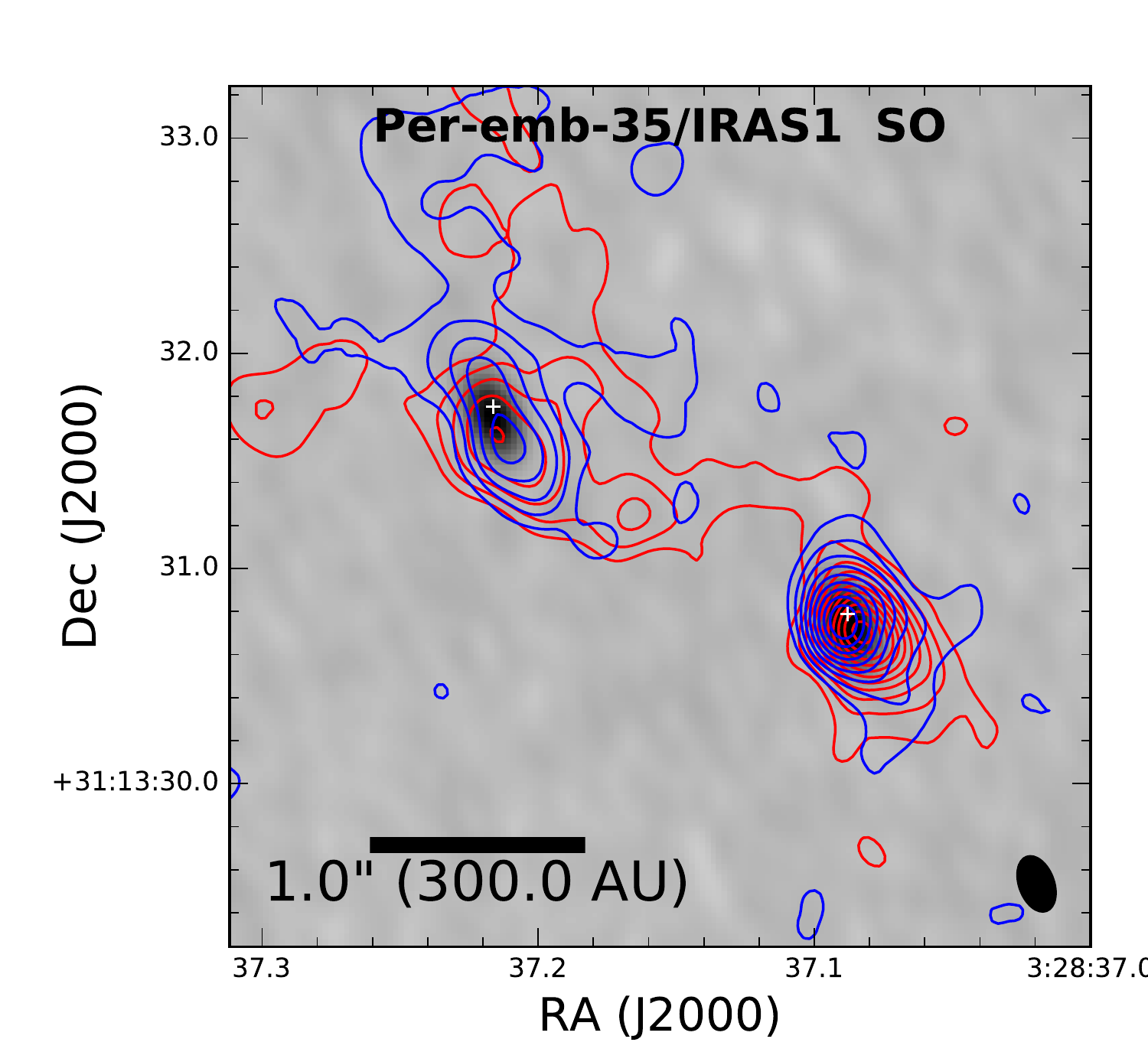}
\includegraphics[scale=0.425]{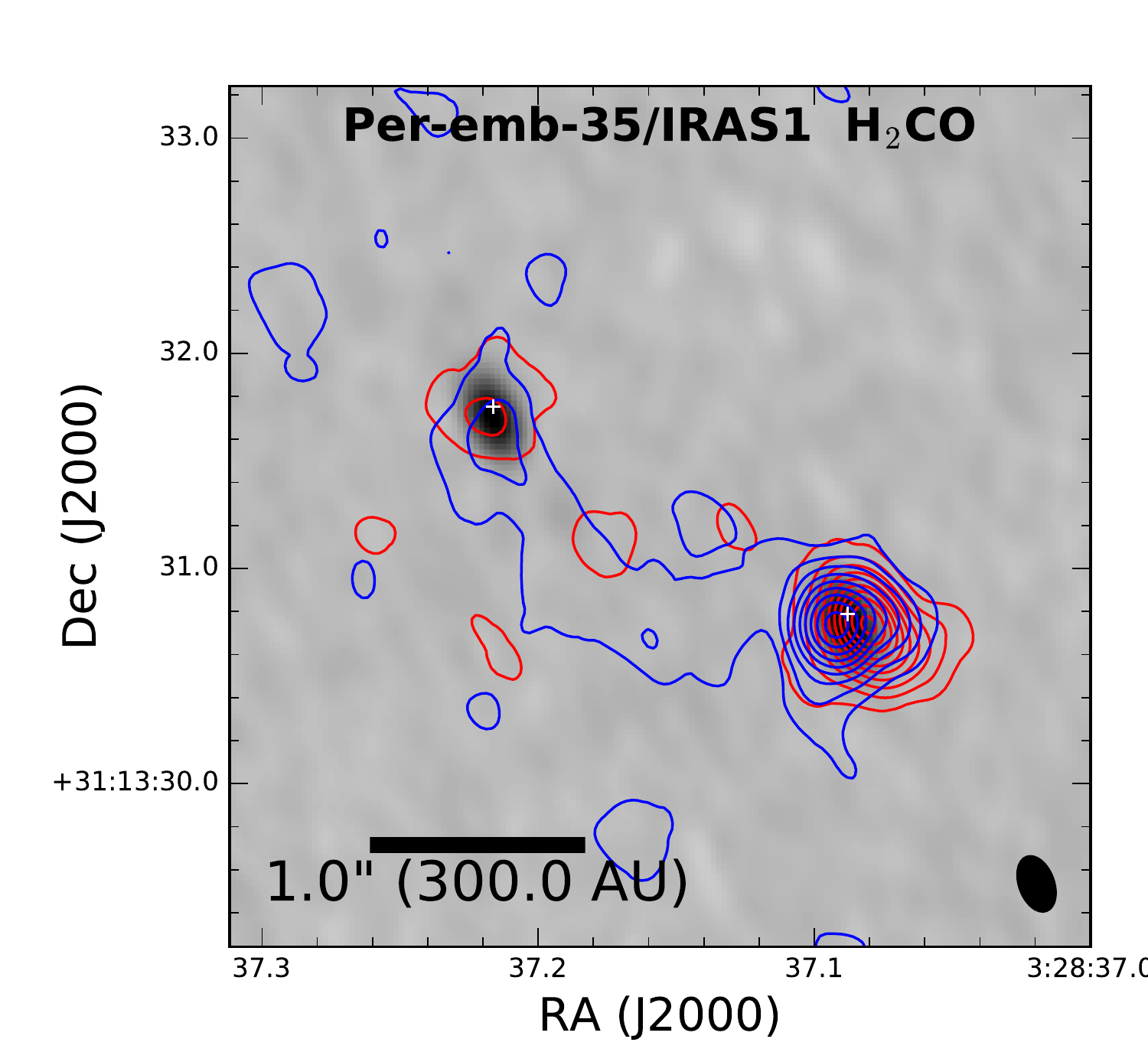}
\end{center}

\caption{Integrated intensity maps of \thco\ (top left panel), \cateo\ (top right panel), SO (bottom left panel), and H$_2$CO (bottom right panel)
toward Per-emb-35 (NGC 1333 IRAS1). The integrated intensity maps are displayed as red and blue contours
corresponding to the integrated intensity of line emission red and blue-shifted with respect
to the system velocity. The contours are overlaid on the 1.3~mm continuum image. The
line emission shows evidence for a velocity gradient consistent with rotation.
The red-shifted contours start at (3,3,3,3)$\sigma$ and increase in (2,1,2,2)$\sigma$ increments and
the blue-shifted contours start at (3,3,3,3)$\sigma$ and increase in (2,1,2,2)$\sigma$ increments, respectively. The values inside the parentheses in the previous sentence
correspond to the \thco, \cateo, SO, and H$_2$CO integrated intensity maps, respectively. The
values for $\sigma_{red}$ and $\sigma_{blue}$ and velocity ranges over which the line
emission was summed can be found in Table 3.
The beam in the images is approximately 0\farcs36$\times$0\farcs26.}
\label{per-emb-35-lines}
\end{figure}
\clearpage

\begin{figure}
\begin{center}
\includegraphics[scale=0.425]{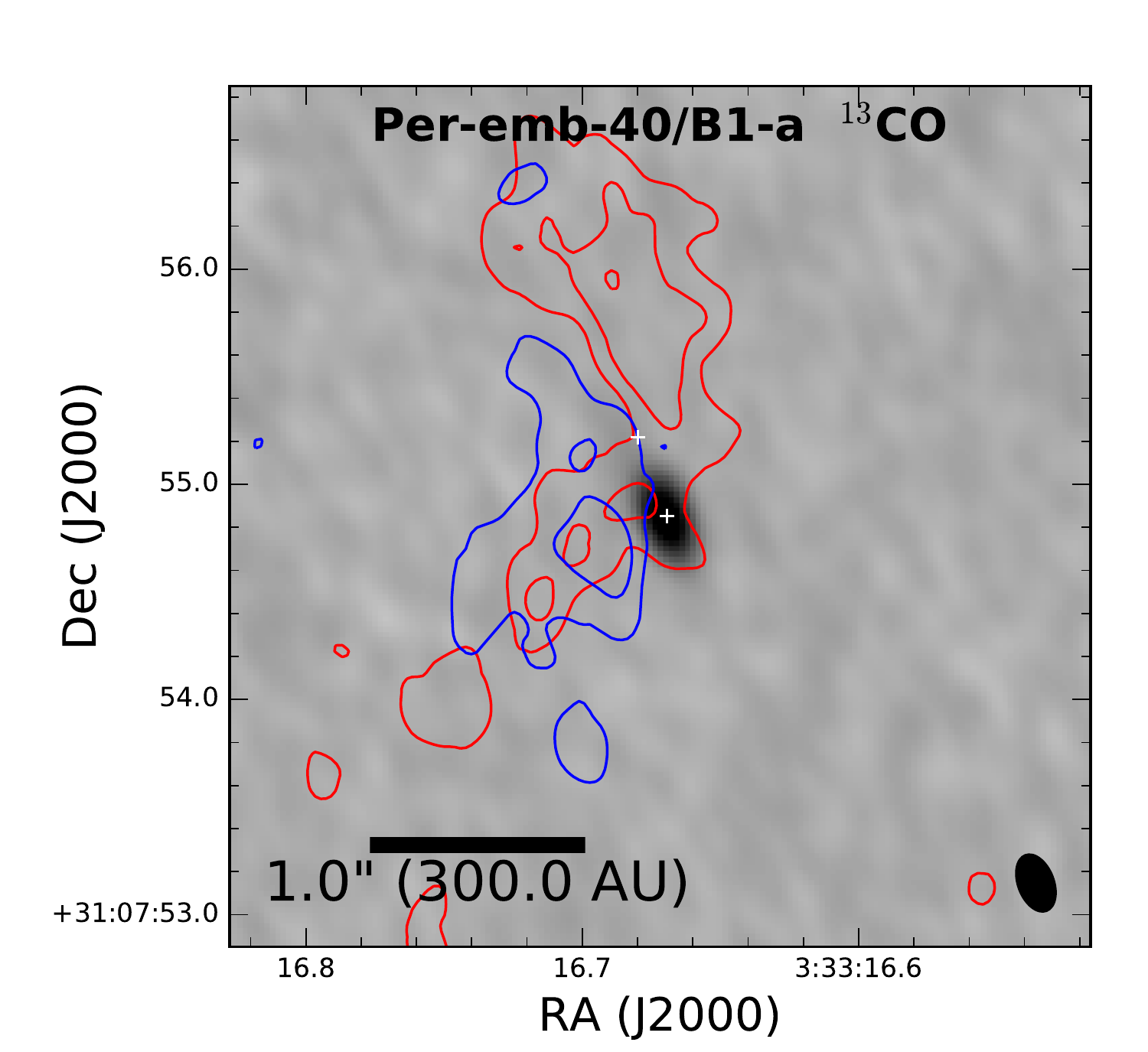}
\includegraphics[scale=0.425]{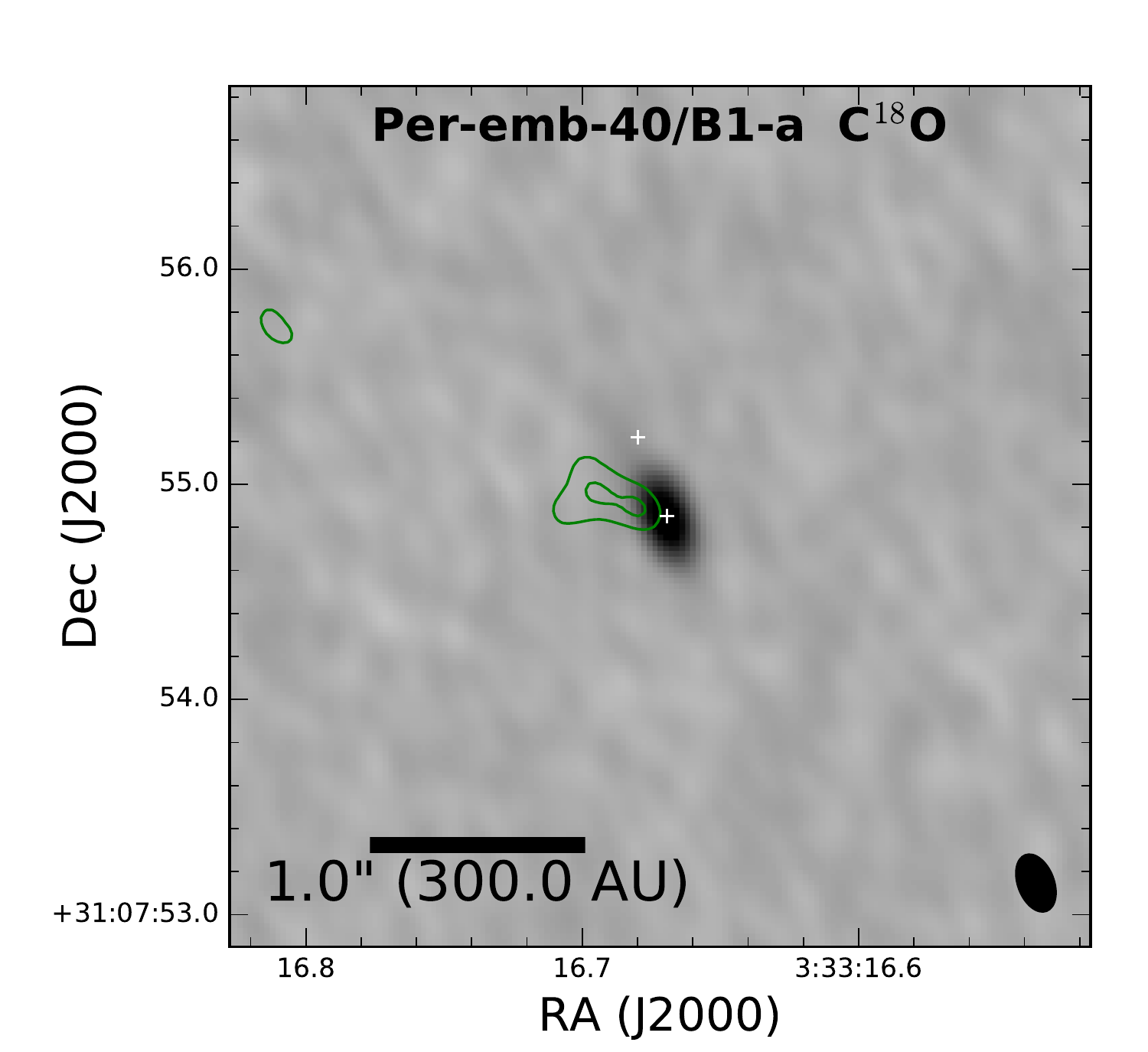}
\end{center}
\caption{Integrated intensity maps of \thco\ (top left panel), \cateo\ (top right panel), SO (bottom left panel), and H$_2$CO (bottom right panel)
toward Per-emb-40 (B1-a). The integrated intensity maps are displayed as red and blue contours
corresponding to the integrated intensity of line emission red and blue-shifted with respect
to the system velocity. The contours are overlaid on the 1.3~mm continuum image. The
line emission shows evidence for a velocity gradient consistent with rotation.
The red-shifted contours start at 3$\sigma$ and increase in 2$\sigma$ increments, and
the blue-shifted contours start at 3$\sigma$ and increase in 2$\sigma$ increments. For
\cateo\ only the line center velocities are plotted, starting at 4$\sigma$ and increasing
by 1$\sigma$ intervals. The values for $\sigma_{red}$ and $\sigma_{blue}$ and velocity ranges over which the line
emission was summed can be found in Table 3.
The beam in the images is approximately 0\farcs36$\times$0\farcs26.}
\label{per-emb-40-lines}
\end{figure}

\begin{figure}
\begin{center}
\includegraphics[scale=0.425]{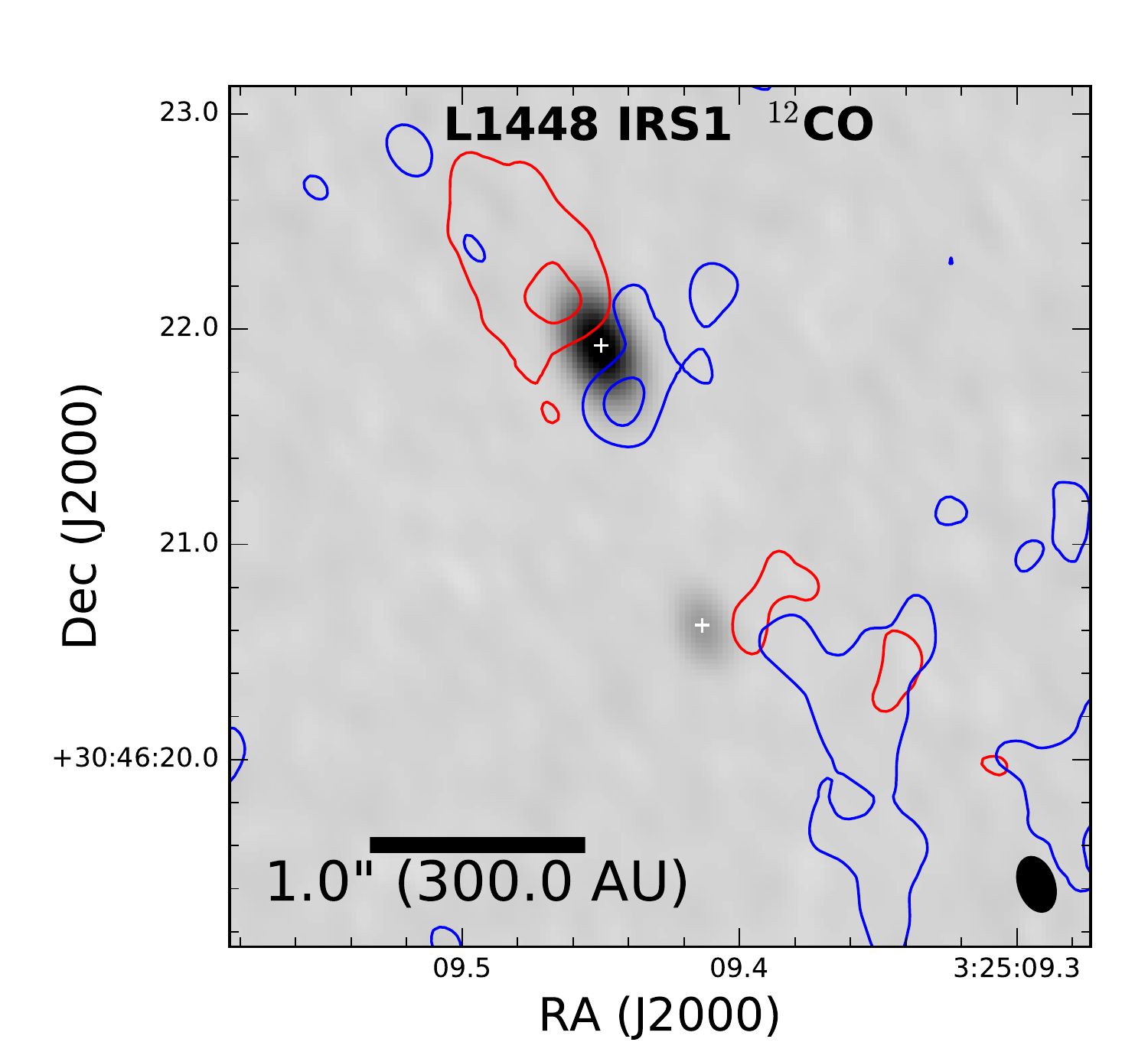}
\end{center}

\caption{Integrated intensity maps of \twco\ 
toward L1448 IRS1. The integrated intensity maps are displayed as red and blue contours
corresponding to the integrated intensity of line emission red and blue-shifted with respect
to the system velocity. The contours are overlaid on the 1.3~mm continuum image. The
line emission shows evidence for a velocity gradient consistent with rotation.
The red-shifted contours start at 3$\sigma$ and increase in 3$\sigma$ increments, and
the blue-shifted contours start at 3$\sigma$ and increase in 3$\sigma$ increments.
The values for $\sigma_{red}$ and $\sigma_{blue}$ and velocity ranges over which the line
emission was summed can be found in Table 3.
The beam in the images is approximately 0\farcs36$\times$0\farcs26.
}
\label{L1448IRS1-lines}
\end{figure}

\begin{figure}
\begin{center}
\includegraphics[scale=0.425]{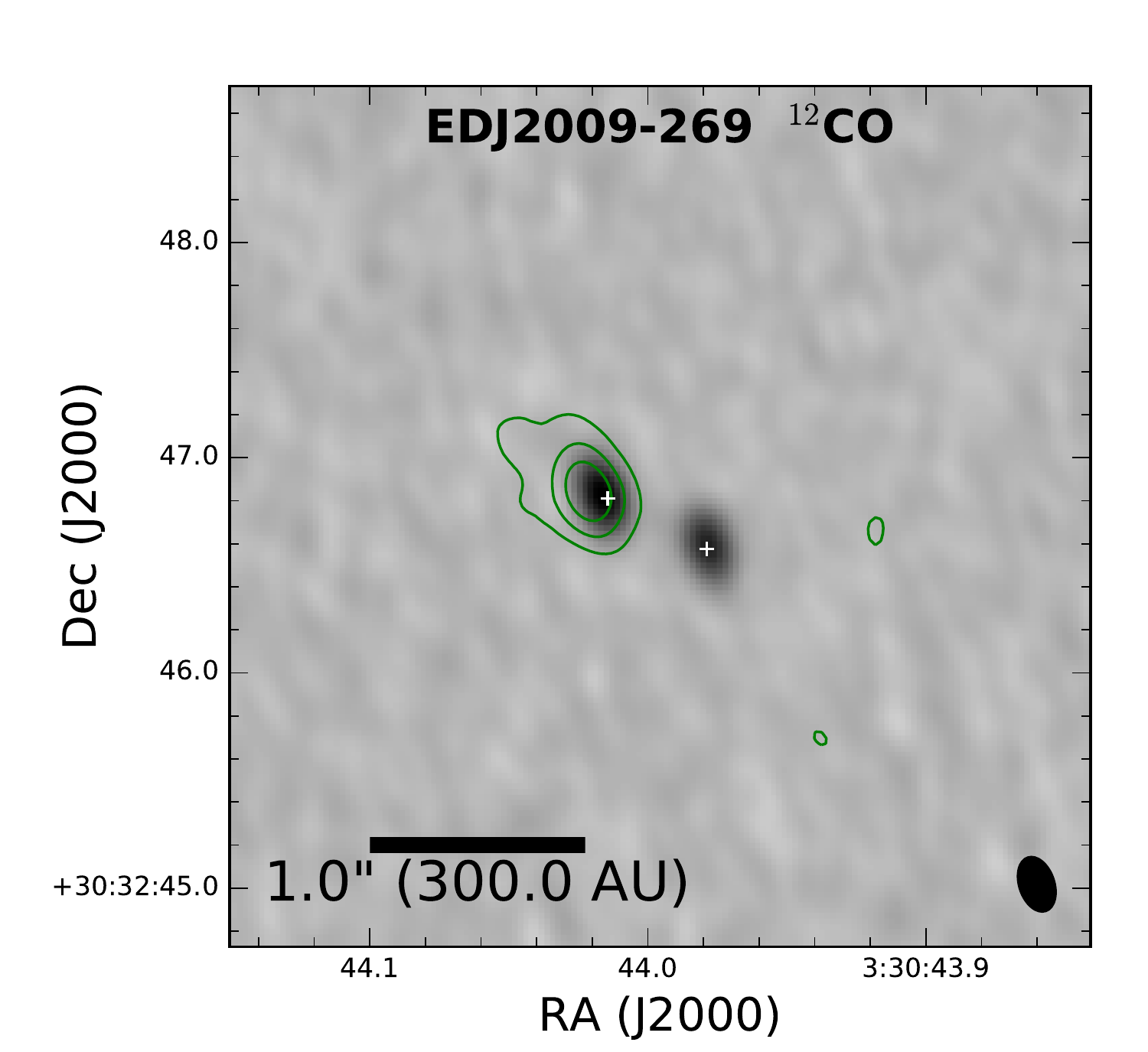}
\end{center}
\caption{Integrated intensity maps of \twco\ 
toward EDJ2009-269. The integrated intensity maps are displayed as red and blue contours
corresponding to the integrated intensity of line emission red and blue-shifted with respect
to the system velocity. The contours are overlaid on the 1.3~mm continuum image. The
line emission shows evidence for a velocity gradient consistent with rotation.
The integrated intensity contours start at 3$\sigma$ and increase in 3$\sigma$ increments.
The values for $\sigma$ and the velocity range over which the line
emission was summed can be found in Table 3.
The beam in the images is approximately 0\farcs36$\times$0\farcs26.}
\label{EDJ2009-269-lines}
\end{figure}

\clearpage
\begin{figure}
\begin{center}
\includegraphics[scale=0.425]{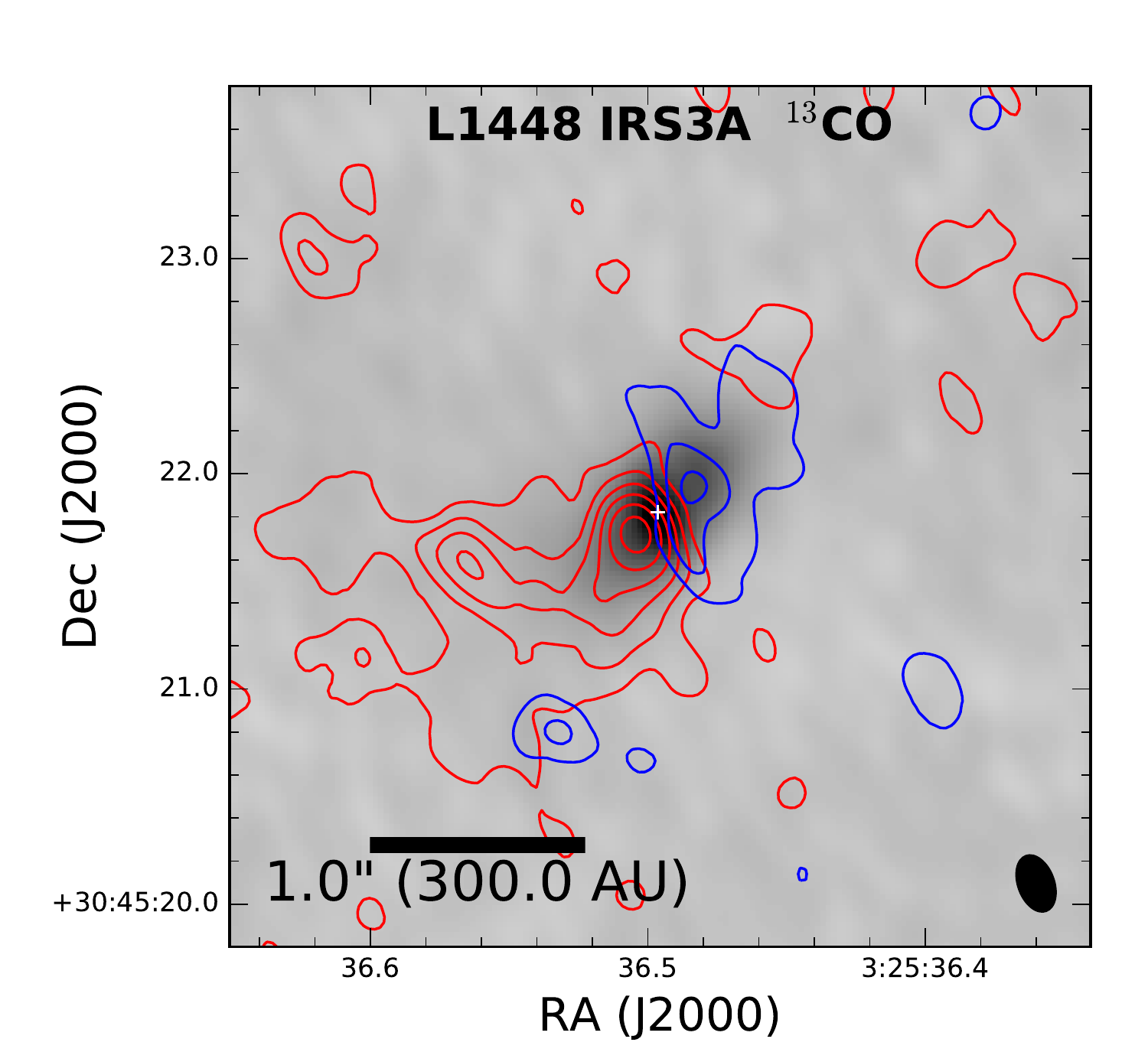}
\includegraphics[scale=0.425]{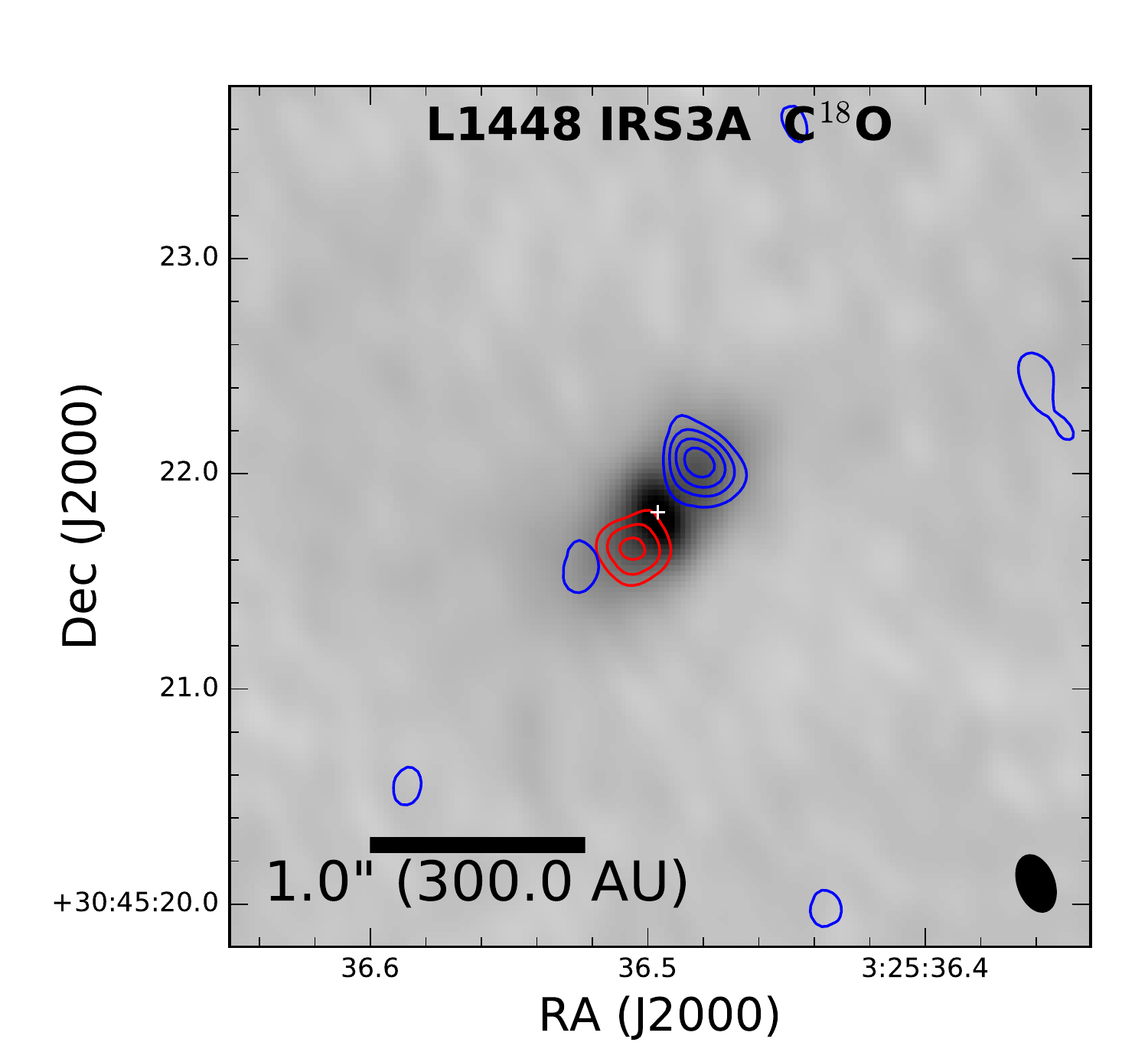}
\includegraphics[scale=0.425]{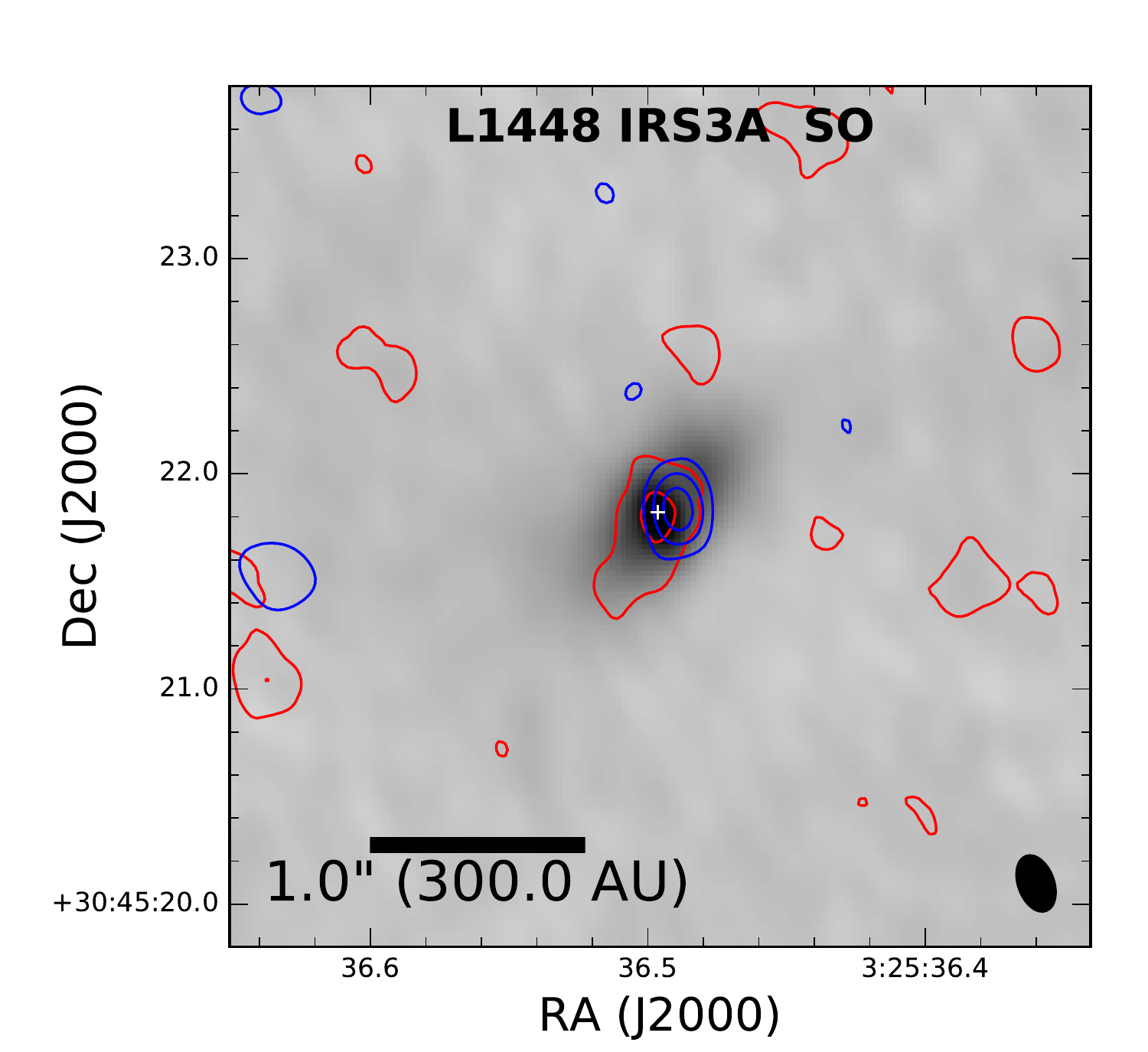}
\includegraphics[scale=0.425]{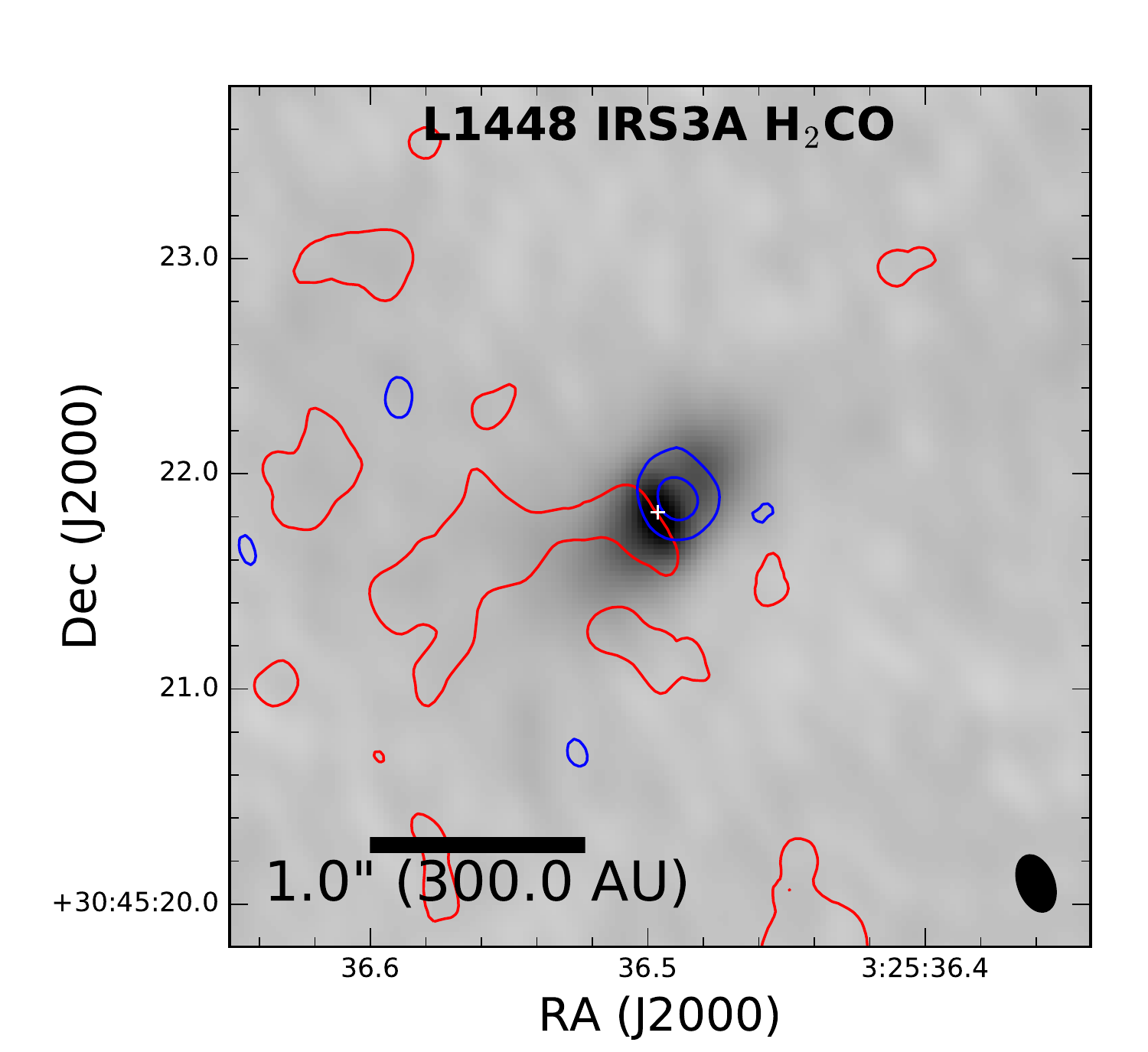}
\end{center}
\caption{Integrated intensity maps of \thco\ (top left panel), \cateo\ (top right panel), SO (bottom left panel), and H$_2$CO (bottom right panel)
toward L1448 IRS3A. The integrated intensity maps are displayed as red and blue contours
corresponding to the integrated intensity of line emission red and blue-shifted with respect
to the system velocity. The contours are overlaid on the 1.3~mm continuum image. The
line emission shows evidence for a velocity gradient consistent with rotation.
The red-shifted contours start at (3,4,3,3)$\sigma$ and increase in (2,1,2,3)$\sigma$ increments, and
the blue-shifted contours start at (3,3,3,3)$\sigma$ and increase in (2,1,2,3)$\sigma$ increments.
The values inside the parentheses in the previous sentence correspond to the \thco, \cateo, SO, and H$_2$CO integrated intensity maps, respectively. The
values for $\sigma_{red}$ and $\sigma_{blue}$ and velocity ranges over which the line
emission was summed can be found in Table 3.
The beam in the images is approximately 0\farcs36$\times$0\farcs26.}
\label{L1448IRS3A-lines}
\end{figure}

\clearpage
\begin{figure}
\begin{center}
\includegraphics[scale=0.425]{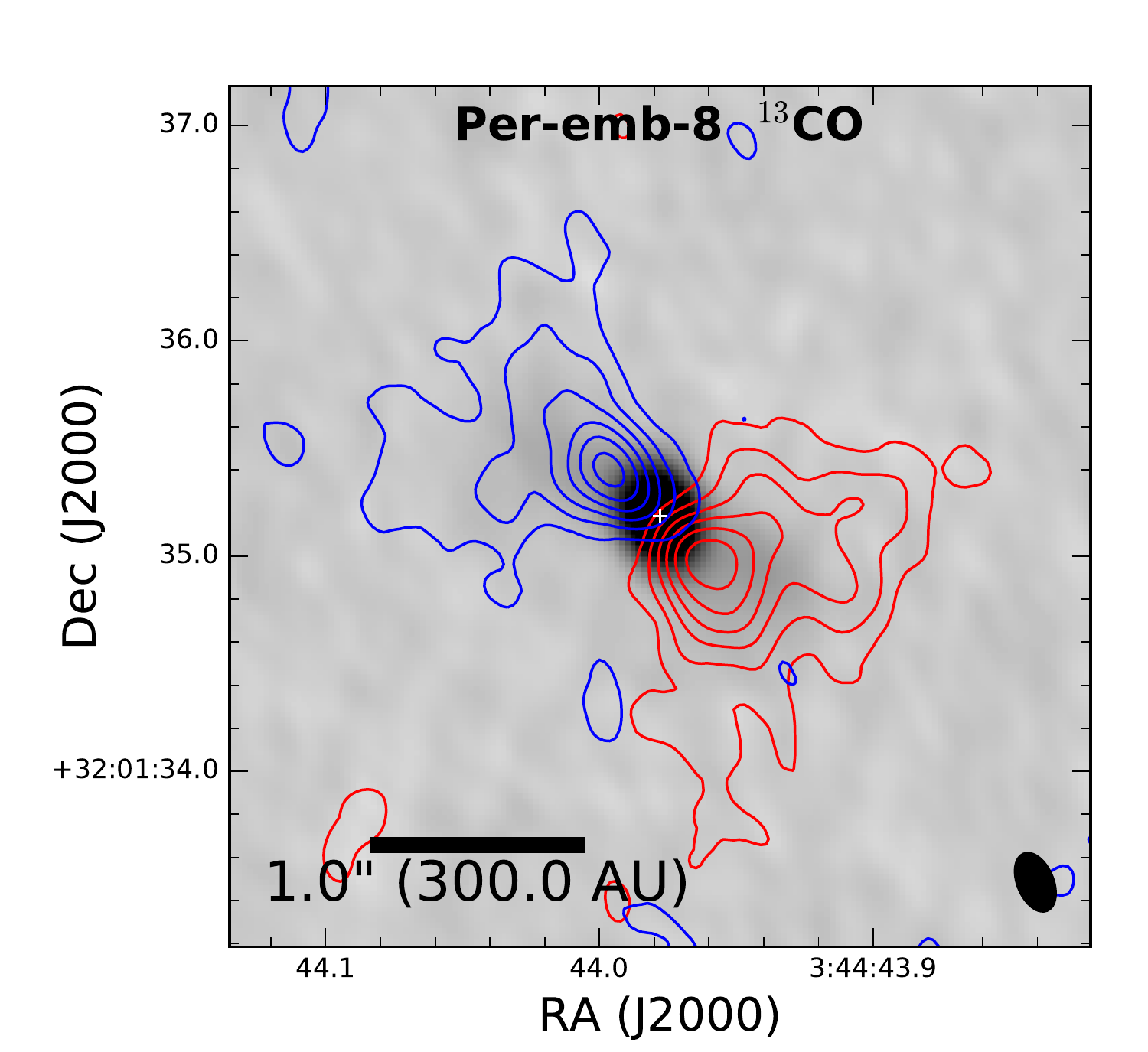}
\includegraphics[scale=0.425]{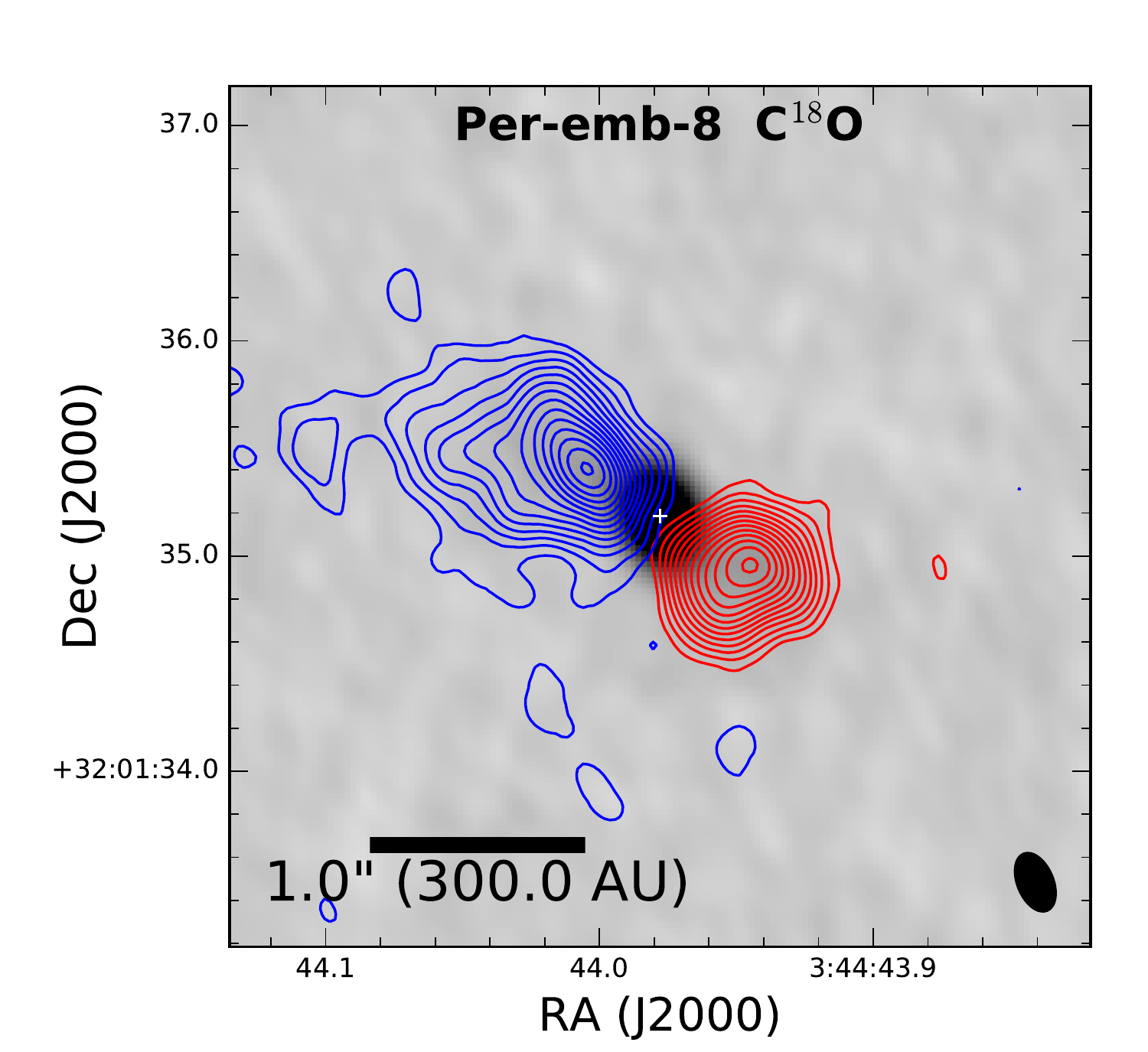}
\includegraphics[scale=0.425]{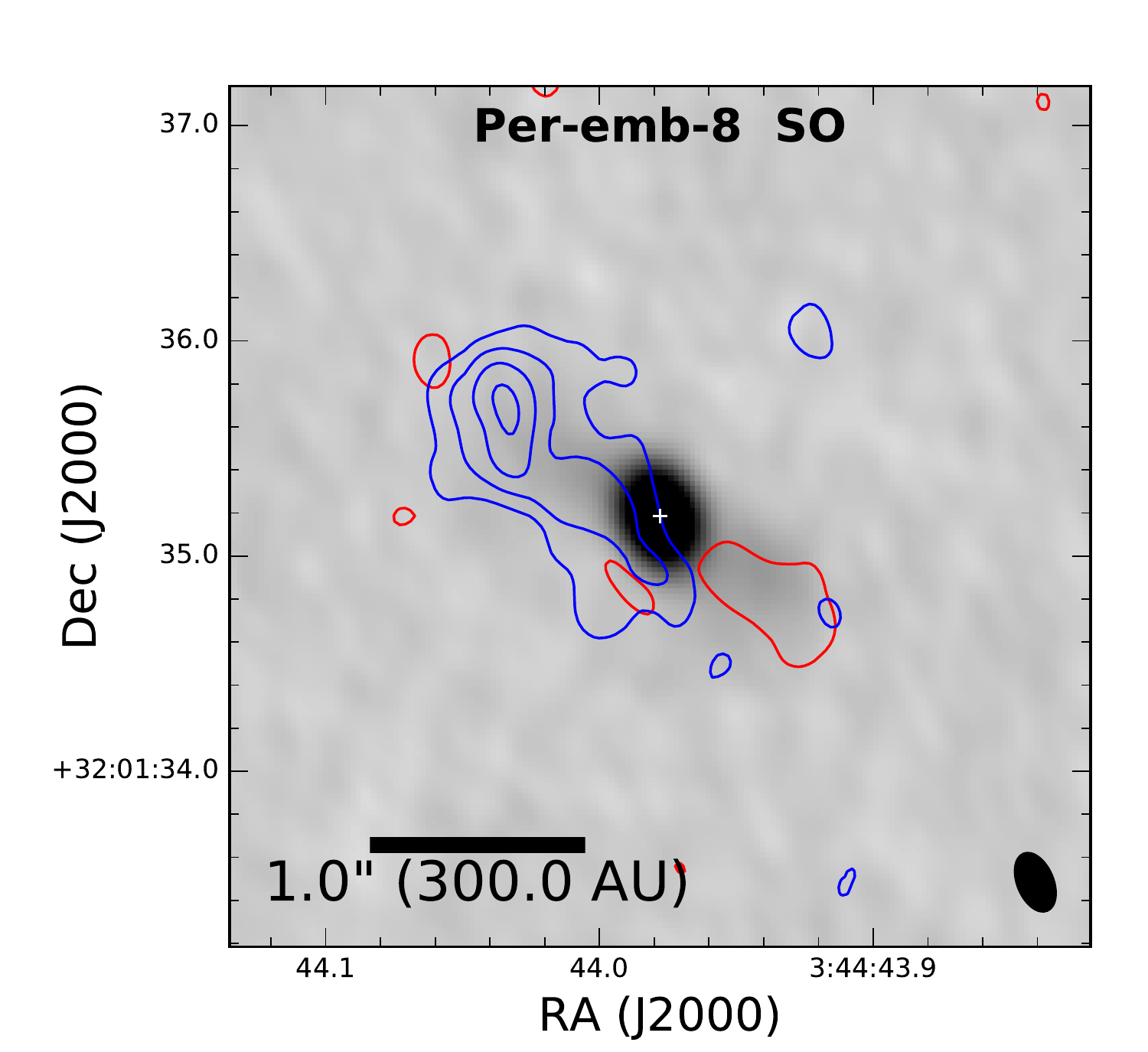}
\includegraphics[scale=0.425]{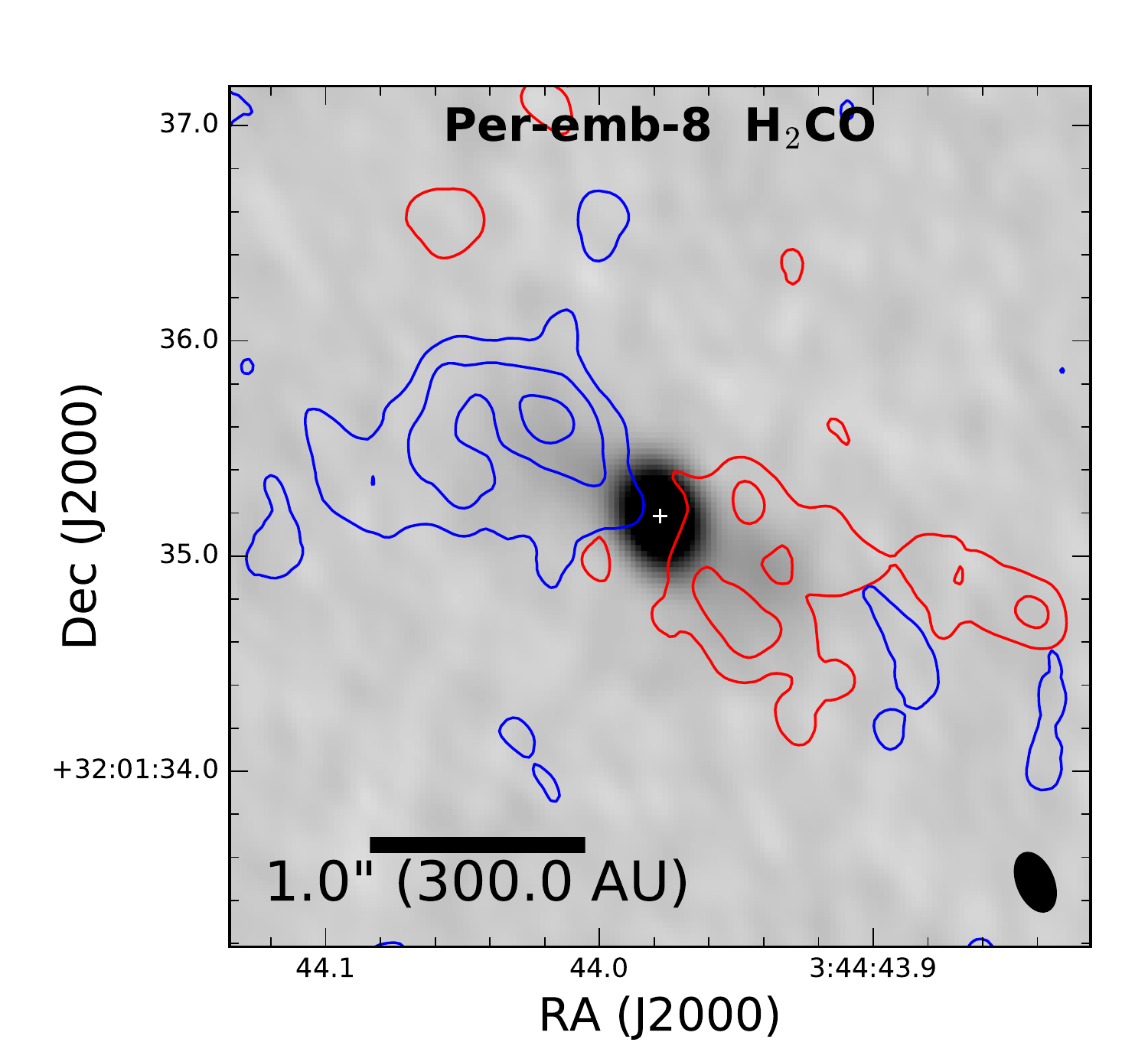}
\end{center}
\caption{Integrated intensity maps of \thco\ (top left panel), \cateo\ (top right panel), SO (bottom left panel), and H$_2$CO (bottom right panel)
toward Per-emb-8. The integrated intensity maps are displayed as red and blue contours
corresponding to the integrated intensity of line emission red and blue-shifted with respect
to the system velocity. The contours are overlaid on the 1.3~mm continuum image. The
line emission shows evidence for a velocity gradient consistent with rotation.
The red-shifted contours start at (3,4,3,3)$\sigma$ and increase in (2,1,2,2)$\sigma$ increments, and
the blue-shifted contours start at (3,3,3,3)$\sigma$ and increase in (2,1,2,2)$\sigma$ increments.
The values inside the parentheses in the previous sentence correspond to the \thco, \cateo, SO, and H$_2$CO integrated intensity maps, respectively. The
values for $\sigma_{red}$ and $\sigma_{blue}$ and velocity ranges over which the line
emission was summed can be found in Table 3.
The beam in the images is approximately 0\farcs36$\times$0\farcs26.}
\label{per-emb-8-lines}
\end{figure}

\clearpage
\begin{figure}
\begin{center}
\includegraphics[scale=0.425]{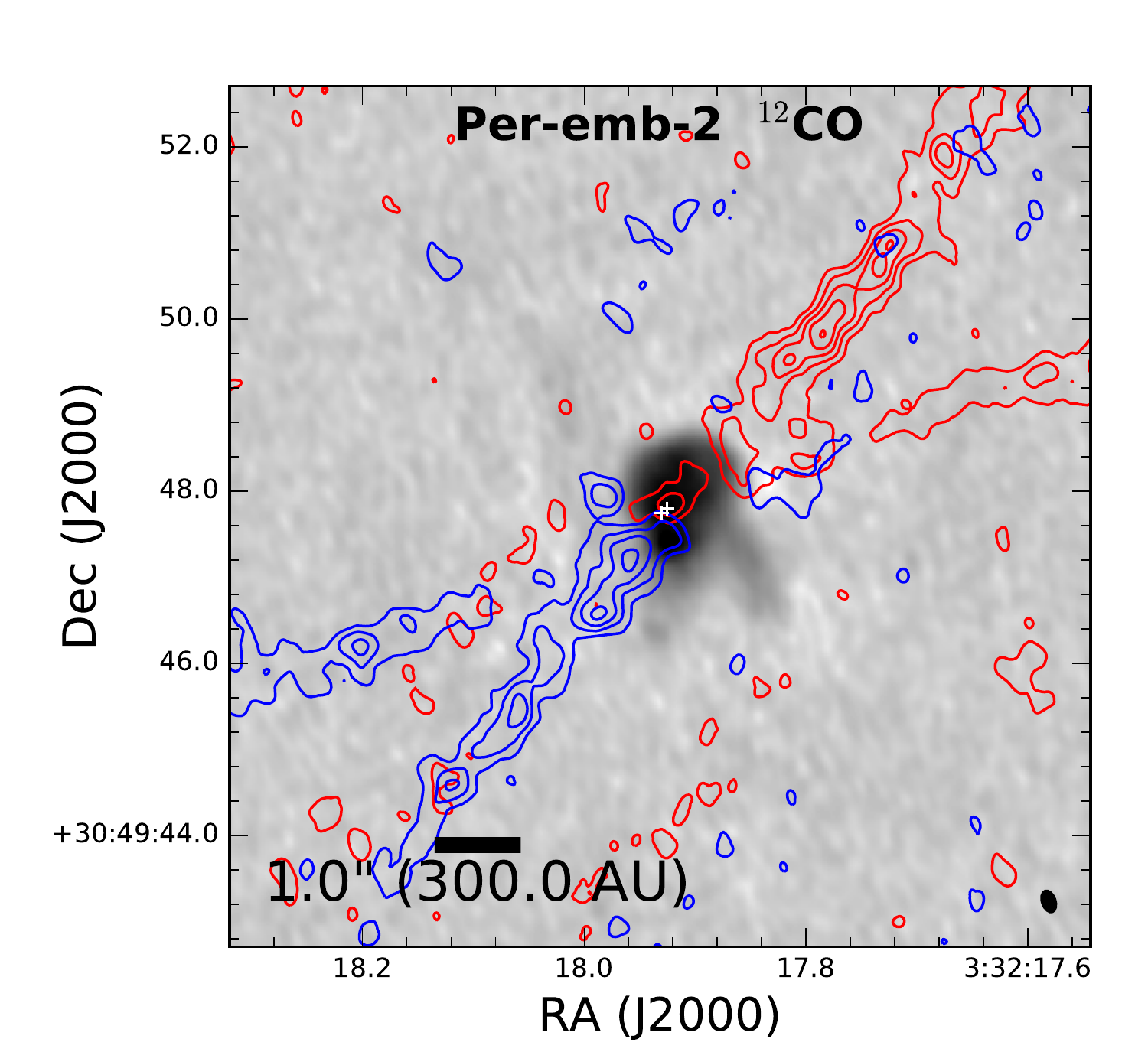}
\includegraphics[scale=0.425]{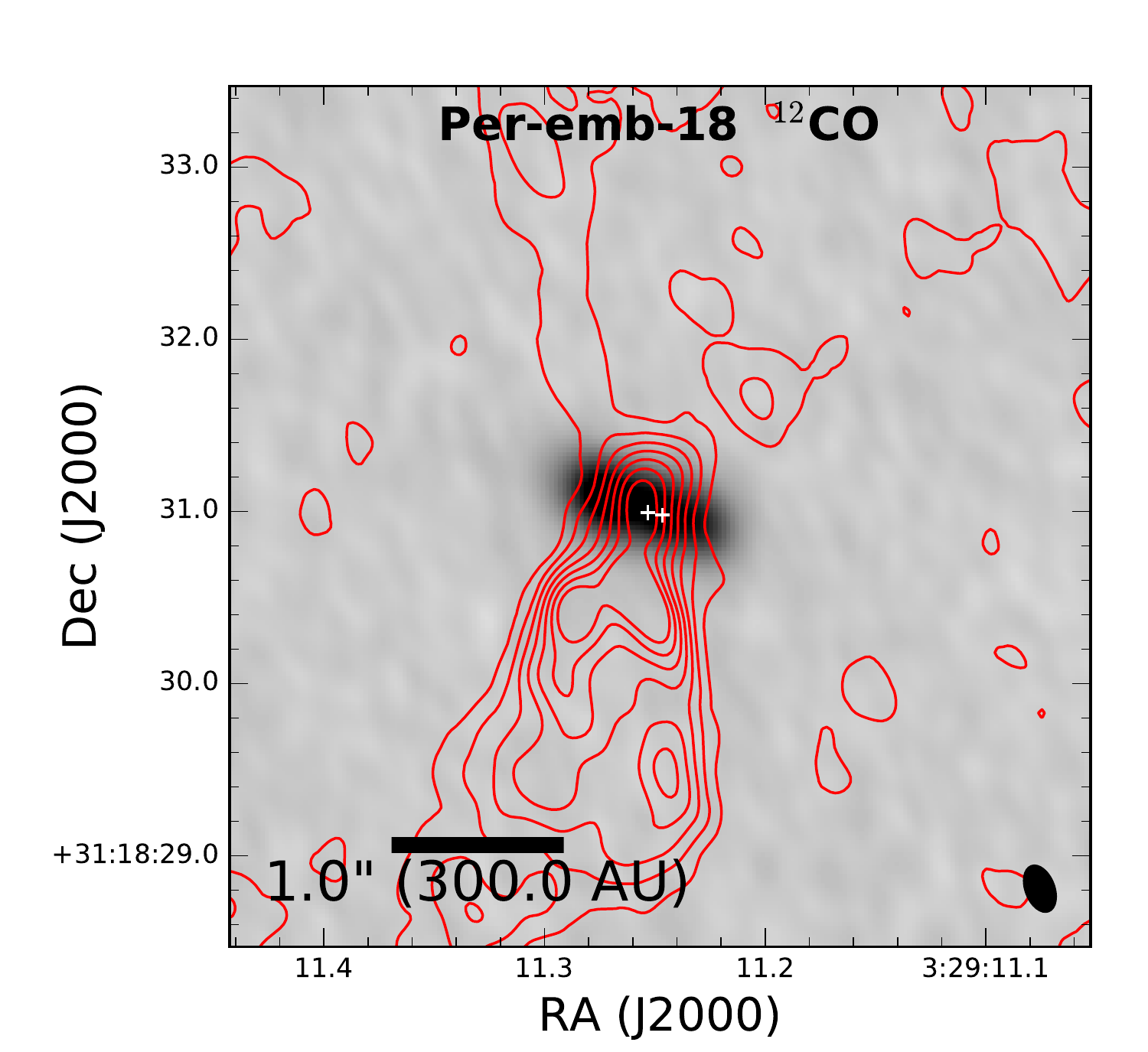}
\includegraphics[scale=0.425]{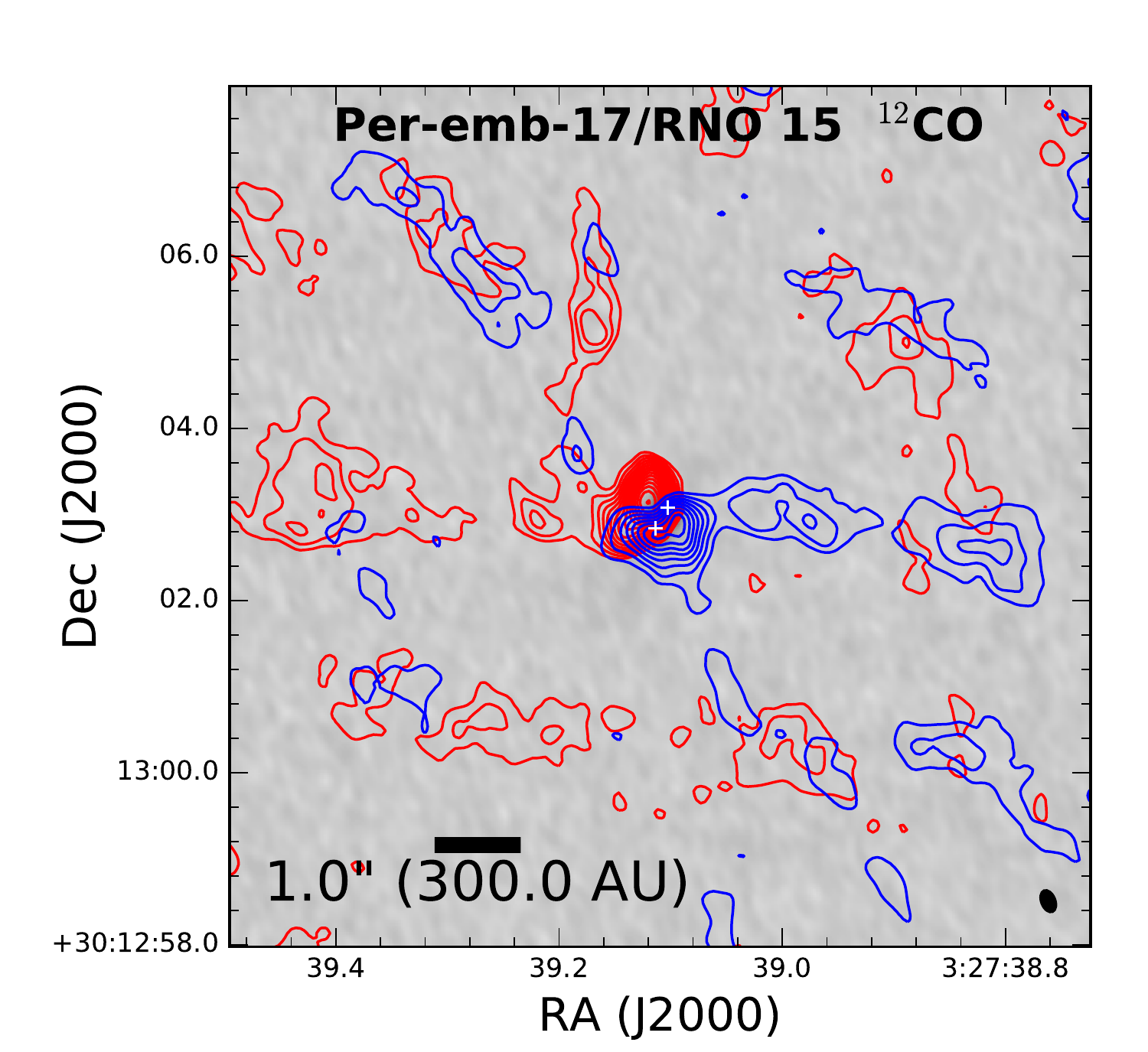}
\includegraphics[scale=0.425]{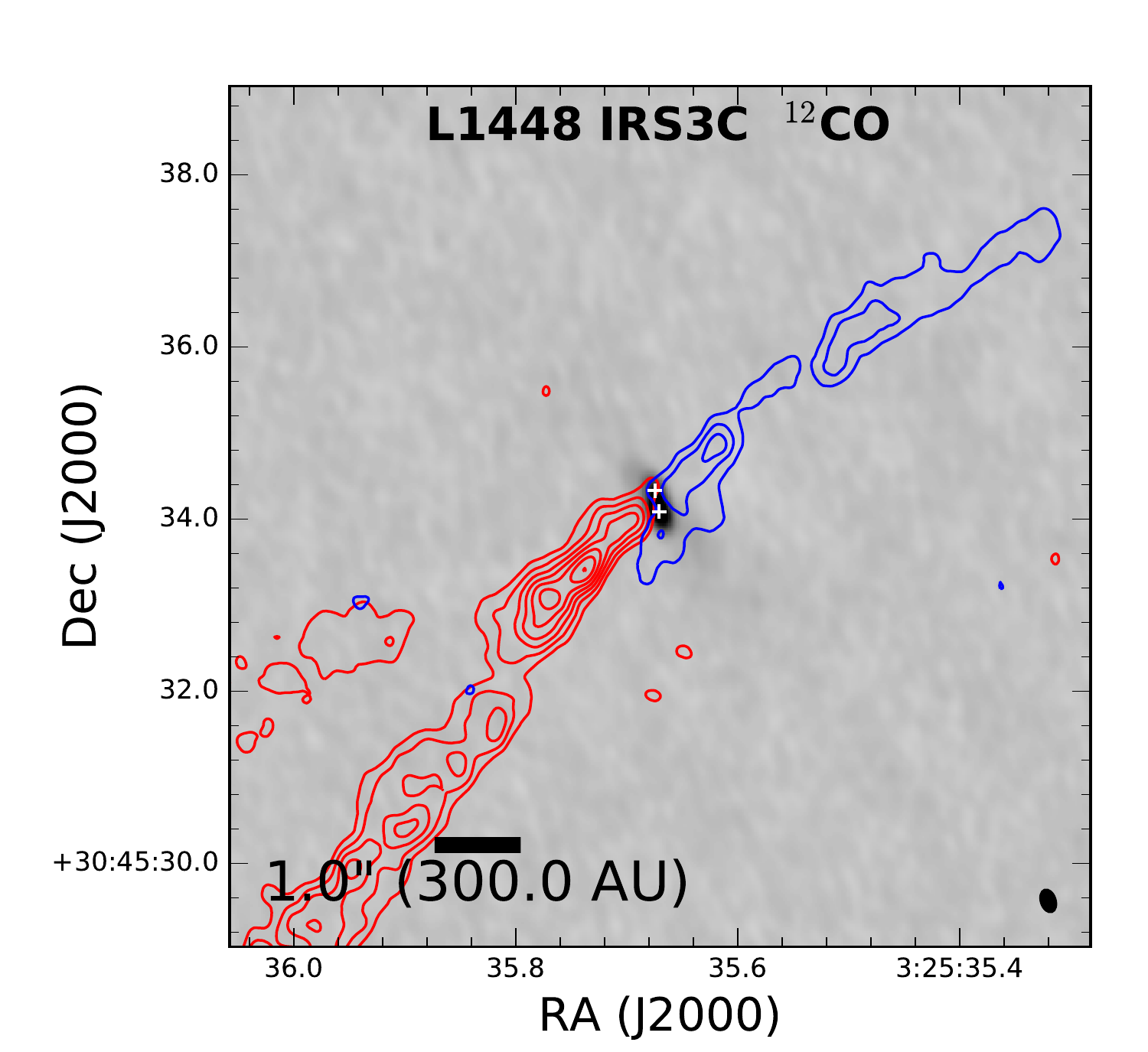}
\includegraphics[scale=0.425]{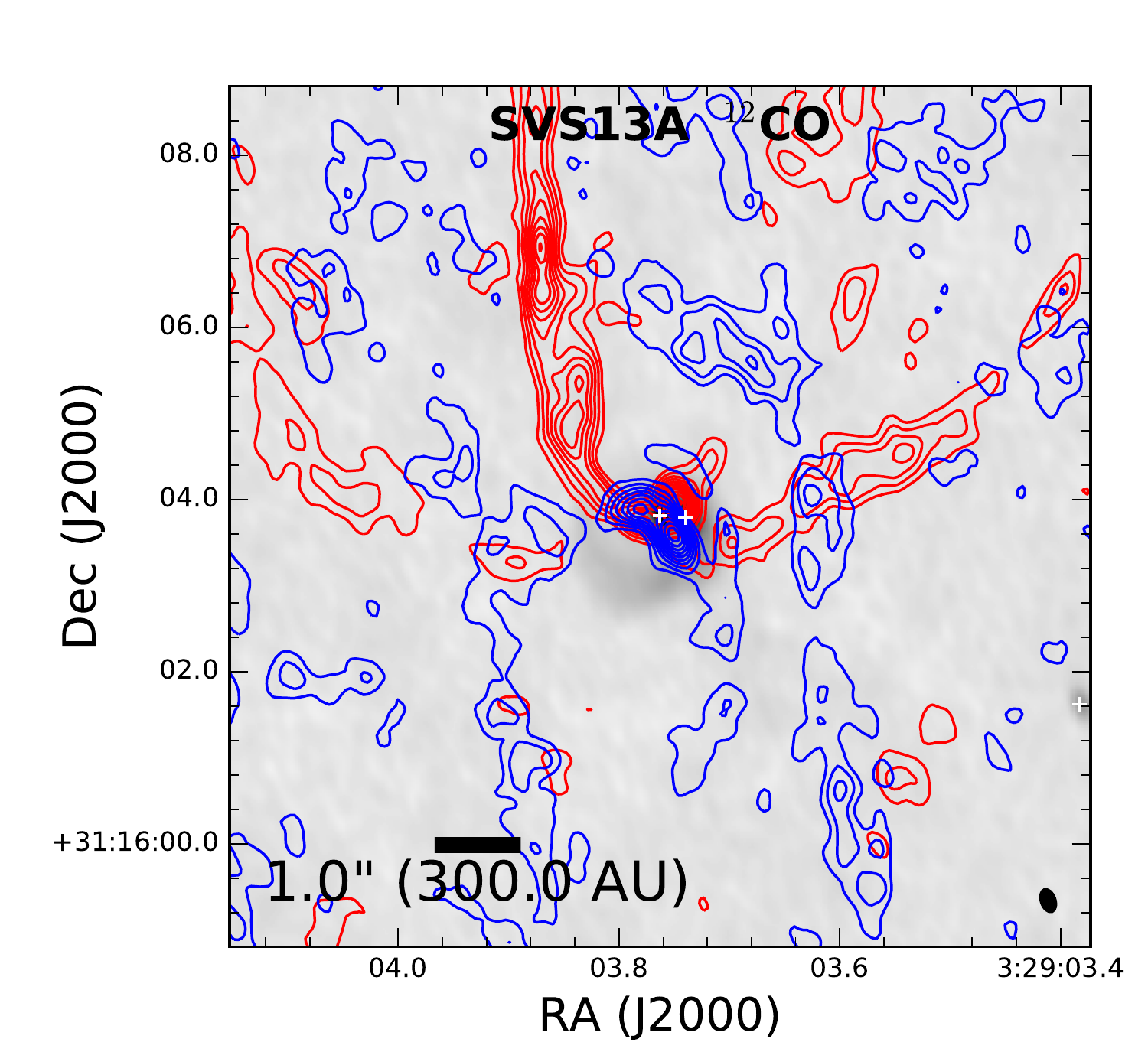}
\includegraphics[scale=0.425]{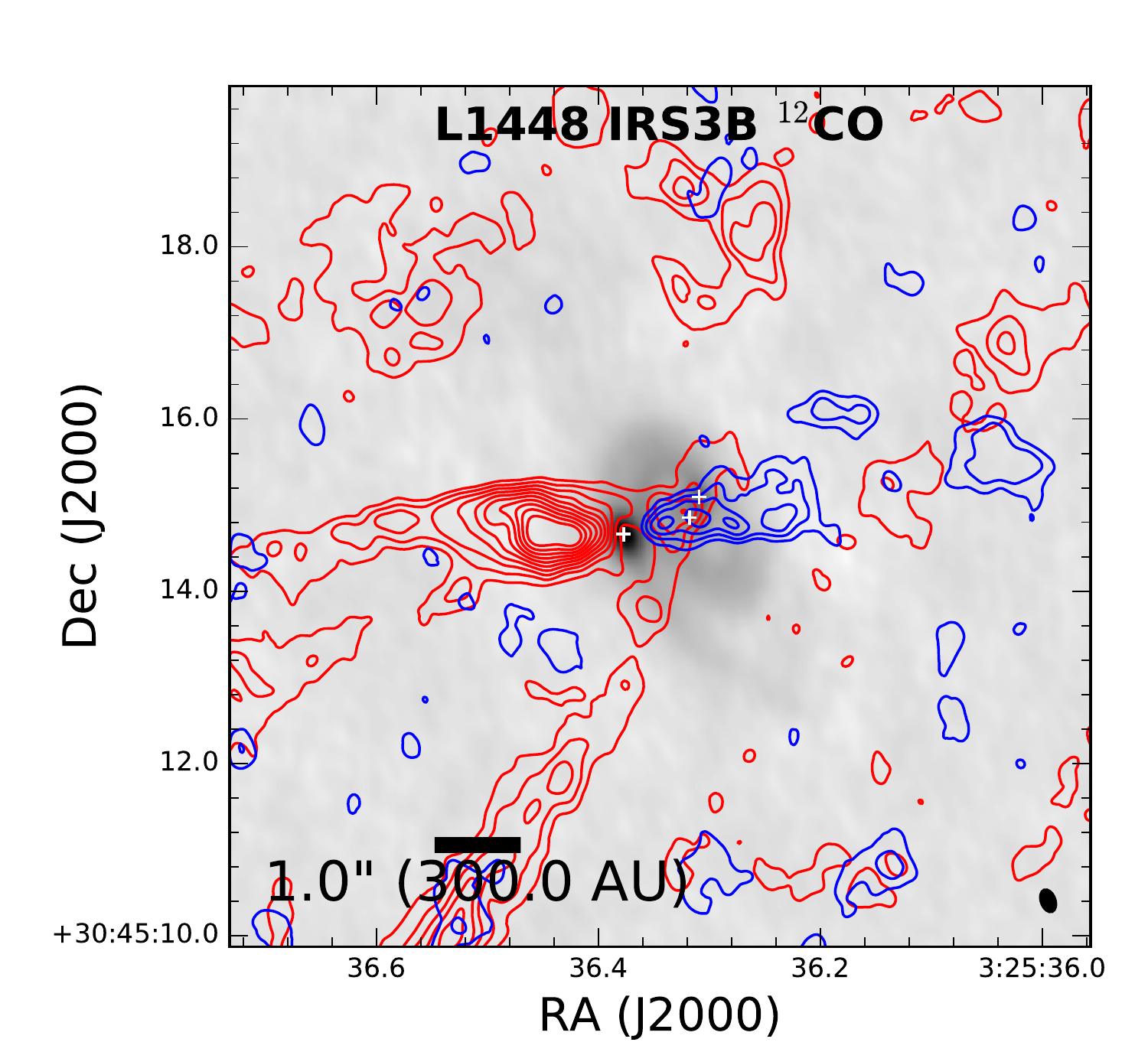}
\end{center}
\caption{ALMA \twco\ integrated intensity maps toward all protostars where \twco\ appears to trace outflowing
gas from the protostellar system. The integrated intensity maps are displayed as red and blue contours
corresponding to the integrated intensity of line emission red and blue-shifted with respect
to the system velocity. The contours are overlaid on the 1.3~mm continuum image. The
line emission shows evidence for a velocity gradient consistent with rotation.
The red-shifted contours start at X$\sigma$ and increase in X$\sigma$ increments, and
the blue-shifted contours start at X$\sigma$ and increase in X$\sigma$ increments.
The values for $\sigma_{red}$ and $\sigma_{blue}$ and velocity ranges over which the line
emission was summed can be found in Table 3.
The beam in the images is approximately 0\farcs36$\times$0\farcs26.}
\label{outflows-1}
\end{figure}
\clearpage

\begin{figure}
\begin{center}
\includegraphics[scale=0.425]{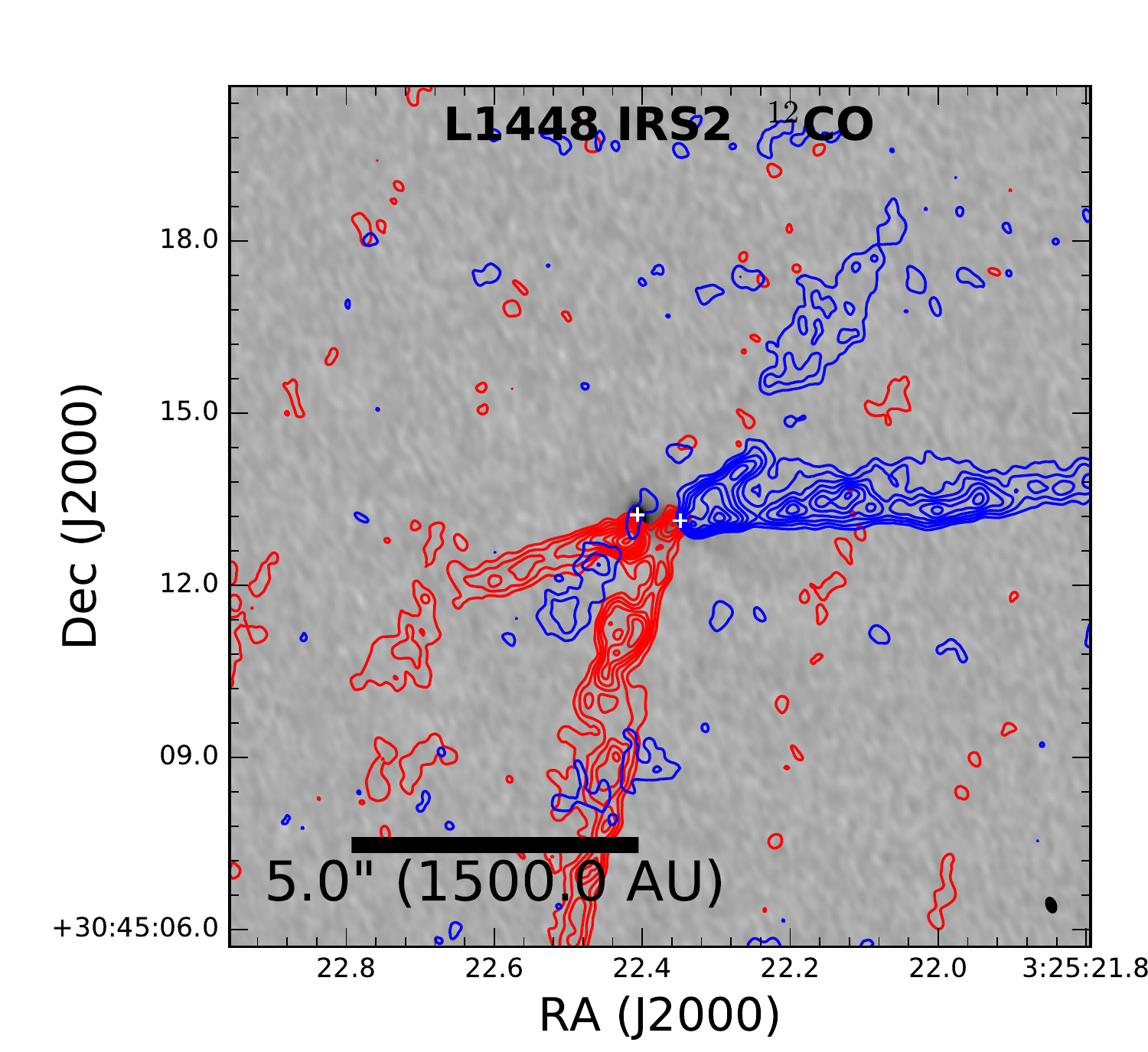}
\includegraphics[scale=0.425]{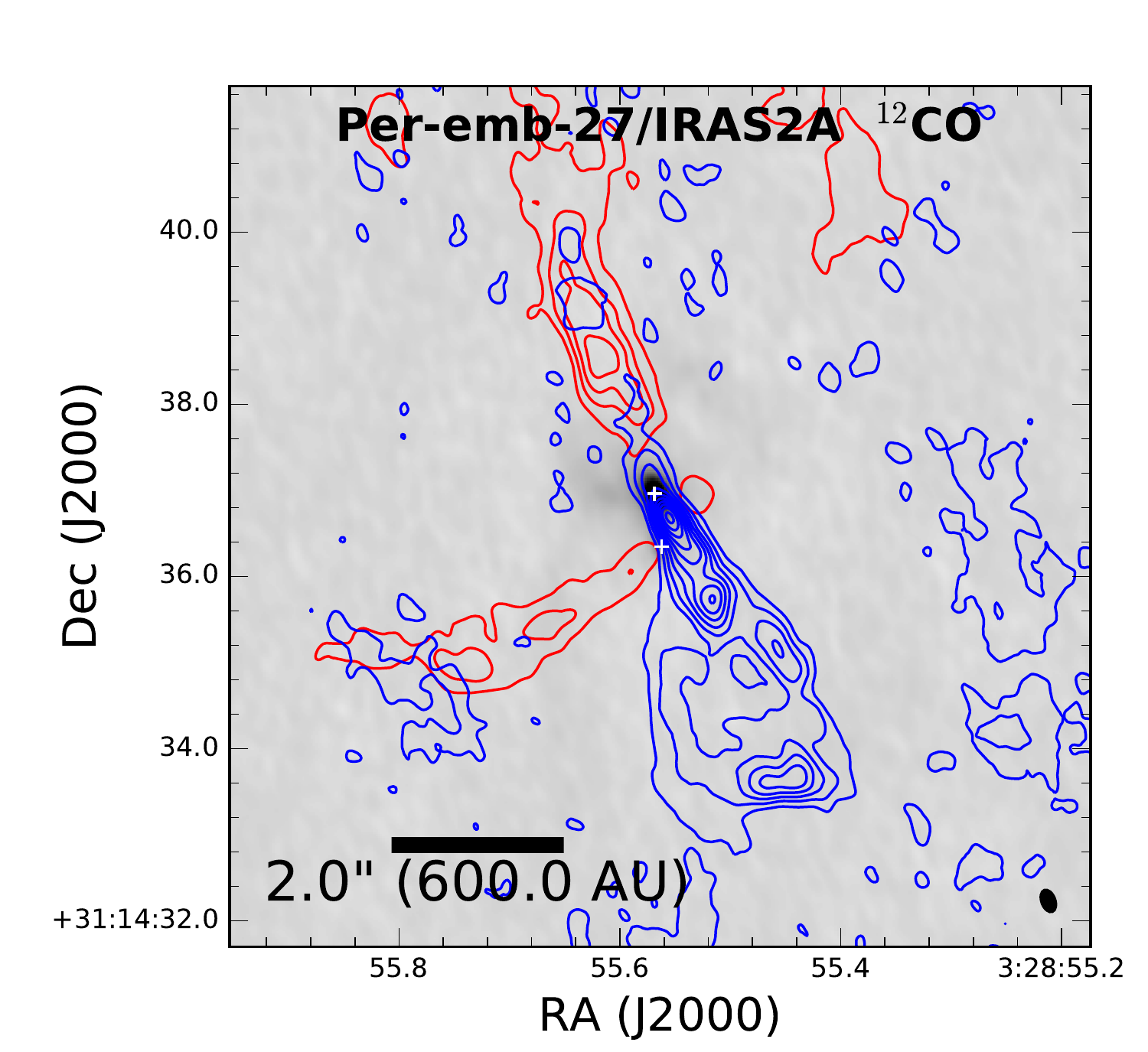}
\includegraphics[scale=0.425]{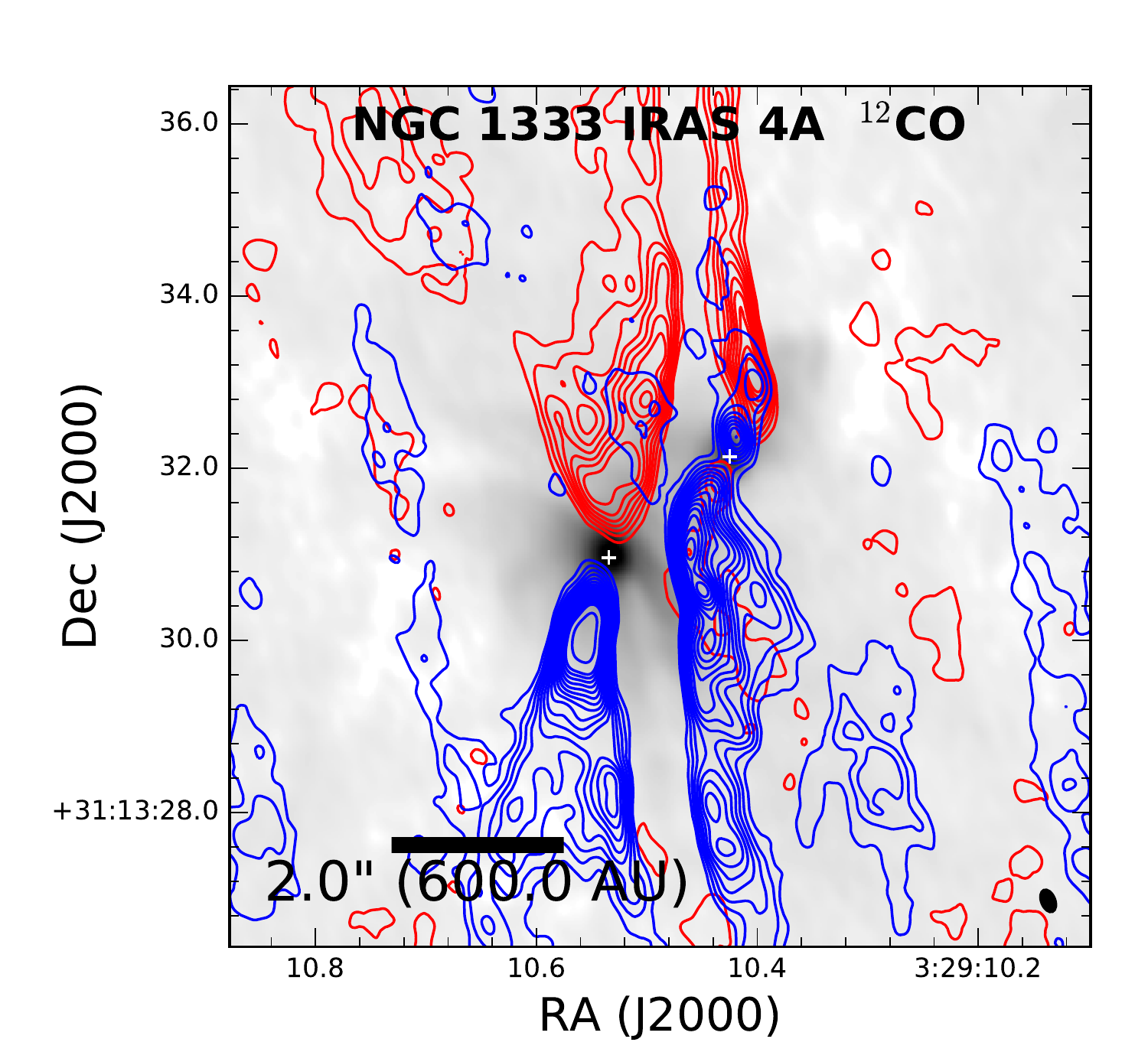}
\includegraphics[scale=0.425]{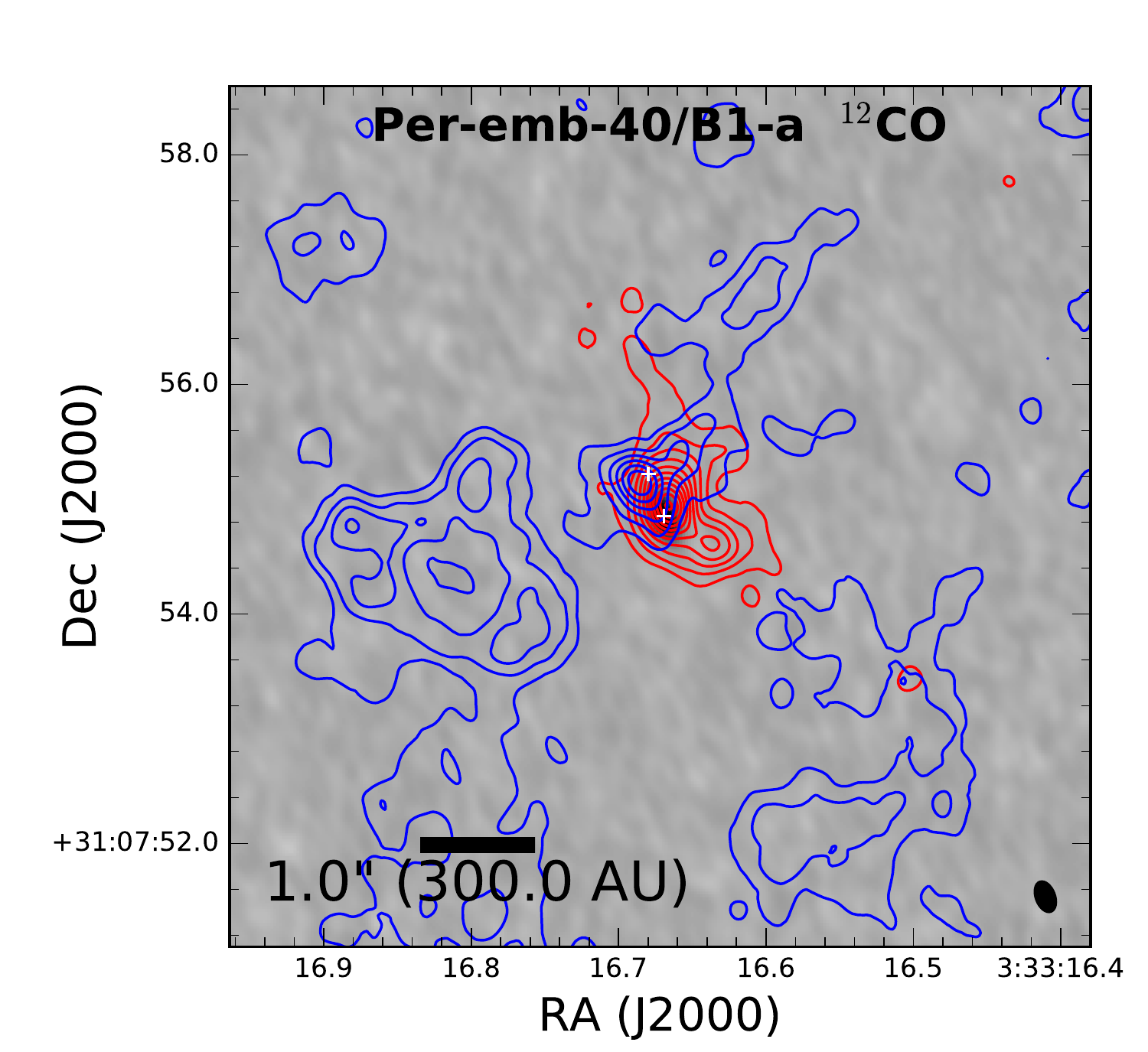}
\includegraphics[scale=0.425]{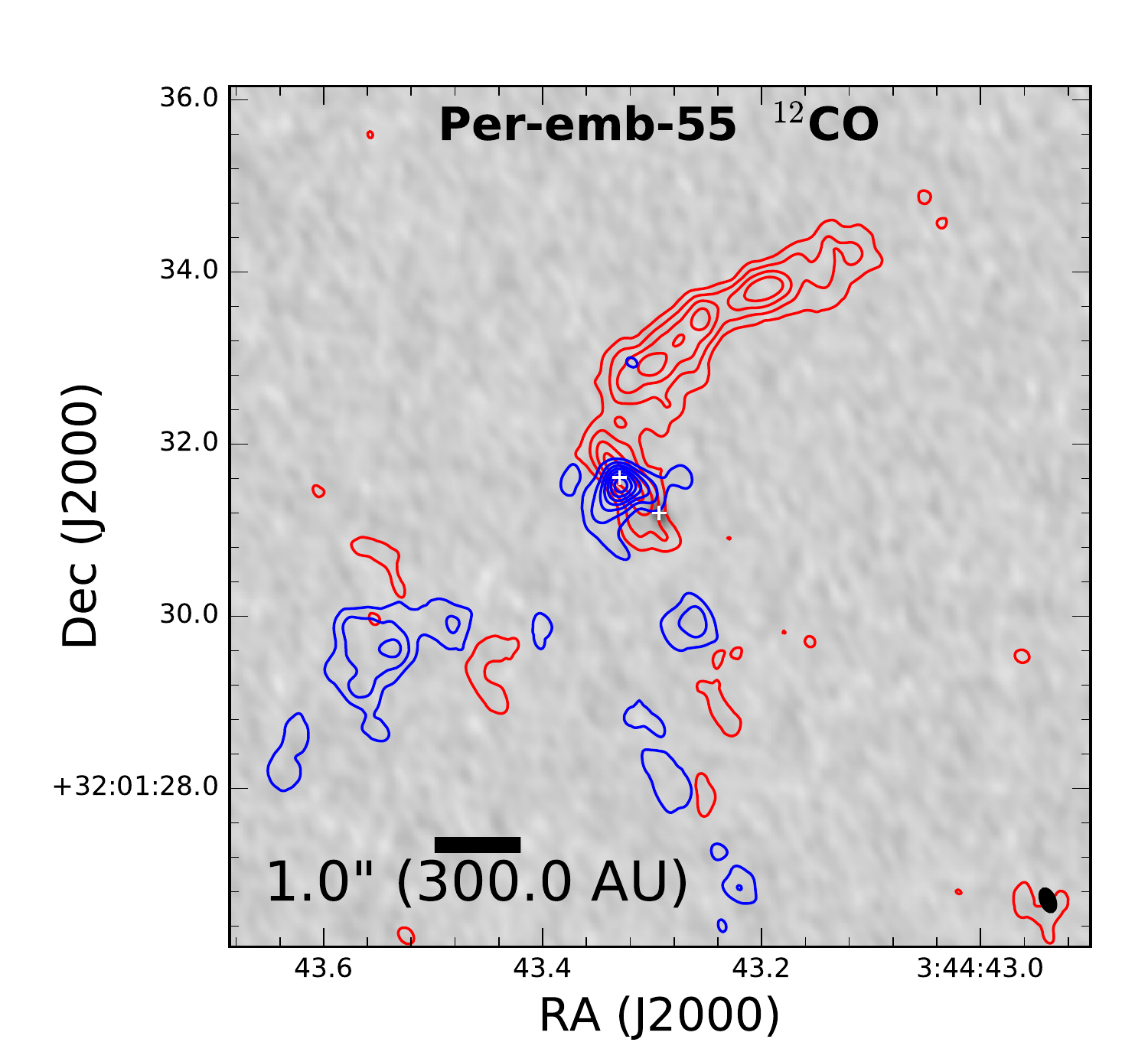}
\includegraphics[scale=0.425]{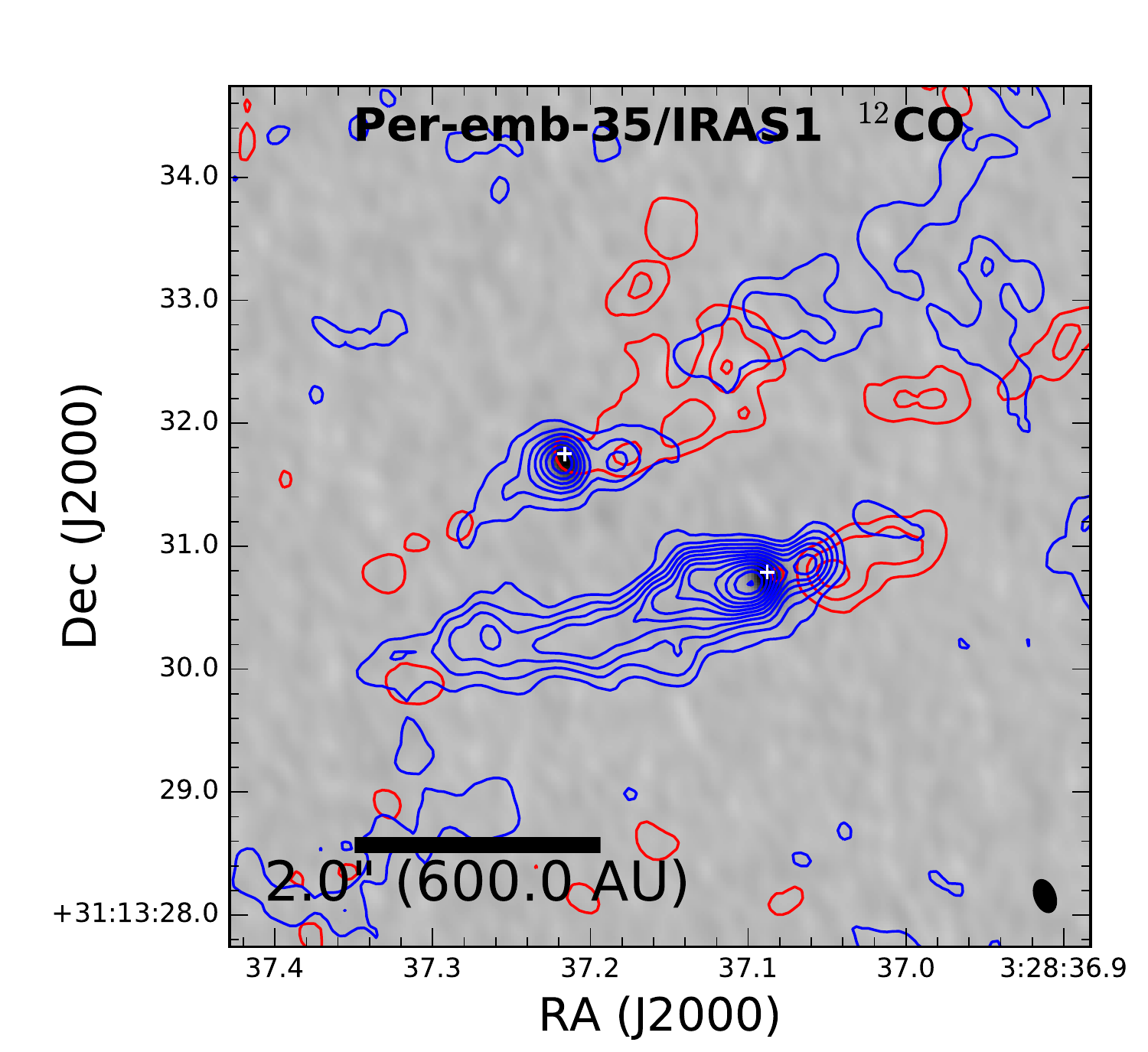}
\end{center}
\caption{Same as Figure \ref{outflows-1}.}
\label{outflows-2}
\end{figure}

\begin{figure}
\begin{center}
\includegraphics[scale=0.425]{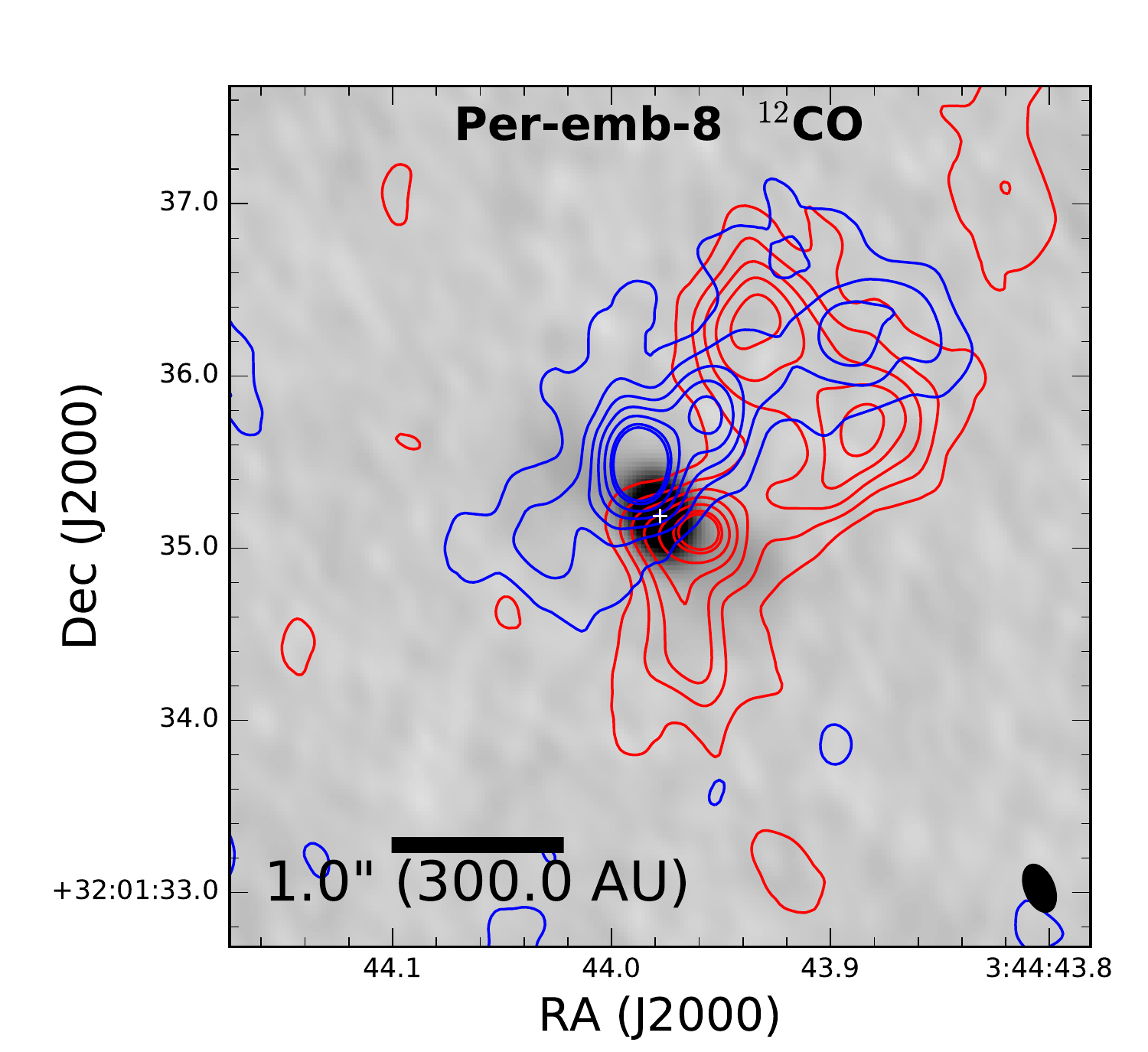}
\includegraphics[scale=0.425]{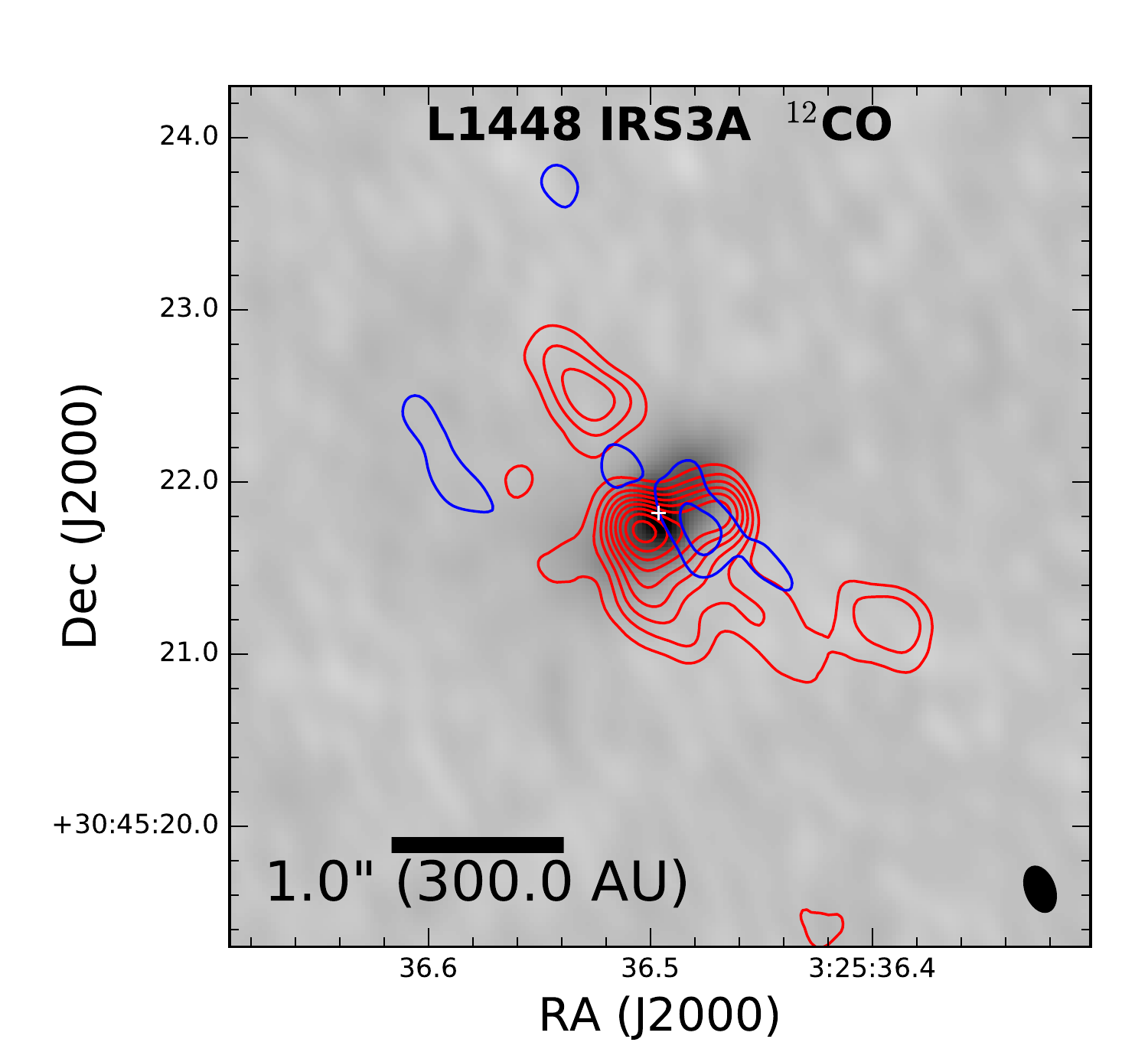}
\end{center}
\caption{Same as Figure \ref{outflows-1}.}
\label{outflows-3}
\end{figure}

\begin{figure}
\begin{center}
\includegraphics[scale=0.35]{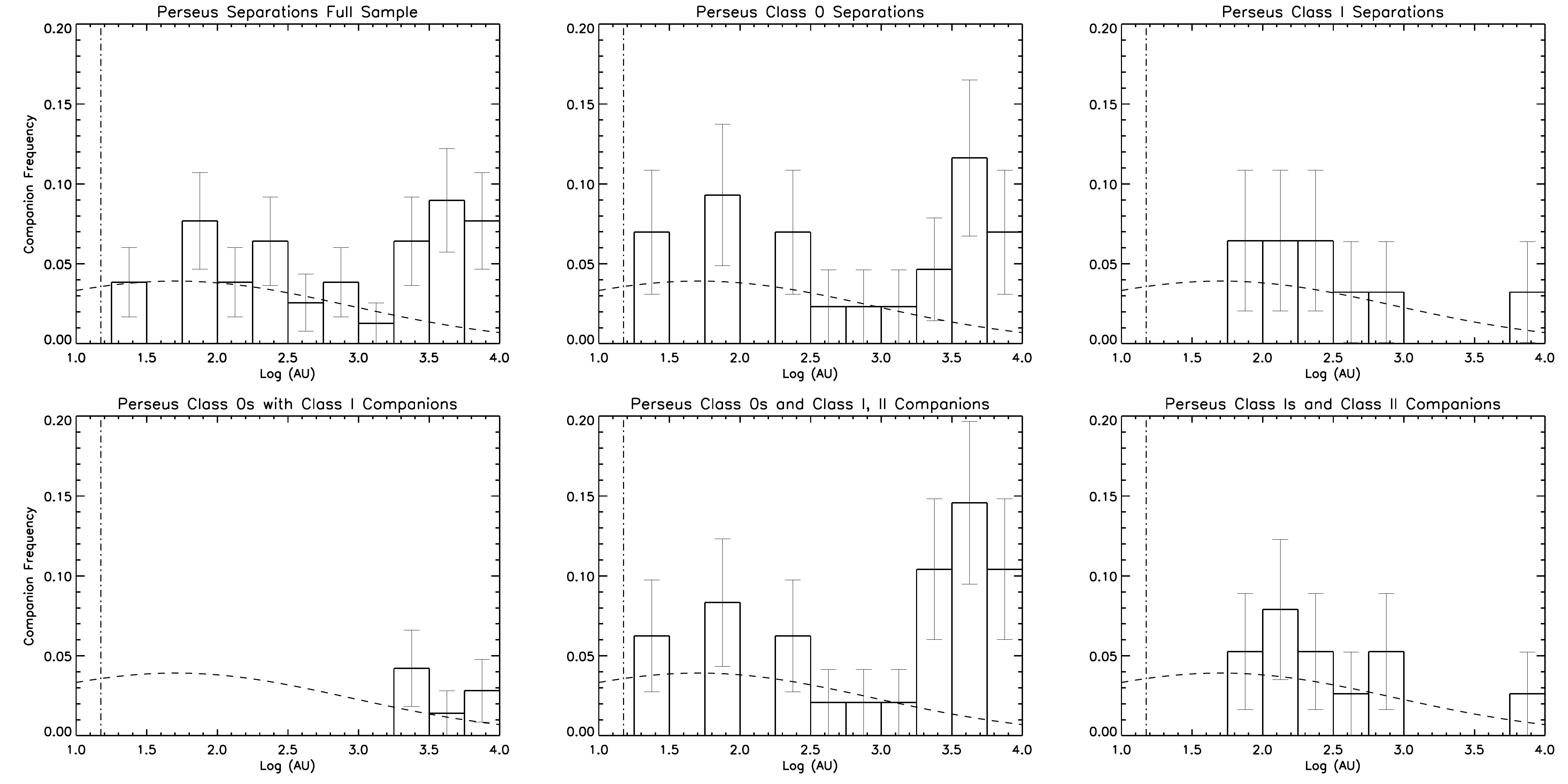}

\end{center}
\caption{Histograms of companion frequency versus separation for 
multiple sources in Perseus; this is an updated version of the plot shown in 
\citet{tobin2016a}, but with the average distance to Perseus revised to 300~pc. The top left panel shows the distribution
for all sources in the sample; the top middle and top right panels break the distribution
into sources that are only comprised of Class 0 protostars and Class I protostars, respectively.
The bottom left panel shows only the multiple systems comprised of Class 0 and I sources, the
bottom middle shows the separation distribution of all systems with a Class 0 primary source and
the bottom left panel shows the same, but with a Class I primary. The systems comprised of
a Class 0 and Class I protostar are not included in the Class I plot in the bottom right.
 Note the apparent bi-modal distribution
for the full sample and Class 0 samples and the apparent deficit of wide companions for the Class I 
systems. In all plots, the dashed
curve is the Gaussian fit to the field star separation distribution 
from \citet{raghavan2010} and the vertical dot-dashed line
corresponds to the approximate resolution limit of 20 AU.}
\label{separations}
\end{figure}

\floattable

\begin{deluxetable}{lllllllllllll}
\rotate
\tablewidth{0pt}
\tabletypesize{\tiny}
\tablecaption{Source List and Gaussian Fitting Results}
\tablehead{\colhead{Source} & \colhead{RA} & \colhead{Decl.}  & \colhead{Int. Flux}  & \colhead{Peak Flux}  & \colhead{$\theta_{maj}$~$\times$~$\theta_{min}$} & \colhead{PA}  & \colhead{Deconv. $\theta_{maj}$} & \colhead{Deconv. $\theta_{min}$} & \colhead{Deconv. PA}  & \colhead{Extended Flux} & \colhead{1.3~mm - 9.1~mm\tablenotemark{a}}\\ 
\colhead{} & \colhead{(J2000)} & \colhead{(J2000)}  & \colhead{(mJy)}  & \colhead{(mJy~beam$^{-1}$)}  & \colhead{(\arcsec~$\times$~\arcsec)} & \colhead{(\degr)}  & \colhead{(arcsec)} & \colhead{(arcsec)} & \colhead{(\degr)}  & \colhead{(mJy)} &  \colhead{Spectral Index}\\
}
\startdata
Per-emb-12-A & 3:29:10.53  & 31:13:30.99     & 1503.7~$\pm$~29.2 & 180.7~$\pm$~3.2 & 0.77~$\times$~0.64 & 50.9  & 0.73~$\pm$~0.02   & 0.61~$\pm$~0.02   & 58.2~$\pm$~6.1   & 3270~$\pm$~140.0 & 2.9 - 2.7 $\pm$ 0.07\\
Per-emb-12-B & 3:29:10.43  & 31:13:32.15     &  434.2~$\pm$~7.9 &  128.6~$\pm$~3.2  & 0.49~$\times$~0.43 & 119.9 & 0.46~$\pm$~0.02  & 0.33~$\pm$~0.02   & 116.4~$\pm$~9.8  & -99.0~$\pm$~-99.0 & 3.7 - 3.3 $\pm$ 0.08\\
Per-emb-17-A & 3:27:39.10  & 30:13:03.08     &   26.1~$\pm$~0.5 &   22.9~$\pm$~0.2 & 0.28~$\times$~0.17 & 21.0  & 0.11~$\pm$~0.01   & 0.062~$\pm$~0.01  & 15.0~$\pm$~8.8   & 55.0~$\pm$~7.0 & 2.2 - 2.1 $\pm$ 0.07\\
Per-emb-17-B & 3:27:39.11  & 30:13:02.85     &   20.4~$\pm$~0.5 &   15.9~$\pm$~0.2 & 0.31~$\times$~0.2  & 18.6  & 0.16~$\pm$~0.01   & 0.11~$\pm$~0.01   & 7.1~$\pm$~11.3   & -99.0~$\pm$~-99.0 & 2.8 $\pm$ 0.09\\
Per-emb-18 & 3:29:11.26 & 31:18:31.03        &  145.2~$\pm$~1.4 &   29.0~$\pm$~0.3 & 0.74~$\times$~0.31 & 72.3  & 0.70~$\pm$~0.01   & 0.19~$\pm$~0.01   & 75.4~$\pm$~0.3   & 150.0~$\pm$~9.0 & 3.0 - 2.8 $\pm$ 0.07\\
Per-emb-21 & 3:29:10.66  & 31:18:20.10       &   49.3~$\pm$~0.5 &   44.3~$\pm$~0.3 & 0.29~$\times$~0.18 & 21.7  & 0.083~$\pm$~0.01  & 0.065~$\pm$~0.01  & 32.4~$\pm$~17.8  & 53.0~$\pm$~4.0 & 2.4  $\pm$ 0.08\\
Per-emb-22-A & 3:25:22.40  & 30:45:13.21     &   33.6~$\pm$~0.7 &   18.8~$\pm$~0.2 & 0.36~$\times$~0.23 & 35.9  & 0.27~$\pm$~0.01   & 0.16~$\pm$~0.01   & 49.7~$\pm$~3.4   & 74.0~$\pm$~12.0 & 2.6 - 2.2 $\pm$ 0.08\\
Per-emb-22-B & 3:25:22.35  & 30:45:13.14     &   10.1~$\pm$~0.2 &    8.5~$\pm$~0.2 & 0.26~$\times$~0.15 & 20.6  & 0.0~$\pm$~0.0     & 0.0~$\pm$~0.0     & 0.0~$\pm$~0.0    & -99.0~$\pm$~-99.0 & 2.7 - 2.2 $\pm$ 0.09\\
Per-emb-27-A & 3:28:55.57  & 31:14:36.95     &  162.5~$\pm$~2.1 &  108.5~$\pm$~0.8 & 0.31~$\times$~0.21 & 24.1  & 0.17~$\pm$~0.01   & 0.14~$\pm$~0.01   & 46.5~$\pm$~10.0  & 290.0~$\pm$~27.0 & 2.5 - 2.4 $\pm$ 0.07\\
Per-emb-27-B & 3:28:55.56  & 31:14:36.38     &   26.8~$\pm$~2.5 &   15.4~$\pm$~0.8 & 0.40~$\times$~0.22 & 25.4  & 0.29~$\pm$~0.05   & 0.15~$\pm$~0.02   & 29.0~$\pm$~10.8  & -99.0~$\pm$~-99.0 & 2.4 $\pm$ 0.09\\
Per-emb-2 & 3:32:17.92 & 30:49:47.85         &  571.8~$\pm$~9.7 &   20.5~$\pm$~0.7 & 1.22~$\times$~0.83 & 173.8 & 1.20~$\pm$~0.02   & 0.81~$\pm$~0.01   & 172.6~$\pm$~1.8  & 570.0~$\pm$~34.0 & 2.9 $\pm$ 0.07\\
Per-emb-33-A & 3:25:36.32  & 30:45:14.84     &    6.5~$\pm$~0.8 &   22.4~$\pm$~0.5 & 0.25~$\times$~0.15 & 31.5  & 0.0~$\pm$~0.0     & 0.0~$\pm$~0.0     & 0.0~$\pm$~0.0    & 740.0~$\pm$~32.0& 2.1 - 1.7 $\pm$ 0.1\\
Per-emb-33-B & 3:25:36.31  & 30:45:15.09     &   31.9~$\pm$~1.4 &   26.5~$\pm$~0.5 & 0.39~$\times$~0.21 & 11.9  & 0.28~$\pm$~0.02   & 0.13~$\pm$~0.01   & 5.7~$\pm$~4.6    & -99.0~$\pm$~-99.0 & 2.3  $\pm$ 0.08\\
Per-emb-33-C & 3:25:36.38  & 30:45:14.64     &  163.0~$\pm$~1.8 &   79.4~$\pm$~0.5 & 0.39~$\times$~0.28 & 21.2  & 0.27~$\pm$~0.01   & 0.22~$\pm$~0.01   & 23.0~$\pm$~2.8   & -99.0~$\pm$~-99.0 & 2.9 $\pm$ 0.08\\
L1448NW-A & 3:25:35.67  & 30:45:34.08        &   49.8~$\pm$~0.5 &   39.7~$\pm$~0.2 & 0.31~$\times$~0.19 & 23.1  & 0.16~$\pm$~0.01   & 0.075~$\pm$~0.01  & 33.6~$\pm$~1.9   & 82.0~$\pm$~7.0 & 2.5 - 2.3 $\pm$ 0.07\\
L1448NW-B & 3:25:35.68  & 30:45:34.34        &   14.8~$\pm$~0.2 &   19.2~$\pm$~0.2 & 0.27~$\times$~0.17 & 18.1  & 0.0~$\pm$~0.0     & 0.037~$\pm$~0.0   & 18.2~$\pm$~0.0   & -99.0~$\pm$~-99.0 & 1.9 $\pm$ 0.08\\
Per-emb-44-A & 3:29:03.76  & 31:16:03.81     &   99.0~$\pm$~1.1 &   80.5~$\pm$~0.5 & 0.32~$\times$~0.19 & 20.7  & 0.16~$\pm$~0.01   & 0.093~$\pm$~0.01  & 23.9~$\pm$~2.3   & 400.0~$\pm$~19.0 & 2.4 - 2.2 $\pm$ 0.07\\
Per-emb-44-B & 3:29:03.74  & 31:16:03.80     &  158.1~$\pm$~1.8 &   74.9~$\pm$~0.5 & 0.42~$\times$~0.26 & 27.3  & 0.32~$\pm$~0.01   & 0.20~$\pm$~0.01   & 33.6~$\pm$~1.6   & -99.0~$\pm$~-99.0 & 3.1 - 3.0 $\pm$ 0.08\\
SVS13B & 3:29:03.07  & 31:15:51.72           &  222.2~$\pm$~2.2 &   73.5~$\pm$~0.5 & 0.43~$\times$~0.35 & 53.0  & 0.36~$\pm$~0.01   & 0.26~$\pm$~0.01   & 75.2~$\pm$~2.5   & 270.0~$\pm$~28.0 & 2.7  $\pm$ 0.07\\
SVS13A2 & 3:29:03.38  & 31:16:01.62          &   13.9~$\pm$~1.0 &   12.8~$\pm$~0.5 & 0.29~$\times$~0.17 & 18.1  & 0.11~$\pm$~0.05   & 0.044~$\pm$~0.03  & 10.5~$\pm$~79.8  & 1.3~$\pm$~4.3 & 2.2 - 2.0 $\pm$ 0.09\\
RAC1999 VLA20 & 3:29:04.25  & 31:16:09.14          &    4.4~$\pm$~1.0 &    3.7~$\pm$~0.5 & 0.28~$\times$~0.19 & 13.7  & 0.11~$\pm$~0.1    & 0.026~$\pm$~0.1   & 136.5~$\pm$~85.1 & 4.5~$\pm$~3.0 & 1.7 $\pm$ 0.2\\
Per-emb-8 & 3:44:43.98 & 32:01:35.15         & 102.8~$\pm$~2.7 &   64.4~$\pm$~0.3 & 0.32~$\times$~0.24 & 24.5  & 0.17~$\pm$~0.01 & 0.14~$\pm$~0.01     & 103.8~$\pm$~4.7   & 120.0~$\pm$~9.3 & 2.8 - 2.4 $\pm$ 0.07\\
Per-emb-8 Inner Disk & 3:44:43.98 & 32:01:35.18       &   89.1~$\pm$~0.7 &   64.4~$\pm$~0.3 & 0.30~$\times$~0.23 & 19.5  & 0.16~$\pm$~0.01 & 0.094~$\pm$~0.01    & 120.9~$\pm$~2.6   & 120.0~$\pm$~9.3 & 2.7 - 2.3 $\pm$ 0.07\\
Per-emb-8 Outer Disk & 3:44:43.98  & 32:01:35.15      &   34.0~$\pm$~2.2 &   64.4~$\pm$~0.3 & 1.25~$\times$~0.38 & 56.5  & 1.22~$\pm$~0.09 & 0.32~$\pm$~0.03     & 57.5~$\pm$~1.6    & -99.0~$\pm$~-99.0 & \nodata \\
Per-emb-35-A & 3:28:37.09  & 31:13:30.74     &   22.6~$\pm$~0.4 &   20.5~$\pm$~0.2 & 0.29~$\times$~0.17 & 21.6  & 0.085~$\pm$~0.01  & 0.054~$\pm$~0.01  & 36.4~$\pm$~19.6  & -99.0~$\pm$~-99.0 & 2.8 - 2.1 $\pm$ 0.08\\
Per-emb-35-B & 3:28:37.22  & 31:13:31.70     &   16.8~$\pm$~0.4 &   15.8~$\pm$~0.2 & 0.28~$\times$~0.16 & 22.3  & 0.078~$\pm$~0.02  & 0.021~$\pm$~0.02  & 40.3~$\pm$~11.0  & -99.0~$\pm$~-99.0 & 2.3 $\pm$ 0.08\\
Per-emb-36-A & 3:28:57.37  & 31:14:15.79     &  129.2~$\pm$~0.9 &   88.6~$\pm$~0.4 & 0.29~$\times$~0.23 & 22.8  & 0.16~$\pm$~0.002  & 0.091~$\pm$~0.01  & 105.7~$\pm$~2.0  & 140.~$\pm$~5.0 & 2.4 - 2.3 $\pm$ 0.07\\
Per-emb-36-B & 3:28:57.37  & 31:14:16.11     &   12.3~$\pm$~0.7 &   29.6~$\pm$~0.4 & 0.27~$\times$~0.18 & 18.9  & 0.0~$\pm$~0.01    & 0.0~$\pm$~0.07    & 0.0~$\pm$~0.0    & -99.0~$\pm$~-99.0 & 2.0 $\pm$ 0.08\\
Per-emb-40-A & 3:33:16.67  & 31:07:54.84     &   15.9~$\pm$~0.3 &   14.1~$\pm$~0.5 & 0.30~$\times$~0.17 & 21.5  & 0.12~$\pm$~0.01   & 0.04~$\pm$~0.01   & 24.1~$\pm$~3.6   & 17.0~$\pm$~2.0 & 2.4 - 2.0 $\pm$ 0.08\\
Per-emb-40-B & 3:33:16.69  & 31:07:55.20     &    0.9~$\pm$~0.1 &    2.2~$\pm$~0.5 & 0.28~$\times$~0.16 & 20.8  & 0.0~$\pm$~0.0     & 0.0~$\pm$~0.0     & 0.0~$\pm$~0.0    & -99.0~$\pm$~-99.0 & 1.6 $\pm$ 0.15\\
Per-emb-48-A & 3:27:38.28 & 30:13:58.60      &    4.0~$\pm$~0.4 &    2.7~$\pm$~0.1 & 0.36~$\times$~0.22 & 21.4  & 0.23~$\pm$~0.04   & 0.13~$\pm$~0.02   & 23.5~$\pm$~18.7  & 4.5~$\pm$~2.0 & 2.0 $\pm$ 0.14\\
Per-emb-48-B & \nodata & \nodata             &    $<$0.4        &  $<$0.4~$\pm$~0.1 & 0.36~$\times$~0.22 & 21.4  & 0.23~$\pm$~0.04   & 0.13~$\pm$~0.02   & 23.5~$\pm$~18.7  & 4.5~$\pm$~2.0 & $<$1.3 \\
Per-emb-49-A & 3:29:12.96  & 31:18:14.33     &   18.3~$\pm$~0.4 &   15.5~$\pm$~0.2 & 0.28~$\times$~0.18 & 14.6  & 0.1~$\pm$~0.01    & 0.052~$\pm$~0.03  & 146.2~$\pm$~12.3  & 20.0~$\pm$~4.3 & 2.1 $\pm$ 0.08\\
Per-emb-49-B & 3:29:12.98  & 31:18:14.44     &    4.9~$\pm$~0.3 &    5.0~$\pm$~0.2 & 0.26~$\times$~0.16 & 20.8  & 0.0~$\pm$~0.0     & 0.0~$\pm$~0.0     & 0.0~$\pm$~0.0     & -99.0~$\pm$~-99.0 & 2.0 - 1.9 $\pm$ 0.1\\
Per-emb-55-A & 3:44:43.29  & 32:01:31.19     &    3.3~$\pm$~0.5 &    3.2~$\pm$~0.3 & 0.29~$\times$~0.18 & 25.3  & 0.078~$\pm$~0.09  & 0.027~$\pm$~0.06  & 79.2~$\pm$~ 70.5  & -99.0~$\pm$~-99.0 & 1.8 $\pm$ 0.11\\
Per-emb-55-B & 3:44:43.33  & 32:01:31.65     &    0.5~$\pm$~0.3 &    0.5~$\pm$~0.3 & 0.29~$\times$~0.17 & 22.6  & 0.052~$\pm$~0.0   & 0.04~$\pm$~0.0    & 17.5~$\pm$~0.0    & -99.0~$\pm$~-99.0 & 1.7 - 1.0 $\pm$ 0.32\\
L1448IRS1-A & 3:25:09.45 & 30:46:21.92       &   73.4~$\pm$~0.9 &   48.4~$\pm$~0.4 & 0.33~$\times$~0.2  & 23.4  & 0.21~$\pm$~0.01   & 0.10~$\pm$~0.01   & 29.3~$\pm$~1.6   & 78.7~$\pm$~14.0 & 2.5 - 2.3 $\pm$ 0.07\\
L1448IRS1-B & 3:25:09.41  & 30:46:20.61      &    5.5~$\pm$~0.7 &    5.5~$\pm$~0.4 & 0.25~$\times$~0.17 & 19.5  & 0.0~$\pm$~0.0     & 0.0~$\pm$~0.0     & 0.0~$\pm$~0.0    & -99.0~$\pm$~-99.0 & 2.2 $\pm$ 0.13\\
EDJ2009-269-A & 3:30:44.02  & 30:32:46.82    &   11.1~$\pm$~0.3 &   10.6~$\pm$~0.2 & 0.26~$\times$~0.17 & 18.5  & 0.043~$\pm$~0.03  & 0.039~$\pm$~0.01  & 147.3~$\pm$~67.5 & -99.0~$\pm$~-99.0 & 2.1 - 2.0 $\pm$ 0.1\\
EDJ2009-269-B & 3:30:43.98  & 30:32:46.59    &    7.9~$\pm$~0.3 &    7.5~$\pm$~0.2 & 0.26~$\times$~0.17 & 19.2  & 0.0~$\pm$~0.07    & 0.0~$\pm$~0.05    & 0.0~$\pm$~0.0    & -99.0~$\pm$~-99.0 & 1.9 $\pm$ 0.1\\
L1448IRS3A & 3:25:36.50  & 30:45:21.84       &  101.0~$\pm$~2.5 &   29.6~$\pm$~0.5 & 0.75~$\times$~0.39 & 136.5 & 0.73~$\pm$~0.03 & 0.30~$\pm$~0.02     & 134.1~$\pm$~2.0   & 110.0~$\pm$~9.0 & 2.7 - 2.4 $\pm$ 0.07\\
L1448IRS3A Inner Disk & 3:25:36.50 & 30:45:21.78      &  165.4~$\pm$~0.8 &   29.6~$\pm$~0.5 & 0.26~$\times$~0.16 & 20.0  & 0.0~$\pm$~0.0     & 0.0~$\pm$~0.0     & 0.0~$\pm$~0.0    & 110.0~$\pm$~9.0 & 3.0 - 2.6 $\pm$ 0.07\\
L1448IRS3A Outer Disk & 3:25:36.50 & 30:45:21.84      &   84.4~$\pm$~3.4 &   29.6~$\pm$~0.5 & 0.75~$\times$~0.39 & 136.5 & 0.73~$\pm$~0.03   & 0.30~$\pm$~0.02   & 134.1~$\pm$~2.0  & -99.0~$\pm$~-99.0 & \nodata\\
\enddata
\tablenotetext{a}{The spectral index is calculated with respect to the 9.1~mm flux density given in Table 7 of 
\citet{tychoniec2018} under the assumption that F$_{\nu}$~$\propto$~$\nu^{\alpha}$. The range of spectral indices for most sources reflects the 9.1~mm flux density
that is corrected and not corrected for free-free emission, respectively, in that study. The sources with
only one spectral index listed could not be corrected for free-free emission. The formal uncertainly of the spectral
index is quite low, $\sim$0.1, even if 10\% systematic uncertainty in the flux density is assumed due to the
large difference in wavelength from 1.3~mm to 9.1~mm. We calculated the uncertainty in the spectral index
following \citet{chiang2012} Equation A5.}
\end{deluxetable}

\begin{deluxetable}{lll}
\tabletypesize{\scriptsize}
\tablewidth{0pt}
\tablecaption{Compact Masses vs. Extended Masses}
\tablehead{\colhead{Source} & \colhead{Gaussian mass} & \colhead{Extended Mass} \\
\colhead{} & \colhead{(M$_{\sun}$)} & \colhead{(M$_{\sun}$)}
}
\startdata
Per-emb-2     & 0.69~$\pm$~0.01 & 0.68~$\pm$~0.04 \\
Per-emb-12    & 2.3~$\pm$~0.04   & 3.9~$\pm$~0.2 \\
Per-emb-12-A  & 1.8~$\pm$~0.04   & \nodata \\
Per-emb-12-B  & 0.53~$\pm$~0.02   & \nodata \\
Per-emb-17    & 0.054~$\pm$~0.001 & 0.066~$\pm$~0.01 \\
Per-emb-17-A  & 0.031~$\pm$~0.0005 & \nodata \\
Per-emb-17-B  & 0.024~$\pm$~0.0007 & \nodata \\
Per-emb-18    & 0.17~$\pm$~0.002 & 0.18~$\pm$~0.01 \\
Per-emb-21    & 0.06~$\pm$~0.0005 & 0.063~$\pm$~0.005 \\
Per-emb-22    & 0.053~$\pm$~0.001 & 0.088~$\pm$~0.014 \\
Per-emb-22-A  & 0.041~$\pm$~0.0009 & \nodata \\
Per-emb-22-B  & 0.012~$\pm$~0.0002 & \nodata \\
Per-emb-27    & 0.24~$\pm$~0.003 & 0.35~$\pm$~0.03 \\
Per-emb-27-A  & 0.204~$\pm$~0.0024  & \nodata \\
Per-emb-27-B  & 0.032~$\pm$~0.003 & \nodata \\
Per-emb-33    & 0.33~$\pm$~0.003  & 0.89~$\pm$~0.04 \\
Per-emb-33-A  & 0.009~$\pm$~0.001 & \nodata \\
Per-emb-33-B  & 0.04~$\pm$~0.002 & \nodata \\
Per-emb-33-C  & 0.2~$\pm$~0.002  & \nodata \\
L1448NW       & 0.08~$\pm$~0.0009 & 0.1~$\pm$~0.009 \\
L1448NW-A     & 0.060~$\pm$~0.0007 & \nodata \\
L1448NW-B     & 0.018~$\pm$~0.0003 & \nodata \\
Per-emb-44    & 0.31~$\pm$~0.002 & 0.48~$\pm$~0.02 \\
Per-emb-44-A  & 0.12~$\pm$~0.001 & \nodata \\
Per-emb-44-B  & 0.19~$\pm$~0.002  & \nodata \\
SVS13B        & 0.27~$\pm$~0.003  & 0.33~$\pm$~0.03 \\
SVS13A2       & 0.017~$\pm$~0.002  & 0.002~$\pm$~0.005 \\
RAC1999 VLA20       & 0.005~$\pm$~0.001  & 0.003~$\pm$~0.002 \\
Per-emb-8     & 0.15~$\pm$~0.003  & 0.15~$\pm$~0.01\\
Per-emb-8 Inner Disk & 0.11~$\pm$~0.0009 & \nodata \\
Per-emb-8 Outer Disk & 0.041~$\pm$~0.003  & \nodata \\
Per-emb-35-A & 0.027~$\pm$~0.0005 & \nodata \\
Per-emb-35-B & 0.020~$\pm$~0.0005 & \nodata \\
Per-emb-36   & 0.17~$\pm$~0.001 & 0.17~$\pm$~0.005 \\
Per-emb-36-A & 0.15~$\pm$~0.001 & \nodata \\
Per-emb-36-B & 0.015~$\pm$~0.0009 & \nodata \\
Per-emb-40-A & 0.019~$\pm$~0.0003 & \nodata\\
Per-emb-40-B & 0.0017~$\pm$~0.0002 & \nodata \\
Per-emb-48-A   & 0.005~$\pm$~0.0005 & 0.005~$\pm$~0.002 \\
Per-emb-48-B   & $<$0.0005 & \nodata \\
Per-emb-49-A & 0.022~$\pm$~0.0005 & \nodata \\
Per-emb-49-B & 0.005~$\pm$~0.0003 & \nodata \\
Per-emb-55-A & 0.003~$\pm$~0.0007  & \nodata \\
Per-emb-55-B & 0.0007~$\pm$~0.0003 & \nodata\\
L1448IRS1-A  & 0.088~$\pm$~0.001 & \nodata \\
L1448IRS1-B  & 0.007~$\pm$~0.0009 & \nodata \\
L1448IRS3A   & 0.12~$\pm$~0.005  & 0.13~$\pm$~0.01 \\
L1448IRS3A Inner Disk  & 0.02~$\pm$~0.001 & \nodata \\
L1448IRS3A Outer Disk  & 0.1~$\pm$~0.004 & \nodata \\
EDJ2009-269-A & 0.014~$\pm$~0.0003 & \nodata \\
EDJ2009-269-B & 0.01~$\pm$~0.0003 & \nodata \\
\enddata
\end{deluxetable}

\begin{deluxetable}{lllllllllll}
\rotate
\tabletypesize{\tiny}
\tablewidth{0pt}
\tablecaption{Line Integration Intervals and Resultant Noise}
\tablehead{\colhead{Source} & \colhead{$^{13}$CO intervals} & \colhead{$^{13}$CO rms} & \colhead{C$^{18}$O intervals} & \colhead{C$^{18}$O rms} & \colhead{SO intervals} & \colhead{SO rms} & \colhead{H$_2$CO intervals}& \colhead{H$_2$CO rms}  & \colhead{$^{12}$CO intervals}& \colhead{$^{12}$CO rms}\\
\colhead{} & \colhead{(blue, red)}     & \colhead{(blue, red)}     & \colhead{(blue, red)}   & \colhead{(blue, red)}     & \colhead{(blue, red)}   & \colhead{(blue, red)}     & \colhead{(blue, red)}   & \colhead{(blue, red)} & \colhead{(blue, red)}   & \colhead{(blue, red)}\\
\colhead{} & \colhead{(km~s$^{-1}$)}   & \colhead{(K~km~s$^{-1}$)} & \colhead{(km~s$^{-1}$)} & \colhead{(K~km~s$^{-1}$)} & \colhead{(km~s$^{-1}$)} & \colhead{(K~km~s$^{-1}$)} & \colhead{(km~s$^{-1}$)} & \colhead{(K~km~s$^{-1}$)} & \colhead{(km~s$^{-1}$)} & \colhead{(K~km~s$^{-1}$)}
}
\startdata
Per-emb-2     & 5.0-6.0, 8.0-9.5 &3.33, 3.95 & 7.5-9.0, 5.0-6.25 & 2.16, 1.65 &\nodata &\nodata & \nodata &\nodata & 1.5-6.5, 9.0-15.0 & 5.8, 6.15\\
Per-emb-12    & 1.25-6.5, 6.5-9.75 & 5.59, 6.65 & 4.25-6.75, 7.75-9.0 & 3.34, 2.46 & 1.0-7.0, 8.0-11.75 & 6.24, 4.99 & 3.75-7.0, 8.0-11.0 & 3.64, 3.51 & -1.5-6.5, 10.0-15.5 & 7.21, 6.05\\
Per-emb-17    & 0.5-4.5, 6.5-9.75 & 6.08, 5.52 & 3.0-4.75, 5.5-7.75 & 2.19, 2.19 & -1.5-5.0, 5.25-10.25 & 6.6, 5.8 & 0.0-5.0, 5.25-9.75 & 4.54, 4.32 & 0.0-4.0, 7.5-14.0 & 6.56, 5.26\\
Per-emb-18    & 4.75-7.5, 9.25-11.5 & 4.79, 4.54 & 4.75-7.25, 8.75-11.5 & 3.22, 3.37 & 4.5-8.0, 8.25-11.5 & 4.83, 4.67 & 4.5-7.5, 9.25,11.25 & 3.51, 3.08 & 3.0-5.0, 10.0~13.0 & 5.26, 4.41\\
Per-emb-21    & \nodata & \nodata & \nodata & \nodata & \nodata & \nodata & \nodata & \nodata & \nodata & \nodata \\
Per-emb-22    & 2.25-3.75, 4.75-5.75  & 3.94, 3.33 & 2.75-4.0, 4.5~5.75 & 2.58, 2.58 & 1.75-5.75 & 5.21 & 3.0-4.0, 4.25-6.0 & 2.2, 2.79 & -8.0-2.0, 7.5-16.0 & 8.1, 7.5 \\
Per-emb-27    & 0.25-6.25, 9.5-10.5  & 7.03, 3.14 & 5.75-7.25, 7.5-9.0 & 2.58, 2.58 & 0.25-7.0, 7.25-11.0 & 7.0, 5.29 & 3.0-6.75, 7.75-10.0 & 3.91, 3.09 & -28.5-6.0, 9.5-20.0 & 13.56, 13.58 \\
Per-emb-33    & 1.25-4, 5.5-7.0  & 4.99, 3.20 & 1.25-4.0, 5.5-7.0 & 2.25, 1.65 & 1.25-4.5, 4.5-8.0 & 4.74, 4.91 & 2.75-4.0, 5.25-6.25 & 2.25, 2.05 & -5.5-1.5, 4.5-8.0 & 6.88, 5.02 \\
L1448NW       & 0.75-3.5, 5.25-7.75 & 4.95, 4.74 & 1.0-3.75, 4.0, 7.25 & 3.66, 3.39 & 2.25-5.0 & 4.41 & 1.5-6.25 & 4.12 & -9.0-3.5, 6.0-16.0 & 8.4, 7.54 \\
Per-emb-44    & 6.75-7.5, 9.0, 11.0 & 2.78, 4.17 & 6.0-8.5, 8.75, 10.0 & 3.63, 2.34 & 4.75-8.5, 8.75-11.75 & 6.36, 5.73 & 4.25-8.5, 8.75-11.75 & 3.63, 2.34 & -25.0-0.0, 10.0-20.0 & 4.42, 4.22 \\
Per-emb-8    & 5.75-8.75, 12.0-14.5 & 5.09, 4.68 & 6.0-9.25, 11.75-14.0 & 3.57, 3.02 & 7.5-10.0, 10.75-12.0 & 4.04, 2.99 & 6.75-10.25, 11.0-13.5 & 3.42, 2.93 & 4.5-8.5, 13.0-16.5 & 4.89, 4.61 \\
Per-emb-55   & \nodata & \nodata & \nodata & \nodata & \nodata & \nodata & \nodata & \nodata & 4.5-8.5, 11.5-16.0 & 4.90, 5.16 \\
Per-emb-35   & 5.75-6.25, 6.5-7.25 & 2.52, 2.91 & 5.25-8.0 & 4.31 & 5.5-7.25, 7.5-9.5 & 3.35, 3.55 & 5.5-7.5, 7.75-10.5 & 2.87, 3.63 & 2.0-7.5, 10.0-14.0 & 6.0, 5.2 \\
Per-emb-36   & 3.5-5.75, 9.0-11.0 & 4.68, 4.17 & 4.5-6.5 & 3.14 & 4.5-7.0, 7.25-9.0 & 4.29, 3.65 & 4.0-6.25 & 2.87 & -3.0-5.0, 9.5-16.0 & 7.81, 7.1 \\
Per-emb-40   & 5.0-6.5 & 3.91, 4.18 & 5.0-7.75 & 3.38 & \nodata & \nodata & \nodata & \nodata & -3.5-5.0, 8.5-13.5 & 8.03, 6.28 \\
Per-emb-48   & \nodata & \nodata & \nodata & \nodata & \nodata & \nodata & \nodata & \nodata & \nodata & \nodata \\
Per-emb-49   & \nodata & \nodata & \nodata & \nodata & \nodata & \nodata & \nodata & \nodata & \nodata & \nodata \\
L1448IRS1    & \nodata & \nodata & \nodata & \nodata & \nodata & \nodata & \nodata & \nodata & 0.0-2.0, 5.0-7.0 & 3.65, 3.65 \\
L1448IRS3A   & 2.5-3.75, 6.25-9.0 & 2.96, 4.19 & 2.25-3.75, 6.25-9.0 & 2.42, 3.18 & 3.0-4.25, 4.25-8.0 & 3.1, 5.07 & 3.0-3.75, 5.25-7.0 & 1.83, 2.6 & -1.0-1.5, 6.5-11.0 & 4.35, 5.61 \\
EDJ2009-269   & \nodata & \nodata & \nodata & \nodata & \nodata & \nodata & \nodata & \nodata & 5.5-8.0 & 5.04 \\
\enddata
\end{deluxetable}

\begin{deluxetable}{llllllllllll}
\rotate
\tabletypesize{\tiny}
\tablewidth{0pt}
\tablecaption{Molecules and Rotation Characteristics}
\tablehead{\colhead{Source} & \colhead{Class} & \colhead{Separation} & \colhead{Outflow PA\tablenotemark{a}} & \colhead{Gradient PA\tablenotemark{b}} & \colhead{$^{12}$CO} & \colhead{$^{13}$CO} & \colhead{C$^{18}$O} & \colhead{H$_2$CO} & \colhead{SO} & \colhead{Rotation vs. Outflow} & \colhead{Disk Frag. Possible?}\\
 & & \colhead{(\arcsec, AU)} & \colhead{(\degr)} & \colhead{(\degr)} & &  &  &  &  & \
}
\startdata
Per-emb-2   & 0 & 0.080, 24.0  & 128 & 232 (104)    & Outflow & Rotation & Rotation & \nodata & \nodata & Perpendicular & Yes\\
Per-emb-12  & 0 & 1.830, 548.9 & 175, 205  & \nodata & Outflow & Indistinct & Indistinct & Indistinct & Indistinct & Indistinct & No\\
Per-emb-17  & 0 & 0.278, 83.3  & 240  & 180/133 (60/107) & Outflow & Rotation & Rotation & Outflow+Rotation & Outflow+Rotation & Perpendicular & Yes\\
Per-emb-18  & 0 & 0.085, 25.6  & 169 & 78 (91)      & Outflow & Rotation & Rotation & Rotation & Rotation & Perpendicular & Yes\\
Per-emb-22  & 0 & 0.751, 225.4 & 306 & 246 (60)      & Outflow & Rotation & Rotation & Indistinct & Indistinct & Perpendicular & Yes\\
Per-emb-27  & 0 & 0.620, 186.0 & 204, 285 & \nodata & Outflow & Indistinct & Indistinct & Outflow/Indistinct & Outflow & Indistinct & No\\
Per-emb-33  & 0 & 0.264, 79.2 & 301, 284 & 212 (89) & Outflow & Rotation & Rotation & Rotation & Outflow? & Perpendicular & Yes\\
            & 0 & 0.795, 238.4 \\
L1448NW     & 0 & 0.251, 75.3 & 305 & 224 (81)      & Outflow & Rotation & Rotation & Indistinct & Indistinct & Perpendicular & Yes\\
Per-emb-44  & 0/I & 0.300, 90.0 & 147 & 220 (73)    & Outflow & Rotation & Rotation & Rotation? & Rotation? & Perpendicular & Yes\\
Per-emb-35  & I & 1.908, 572.3 292, 279 & \nodata & Outflow & Outflow? & Indistinct & Indistinct & Indistinct & Indistinct & No?\\
Per-emb-36  & I & 0.311, 93.4 & 24  & \nodata       & Outflow & Indistinct & Indistinct & Rotation? & Outflow? & Indistinct & No?\\
Per-emb-40  & I & 0.391, 117.4 & 123? & \nodata      & Outflow/Rotation & Indistinct & Indistinct & \nodata & \nodata & \nodata & \nodata \\
Per-emb-48  & I & 0.346, 103.7 & \nodata & \nodata   & Indistinct & \nodata  & \nodata  & \nodata & \nodata & \nodata & \nodata \\
Per-emb-49  & I & 0.313, 93.8 & \nodata & \nodata   & \nodata  & \nodata  & \nodata  & \nodata & \nodata & \nodata & \nodata \\
Per-emb-55  & I & 0.618, 185.3 & 121? & \nodata     & Outflow & \nodata  & \nodata  & \nodata  & \nodata  & \nodata & \nodata \\
L1448IRS1   & I & 1.424, 427.0 & 113 & 212 (99)     & Rotation & \nodata  & \nodata  & \nodata  & \nodata  & Perpendicular & Yes?\\
EDJ2009-269 & II & 0.524, 157.3 & \nodata & \nodata & Indistinct & \nodata  & \nodata  & \nodata  & \nodata  & \nodata & \nodata \\
\\
Per-emb-35-A & I & & 292 & 201 (91) & Outflow & Outflow? & Indistinct & Rotation? & Indistinct & Perpendicular? \\
Per-emb-35-B & I & & 279 & 233 (46) & Outflow & Outflow? & Indistinct & Rotation? & Indistinct & Perpendicular? \\
L1448IRS3A   & I & & 38  & 313 (85) & Outflow? & Rotation & Rotation & Rotation & Rotation & Perpendicular\\
Per-emb-8    & 0 & & 314 & 44 (90)  & Outflow/Rotation & Rotation & Rotation & Rotation & Rotation & Perpendicular\\
Per-emb-21   & 0 &  & 80? & \nodata & Outflow & \nodata  & \nodata  & \nodata  & \nodata  & \nodata\\
SVS13B       & 0 &  & 160 & \nodata & Outflow & \nodata  & \nodata  & \nodata  & \nodata  & \nodata\\
\enddata
\tablecomments{This table examines what kinematics are being displayed by each molecular line. The ellipsis symbols (...) correspond to non-detections. The question marks (?)
 following a descriptor indicate that the line might display a certain 
kinematic signature, but that ambiguity remains. 
We expect that the outflow position angle (PA)
and gradient PA to have a typical uncertainty of 10\degr.}
\tablenotetext{a}{This is measured from the \twco\ data presented in this paper except
for Per-emb-36 and L1448 IRS1. To measure the outflow PA, we draw a line that best bisects
the outflow cavity. We adopt the outflow PA value from \citet{plunkett2013}
for Per-emb-36 because of how wide the outflow is. For L1448 IRS1, \twco\ is tracing
rotation, thus we use the scattered light nebula in the image from \citet{foster2006}.}
\tablenotetext{b}{The PA of the velocity gradient is measured by drawing a line
from the peaks of the blue- and red-shifted integrated intensity maps, generally \cateo.}
\end{deluxetable}

\begin{deluxetable}{llllll}
\tablewidth{0pt}
\tabletypesize{\scriptsize}
\tablecaption{Class 0 Multiple Systems}
\tablehead{
  \colhead{Source}  &   \colhead{Separation} & \colhead{Separation} & \colhead{Flux Difference} & \colhead {Type}  \\
  \colhead{}  &   \colhead{(\arcsec)} &\colhead{(AU)} & \colhead{(Log [F$_1$/F$_2$])} &}
\startdata
Per-emb-2 & 0.080 $\pm$ 0.006 & 24.0 $\pm$ 1.7 & 0.17 $\pm$ 0.15 & Class 0\\
Per-emb-18 & 0.085 $\pm$ 0.004 & 25.6 $\pm$ 1.2 & -0.02 $\pm$ 0.12 & Class 0\\
Per-emb-5 & 0.097 $\pm$ 0.006 & 29.1 $\pm$ 1.8 & 0.26 $\pm$ 0.07 & Class 0\\
L1448NW & 0.251 $\pm$ 0.004 & 75.3 $\pm$ 1.7 & 0.17 $\pm$ 0.03 & Class 0\\
Per-emb-33 & 0.264 $\pm$ 0.008 & 79.2 $\pm$ 2.3 & 0.12 $\pm$ 0.05 & Class 0\\
Per-emb-17 & 0.278 $\pm$ 0.014 & 83.3 $\pm$ 4.0 & 0.75 $\pm$ 0.08 & Class 0\\
Per-emb-44 & 0.300 $\pm$ 0.003 & 90.0 $\pm$ 0.9 & 0.48 $\pm$ 0.02 & Class 0\\
Per-emb-27 & 0.620 $\pm$ 0.003 & 186.0 $\pm$ 0.9 & 0.75 $\pm$ 0.03 & Class 0\\
Per-emb-22 & 0.751 $\pm$ 0.004 & 225.4 $\pm$ 1.3 & 0.51 $\pm$ 0.05 & Class 0\\
Per-emb-33 & 0.795 $\pm$ 0.004 & 238.4 $\pm$ 1.3 & 0.33 $\pm$ 0.04 & Class 0\\
Per-emb-12 & 1.830 $\pm$ 0.002 & 548.9 $\pm$ 0.5 & 1.04 $\pm$ 0.02 & Class 0\\
Per-emb-11 & 2.951 $\pm$ 0.008 & 885.4 $\pm$ 2.2 & 0.93 $\pm$ 0.08 & Class 0\\
Per-emb-44+SVS13A2 & 5.314 $\pm$ 0.004 & 1594.2 $\pm$ 1.2 & 1.78 $\pm$ 0.03 & Class 0/I\\
Per-emb-32 & 6.066 $\pm$ 0.022 & 1820.0 $\pm$ 6.5 & -0.16 $\pm$ 0.27 & Class 0/I\\
Per-emb-33+L1448IRS3A & 7.317 $\pm$ 0.004 & 2195.2 $\pm$ 1.2 & -0.24 $\pm$ 0.02 & Class 0-Class I\\
Per-emb-26+Per-emb-42 & 8.104 $\pm$ 0.005 & 2431.3 $\pm$ 1.6 & 1.97 $\pm$ 0.03 & Class 0-Class I\\
Per-emb-11 & 9.469 $\pm$ 0.025 & 2840.6 $\pm$ 7.6 & 0.76 $\pm$ 0.13 & Class 0\\
Per-emb-8+Per-emb-55 & 9.557 $\pm$ 0.013 & 2867.2 $\pm$ 3.8 & 2.38 $\pm$ 0.10 & Class 0-Class I\\
Per-emb-37+EDJ2009+235 & 10.556 $\pm$ 0.009 & 3166.7 $\pm$ 2.9 & 1.53 $\pm$ 0.11 & Class 0-Class II\\
Per-emb-13+IRAS4B' & 10.654 $\pm$ 0.005 & 3196.2 $\pm$ 1.6 & 0.81 $\pm$ 0.02 & Class 0-Class 0\\
Per-emb-21+Per-emb-18 & 13.252 $\pm$ 0.004 & 3975.7 $\pm$ 1.3 & -0.44 $\pm$ 0.02 & Class 0-Class 0\\
B1-bS+Per-emb-41 & 13.957 $\pm$ 0.014 & 4187.1 $\pm$ 4.2 & 2.08 $\pm$ 0.09 & Class 0-Class 0/I\\
Per-emb-44+SVS13B & 14.932 $\pm$ 0.002 & 4479.7 $\pm$ 0.7 & 0.40 $\pm$ 0.01 & Class 0/I-Class 0\\
Per-emb-16+Per-emb-28 & 16.063 $\pm$ 0.037 & 4818.9 $\pm$ 11.1 & -0.05 $\pm$ 0.17 & Class 0-Class 0\\
B1-bN+B1-bS & 17.395 $\pm$ 0.009 & 5218.4 $\pm$ 2.6 & 0.31 $\pm$ 0.03 & Class 0-Class 0\\
Per-emb-33+L1448NW & 21.503 $\pm$ 0.004 & 6450.8 $\pm$ 1.3 & -0.23 $\pm$ 0.02 & Class 0-Class 0\\
Per-emb-18+Per-emb-49 & 27.474 $\pm$ 0.007 & 8242.3 $\pm$ 2.0 & 0.33 $\pm$ 0.03 & Class 0-Class I\\
Per-emb-12+Per-emb-13 & 29.739 $\pm$ 0.002 & 8921.7 $\pm$ 0.7 & 1.04 $\pm$ 0.01 & Class 0-Class 0\\
Per-emb-36+Per-emb-27 & 31.420 $\pm$ 0.001 & 9426.0 $\pm$ 0.4 & 0.15 $\pm$ 0.01 & Class 0-Class I\\
Per-emb-6+Per-emb-10 & 31.947 $\pm$ 0.005 & 9584.2 $\pm$ 1.6 & -0.28 $\pm$ 0.03 & Class 0-Class 0\\
Per-emb-37+EDJ2009+233 & 33.704 $\pm$ 0.006 & 10111.3 $\pm$ 1.7 & -0.62 $\pm$ 0.04 & Class 0-Class II\\
Per-emb-44+SVS13C & 34.528 $\pm$ 0.001 & 10358.5 $\pm$ 0.4 & -0.21 $\pm$ 0.01 & Class 0/I-Class 0\\
Per-emb-32+EDJ2009+366 & 36.605 $\pm$ 0.010 & 10981.6 $\pm$ 3.1 & -1.31 $\pm$ 0.11 & Class 0/I-Class II\\
\enddata
\tablecomments{This table includes Class 0 + Class 0, Class 0 + Class I, and Class 0 + Class II multiple systems. The flux difference
is calculated in the 9 mm band. The distance has been revised to 300~pc for the physical
separation distance.}
\end{deluxetable}

\begin{deluxetable}{llll}
\tablewidth{0pt}
\tabletypesize{\scriptsize}
\tablecaption{Class I Multiple Systems}
\tablehead{
  \colhead{Source}  &   \colhead{Separation} &\colhead{Separation} & \colhead{Flux Difference}  \\
  \colhead{}  &   \colhead{(\arcsec)} & \colhead{(AU)} & \colhead{(Log [F$_1$/F$_2$])}}
\startdata
Per-emb-36 & 0.311 $\pm$ 0.005 & 93.4 $\pm$ 1.6 & 0.80 $\pm$ 0.02\\
Per-emb-49 & 0.313 $\pm$ 0.009 & 93.8 $\pm$ 2.6 & 0.46 $\pm$ 0.11\\
Per-emb-48 & 0.346 $\pm$ 0.019 & 103.7 $\pm$ 5.7 & 0.06 $\pm$ 0.17\\
Per-emb-40 & 0.391 $\pm$ 0.022 & 117.4 $\pm$ 6.7 & 0.94 $\pm$ 0.19\\
Per-emb-55 & 0.618 $\pm$ 0.009 & 185.3 $\pm$ 2.6 & 0.14 $\pm$ 0.07\\
EDJ2009-183 & 1.025 $\pm$ 0.028 & 307.6 $\pm$ 8.3 & 0.45 $\pm$ 0.13\\
L1448IRS1 & 1.424 $\pm$ 0.015 & 427.0 $\pm$ 4.6 & 1.02 $\pm$ 0.09\\
Per-emb-35 & 1.908 $\pm$ 0.003 & 572.3 $\pm$ 0.9 & 0.23 $\pm$ 0.04\\
EDJ2009-156 & 3.107 $\pm$ 0.011 & 932.1 $\pm$ 3.3 & 0.17 $\pm$ 0.13\\
Per-emb-58+Per-emb-65 & 28.878 $\pm$ 0.023 & 8663.3 $\pm$ 6.9 & -0.44 $\pm$ 0.13\\
\enddata
\tablecomments{This table includes only Class I + Class I multiple systems. The flux difference
is calculated in the 9 mm band. The distance has been revised to 300~pc for the physical
separation distance.}
\end{deluxetable}

\begin{deluxetable}{lllll}
\tablewidth{0pt}
\tabletypesize{\scriptsize}
\tablecaption{Class II Multiple Systems}
\tablehead{
  \colhead{Source}  &   \colhead{Separation} &\colhead{Separation} & \colhead{Flux Difference}   \\
  \colhead{}  &   \colhead{(\arcsec)} & \colhead{(AU)} & \colhead{(Log [F$_1$/F$_2$])}}
\startdata
EDJ2009-269 & 0.524 $\pm$ 0.007 & 157.3 $\pm$ 2.1 & 0.12 $\pm$ 0.08\\
EDJ2009-156 & 3.107 $\pm$ 0.011 & 932.1 $\pm$ 3.3 & 0.17 $\pm$ 0.13\\
\enddata
\tablecomments{This table includes only Class II + Class II multiple systems. The flux difference
is calculated in the 9 mm band. The distance has been revised to 300~pc for the physical
separation distance.}
\end{deluxetable}

\end{document}